\pdfoutput=1
 
\documentclass[cernpreprint, atlasdraft=false, texlive=2020, UKenglish, texmf, orcidlogo]{atlasdoc}
 
\makeatletter
\DeclareOldFontCommand{\rm}{\normalfont\rmfamily}{\mathrm}
\DeclareOldFontCommand{\sf}{\normalfont\sffamily}{\mathsf}
\DeclareOldFontCommand{\tt}{\normalfont\ttfamily}{\mathtt}
\DeclareOldFontCommand{\bf}{\normalfont\bfseries}{\mathbf}
\DeclareOldFontCommand{\it}{\normalfont\itshape}{\mathit}
\DeclareOldFontCommand{\sl}{\normalfont\slshape}{\@nomath\sl}
\DeclareOldFontCommand{\sc}{\normalfont\scshape}{\@nomath\sc}
\makeatother
 
\usepackage[subfigure=true]{atlaspackage}
\usepackage{atlasbiblatex}
 
\usepackage{atlasphysics}
 
\graphicspath{{logos/}{figures/}}
 
\usepackage{ANA-TOPQ-2018-11-PAPER-defs}
 
\usepackage{nicefrac}
\usepackage{multirow}
\usepackage{color, colortbl}

\AtlasTitle{Differential $t\bar{t}$ cross-section measurements using boosted top quarks in the
all-hadronic final state with 139~fb$^{-1}$ of ATLAS data}

\AtlasAbstract{
Measurements of single-, double-, and triple-differential cross-sections are presented for
boosted top-quark pair-production in 13~$\text{TeV}$ proton--proton
collisions recorded by the ATLAS detector at the LHC.
The top quarks are observed through their hadronic decay and reconstructed as large-radius jets with
the leading jet having
transverse momentum ($p_{\text{T}}$) greater than 500~GeV.
The observed data are unfolded to remove detector effects.
The particle-level cross-section, multiplied by the $t\bar{t} \rightarrow W W b \bar{b}$ branching fraction and measured in a
fiducial phase space defined by requiring the leading and second-leading jets to have
$p_{\text{T}} > 500$~GeV and $p_{\text{T}} > 350$~GeV, respectively,
is $331 \pm 3 \text{(stat.)} \pm 39 \text{(syst.)}$~fb.
This is approximately 20\% lower than the prediction of $398^{+48}_{-49}$~fb
by \textsc{Powheg+Pythia}\,8 with next-to-leading-order (NLO) accuracy but
consistent within the theoretical uncertainties.
Results are also presented at the parton level, where the effects of top-quark decay, parton showering,
and hadronization are removed  such that they can be compared with fixed-order
next-to-next-to-leading-order (NNLO) calculations.
The parton-level cross-section, measured in a fiducial phase space similar to that at particle level, is
$1.94 \pm 0.02 \text{(stat.)} \pm 0.25 \text{(syst.)}$~pb.
This agrees with the NNLO prediction of $1.96^{+0.02}_{-0.17}$~pb.
Reasonable agreement with the differential cross-sections is found for most NLO models, while the
NNLO calculations are generally in better agreement with the data.
The differential cross-sections are interpreted using a Standard Model effective field-theory formalism and
limits are set on Wilson coefficients of several four-fermion operators.
}
 
\author{The ATLAS Collaboration}
 
\AtlasRefCode{TOPQ-2018-11}
 
\AtlasJournalRef{\JHEP 04 (2023) 80}
\AtlasDOI{10.1007/JHEP04(2023)080}
 
\PreprintIdNumber{CERN-EP-2022-026}

\hypersetup{pdftitle={ATLAS document},pdfauthor={The ATLAS Collaboration}}
 
\begin{document}
 
\maketitle
 
\tableofcontents

\section{Introduction}
\label{sec:intro}

The large top-quark pair-production cross-section at the Large Hadron Collider (LHC) results in
high-statistics samples of top-quark--top-antiquark (\ttbar) pairs that enable unique tests of the Standard
Model (SM) and searches for new phenomena that affect \ttbar\ production.
A focus on final states with highly boosted top quarks probes the QCD \ttbar\ production processes at the \TeV\ scale,
a kinematic region where theoretical calculations based on the SM still have large
uncertainties~\cite{Czakon:2017wor,NNLO_calc,nnloMtt}.
High-precision measurements constrain these predictions, especially at \ttbar\ invariant masses of 2~\TeV\ or more.
Furthermore, effects beyond the SM may appear as deviations of $\ttbar{}$ differential
distributions from the SM prediction~\cite{Hill:1993hs,Frederix:2009,top_bkg_interference}.
 
In the SM, the top quark decays almost exclusively into a~\Wboson{} boson and a~$b$-quark.
The signature of a~\ttbar\ final state is therefore determined by the \Wboson{} boson decay modes.
The
ATLAS~\cite{TOPQ-2011-07,TOPQ-2012-08,TOPQ-2013-07,TOPQ-2014-15,TOPQ-2015-06,TOPQ-2015-07,TOPQ-2016-04,TOPQ-2016-01,TOPQ-2016-09}\ and
CMS~\cite{CMS-TOP-11-013,CMS-TOP-12-028,CMS-TOP-14-018,CMS-TOP-14-012,CMS-TOP-16-008,CMS-TOP-14-013,CMS-TOP-15-015,CMS-TOP-16-007,CMS-TOP-16-014,CMS-TOP-17-002,CMS-TOP-18-004,CMS-TOP-19-005,CMS-TOP-18-013,CMS-TOP-20-001}\
collaborations have published measurements of
the $\ttbar{}$ differential cross-sections at centre-of-mass energies of $\sqrt{s}=7$~\TeV{},
8~\TeV{}, and 13~\TeV{}
in proton--proton ($pp$) collisions using final states containing leptons and jets,
with most measurements employing the lepton+jets or dilepton channels.
The analysis presented here makes use of the all-hadronic \ttbar{} decay mode, where
only top-quark candidates with high transverse momentum (\pt{})
being reconstructed as jets are selected.
This highly boosted topology has
advantages over other final states since the Lorentz boost of the
top quark collimates its decay products such that they can be collected into a large-radius (\largeR) jet.
Although this final state has the largest branching ratio
and the absence of neutrinos from W decays allows direct detection of all decay products,
it is also less studied given the large backgrounds coming from multijet
production~\cite{TOPQ-2016-09,CMS-TOP-18-013}.
These features make measurements of the all-hadronic final state complementary to studies of the
lepton+jets and dilepton channels.
 
This analysis is performed by targeting events where the leading top-quark jet has $\pTtone > 500$~\GeV\ and the second-leading top-quark jet has $\pTttwo > 350$~\GeV.
These jets are reconstructed from calorimeter energy deposits and tagged as top-quark candidates to separate
the \ttbar\ final state from background sources.
The top-quark-tagging algorithm uses high-level jet-substructure information as input to a deep neural
network (DNN) that efficiently discriminates
between top-quark jets resulting from hadronic top-quark decay and the various
backgrounds~\cite{JETM-2018-03,ATL-PHYS-PUB-2020-017}.
Moreover, the jets containing $b$-hadrons coming from hadronization of $b$-quarks from top-quark decays are identified using another DNN
that exploits information from large-impact-parameter tracks, the topological decay chain and the displaced vertices of b-hadron
decays~\cite{FTAG-2018-01, ATL-PHYS-PUB-2017-013}.
The event selection and background estimation follows the approach used in Ref.~\cite{TOPQ-2012-15}, but with
a fourfold increase in sample size,
improved tagging methods~\cite{JETM-2018-03,ATL-PHYS-PUB-2020-017,ATL-PHYS-PUB-2017-013,FTAG-2018-01,Kogler:2018hem}, and
more precise background estimates.
 
These measurements utilise data collected by the ATLAS detector between 2015 and 2018 from $pp$ collisions at
$\sqrt{s}=13\,\TeV$, referred to as Run~2, corresponding to an integrated luminosity of \lumitot.
The \ttbar\ differential cross-sections are measured by unfolding the detector-level distributions
to a particle-level fiducial phase-space region and a parton-level fiducial phase-space region.
The particle-level criteria intend to match the kinematic requirements used for the detector-level selection of the \ttbar\ events
while the parton-level region is defined by making the same cuts on the \pt\ of
the leading and second-leading top quark as at the detector-level.
Unfolding the observed distributions to distributions of variables
directly related to the detector observables in a particle-level fiducial phase space
allow for precision tests of QCD,
as the particle-level results can be compared with Monte Carlo (MC) generator predictions
that implement
matrix-element calculations at next-to-leading order (NLO) in the strong coupling constant \alphas,
leading-order (LO) models for top-quark decay, and parton-shower and hadronization models.
This procedure avoids model-dependent extrapolation
of the measurements to a phase-space region outside the detector acceptance.
The size of the phase space nonetheless is large enough to allow robust tests of the QCD predictions.
Parton-level differential cross-sections are also presented, where the detector-level
distributions are unfolded to measure the  top-quark kinematics at the parton level in a
larger phase-space region, allowing measurement of the QCD process factoring out the parton
showering and hadronization of the quarks and gluons.
These allow comparisons with next-to-next-to-leading-order (NNLO) predictions for which matching to a parton-shower
algorithm is not yet available.
Here and in what follows, the cross-sections and distributions created using MC generators are referred to as
calculations or predictions, while distributions created using  MC events
passed through the detector simulation and event reconstruction are referred to as simulations.
 
The modelling of differential cross-sections are affected by
the models for initial- and final-state radiation (ISR and FSR), parton distribution functions (PDFs),
and the scheme for matching matrix-element calculations to parton-shower models.
Their measurement therefore tests the NLO QCD predictions for these aspects.
 
Measurements of the differential cross-sections for the leading and second-leading top quarks are made as a function of
their \pT\ and rapidity $y$.
In addition, the differential cross-sections for the transverse momentum ($\pt^{t}$) and
absolute rapidity ($|y^t|$) of a top quark chosen at random from the event are measured.
These are equivalent to the average of the top-quark and top-antiquark distributions and are
typically easier to compare with fixed-order predictions and measurements
in other channels than kinematic distributions of the leading or second-leading jet.
The rapidities of the leading and second-leading top quarks in the laboratory frame are denoted by $\ytone$ and $\yttwo$, respectively,
while their rapidities in the $\ttbar{}$ centre-of-mass frame are $\ystar = \nicefrac{1}{2}\left(\ytone - \yttwo\right)$ and $-\ystar$.
These allow the construction of the observable $\chittbar=\exp(2|\ystar|)$, which is of particular interest
since many phenomena not included in the SM, such as quark substructure, are predicted to peak at low
values of \chittbar{}~\cite{EXOT-2014-15}.
The longitudinal motion of the \ttbar\ system in the laboratory frame is described by the rapidity
boost $\boostttbar=\nicefrac{1}{2} \left( \ytone + \yttwo \right) $\ and is sensitive to PDFs.
The unfolded distributions for the \ttbar\ invariant mass (\mttbar), transverse momentum (\ptttbar), and absolute value
of rapidity (\absyttbar) are constructed, as these test
QCD predictions and are sensitive to processes beyond the SM (BSM processes).
Measurements of the differential cross-sections are also performed as a function of
the absolute value of the azimuthal angle between the two top quarks, \absdeltaPhittbar;
the absolute value of the out-of-plane momentum, \absPoutttbar\ (i.e. the projection of the three-momentum of the
second-leading top quark onto the direction perpendicular to a plane defined by the leading top quark and
the beam axis ($z$) in the laboratory frame \cite{PhysRevLett.81.2642});
the cosine of the production angle in the Collins--Soper
reference frame~\cite{Collins:1977iv}, \cosThetaStar; and
the scalar sum of the transverse momenta of the two top quarks, \HTttbar\ \cite{Denner:2010jp,Bevilacqua:2010qb}.
Several double- and triple-differential cross-sections employing pairs and triplets of
these observables are measured to provide information about
correlations. These are particularly sensitive to QCD modelling and have been
shown to constrain PDFs~\cite{CMS-TOP-14-013,CMS-TOP-18-004}.
 
These measurements are compared with different QCD predictions.
Direct comparisons of the
differential cross-sections incorporating statistical and systematic uncertainties
identify the predictions that are in best agreement with the data.
Measurements unfolded to the particle-level fiducial phase space and parton-level fiducial phase space
are compared with QCD predictions at NLO produced by the
\POWHEGBOX[v2]~\cite{Frixione:2007nw,Nason:2004rx,Frixione:2007vw,Alioli:2010xd}\ and
\MGNLO{}~\cite{Alwall:2014hca}\ programs.
Measurements unfolded to the parton-level fiducial phase space are compared with a
NNLO calculation implemented in the \MATRIX program~\cite{Grazzini:2017mhc,Catani:2019iny,Catani:2019hip}.
Other NNLO calculations for \ttbar\ production exist~\cite{Czakon:2015owf,Czakon:2016dgf,Czakon:2017wor,Behring:2019iiv}\
but are not publicly available to make predictions
in the phase space and/or final state employed in these measurements.
 
The unfolded distributions are also used to set constraints on the magnitude of particle couplings beyond the SM using the SM effective field theory (SMEFT) that extends the SM by adding higher-order dimension operators suppressed
by the scale of the new physics that respect the Lorentz and gauge invariance and other assumed basic symmetries~\cite{Buchmuller:1985jz}.
Using a LO SMEFT model that incorporates a full list of dimension-6 operators involved in top-quark
interactions~\cite{AguilarSaavedra:2018nen}, limits are set on a subset of Wilson coefficients
and on selected pairs of these coefficients.
These can be related to the couplings and production of massive particles beyond the SM.
 
The paper is organized as follows. Section~\ref{sec:detector} describes the ATLAS detector,
while Section~\ref{sec:datasets} describes the data and simulation samples used in the measurements.
The reconstruction of lepton and jet candidates, and the event selection based on these,
is explained in Section~\ref{sec:selection}\
and the background estimates
are discussed in Section~\ref{sec:backgrounds}.
The detector-level results are presented in Section~\ref{sec:reco_level}.
The procedure for particle-level and parton-level unfolding is described in Section~\ref{sec:unfolding}.
The systematic uncertainties affecting the measurements are summarized in Section~\ref{sec:systematics}.
The results of the measurements are presented in Section~\ref{sec:measurements} and the discussion and the interpretation of these results is made in Section~\ref{sec:comparisons}.
The results of the analysis using the SMEFT formalism are presented in Section~\ref{sec:eft}.
Conclusions of this study are summarized in Section~\ref{sec:conclusion}.


\section{ATLAS detector}
\label{sec:detector}

 
\newcommand{\AtlasCoordFootnote}{
ATLAS uses a right-handed coordinate system with its origin at the nominal interaction point (IP)
in the centre of the detector and the \(z\)-axis along the beam pipe.
The \(x\)-axis points from the IP to the centre of the LHC ring,
and the \(y\)-axis points upwards.
Cylindrical coordinates \((r,\phi)\) are used in the transverse plane,
\(\phi\) being the azimuthal angle around the \(z\)-axis.
The pseudorapidity is defined in terms of the polar angle \(\theta\) as \(\eta = -\ln \tan(\theta/2)\).
Angular distance is measured in units of \(\Delta R \equiv \sqrt{(\Delta\eta)^{2} + (\Delta\phi)^{2}}\).}
 
The ATLAS detector~\cite{PERF-2007-01} at the LHC covers nearly the entire solid angle around the collision point.\footnote{\AtlasCoordFootnote}
It consists of an inner tracking detector surrounded by a thin superconducting solenoid, electromagnetic and hadron calorimeters,
and a muon spectrometer incorporating large superconducting air-core toroidal magnets.
 
The inner-detector system (ID) is immersed in a \SI{2}{\tesla} axial magnetic field
and provides charged-particle tracking in the range \(|\eta| < 2.5\).
The high-granularity silicon pixel detector covers the vertex region and typically provides four measurements per track,
the first hit normally being in the insertable B-layer installed before Run~2~\cite{ATLAS-TDR-19,PIX-2018-001}.
It is followed by the silicon microstrip tracker, which usually provides eight measurements per track.
These silicon detectors are complemented by the transition radiation tracker,
which enables radially extended track reconstruction up to \(|\eta| = 2.0\).
 
The calorimeter system covers the pseudorapidity range \(|\eta| < 4.9\).
Within the region \(|\eta|< 3.2\), electromagnetic calorimetry is provided by barrel and
endcap high-granularity lead/liquid-argon (LAr) calorimeters,
with an additional thin LAr presampler covering \(|\eta| < 1.8\)
to correct for energy loss in material upstream of the calorimeters.
Hadron calorimetry is provided by the steel/scintillator-tile calorimeter,
segmented into three barrel structures within \(|\eta| < 1.7\), and two copper/LAr hadron endcap calorimeters.
The solid angle coverage is completed with forward copper/LAr and tungsten/LAr calorimeter modules
optimized for electromagnetic and hadronic energy measurements, respectively.
 
The muon spectrometer (MS) comprises separate trigger and
high-precision tracking chambers measuring the deflection of muons in a magnetic field generated by the superconducting air-core toroidal magnets.
The field integral of the toroids ranges between \num{2.0} and \SI{6.0}{\tesla\metre}
across most of the detector.
A set of precision chambers covers the region \(|\eta| < 2.7\) with three layers of monitored drift tubes,
complemented by cathode-strip chambers in the forward region, where the background is highest.
The muon trigger system covers the range \(|\eta| < 2.4\) with resistive-plate chambers in the barrel, and thin-gap chambers in the endcap regions.
 
Interesting events are selected by the first-level trigger system implemented in custom hardware,
followed by selections made by algorithms implemented in software in the high-level trigger~\cite{TRIG-2016-01}.
The first-level trigger accepts events from the \SI{40}{\MHz} bunch crossings at a rate below \SI{100}{\kHz},
which the high-level trigger further reduces in order to record events to disk at about \SI{1}{\kHz}.
An extensive software suite~\cite{ATL-SOFT-PUB-2021-001} is used in the reconstruction
and analysis of real and simulated data, in detector operations, and in the trigger and data acquisition systems of the experiment.


\section{Data and simulated event samples}
\label{sec:datasets}

The data used for this analysis were recorded with the ATLAS detector at a $pp$\
centre-of-mass energy of 13~\TeV\ between 2015 and 2018 and correspond to an  integrated luminosity of \lumitot. Only data taken under stable beam conditions with fully operational subdetectors are considered~\cite{DAPR-2018-01}.
The events for this analysis were collected using an inclusive
jet trigger employing anti-$k_{t}$~\cite{Cacciari:2008gp}\ reconstruction with radius parameter $R=1.0$\
and nominal \pT\ thresholds of 360~\GeV\ in 2015, 420~\GeV\ in 2016, and 460~\GeV\ in 2017 and 2018.
Moreover, single-jet and double-jet triggers with lower \pt thresholds and
jet-mass requirements of ${>}40$~\GeV\ and ${>}35$~\GeV\ were used in 2017 and 2018, respectively.
These triggers are fully efficient for jets with $\pT > 500$~\GeV~\cite{TRIG-2016-01}.
 
The signal and several background processes were modelled using MC generators.
The effect of multiple interactions in the same and neighbouring bunch
crossings (\pileup) was modelled by overlaying the simulated hard-scattering event with inelastic $pp$\ events generated with
\PYTHIA[8.186]~\cite{Sjostrand:2007gs}
using the \NNPDF[2.3lo] set of PDFs~\cite{Ball:2012cx} and the
A3 set of tuned parameters~\cite{ATL-PHYS-PUB-2016-017}.
The detector response was simulated using the \GEANT\ framework~\cite{Agostinelli:2002hh,SOFT-2010-01}.
The data and MC events are reconstructed with the same software algorithms.
 
Several NLO calculations of the \ttbar\
process are used to generate the simulated events and in comparisons with the measured differential cross-sections. The \POWHEGBOX[v2]~\cite{Frixione:2007nw,Nason:2004rx,Frixione:2007vw,Alioli:2010xd} and
\MGNLO{}~\cite{Alwall:2014hca}\ (hereafter referred to as \AMCatNLO)
MC generators encode different approaches to the matrix-element calculation and different matching schemes between the NLO QCD matrix-element (ME) calculation and the parton-shower (PS) algorithm.
Unless explicitly noted below, the following generator set-ups were used.
The employed PDF set is \NNPDF[3.0nlo]~\cite{Ball:2014uwa}.
Parton showering and hadronization was performed with \PYTHIA[8.230]~\cite{Sjostrand:2014zea}
using the A14 set of tuned parameters~\cite{ATL-PHYS-PUB-2014-021} and the \NNPDF[2.3lo] set of PDFs.
The top-quark mass was set to $\mt=172.5$~\GeV\ for all samples with the top quark in the final state and the renormalization and factorization scales were
set to $\mu_{\mathrm{r/f}} = \sqrt{\mtop^{2} + \pTX[2]}$ for all \ttbar\ samples, where \pT\ is the transverse momentum of the top quark.
The decays of bottom and charm hadrons were simulated using the \EVTGEN[1.6.0] program~\cite{Lange:2001uf}.
 
The nominal sample used the \POWHEGBOX[v2] generator at NLO in QCD.
The \hdamp parameter, which controls the matching in the \POWHEG calculation and
effectively regulates the high-\pt\ radiation against which the
\ttbar system recoils, was set to 1.5\,\mtop~\cite{ATL-PHYS-PUB-2016-020}.
To increase the available statistics for events with high-\pT\ top quarks, multiple samples were
generated
with different ranges of the total scalar sum of \pT in the event.
 
An alternative matrix-element calculation and matching with the parton shower was realized with the
\AMCatNLO[2.6.0] generator.
Top quarks were decayed at LO using the
\MADSPIN\ program~\cite{Frixione:2007zp,Artoisenet:2012st} to preserve spin correlations.
The parton-shower starting scale has the functional form $\muQ = H_{\mathrm T}/2$~\cite{ATL-PHYS-PUB-2017-007},
where $H_{\mathrm T}$ is defined as the scalar sum of the \pT of all outgoing partons.
An alternative \POWPY[8] sample with the \POWHEG{}\ parameter $h_{\mathrm{damp}}$ set to $3\mt$ was used to assess
part of the ISR systematic uncertainty~\cite{ATL-PHYS-PUB-2018-009}.
An additional \POWPY[8] sample was generated with the matrix-element correction turned off in order
to assess the systematic uncertainty due to this change in the matrix-element calculation.
The effects of using alternative parton-shower and hadronization models were probed by combining the nominal \POWHEG{}\ set-up
with the \HERWIG[7.1.3] parton-shower and hadronization model~\cite{Bahr:2008pv,Bellm:2015jjp}, using the
\HERWIG[7.1] default set of tuned parameters~\cite{Bellm:2015jjp,Bellm:2017jjp} and the \MMHT[lo] PDF set~\cite{Harland-Lang:2014zoa}.
 
Single-top-quark production in association with a \Wboson~boson ($tW$) was
modelled by the \POWHEGBOX[v2] generator at NLO using the five-flavour scheme.
The diagram-removal scheme~\cite{Frixione:2008yi} was used to
remove interference and overlap with \ttbar\ production.
Electroweak $t$-channel single-top-quark events were modelled using \POWHEGBOX[v2]~\cite{Re:2010bp} at NLO in the four-flavour
scheme.
The electroweak $s$-channel single-top-quark process was not modelled explicitly, owing to
its small cross-section and its negligible contribution after the event selection described
in Section~\ref{sec:selection}.
Its remaining contribution is largely accounted for in the data-driven background estimate.
The associated production of \ttbar\ and a Higgs boson (\ttH) was modelled by the
\POWHEGBOX[v2]~\cite{Hartanto:2015uka} generator at NLO.
The associated production of \ttbar\ and a \Wboson or \Zzero\ boson (\ttV) was modelled using the \AMCatNLO[2.3.3]
generator at NLO.
Parton showering and hadronization was performed by the \PYTHIA[8.210] generator.
The decays of bottom and charm hadrons were simulated using the \EVTGEN[1.2.0] program.
 
The \ttbar samples are normalized to the cross-section prediction at NNLO
in QCD including the resummation of next-to-next-to-leading logarithmic (NNLL) soft-gluon terms calculated using
the \TOPpp[2.0] program~\cite{Beneke:2011mq,Cacciari:2011hy,Baernreuther:2012ws,Czakon:2012zr,Czakon:2012pz,Czakon:2013goa,Czakon:2011xx}.
For $pp$\ collisions at a centre-of-mass energy of \rts~=~\SI{13}{\TeV}, this cross-section corresponds to
$\sigma(\ttbar)_{\textrm{NNLO+NNLL}} = \mathrm{832\pm51~pb}$ using a top-quark mass of $\mtop = 172.5\,\GeV$.
The uncertainties in the cross-section due to the PDFs and \alphas are calculated using the \PDFforLHC[15] prescription~\cite{Butterworth:2015oua}
with the \MSTW[nnlo]~\cite{Martin:2009iq,Martin:2009bu}, \CT[10nnlo]~\cite{Lai:2010vv,Gao:2013xoa},
and \NNPDF[2.3lo] PDF sets in the five-flavour scheme, and are added in quadrature to the effect of the scale uncertainty.
 
Calculations of \ttbar{} production at NNLO matched to the parton shower to
produce particle-level predictions are not yet available for the all-hadronic final state.
In order to evaluate the impact of NNLO corrections, the MC set-ups are reweighted at parton level.
The reweighting is performed on the three variables: $\pt^{t}$, $\mttbar$, and $\ptttbar$, using the kinematics of the top-quarks in the MC samples after ISR and FSR.
The predictions for $\pt^{t}$ and $\mttbar$ are calculated at NNLO in QCD with NLO EW corrections~\cite{Czakon:2017wor} with the \NNPDF[3.0QED] PDF set using the dynamic renormalisation and factorisation scales $m_\text{T}(t)/2$ for $\pt^{t}$ and $\HTttbar/4$ for $\mttbar$ as proposed in Ref.~\cite{Czakon:2017wor}.
The prediction for $\ptttbar$ is calculated at NNLO in QCD~\cite{Grazzini:2017mhc,Catani:2019iny,Catani:2019hip} with the \NNPDF[3.0] PDF set using renormalisation and factorisation scales $\HTttbar/4$.
All the predictions use $\mtop = 173.3\,\GeV$, a value 0.8~\GeV\ larger than the nominal samples.
It was verified that the changes in the predicted distributions due to the different choice of \mtop\ are negligible.
The reweighting has been derived iteratively~\cite{Serkin:2021bbn}, such that at the end of the procedure the reweighted MC sample is in good agreement with the high-order prediction for each of the three variables.
These samples are referred to as being reweighted to the NNLO prediction in the remainder of the paper.
 
The single-top-quark $tW$ cross-section is normalized to the calculation
at NLO with NNLL soft-gluon corrections~\cite{Kidonakis:2010ux,Kidonakis:2013zqa}.
The single-top-quark $t$-channel cross-section is normalized to the NLO calculation with the
\HATHOR[2.1] program~\cite{Aliev:2010zk,Kant:2014oha}.
The predicted values at \rts~=~\SI{13}{\TeV} are $136.02^{+5.40}_{-4.57}$~pb, $80.95^{+4.06}_{-3.61}$~pb, and $71.7 \pm 3.8$~pb
for $t$-channel top-quark production, $t$-channel top-antiquark production, and $tW$ production, respectively.
The cross-sections for \ttbar\ production in association with a \Zzero, \Wboson, or Higgs boson are
normalized to the NLO QCD + NLO electroweak predictions as calculated by the \AMCatNLO generator and
reported in Ref.~\cite{deFlorian:2016spz}.
The predicted values at \rts~=~\SI{13}{\TeV} are
$0.88^{+0.09}_{-0.11}$~pb,
$0.60^{+0.08}_{-0.07}$~~pb, and
$0.51^{+0.04}_{-0.05}$~~pb, respectively.
 
Comparisons with the measured differential cross-sections at parton level use a calculation of
\ttbar production at QCD NNLO precision by the
\MATRIX program~\cite{Grazzini:2017mhc,Catani:2019iny,Catani:2019hip},
which provides differential \ttbar\ predictions in the full and fiducial phase space.
For the nominal \MATRIX prediction, the top-quark mass is set to 172.5~\GeV\ and
the \NNPDF[3.1nnlo] PDFs are employed with $\alphas(m_Z) = 0.118$
set in the calculations~\cite{Ball:2017nwa}, together with renormalization and factorization dynamical scales of $\muR = \muF = H_\text{T}/2$,
where $H_\text{T}=\sqrt{\mt^2+p_{{\rm T},t}^2} + \sqrt{\mt^2+p_{{\rm T},\bar{t}}^2}$.
For the alternative predictions, the dynamical scales are defined by
$\muR = \muF = H_\text{T}/4$ and $\muR = \muF = m_{t\bar{t}}/2$, and the \CT[18nnlo]~\cite{Hou:2019qau} and
\MMHT[nnlo]\cite{Harland-Lang:2014zoa} PDFs are used.
A seven-point scale variation is used to obtain the effect of the scale uncertainty by varying the renormalization and factorization scales by
a factor of two around their central value with the constraint $0.5 \le \muF / \muR \le 2$.
The largest upward and downward changes from the central-value result are taken as positive and negative uncertainties, respectively.


\section{Selection of events}
\label{sec:selection}

Fully reconstructed and individually selected jets, electrons, and muons, together with selections on the final-state
topology measured with those, are used when choosing the events considered for this analysis.
The applied selection criteria are summarized in the following subsections.
 
\subsection{Particle and jet selections}
\label{sec:objects_selection}
 
Electron candidates are identified from high-quality ID tracks matched to calorimeter
energy deposits consistent with an electromagnetic shower.
The energy deposits have to form a cluster with $\pt > 25$~\GeV\ and $|\eta| < 2.47$, and
be outside the transition region $1.37 \leq |\eta| \leq 1.52$\ between the barrel and endcap calorimeters. A tight likelihood-based requirement is used to reject fake-electron candidates, and calorimeter- and track-based isolation requirements are imposed~\cite{EGAM-2018-01}.
 
Muon candidates are reconstructed using high-quality inner-detector tracks combined with tracks
reconstructed in the muon spectrometer.
Only muon candidates satisfying `medium' identification criteria~\cite{MUON-2018-03}, with $\pt > 25$~\GeV\
and $|\eta| < 2.5$, are considered.
Isolation criteria similar to those used for electrons are imposed~\cite{MUON-2018-03}.
To reduce the impact of non-prompt leptons, muons within $\Delta R = 0.4$ of a
\smallR\ jet, as defined below, are removed.
 
The \antikt\ \cite{Cacciari:2008gp}\ and variable-$R$\ \cite{VRjets}\ algorithms implemented in the
FastJet package \cite{Fastjet}\ are used to define three types of jets for this analysis:
\largeR\ jets with fixed $R = 1.0$~\cite{JETM-2018-06},
\smallR\ jets with fixed $R=0.4$\ used to investigate the internal kinematics of the \largeR\ jets, and
\VRTrack\ jets with a \pT-dependent variable-radius parameter, ranging between $R = 0.02$
and $R = 0.4$~\cite{VRjets,ATL-PHYS-PUB-2020-019}, which are used to identify $b$-hadrons.
These are reconstructed independently of each other.
The \largeR\ jets are formed from topological clusters in the calorimeter
calibrated using the local calibration method described in Ref.~\cite{PERF-2014-07}, while
the \smallR\ jets are reconstructed from both calorimeter energy clusters and charged-particle tracks.
The \VRTrack\ jets are reconstructed from
inner-detector tracks that are  used as input to the clustering algorithm.
Only \VRTrack\ jets that have $|\eta| < 2.5$ and $\pt > 10$~\GeV\ are considered.
Small-$R$\ jets with $\pT < 60$~\GeV\ are required to have charged-particle tracks matched to the
primary interaction vertex~\cite{PERF-2014-03}.
 
Variable-$R$\ jets containing $b$-hadrons are identified ($b$-tagged) using a DNN that exploits
information from large-impact-parameter tracks, the topological decay chain and the displaced vertices of $b$-hadron decays~\cite{FTAG-2018-01, ATL-PHYS-PUB-2017-013}.
The \VRTrack\ jets are considered $b$-tagged if the value of the discriminant is larger than a threshold
that provides 77\%\ efficiency as measured in inclusive \ttbar\ events.
The $b$-tagging efficiency observed in the boosted top-quark jets employed in this analysis is found to be ${\sim}70$\%,
which arises from the increased collimation and charged-particle track density in the top-quark jets.
The corresponding rejection factors for gluon/light-quark jets and charm-quark jets are approximately 300 and 7, respectively,
as measured in inclusive \ttbar\ events.
The \VRTrack\ jets are associated with the \largeR\ jets using a ghost-matching
algorithm~\cite{ghostmatch, Cacciari:2008gn}, which identifies those \VRTrack\ jets that are contained within
or are in proximity to the \largeR\ jet.
A \largeR\ jet with at least one associated $b$-tagged \VRTrack\ jet is considered $b$-tagged.
 
The \largeR\ and \smallR\ jet energy and mass scales are corrected by using energy- and $\eta$-dependent calibration factors derived from simulation and in situ measurements~\cite{JETM-2018-02, JETM-2018-05}.
The \largeR\ jet candidates are required to have $|\eta| < 2.0$, $200~\GeV < \pt < 3000~\GeV$, and jet mass ${>}50$~\GeV,
where $\eta$\ is used instead of rapidity for selection at the detector level because the jet calibrations
were determined as a function of $\eta$. The 'combined jet mass'~\cite{JETM-2018-02}, which
uses both the information from the calorimeter and the tracking system to measure the
jet mass, is used to define the \largeR-jet mass.
A trimming algorithm~\cite{trimming}\ with parameters $R_{\mathrm{sub}} = 0.2$ and $f_{\mathrm{cut}} = 0.05$\ is applied to
the \largeR\ jets to suppress gluon radiation and mitigate \pileup\ effects.
 
The top-quark tagging of \largeR\ jets relies on a DNN that uses jet-substructure variables
such as the jet mass and measures of energy flow as input~\cite{JETM-2018-03,ATL-PHYS-PUB-2020-017}.
The \pT-dependent requirements on the DNN score provide 80\%\ top-quark-tagging efficiency across the full jet-\pT\ range, as measured in inclusive \ttbar\ events where the
top-quark decay products are contained within the \largeR\ jet.
The algorithm has a light-quark and gluon jet-rejection factor that is \pT-dependent, being
${\sim}15$\ at $\pT=500$~\GeV\ and decreasing to ${\sim}12$\ at $\pT=1$~\TeV, as measured in multijet events.
 
\subsection{Event selection}
\label{sec:events_selection}
 
The event selection targets fully hadronic \ttbar\ events where both top-quark jets have high \pT.
Each event is required to have a primary vertex with at least two associated ID tracks with $\pT > 0.5$~\GeV.
The vertex with the highest $\sum \pt^2$ of the associated tracks is taken as the primary vertex.
In order to reject top-quark events where a top quark has decayed semileptonically,
the events are required to contain no reconstructed electron or muon candidates.
To identify the fully hadronic decay topology, events must have at least two \largeR\ jets with
$\pt > 350$~\GeV, with at least one of these having $\pt > 500$~\GeV.
The first top-quark-candidate jet is selected among all the \largeR\ jets with $\pt > 500$~\GeV\ as
that with the closest mass to the nominal top-quark mass of 172.5~\GeV.
The second top-quark-candidate jet is selected from the remaining \largeR\ jets with $\pt > 350$~\GeV,
using the same mass requirement.
Both top-quark-candidate jets must have a mass within 50~\GeV\ of the top-quark mass.
This preselection results in a sample of 2.2~million events.
 
To reject multijet background events, the two top-quark-candidate jets must satisfy the top-quark-tagging
criteria described in Section~\ref{sec:objects_selection}\ and must be $b$-tagged.
The final-state \ttbar\ candidate's momentum is defined by the sum of the four-momenta of the two top-quark-candidate jets.
 
This selection defines the signal region, which has 17\,261 events.


\section{Background estimation}
\label{sec:backgrounds}

The backgrounds in the selected data sample are events characterized by a number of
high-\pT\ jets that do not arise from a top quark, and events that have at least one top quark decaying
semileptonically, have only one top quark decaying hadronically, or arise from production
of a top-quark pair in association with a $W$, $Z$, or Higgs boson.
The first contribution, referred to as multijet background and where the two leading jets both arise
from gluons or lighter quarks, is found to be the largest
background.
Because the uncertainties in MC predictions of this background are large~\cite{STDM-2013-11,STDM-2011-38}, it is estimated
using a data-driven
approach.
A similar method was used in previous work \cite{TOPQ-2012-15,TOPQ-2016-09}.
The second set of contributions are from processes that can be relatively accurately calculated and
simulated, and so MC calculations are used to estimate them.
 
The estimation of backgrounds from these sources is described in the following subsections.
 
\subsection{Multijet background}
 
\begin{table}[t!]
\begin{center}
\small{
\begin{tabular}{c|c|c|c|c|c|}
\cline{2-6}
\multirow{4}{*}{\rotatebox{90}{2nd large-$R$ jet}}&1t1b & \cellcolor{green}J (10\%) & \cellcolor{cyan}K (29\%) & \cellcolor{cyan}L (45\%) & \cellcolor{red}S \\ \cline{2-6}
&  0t1b & \cellcolor{green}B (1.9\%) & \cellcolor{green}D (6.6\%) & \cellcolor{green}H (12\%) & \cellcolor{cyan}N (56\%) \\ \cline{2-6}
&  1t0b & \cellcolor{green}E (0.7\%) & \cellcolor{green}F (2.7\%) & \cellcolor{green}G (5.9\%) & \cellcolor{cyan}M (35\%) \\ \cline{2-6}
&  0t0b & \cellcolor{green}A (0.1\%) & \cellcolor{green}C (0.7\%) & \cellcolor{green}I (1.6\%) & \cellcolor{green}O (12\%) \\ \cline{2-6}
& &0t0b&1t0b&0t1b&1t1b \\ \cline{2-6}
\multicolumn{1}{c}{}& \multicolumn{5}{c}{1st \largeR\ jet} \\
\end{tabular}
}
\vspace{0.2cm}
\caption{Region labels as a function of tagging states of the leading ('1st \largeR\ jet')  and
the second-leading ('2nd \largeR\ jet') \largeR\ jet.
A top-quark-tagged jet is defined by the tagging algorithm described in Section~\ref{sec:objects_selection},
and denoted by `1t' in the table, while
a jet that is not top-quark-tagged is labelled `0t'.
Jets that are or are not $b$-tagged are denoted by `1b' or `0b', respectively.
The expected proportion of \ttbar\ signal events and MC-predicted background events relative to
the number of data events in each region is given in parentheses.
The regions marked in green, blue, and red correspond to control, validation, and signal regions, respectively.
}
\label{tab:ABCD:16regions}
\end{center}
\end{table}
 
The data-driven multijet-background estimate is made using a set of control regions.
Sixteen separate regions are defined by classifying each event in the preselection
sample according to whether the leading and second-leading jets are top-quark-tagged or $b$-tagged.
Table~\ref{tab:ABCD:16regions}\ shows the 16 regions that are defined in this way, and indicates
the expected proportion of \ttbar\ signal events and MC-predicted background events relative to the
number of data events in each region.
These fractions illustrate the size of the MC-predicted subtractions in each region when calculating the
data-driven multijet-background estimates.
Region S is the signal region, while the regions with at most two tags
that are either top-tags or $b$-tags (A--J, O)
are dominated by multijet background and serve as control regions.
Regions with three tags (K, L, N and M) have an expected contribution from top-quark pairs
of at least 20\%\ of the observed yield and are validation regions.
In other regions, the expected contribution from signal and MC-predicted backgrounds is ${<}10$\%\ of the number of observed events.
The asymmetry in the signal contribution in the regions of the table
where the tagging states of the leading and second-leading jet are interchanged is
due to the interplay of several factors, such as the \pT- and jet-mass dependence of
top-quark-tagging and $b$-tagging efficiencies.
 
The estimated contributions of the \ttbar\ signal and the MC-predicted backgrounds
are subtracted from each control region, ignoring the small $s$-channel single-top-quark contribution.
This provides an estimate of the number of multijet events in each control region.
The number of multijet events in region J divided by the number of multijet events in
region A gives an estimate of the ratio of the number of multijet events in region S to the number of multijet events in region O, since the events in different regions in a given ratio always
differ only by the top-quark-tagging and $b$-tagging state of the second-leading \largeR\ jet.
These relationships are used to estimate the multijet-background rate in region S, i.e.\ $S=(O\times J)/A$,
where $O$, $J$\ and $A$\ are the number of multijet events in each region, while $S$\ is the estimate of
the number of multijet background events in region S.
 
This `ABCD' estimate assumes that the mistagging rate of the leading jet does not depend on how the
second-leading jet is tagged, but in practice there are correlations between the mistagging
rates of the two \largeR\ jet candidates. The primary source of such mistagging correlations is
between top-quark-tagging and $b$-tagging in the same jet.
These correlations are measured in the background-dominated regions, e.g.\ a comparison of the ratio of the numbers of events in regions F and E (representing the leading-jet top-quark-mistagging rate when the
second-leading jet is top-quark-tagged) with
the ratio of the numbers of events in regions C and A (giving the leading-jet top-quark-mistagging rate when the
second-leading jet is not top-quark-tagged) gives the correction factor due to the correlation
between top-quark-mistagging states of the two \largeR\ jets.
This results in a data-driven estimate of the number of multijet background events in region S given by
\begin{eqnarray}
S &=& \dfrac{J \times O}{A}\cdot\dfrac{D\times A}{B\times C}\cdot\dfrac{G\times A}{E\times I}\cdot\dfrac{F\times A}{E\times C}\cdot\dfrac{H\times A}{B\times I} \nonumber \\
&=&
\dfrac{ J\times O \times H \times F \times D\times G\times A^{3}}{\left( B\times E\times C\times I \right)^{2}},
\label{eq:ABCD16}
\end{eqnarray}
where the italic letters again represent the number of observed events in that region after the subtraction
of \ttbar\ signal events and the MC-predicted background events.
The measured correlations result in a ${\sim}15$\%\ increase in the multijet estimate relative to the estimate which does not take into account these measured correlations.
There is no significant jet-\pT\ dependence of these correlations.
 
This background estimate depends on the assumed inclusive \ttbar\ cross-section, which is the NNLO+NNLL
cross-section prediction described in Section~\ref{sec:datasets}.
However, it was found that even a 20\% change in the assumed inclusive \ttbar\ cross-section would cause only
${\sim}1$\% change in the measured fiducial phase-space \ttbar\ cross-section, which is
negligible compared to the uncertainty of the measurement.
 
This background estimate, including the mistagging correlations, is made bin-by-bin in the observed distributions.
The statistical uncertainties reflect the number of events found in the regions used in Eq.~(\ref{eq:ABCD16}) for each bin.
 
\subsection{Non-all-hadronic \ttbar\ background}
 
The \POWHEG{}+\PYTHIA[8] \ttbar\ sample described in Section~\ref{sec:datasets}\ is used to estimate the
number of \ttbar\ events in the
sample that arise from at least one top quark decaying semileptonically.
This estimate includes contributions from decays resulting in $\tau$-leptons, as no attempt is made to
identify $\tau$-lepton candidates and reject them.
 
The proportion of this non-all-hadronic \ttbar\ background is estimated to be only ${\sim}3$\% in the signal region,
primarily due to the lepton-veto requirement
in the preselection and the top-quark-tagging requirements.
However, these \ttbar\ events contribute to a greater degree to the control and validation regions,
where the top-quark-tagging and/or $b$-tagging requirements are relaxed,
therefore affecting the multijet-background estimate.
Although the \ttbar\ production cross-section in the signal region is observed to be lower than the MC prediction, see Section~\ref{sec:reco_level}, the use of a cross-section scaled to
the signal region produces a negligible change in the
multijet-background estimate.
 
\subsection{Backgrounds from other top-quark production processes}
 
Single-top-quark production in the $Wt$- and $t$-channel makes a small contribution to the signal sample,
which is estimated using the MC predictions described in Section~\ref{sec:datasets}.
The $s$-channel single-top-quark process is not explicitly calculated given its expected small contribution
and its inclusion in the multijet-background estimate.
 
Backgrounds from a top-quark pair produced in association with a $W$, $Z$, or Higgs boson are estimated using the
MC predictions also described in Section~\ref{sec:datasets}.
 
The cumulative background from these processes is ${\sim}2$\%\ in the signal region.
 
\subsection{Validation of background calculations}
 
The multijet-background estimate is validated using four validation regions, each with a different ratio of
all-hadronic \ttbar\ events to multijet events.
In all these validation regions, the predicted \ttbar\ contribution was scaled by the same factor of 0.83,
obtained by scaling the \ttbar\ contribution to match the total yield prediction to the data in the signal region.
Regions L and N are estimated to consist of approximately  equal
numbers of \ttbar\ signal events and multijet-background events
while regions K and M are estimated to have a 2:1 ratio of multijet to \ttbar\ events.
In these regions, the multijet background is estimated using different combinations of control regions
along with different corrections for the mistagging correlations due to the modified tagging requirements used to define the validation regions.
 
The number of multijet events in the signal region is calculated by applying
Eq.~(\ref{eq:ABCD16}) to the number of events in the control regions.
This results in an estimate of $2900 \pm 160$\ multijet events in the signal region,
where the uncertainty takes into account the data statistical uncertainties in the signal and control
regions as well as the systematic uncertainties in the MC-based subtraction of top-quark-related
contributions in the regions used in Eq.~(\ref{eq:ABCD16}).
 
The ratios of predicted to observed yields in regions K, L, M, and N are
1.03, 0.99, 1.02, and 0.98, respectively, illustrating
good agreement between the predicted and observed event yields
in these validation regions within statistical and detector-related systematic uncertainties.
 
\begin{figure*}[htbp]
\centering
\subfigure[]{\includegraphics[width=0.45\textwidth]{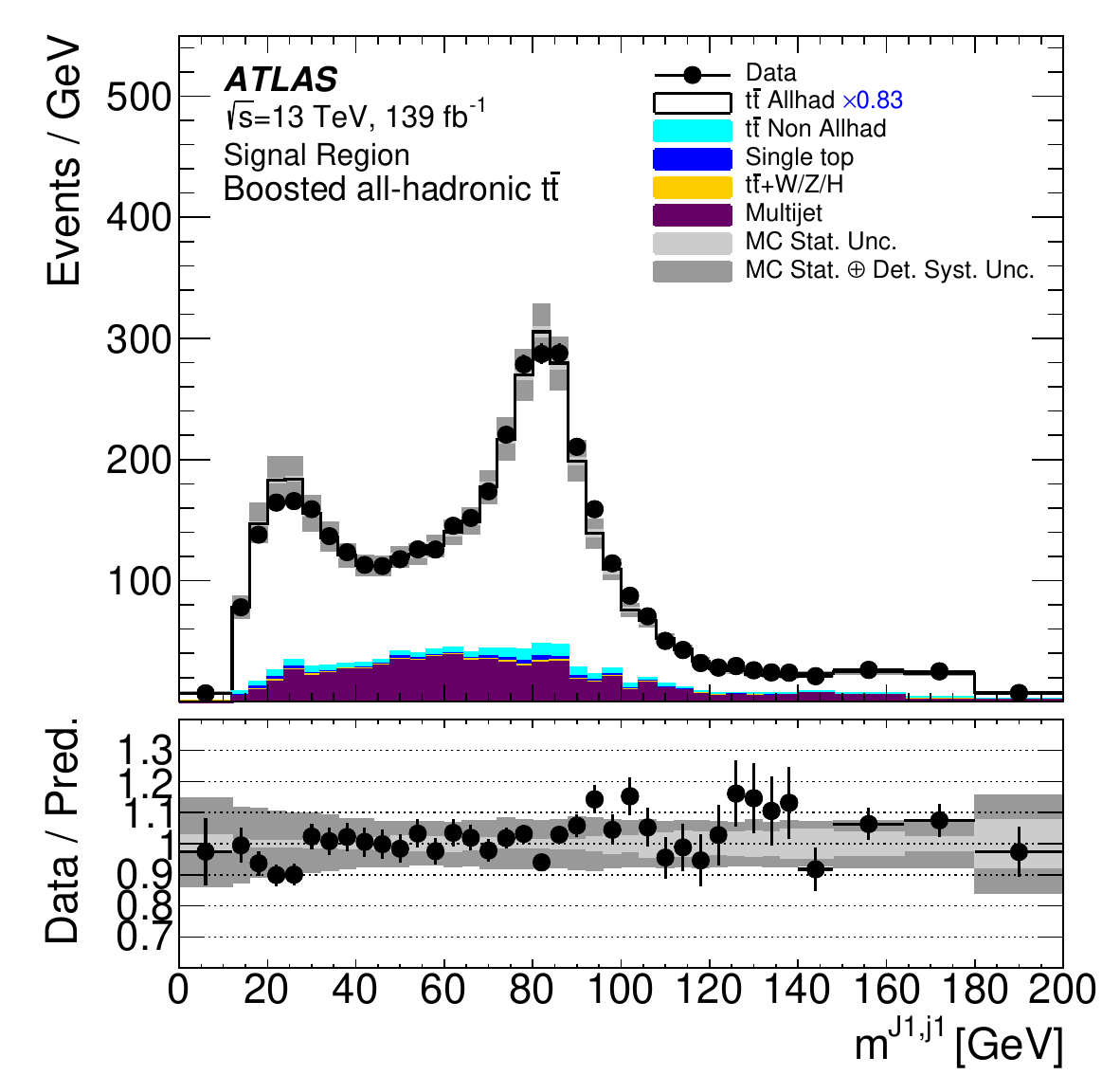}
\label{fig:SR:t1_sjM}}
\subfigure[]{\includegraphics[width=0.45\textwidth]{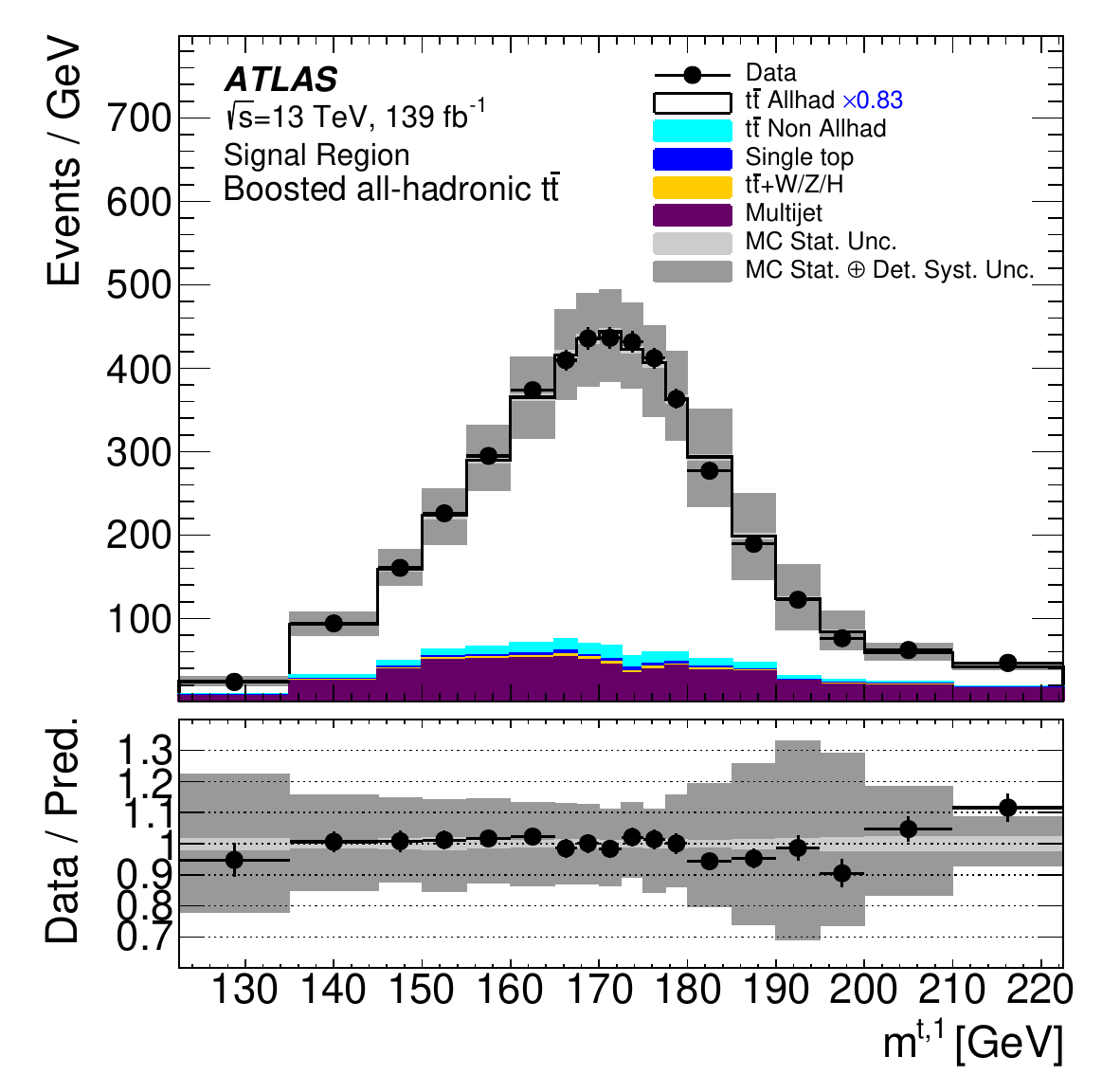}
\label{fig:SR:t1_M}}
\subfigure[]{\includegraphics[width=0.45\textwidth]{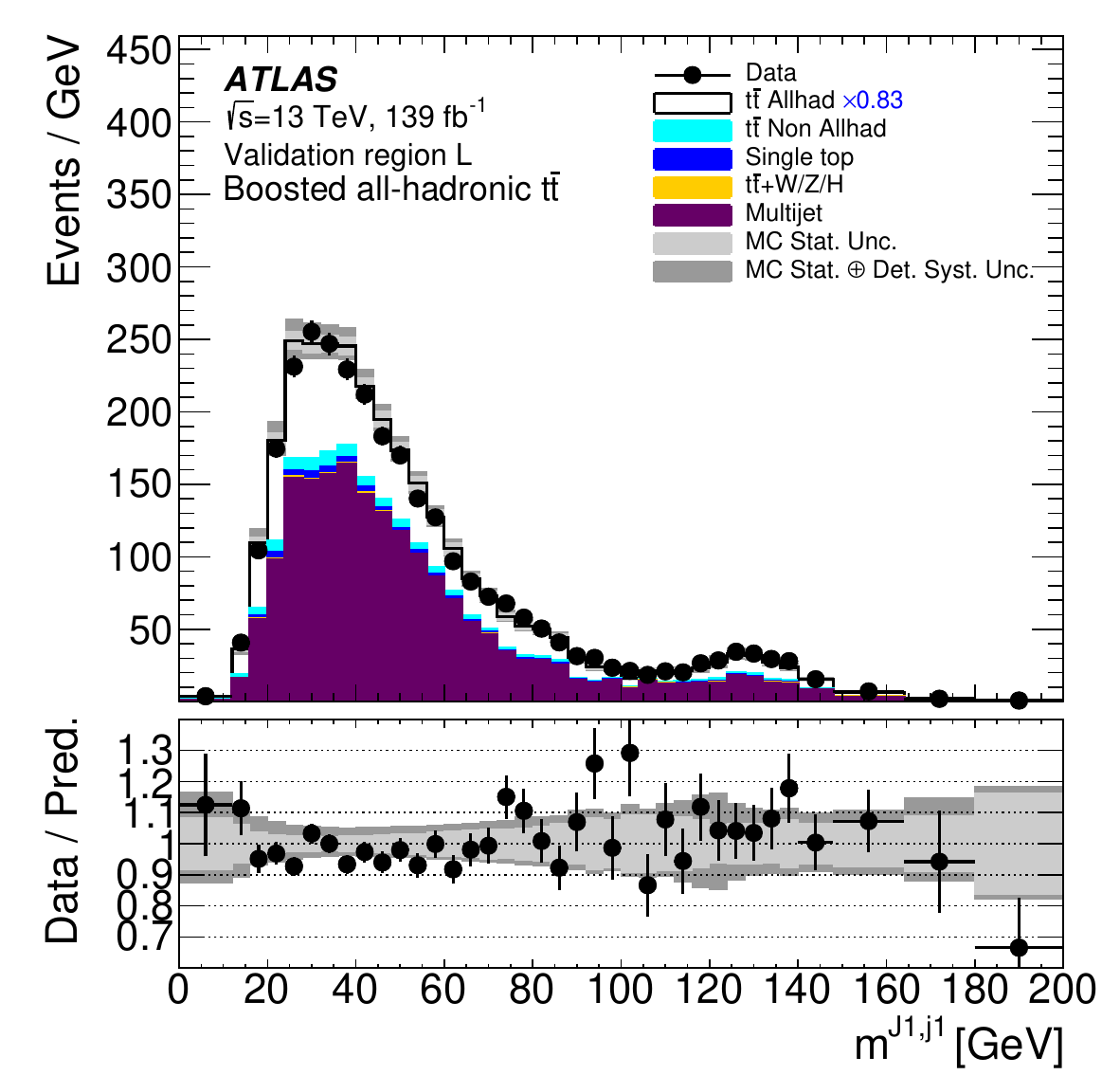}
\label{fig:VR_L:t1_sjM}}
\subfigure[]{ \includegraphics[width=0.45\textwidth]{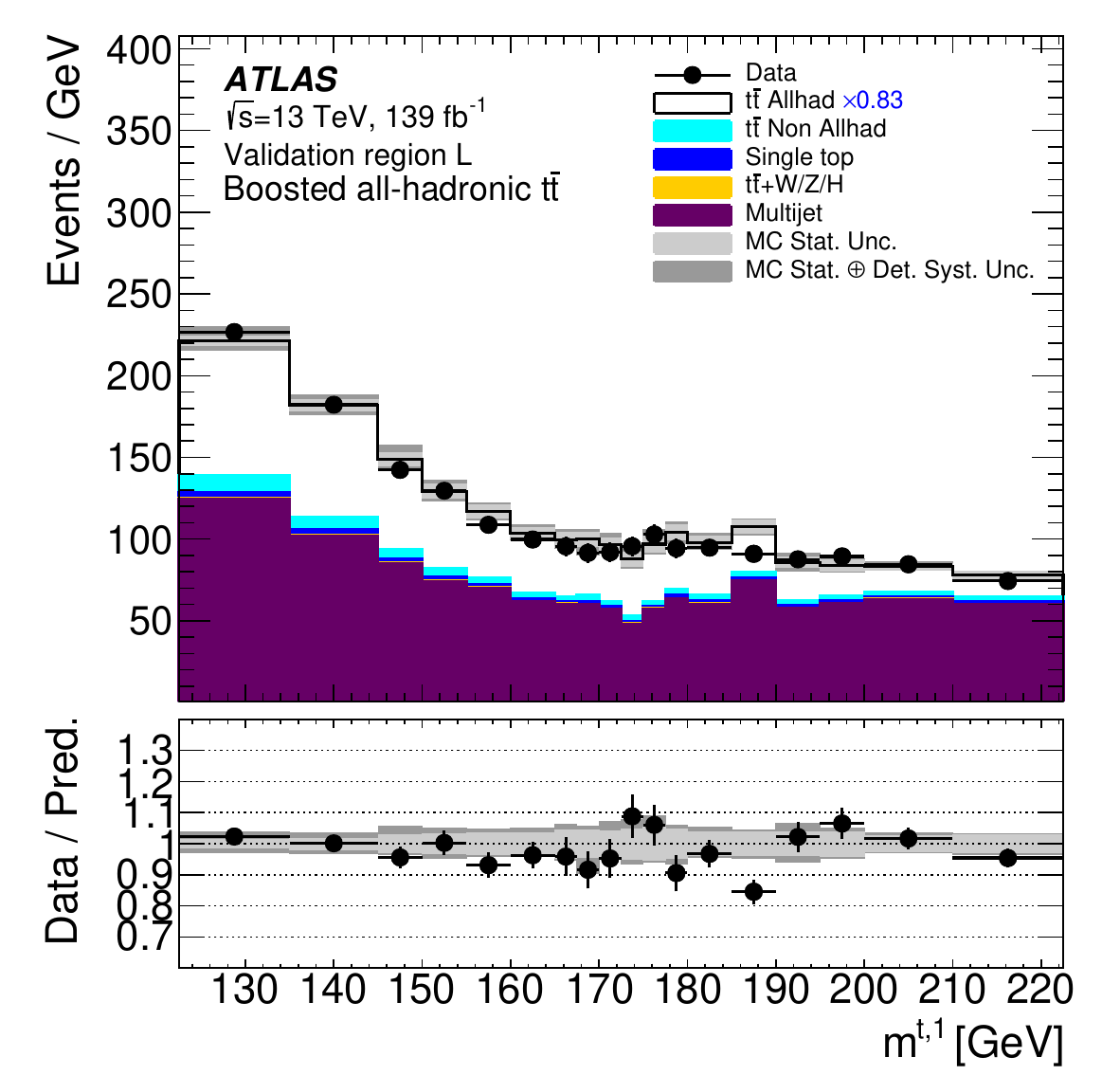}
\label{fig:VR_L:t1_M}}
\caption{Kinematic distributions of the leading top-quark-candidate jets in the signal region and
validation region L.
The mass distributions of the leading $R=0.4$\ \antikt\ jet in the leading \largeR\ jet for events
in the signal region and region L are
shown in \subref{fig:SR:t1_sjM}\ and \subref{fig:VR_L:t1_sjM}, respectively.
The leading \largeR\ jet-mass distributions for the events in the signal region
and validation region L
are shown in \subref{fig:SR:t1_M}\ and \subref{fig:VR_L:t1_M}, respectively.
The signal prediction (open histogram) is based on the \POWHEG{}+\Pythia[8] \ttbar\ calculation
normalized to the observed yield in the signal region.
The background (solid histogram) is the sum of the data-driven multijet estimate and the MC-based
expectation for the non-all-hadronic \ttbar, single-top-quark, and $\ttbar$\,+\,$W/Z/H$ processes.
The light grey bands indicate the statistical uncertainties and the dark grey bands indicate the combined statistical
and detector-related systematic uncertainties defined in Section~\ref{sec:systematics}.
}
\label{fig:VR_LSR:t1_M}
\end{figure*}

\begin{figure*}[htbp]
\centering
\subfigure[]{\includegraphics[width=0.45\textwidth]{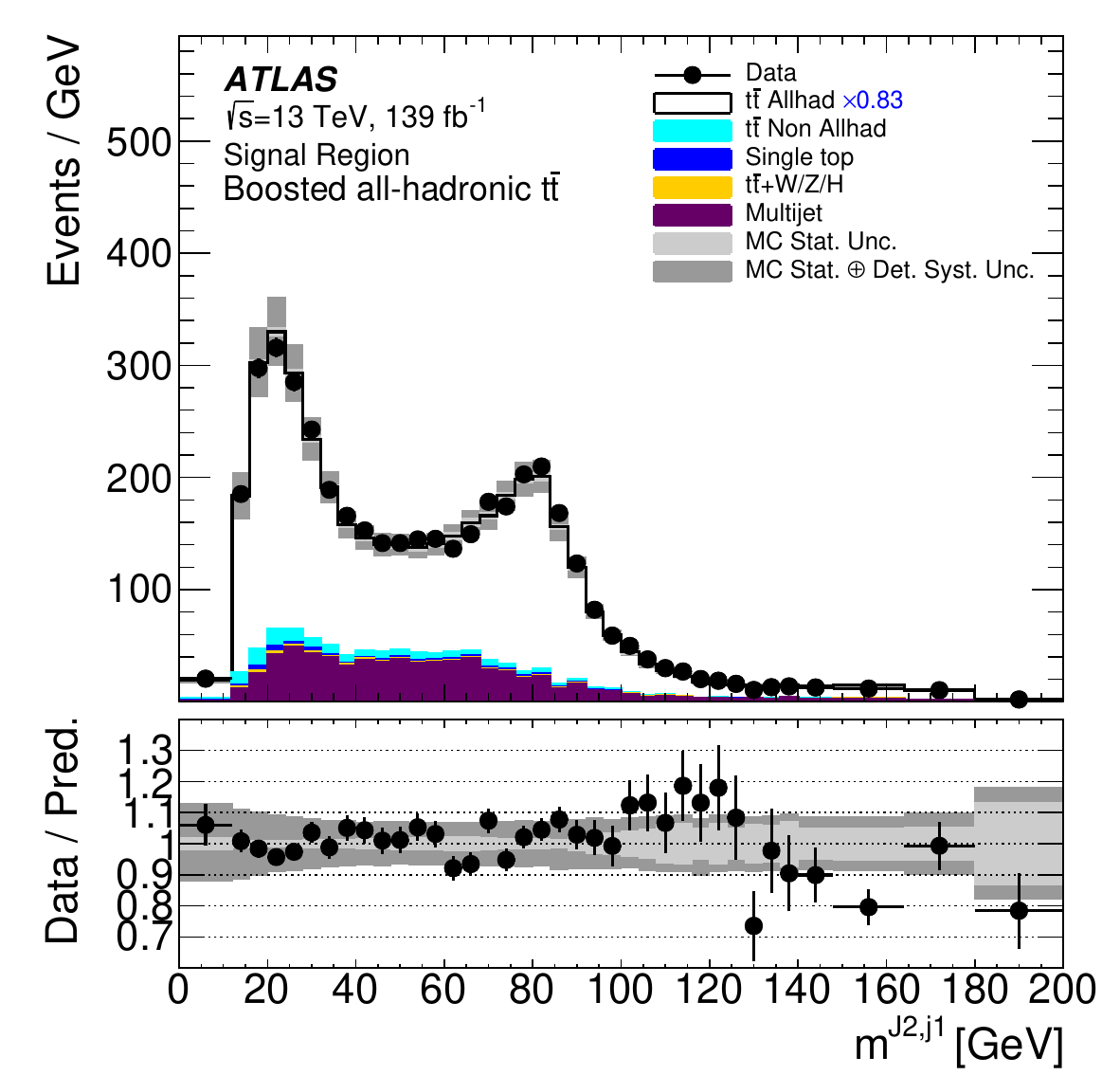}
\label{fig:SR:t2_sjM}}
\subfigure[]{\includegraphics[width=0.45\textwidth]{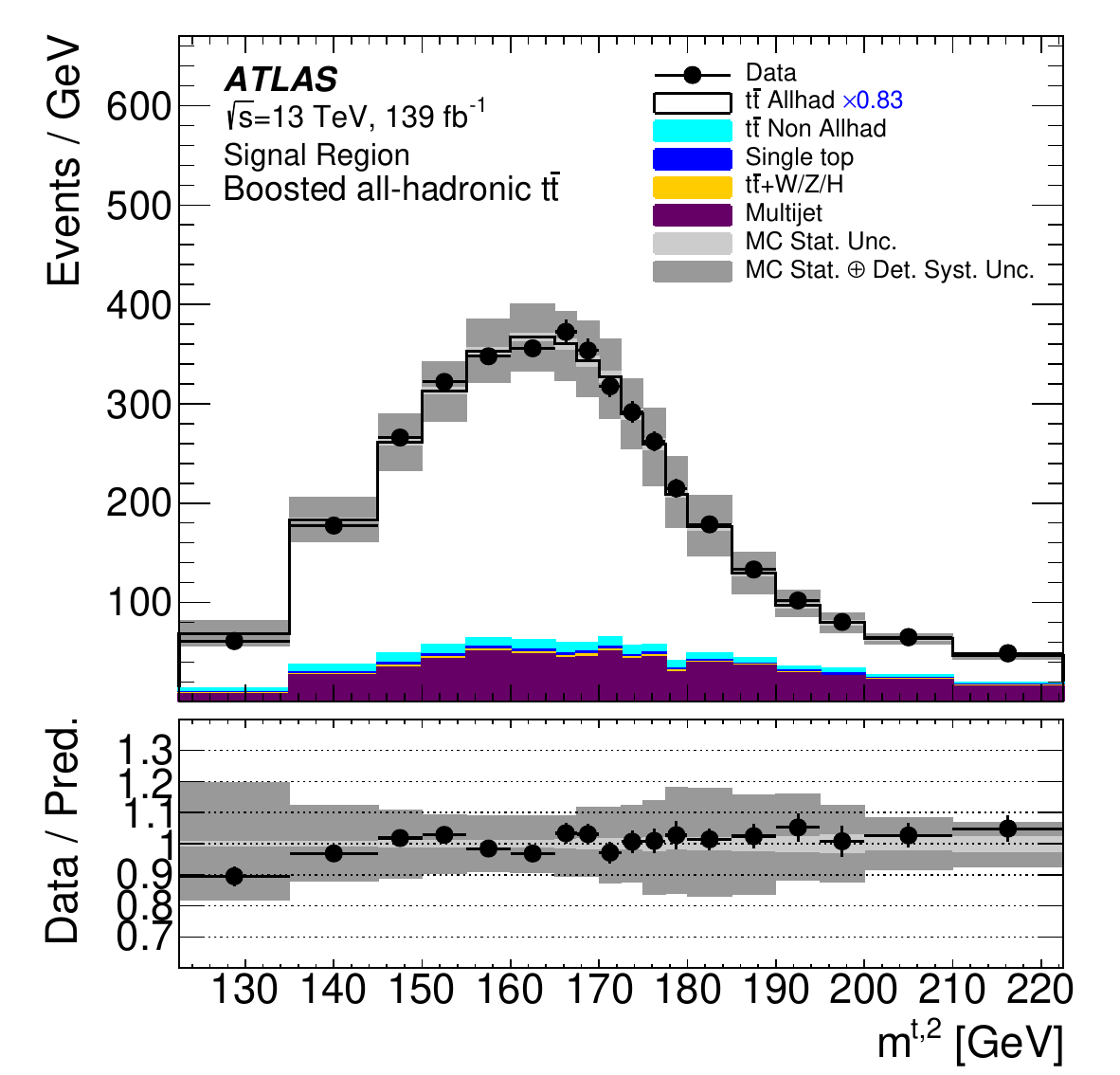}
\label{fig:SR:t2_M}}
\subfigure[]{\includegraphics[width=0.45\textwidth]{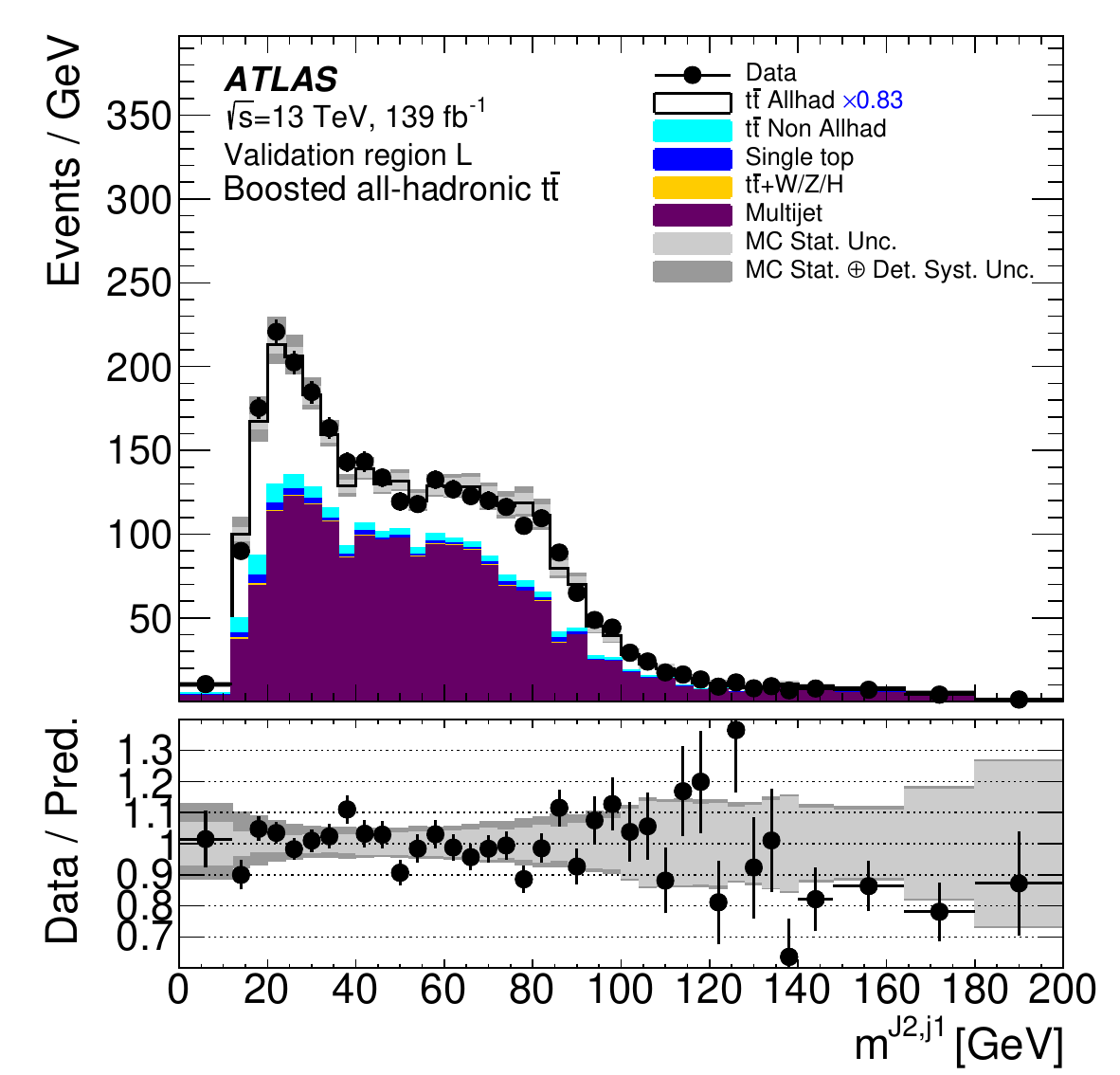}
\label{fig:VR_L:t2_sjM}}
\subfigure[]{ \includegraphics[width=0.45\textwidth]{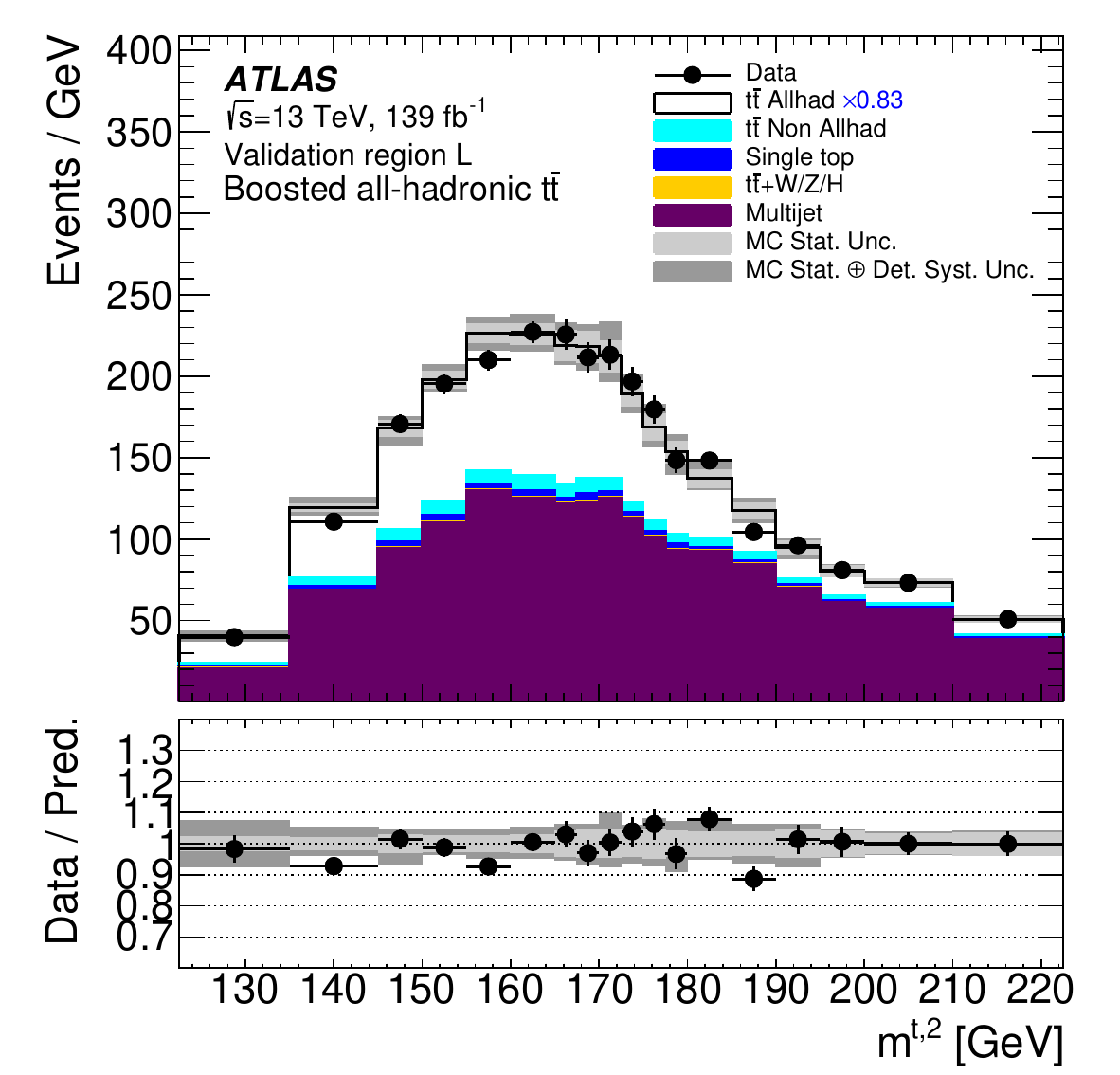}
\label{fig:VR_L:t2_M}}
\caption{Kinematic distributions of the second-leading top-quark-candidate jets in the signal region and
validation region L.
The mass distributions of the leading $R=0.4$\ \antikt\ jet in the second-leading \largeR\ jet for events
in the signal region and region L are
shown in \subref{fig:SR:t2_sjM}\ and \subref{fig:VR_L:t2_sjM}, respectively.
The second-leading \largeR\ jet-mass distributions for the events in the signal region
and validation region L
are shown in \subref{fig:SR:t2_M}\ and \subref{fig:VR_L:t2_M}, respectively.
The signal prediction (open histogram) is based on the \POWHEG{}+\Pythia[8] \ttbar\ calculation
normalized to the observed yield in the signal region.
The background (solid histogram) is the sum of the data-driven multijet estimate and the MC-based
expectation for the contributions of non-all-hadronic \ttbar, single-top-quark, and $\ttbar$\,+\,$W/Z/H$ processes.
The light grey bands indicate the statistical uncertainties and the dark grey bands indicate the combined statistical
and detector-related systematic uncertainties defined in Section~\ref{sec:systematics}.
}
\label{fig:VR_LSR:t2_M}
\end{figure*}

\begin{figure*}[htbp]
\centering
\subfigure[]{ \includegraphics[width=0.45\textwidth]{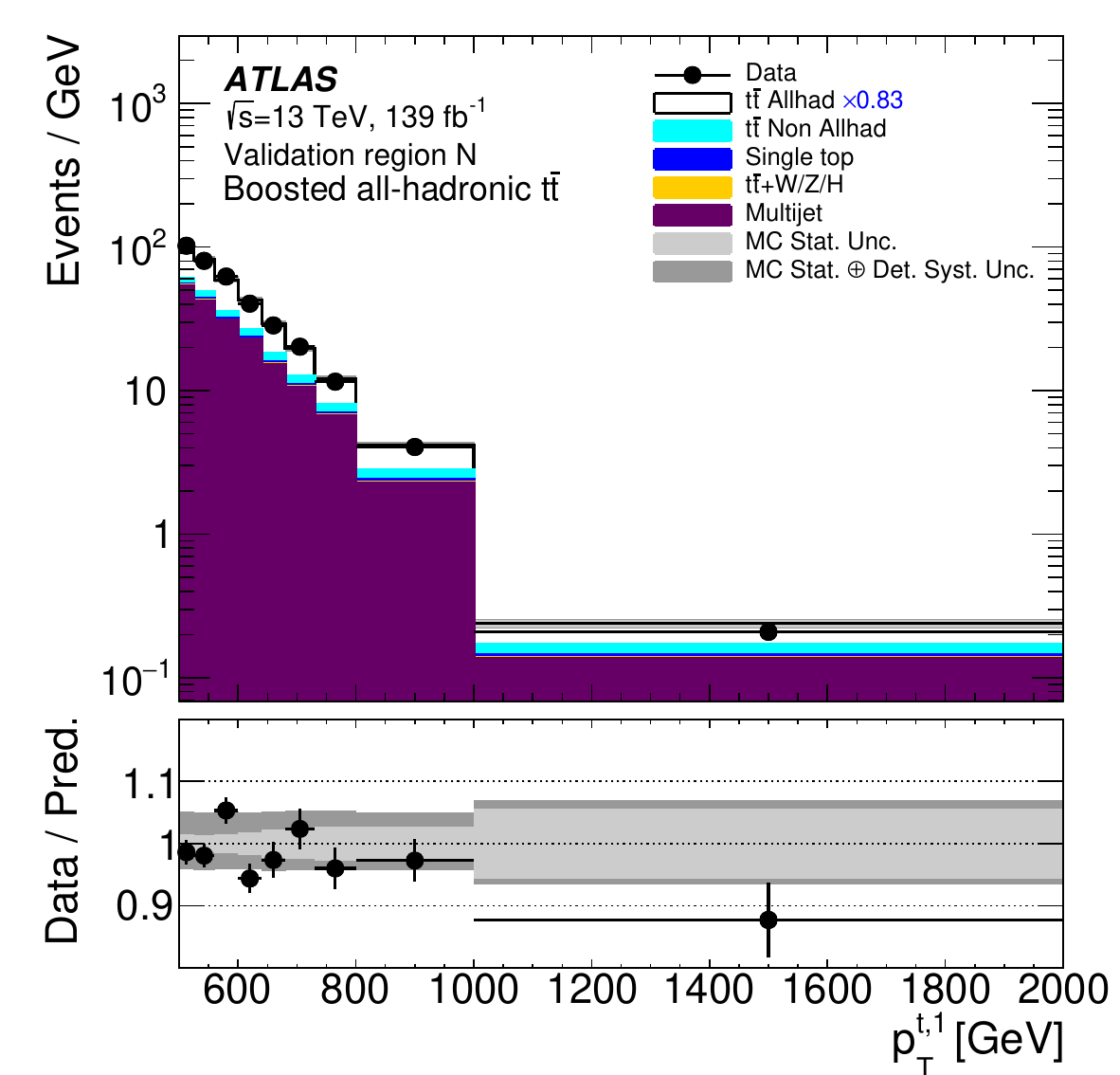}
\label{fig:CP:VR_N:t1_pt}}
\subfigure[]{ \includegraphics[width=0.45\textwidth]{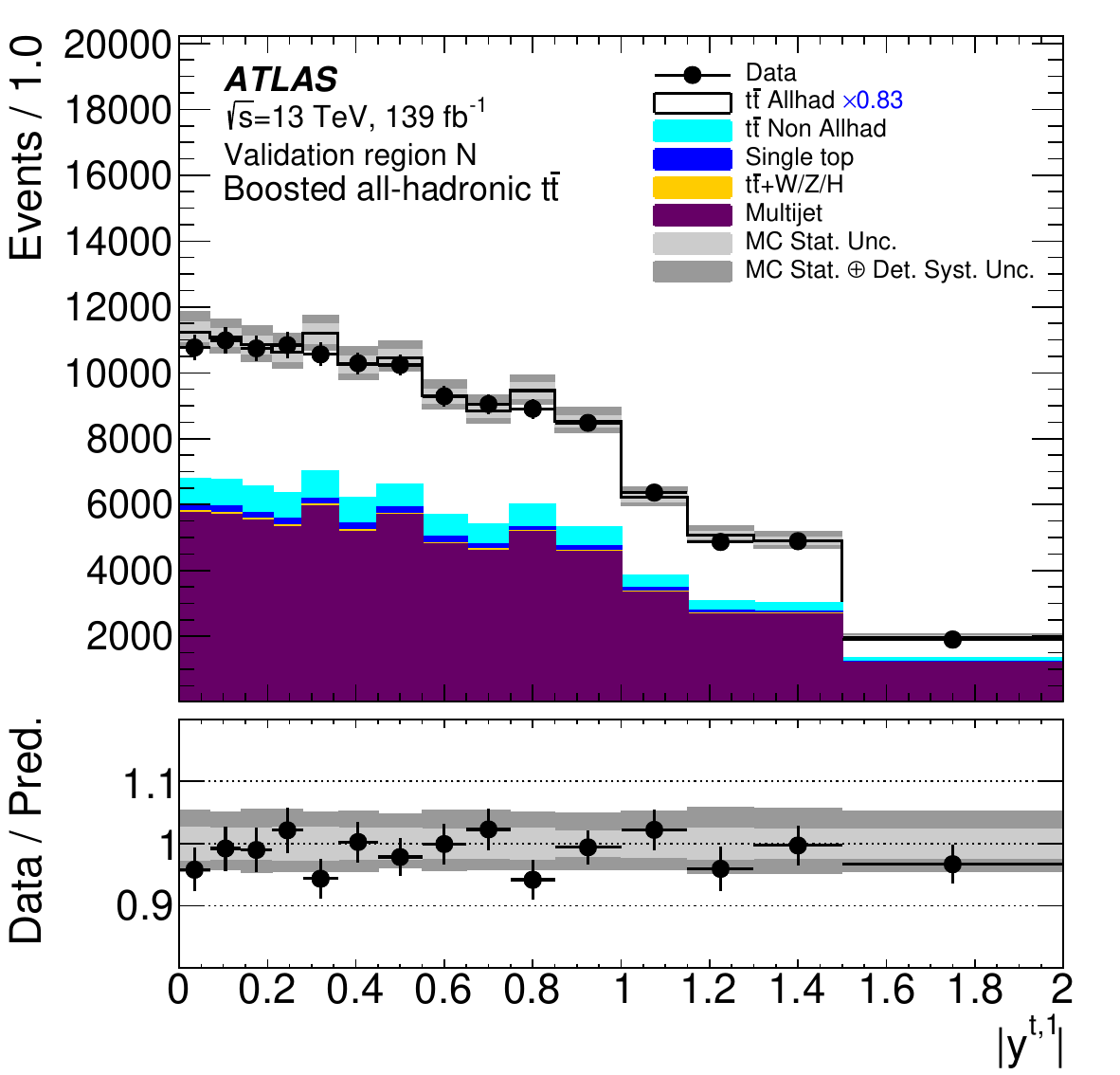}
\label{fig:CP:VR_N:t1_y}}
\subfigure[]{ \includegraphics[width=0.45\textwidth]{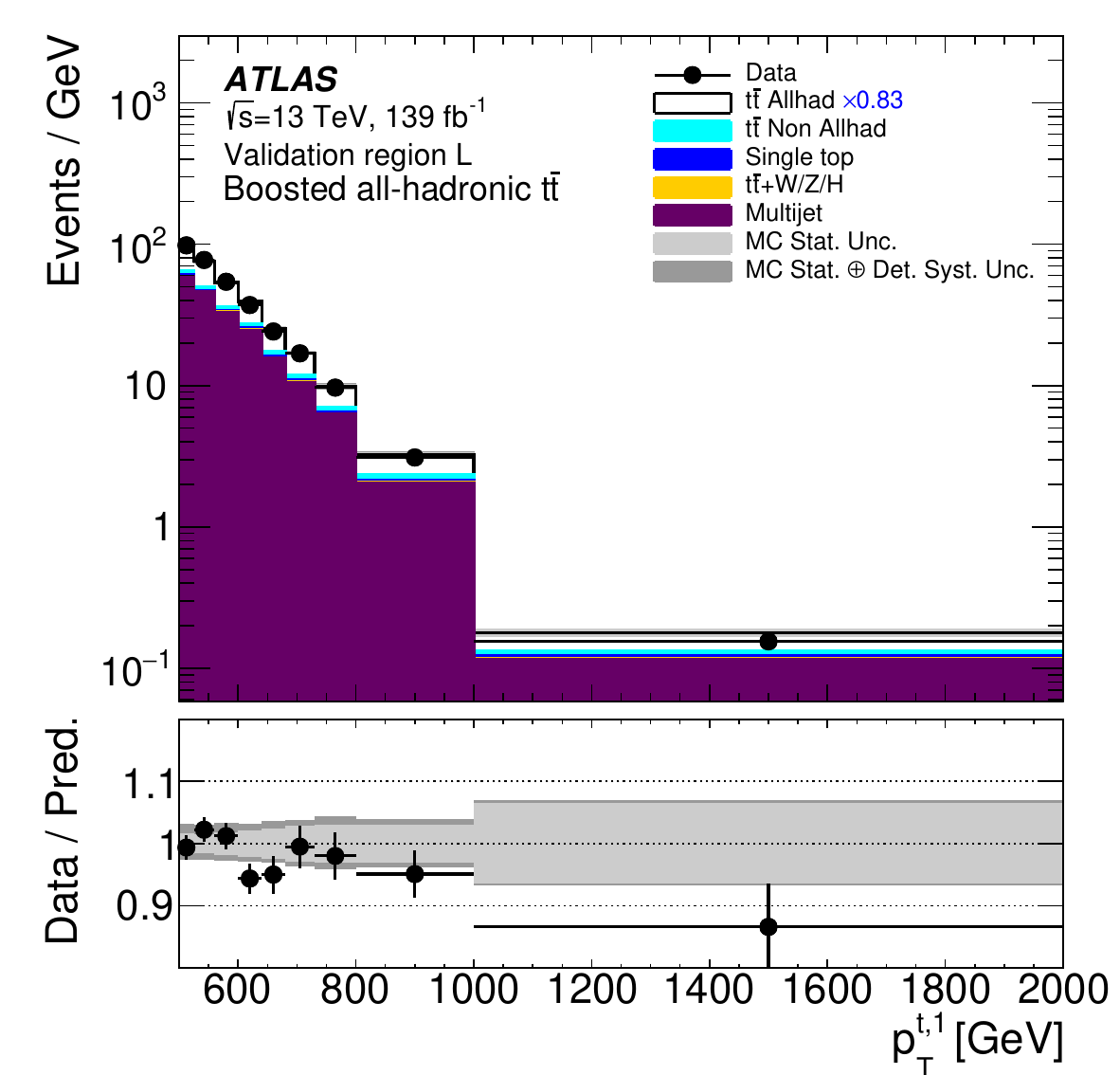}
\label{fig:CP:VR_L:t2_pt}}
\subfigure[]{ \includegraphics[width=0.45\textwidth]{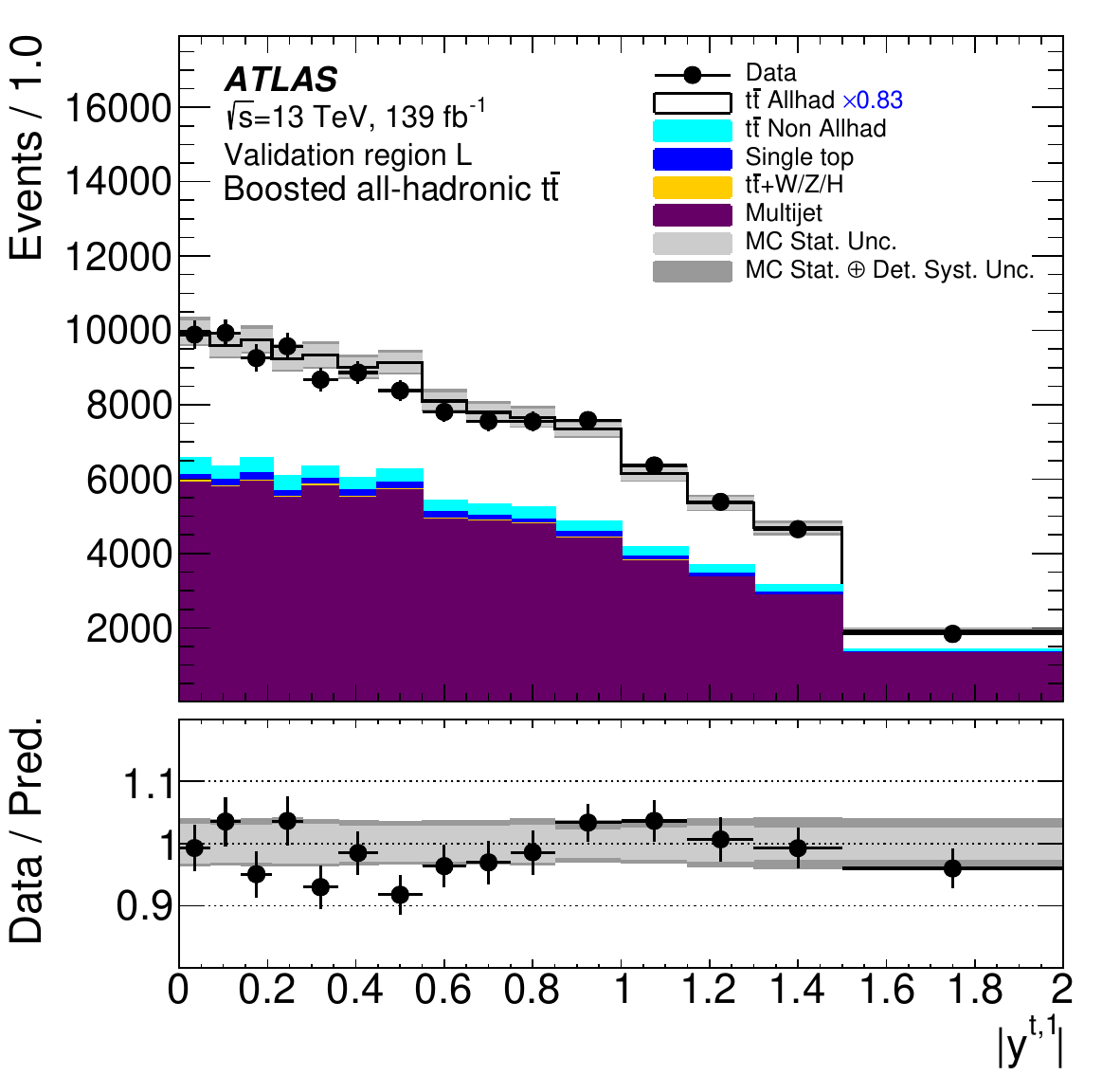}
\label{fig:CP:VR_L:t2_y}}
\caption{Kinematic distributions of the leading top-quark-candidate jets in validation
regions N and L:
\subref{fig:CP:VR_N:t1_pt} and \subref{fig:CP:VR_N:t1_y} are distributions of the
leading \largeR\ jet \pT\ and the absolute value of the
rapidity in region N, respectively, and
\subref{fig:CP:VR_L:t2_pt} and \subref{fig:CP:VR_L:t2_y} are the distributions of the leading \largeR\ jet \pT\ and
absolute value of the rapidity in region L.
The signal prediction (open histogram) is based on the \POWHEG{}+\Pythia[8] \ttbar\ calculation
normalized to the observed yield in a signal region.
The background (solid histogram) is the sum of the data-driven multijet estimate and the MC-based
expectation for the contributions of non-all-hadronic \ttbar, single-top-quark, and $\ttbar$\,+\,$W/Z/H$ processes.
The light grey bands indicate the statistical uncertainties and the dark grey bands indicate the combined statistical
and detector-related systematic uncertainties defined in Section~\ref{sec:systematics}.
}
\label{fig:VRSR:t1_t2_pt}
\end{figure*}

Good agreement between the distribution shapes is illustrated in Figures~\ref{fig:VR_LSR:t1_M}\ and \ref{fig:VR_LSR:t2_M},
which compare
the mass distributions of the highest-\pT\ \smallR\ jet associated with the leading and second-leading \largeR\ jet,
and the leading and second-leading \largeR\ jet-mass distributions,
for events
in the signal region and in region L, where the leading jet is not top-quark-tagged but is $b$-tagged,
while the second-leading jet is both top-quark-tagged and $b$-tagged.
The \smallR\ and  \largeR\ jets are considered to be associated if the angular separation
between the \smallR\ and \largeR\ jet axes satisfies $\Delta R < 1.0$.
The different size of the systematic uncertainties in Figure~\ref{fig:SR:t1_M} and
Figure~\ref{fig:VR_L:t1_M} arise from the relatively large jet-mass systematic uncertainties
of the \largeR\ jets in the \ttbar\ signal and the different proportion of MC-based
signal prediction and the data-driven multijet background.
The distributions of the leading-jet \pt\ and rapidity in regions N and L are shown
in Figure~\ref{fig:VRSR:t1_t2_pt}.
The predictions are in good agreement with the data distributions.
Similar studies of validation regions with varying
correlations between tagging states demonstrate that the multijet-background
estimates are robust.
These distributions can also be compared with the signal-region distributions in
Figure~\ref{fig:CP:SR:top_quarks}, which illustrates the difference in the
kinematics of the leading and second-leading \largeR\ jets.
 
\section{Detector-level results}
\label{sec:reco_level}

 
The event yields in the signal region are summarized in
Table~\ref{tab:yields} for the simulated signal, the background
sources, and the data sample.
The prediction overestimates the data by about 16\%.
The systematic uncertainties apart from signal-modelling uncertainties, as described in detail in Section~\ref{sec:systematics}, are included in the prediction.
 
\begin{figure*}[htbp]
\centering
\subfigure[]{ \includegraphics[width=0.45\textwidth]{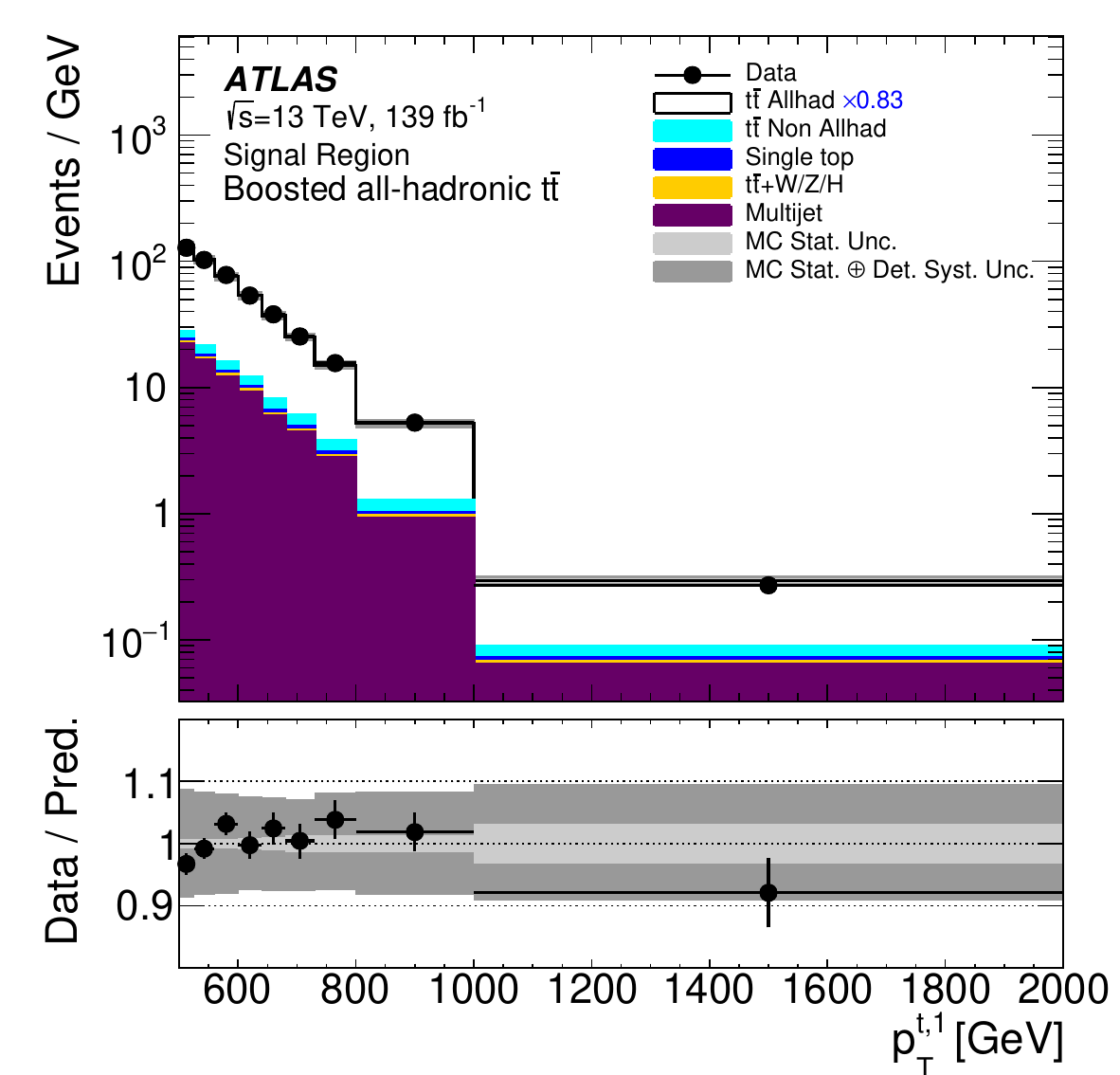}\label{fig:CP:SR:t1_pt}}
\subfigure[]{ \includegraphics[width=0.45\textwidth]{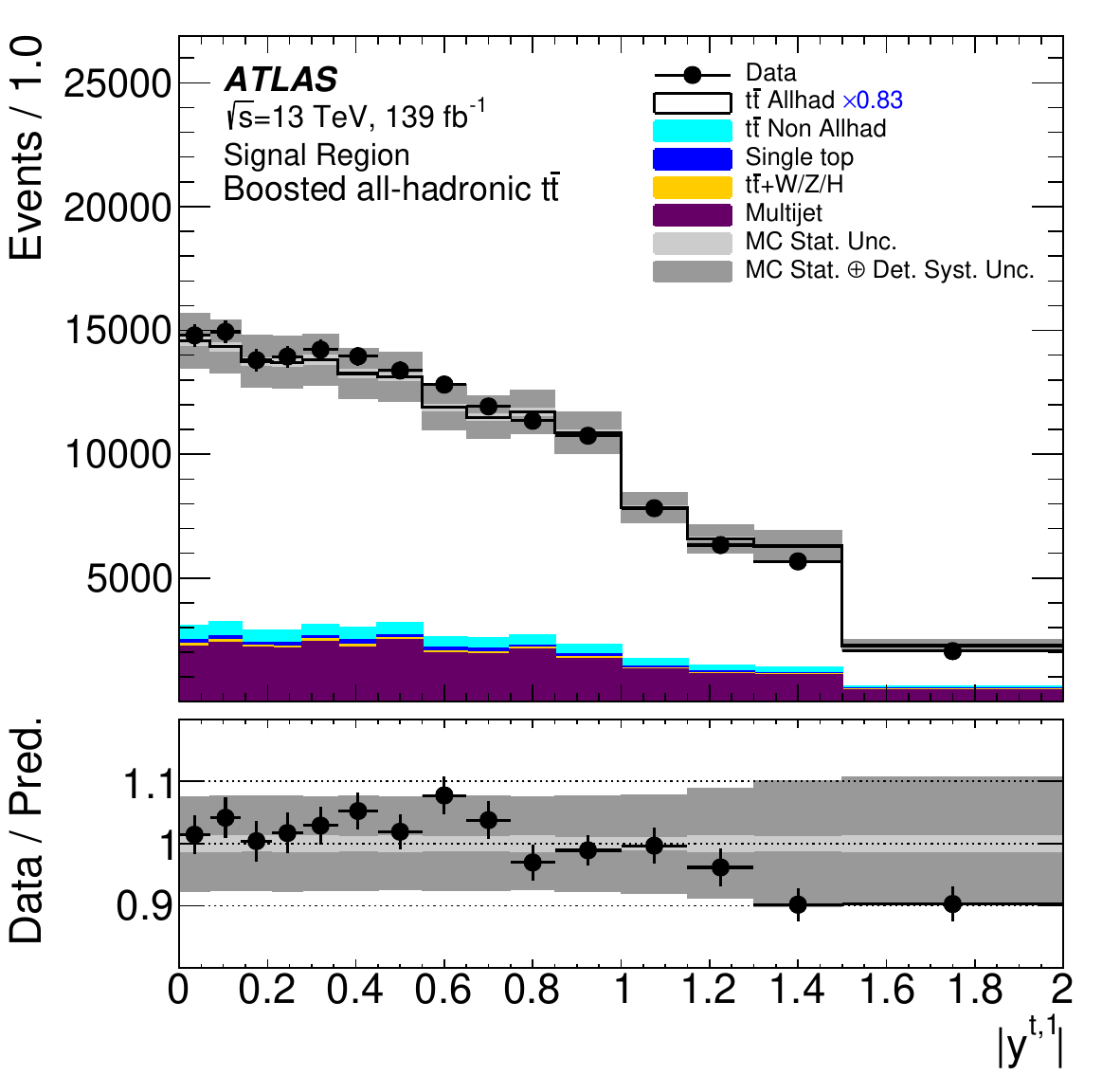} \label{fig:CP:SR:t1_y}}
\subfigure[]{ \includegraphics[width=0.45\textwidth]{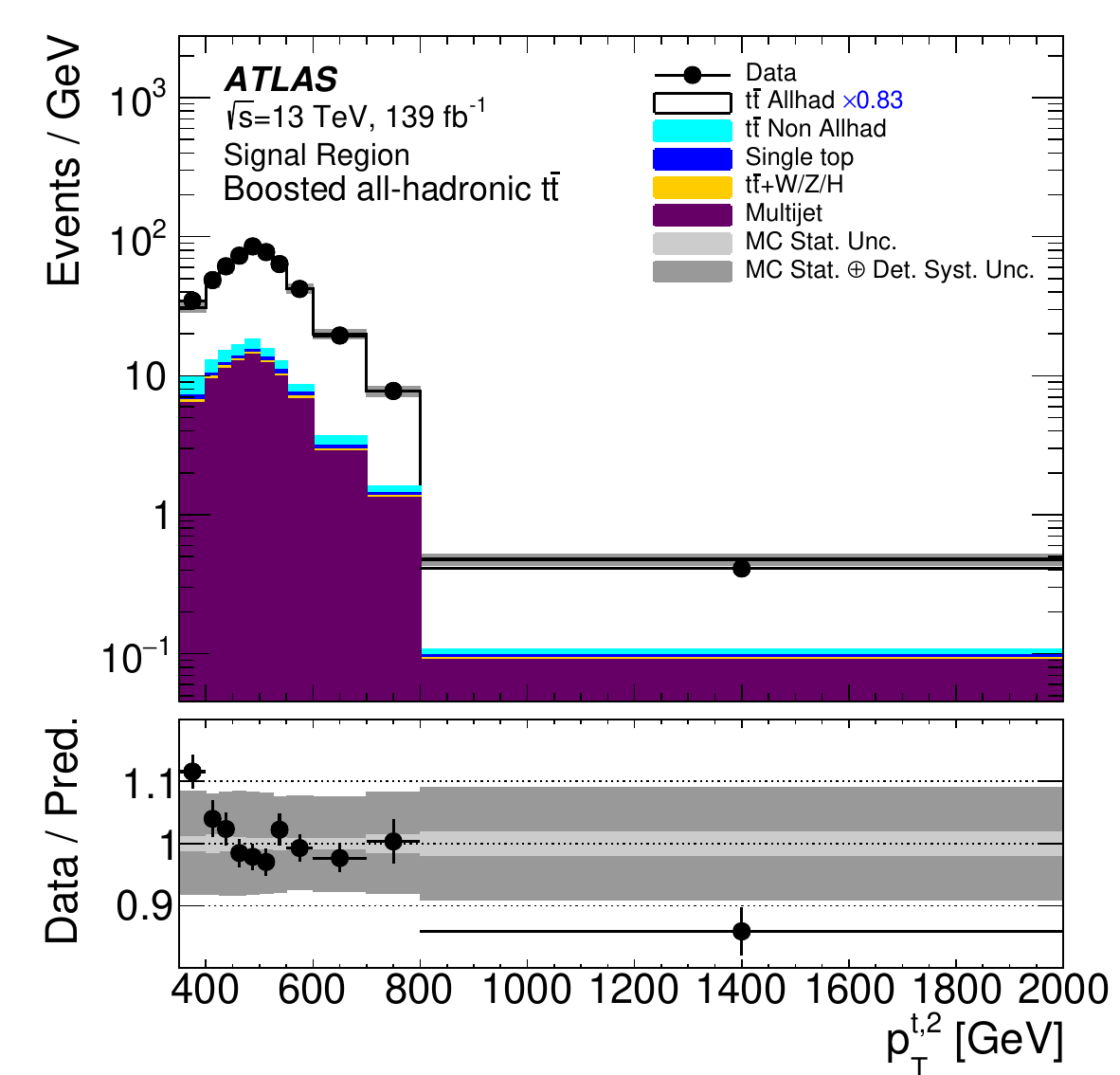}\label{fig:CP:SR:t2_pt}}
\subfigure[]{ \includegraphics[width=0.45\textwidth]{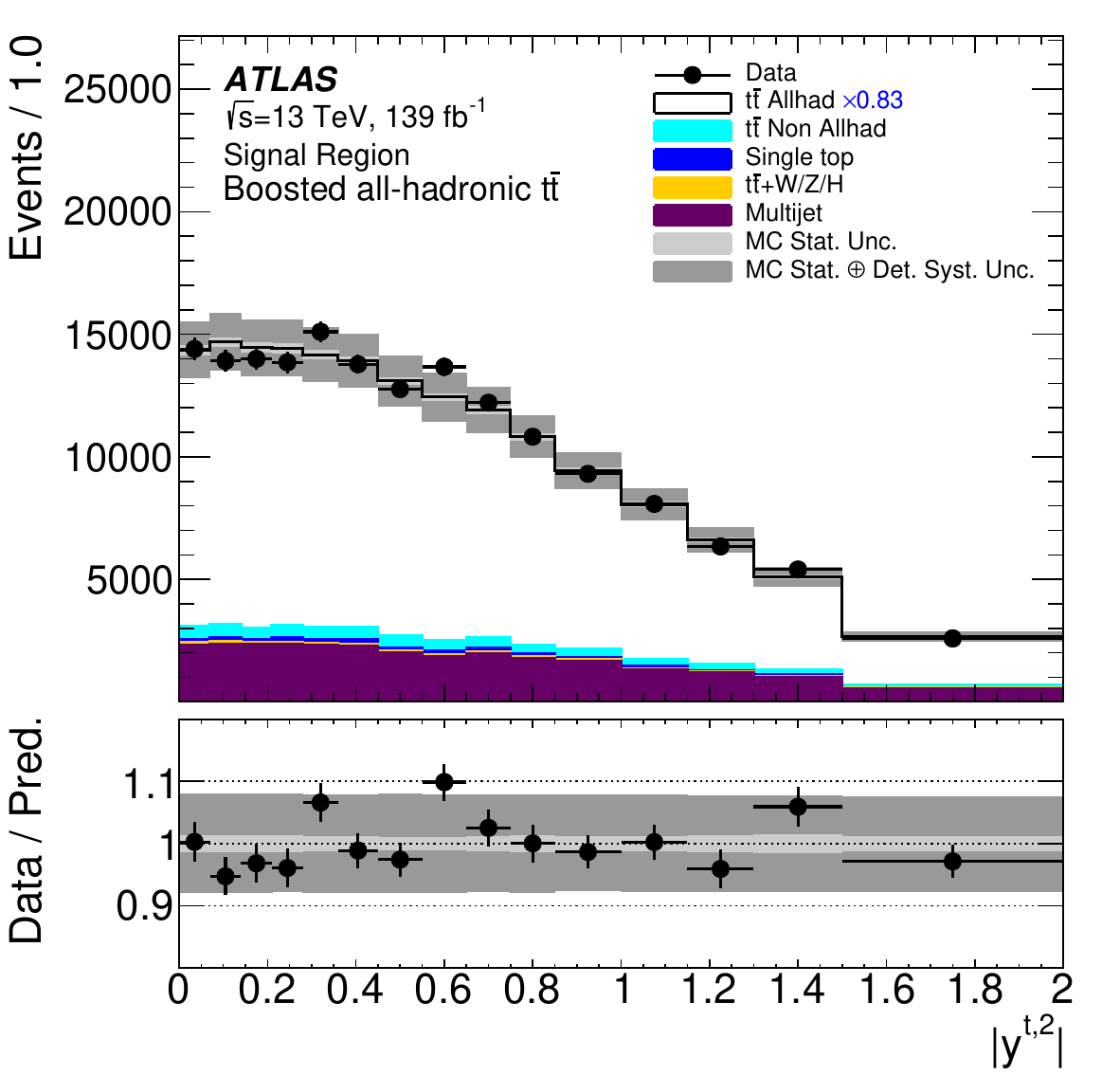} \label{fig:CP:SR:t2_y}}
\caption{Kinematic distributions of top-quark-candidate jets in the signal region:
\subref{fig:CP:SR:t1_pt}~\pT\ and \subref{fig:CP:SR:t1_y} $|y|$\ of the leading jet, and
\subref{fig:CP:SR:t2_pt}~\pT\ and \subref{fig:CP:SR:t2_y} $|y|$\ of the second-leading jet.
The signal prediction (open histogram) is based on the \POWHEG{}+\Pythia[8] \ttbar\ calculation
normalized to the observed yield.
The background (solid histogram) is the sum of the data-driven multijet estimate and the MC-based
expectation for the contributions of non-all-hadronic \ttbar, single-top-quark, and $\ttbar$\,+\,$W/Z/H$ processes.
The light grey bands indicate the statistical uncertainties and the dark grey bands indicate the combined statistical
and detector-related systematic uncertainties defined in Section~\ref{sec:systematics}.
}
\label{fig:CP:SR:top_quarks}
\end{figure*}

\begin{figure*}[htbp]
\centering
\subfigure[]{ \includegraphics[width=0.75\textwidth]{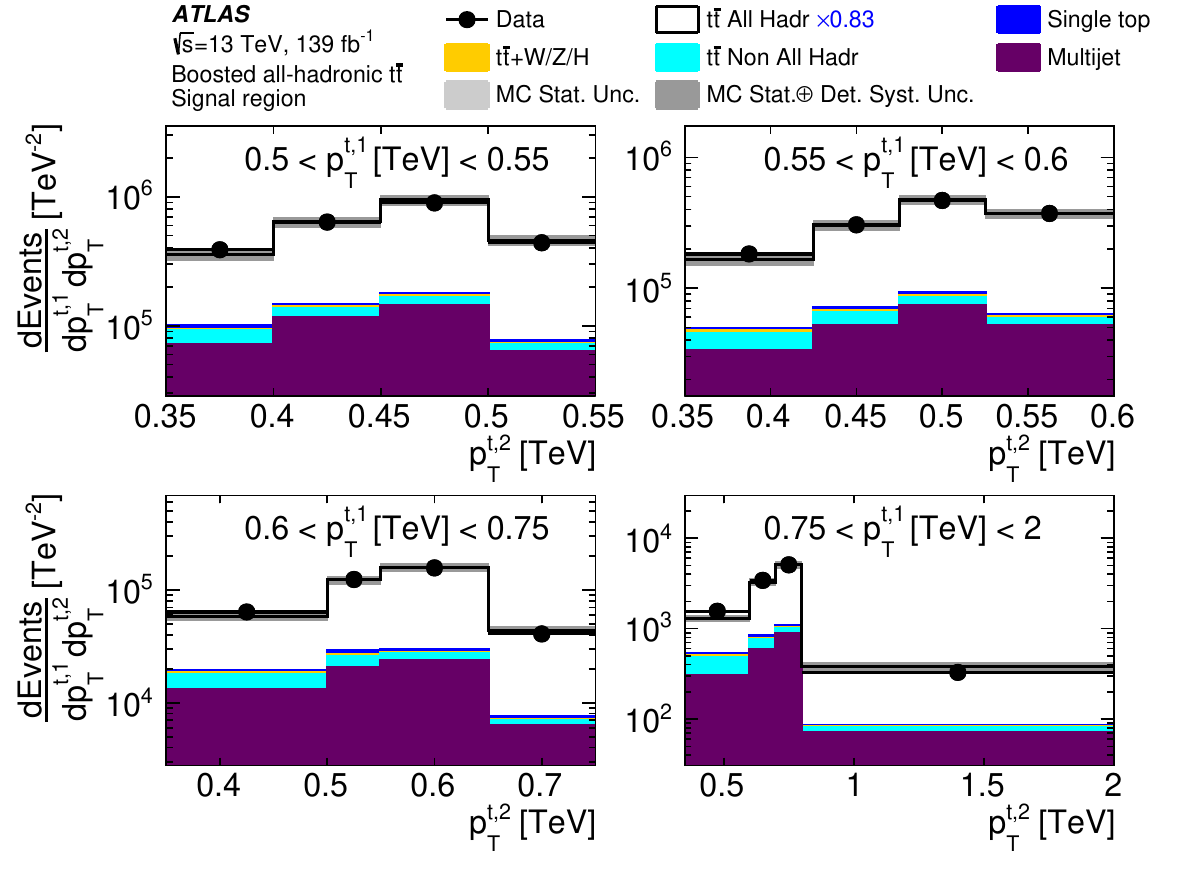}  \label{fig:CP:SR:t1_pt_vs_t2_pt}}
\subfigure[]{ \includegraphics[width=0.75\textwidth]{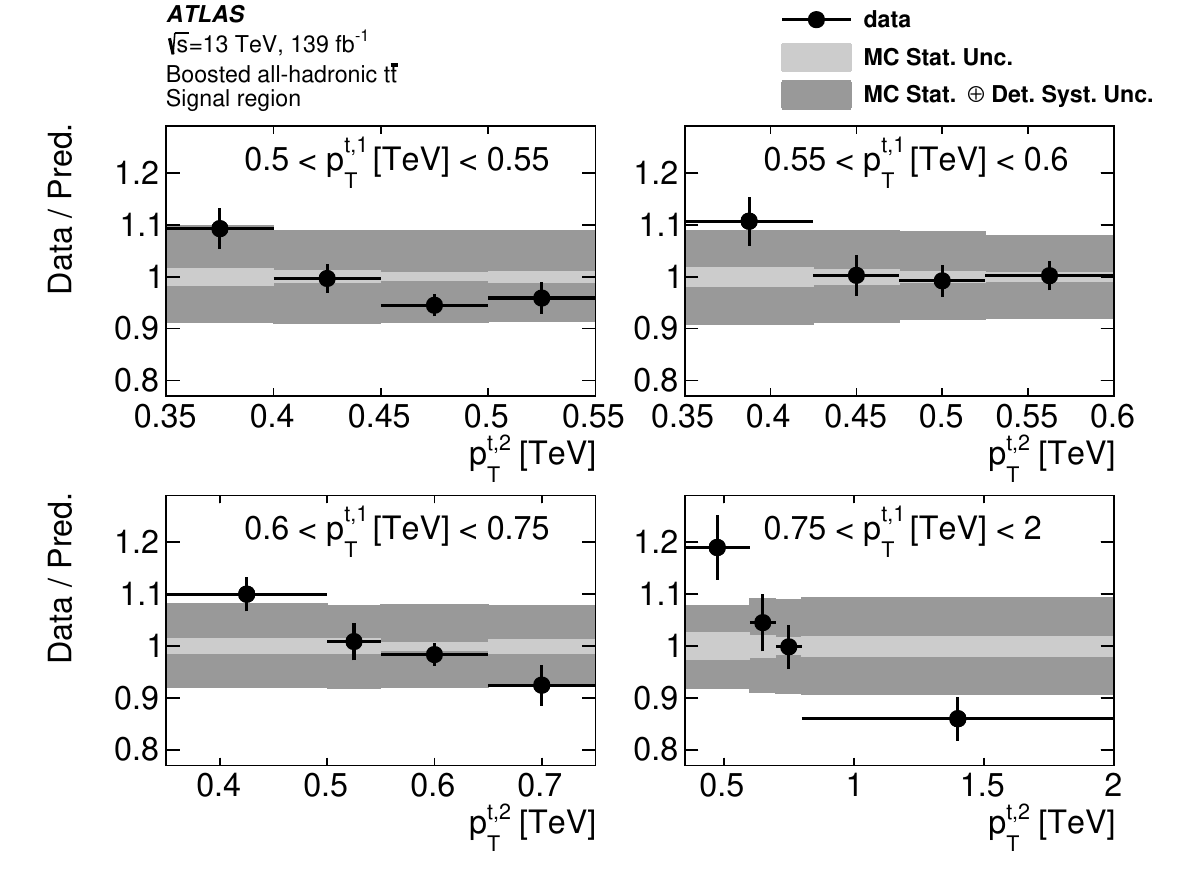}\label{fig:CP:SR:t1_pt_vs_t2_pt_ratio}}
\caption{Two-dimensional kinematic distributions of second-leading
jet \pt versus leading-jet \pt in the signal region:
\subref{fig:CP:SR:t1_pt_vs_t2_pt}~the differential event yield as a function of the
\pT\ of the leading jet and \pT\ of second-leading jet, and
\subref{fig:CP:SR:t1_pt_vs_t2_pt_ratio}~the ratio of the observed and predicted distributions.
The signal prediction (open histogram) is based on the \POWHEG{}+\Pythia[8] \ttbar\ calculation
normalized to the observed yield.
The background (solid histogram) is the sum of the data-driven multijet estimate and the MC-based
expectation for the contributions of non-all-hadronic \ttbar, single-top-quark, and $\ttbar$\,+\,$W/Z/H$ processes.
The light grey bands indicate the statistical uncertainties and the dark grey bands indicate the combined statistical
and detector-related systematic uncertainties defined in Section~\ref{sec:systematics}.
}
\label{fig:CP:SR:2D_top_quarks}
\end{figure*}

The comparisons between predicted and observed distributions in the signal region are shown in
Figures~\ref{fig:CP:SR:top_quarks} and Figure~\ref{fig:CP:SR:2D_top_quarks}. Here, the \ttbar\ MC prediction is scaled by requiring that the predicted and observed event yields match.
The event yield decreases rapidly with increasing \pt\ of the leading and
second-leading jets, which extends
beyond 1~\TeV, while the rapidity distributions fall slowly across the
interval $|y| < 2.0$, as shown in Figure~\ref{fig:CP:SR:top_quarks}.
Good agreement between the observed and predicted distributions also can be seen in
Figure~\ref{fig:CP:SR:top_quarks}.
In the signal region, the maximum observed \pt\ of the leading jet is 1.73~\TeV\ and the maximum
observed \ttbar\ invariant mass is 4.1~\TeV.
 
The distributions of second-leading jet \pT\ as a function of the leading
top-quark-jet \pT\ are shown in Figure~\ref{fig:CP:SR:2D_top_quarks}.
The distributions of the top-quark-jet \pT\ fall more rapidly than the predictions.
 
\begin{table}[t!]
\begin{center}

\sisetup{
group-minimum-digits = 5,
table-format = 4.0,
separate-uncertainty = true,
table-align-uncertainty = true,
}
 
\begin{tabular}{l S}
\toprule
\vspace{.10cm}{Source} & {Event Yields} \\
\hline
\vspace{.10cm}
\ttbar{} (all-hadronic) 		& \num{16200(1400)} \\ 
\vspace{.10cm}
\ttbar{} (non-all-hadronic) 	& \num{625(63)}   \\
\vspace{.10cm}
Single top-quarks 			& \num{179(21)}   \\
\vspace{.10cm}
\ttbar{}\,+\,$W/Z/H$			& \num{114(11)}    \\
\vspace{.10cm}
Multijet events			& \num{2900(160)}  \\ 
\hline
\vspace{.10cm}
All Backgrounds${}^{\phantom{1}^{\phantom{1}}}$  & \num{3820(200)}  \\
\vspace{.10cm}
Prediction${}^{\phantom{1}^{\phantom{1}}}$ & \num{20000(1600)} \\ 
\vspace{.10cm}
Data (\lumitot) 	 		& \num{17261}     \\
\bottomrule
\end{tabular}


\end{center}
\caption{Event yields in the signal region for the expected \ttbar\
signal process and the background processes.
The sum of these is compared with the observed yield.
The uncertainties represent the sum in quadrature of the statistical and systematic
uncertainties for each process, as described in Section~\ref{sec:systematics}.
Neither \ttbar\ modelling uncertainties nor uncertainties in the inclusive \ttbar\ cross-section are included in the systematic uncertainties.
The multijet-background uncertainty includes the statistical
uncertainties in the signal and control regions
as well as the systematic
uncertainties arising from the MC-based subtraction in the
control regions used to make the data-driven estimate.
The column entries do not add up exactly to `All Backgrounds' and `Prediction' due to rounding.
}
\label{tab:yields}
\end{table}
 
\FloatBarrier


\section{Correction procedures}
\label{sec:unfolding}

The observed differential cross-sections reflect the underlying physics processes as well as the
acceptance, efficiency, and resolution of the detector and reconstruction algorithms.
These distributions are unfolded to particle level in a fiducial phase space in order to correct for detector effects.
The correction is made to the particle-level differential cross-sections, i.e.\ the distributions
defined by the stable particles in the MC simulation.
 
Parton-level differential cross-sections are measured in a similar manner.
The parton level is defined in the MC simulation by the top quark after final-state radiation, i.e.\
immediately before its decay.
 
The following subsections describe the particle-level fiducial
phase space, the parton-level fiducial phase space, and the algorithm used for the unfolding.
 
\subsection{Particle-level fiducial phase-space and parton-level fiducial phase-space regions}
 
The particle-level fiducial phase-space definition is intended to match the kinematic requirements used to
select the \ttbar\ process as described in Section~\ref{sec:events_selection}.
Particle-level jets and leptons are defined so as to closely match the detector-level objects.
 
In the MC signal sample, electrons and muons that do not originate from hadron decays are
`dressed' with prompt photons found in a cone of size $\Delta R = 0.1$ around the lepton direction.
The four-momentum of each photon in the cone is added to the four-momentum
of the lepton to produce the dressed lepton.
The leptons within  $\Delta R = 0.4$ of a \smallR\ jet, as defined below, are removed.
 
Jets are clustered using all stable particles (lifetimes $> 30$~ps) except those used in
the definition of dressed electrons, dressed muons,
and neutrinos not from hadron decays, using the anti-$k_t$ algorithm with a radius parameter
$R=1.0$ for \largeR\ jets and $R=0.4$ for \smallR\ jets~\cite{ATL-PHYS-PUB-2015-013}.
The decay products of hadronically decaying $\tau$-leptons are included.
These jets include the particles from the underlying event in the $pp$ collision
but do not include particles from additional interactions in the same $pp$ bunch crossing.
Large-$R$ jets are required to have $\pT>350$~\GeV\ and a mass within
$50$~\GeV\ of the top-quark mass.
Small-$R$ jets are required to have $\pT>25$~\GeV\ and $|\eta|<2.5$.
 
The requirements on particle-level objects in the all-hadronic \ttbar\ MC events define the
particle-level fiducial phase space:
(1) there can be no dressed electrons or muons with $\pt>25$~\GeV\ and $|\eta|<2.5$\ in the event,
(2) there must be at least two \antikt\ $R=1.0$\ jets with $\pt>350$~\GeV, $|\eta|<2.0$,
and jet mass between 122.5 and 222.5~\GeV,
(3) there must be at least one \antikt\ $R=1.0$\ jet with $\pt>500$~\GeV\ and $|\eta|<2.0$,
and
(4) each of the two leading $R=1.0$\ jets must be matched to a $b$-hadron with $\pt > 5$~\GeV\
using a ghost-matching technique~\cite{Cacciari:2008gn}.
The use of $\eta$ instead of rapidity for defining the fiducial phase space was motivated by its
use in the event selection at detector level.
These requirements are used to derive the migration matrices, efficiency corrections,
and acceptance corrections needed for the unfolding procedure.
 
The parton-level fiducial phase space is defined by requiring that the
leading top quark has $\pT > 500$~\GeV\ and the second-leading
top quark has $\pT > 350$~\GeV.
No rapidity or other kinematic requirements are applied.
 
\subsection{Unfolding algorithm}
The iterative Bayesian method~\cite{DAgostini:1995} as
implemented in \textsc{RooUnfold}~\cite{Adye:2011gm} is used to correct the detector-level event
distributions to their corresponding particle- and parton-level differential cross-sections.
The unfolding starts from the observed differential distributions after subtraction of the
estimated backgrounds.
 
\begin{figure*}[htbp]
\centering
\subfigure[]{ \includegraphics[width=0.45\textwidth]{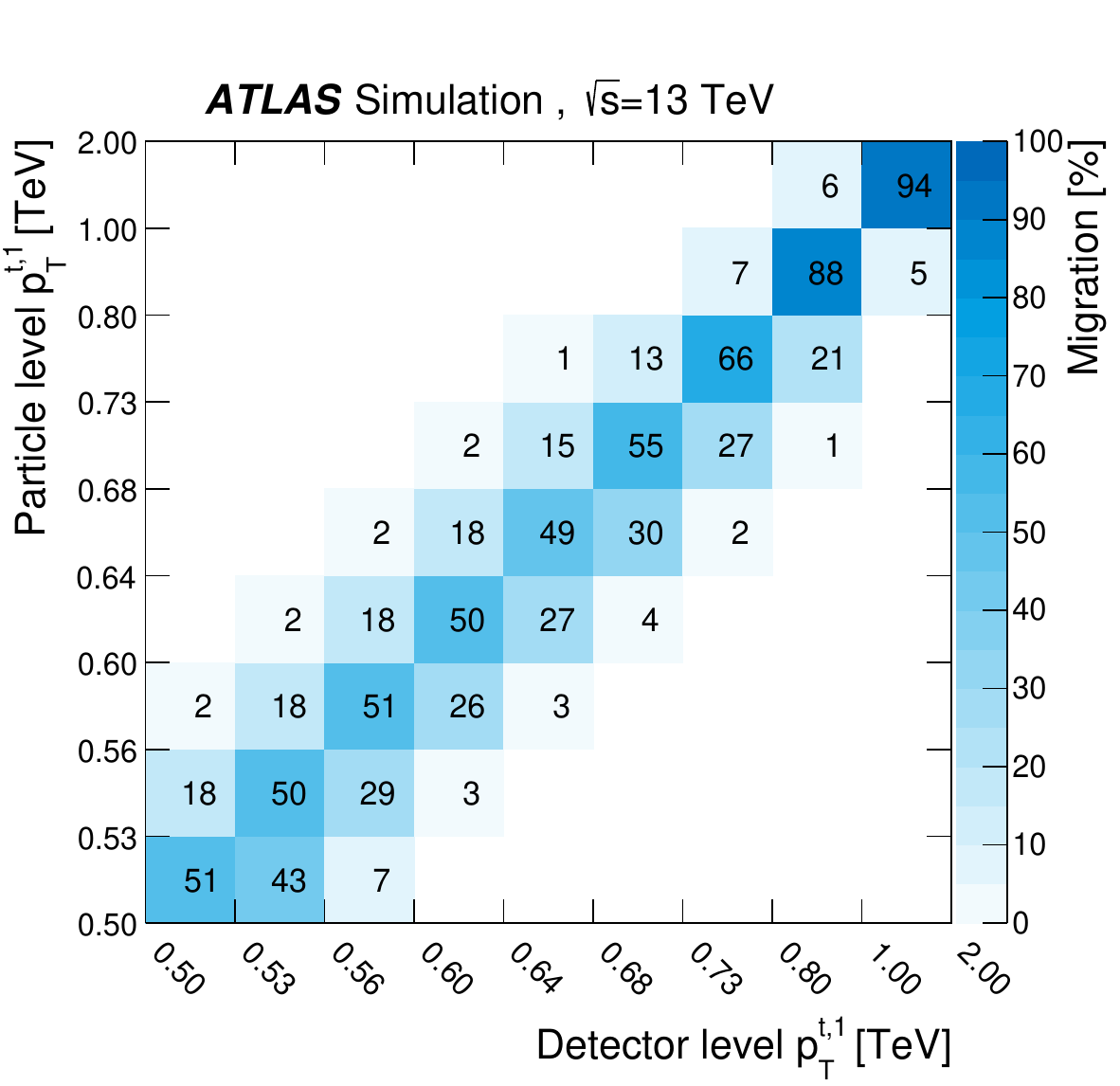}
\label{fig:migrations:particle:t1_pt}}
\subfigure[]{ \includegraphics[width=0.45\textwidth]{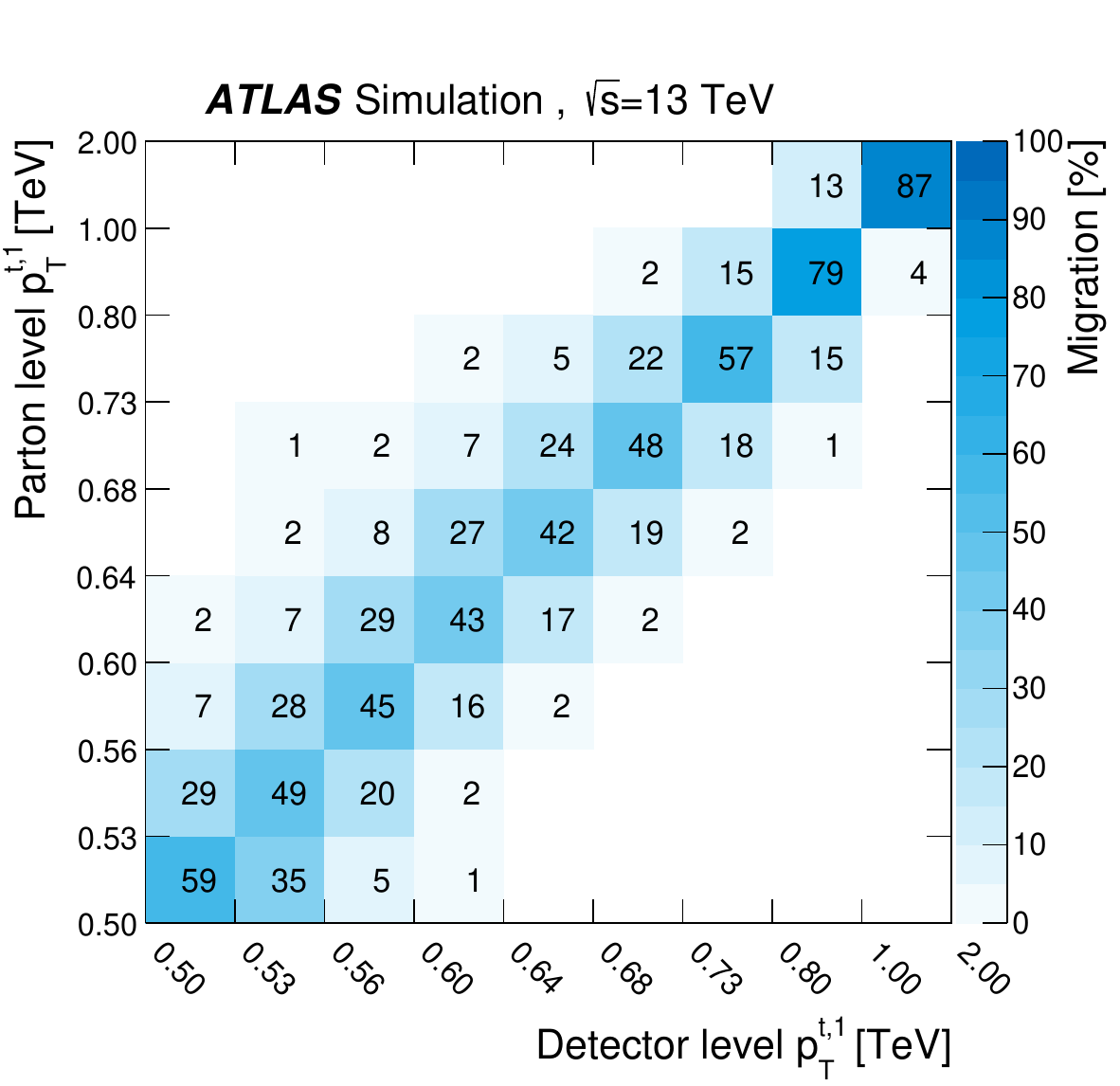}
\label{fig:migrations:parton:t1_pt}}
\subfigure[]{ \includegraphics[width=0.45\textwidth]{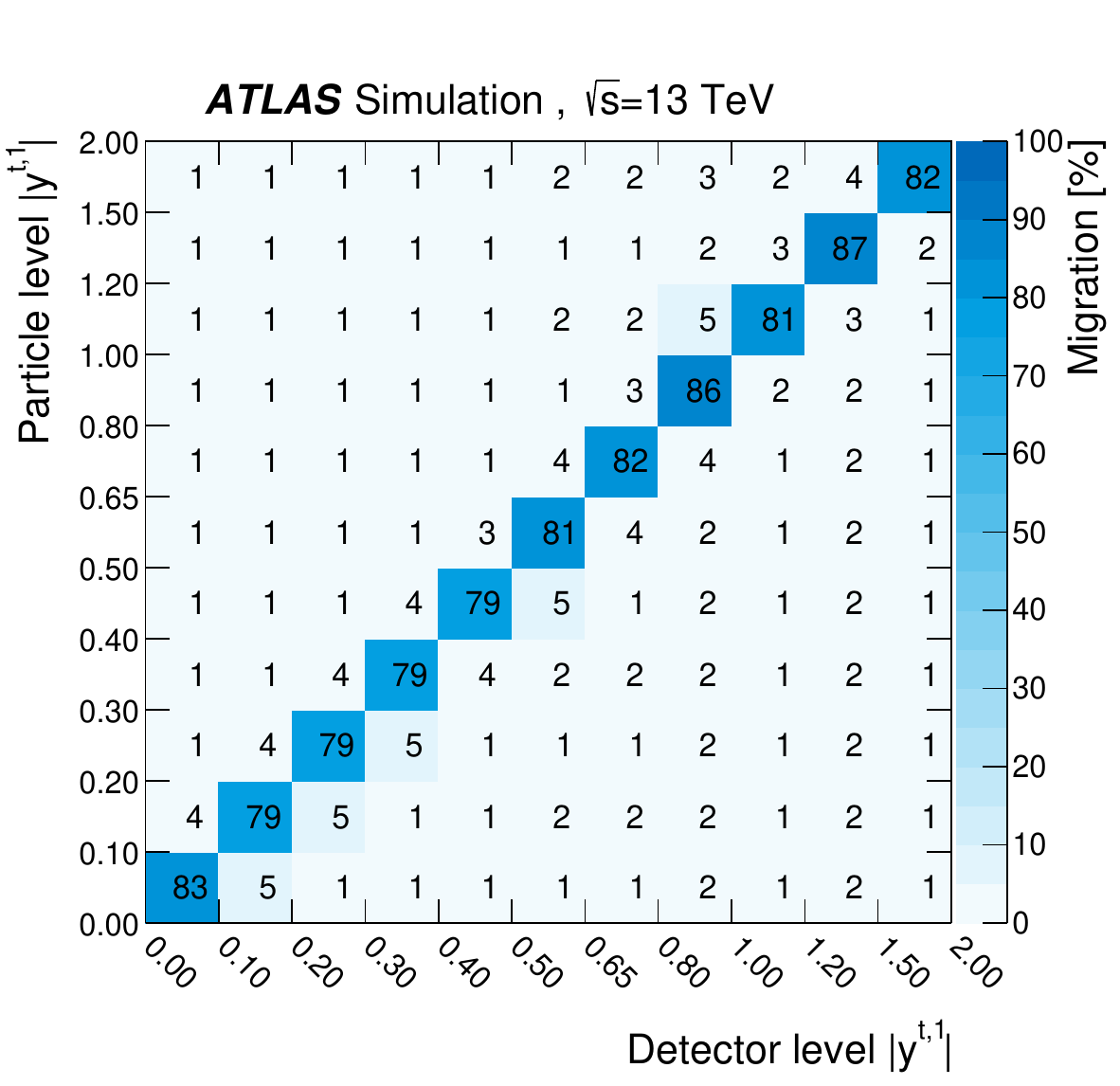}
\label{fig:migrations:particle:t1_y}}
\subfigure[]{ \includegraphics[width=0.45\textwidth]{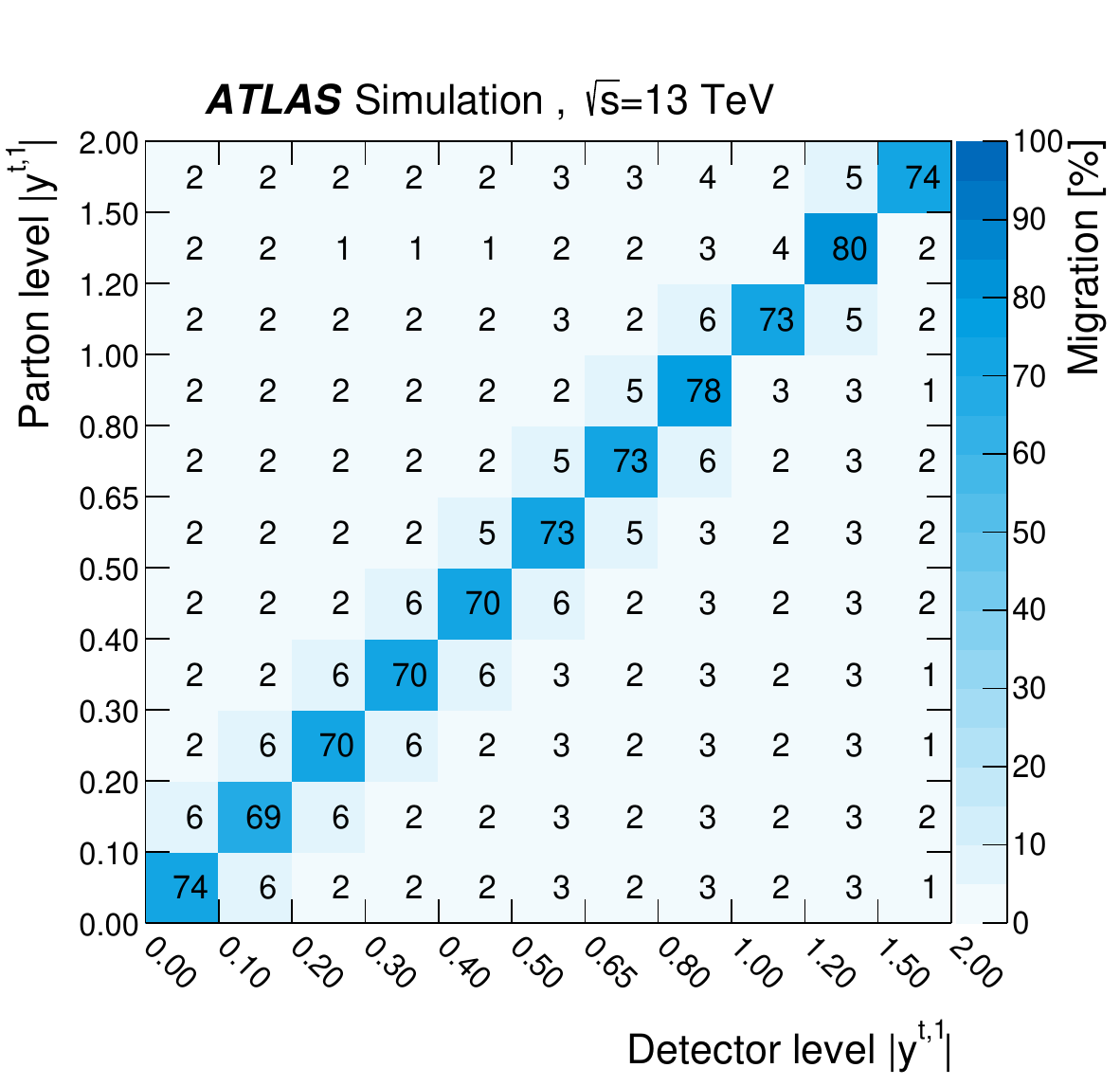}
\label{fig:migrations:parton:t1_y}}
\caption{Migration matrices for \pT\ and $|y|$\ of the leading top-quark jet
for the particle-level fiducial phase space in \subref{fig:migrations:particle:t1_pt}\
and \subref{fig:migrations:particle:t1_y}, respectively,
and for the parton-level fiducial phase space in
\subref{fig:migrations:parton:t1_pt}\ and \subref{fig:migrations:parton:t1_y}, respectively.
Each row is normalized to 100\%.
The \POWHEG{}+\Pythia[8] generator together with the \GEANT\ detector simulation framework is used to determine these matrices.}
\label{fig:migrations:particle:top_quarks}
\end{figure*}

The unfolding step for each observable uses a migration matrix ($\mathcal{M}$) derived from simulated \ttbar\
events by binning the  events in the particle-level
(parton-level) fiducial phase space using the true value for the observable and subdividing the events
in each particle-level (parton-level) bin into bins of the detector-level observable.
The resulting matrix, defined by the detector-level observable bins on the $x$-axis and the
particle-level (parton-level) bins on the $y$-axis, is normalized so that each row sums to unity, as shown in
Figure~\ref{fig:migrations:particle:top_quarks}.
 
The bin widths are chosen by considering the measurement resolution of a given
observable to achieve the migration matrix to be largely diagonal and that the
unfolding procedure is stable, as determined by the stress tests described below.
The migration matrices for the rapidity of the leading and second-leading
top-quark candidates are the exceptions, where there are a small number of
entries in very off-diagonal bins.
This is due to cases where the two \largeR\ jets swap order in \pT when
they evolve from the particle level or parton level to the detector level.
 
\begin{figure*}[htbp]
\centering
\subfigure[]{ \includegraphics[width=0.45\textwidth]{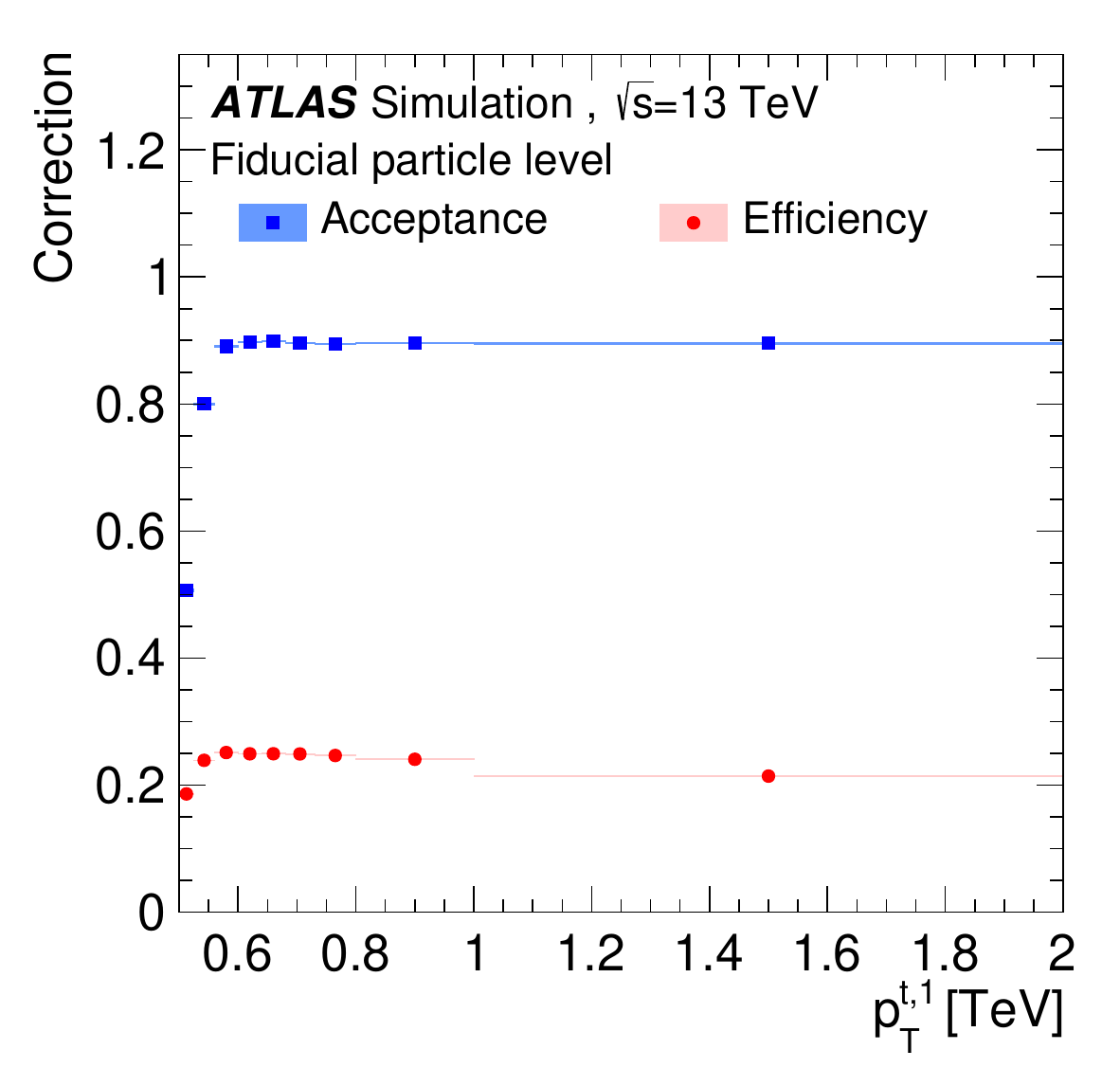}
\label{fig:corrections:particle:t1_pt}}
\subfigure[]{ \includegraphics[width=0.45\textwidth]{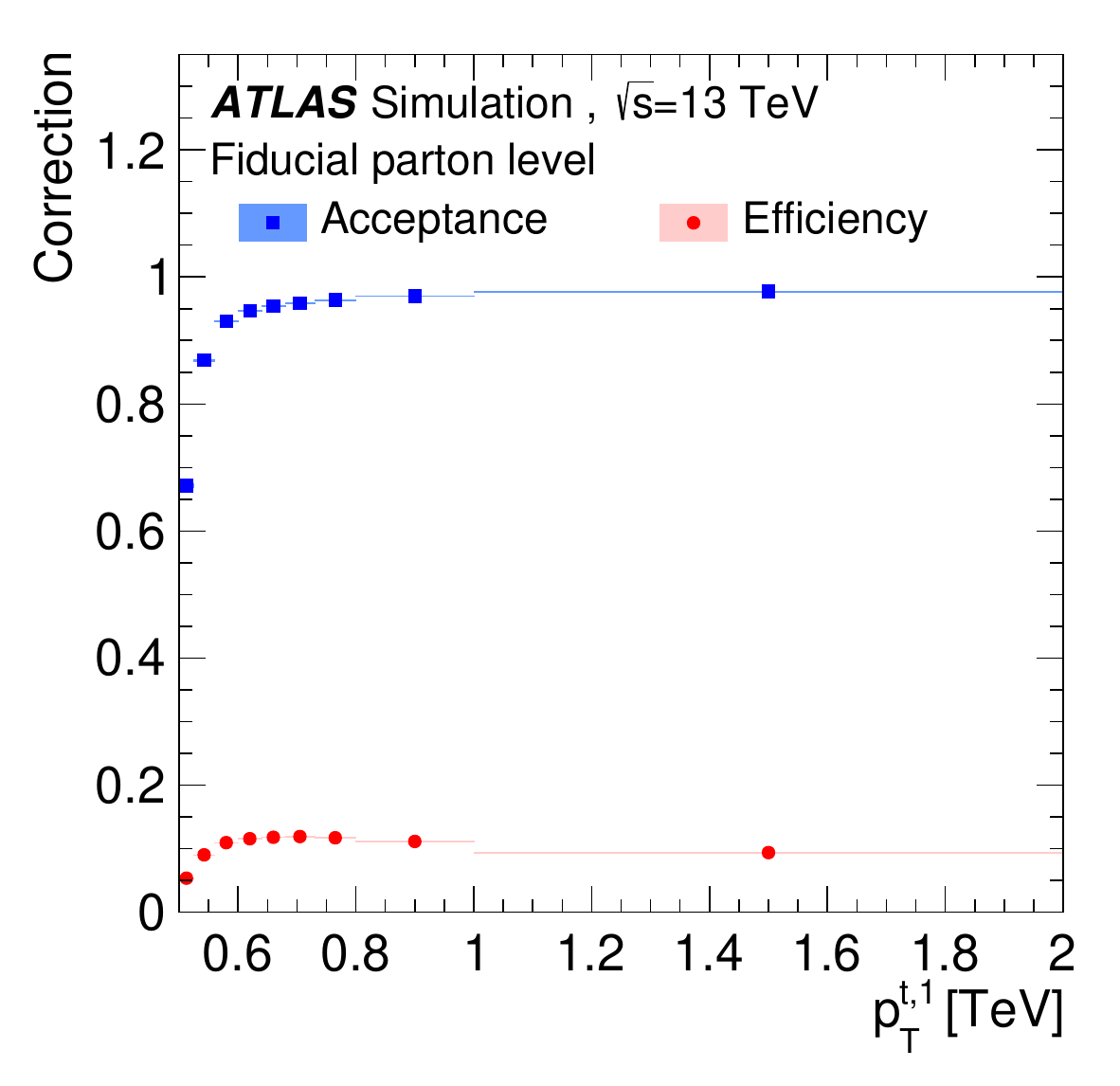}
\label{fig:corrections:parton:t1_pt}}
\subfigure[]{ \includegraphics[width=0.45\textwidth]{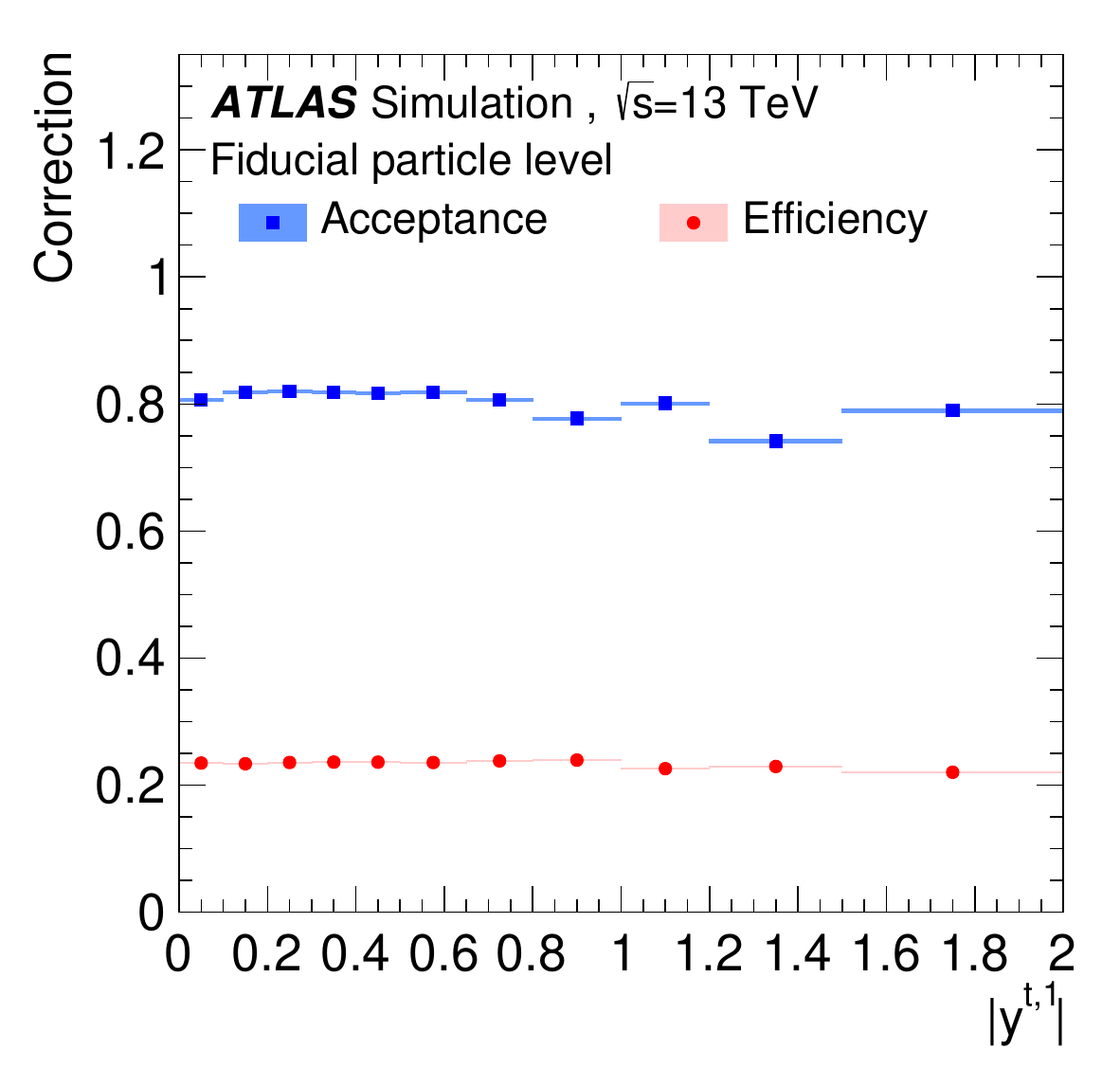}
\label{fig:corrections:particle:t1_y}}
\subfigure[]{ \includegraphics[width=0.45\textwidth]{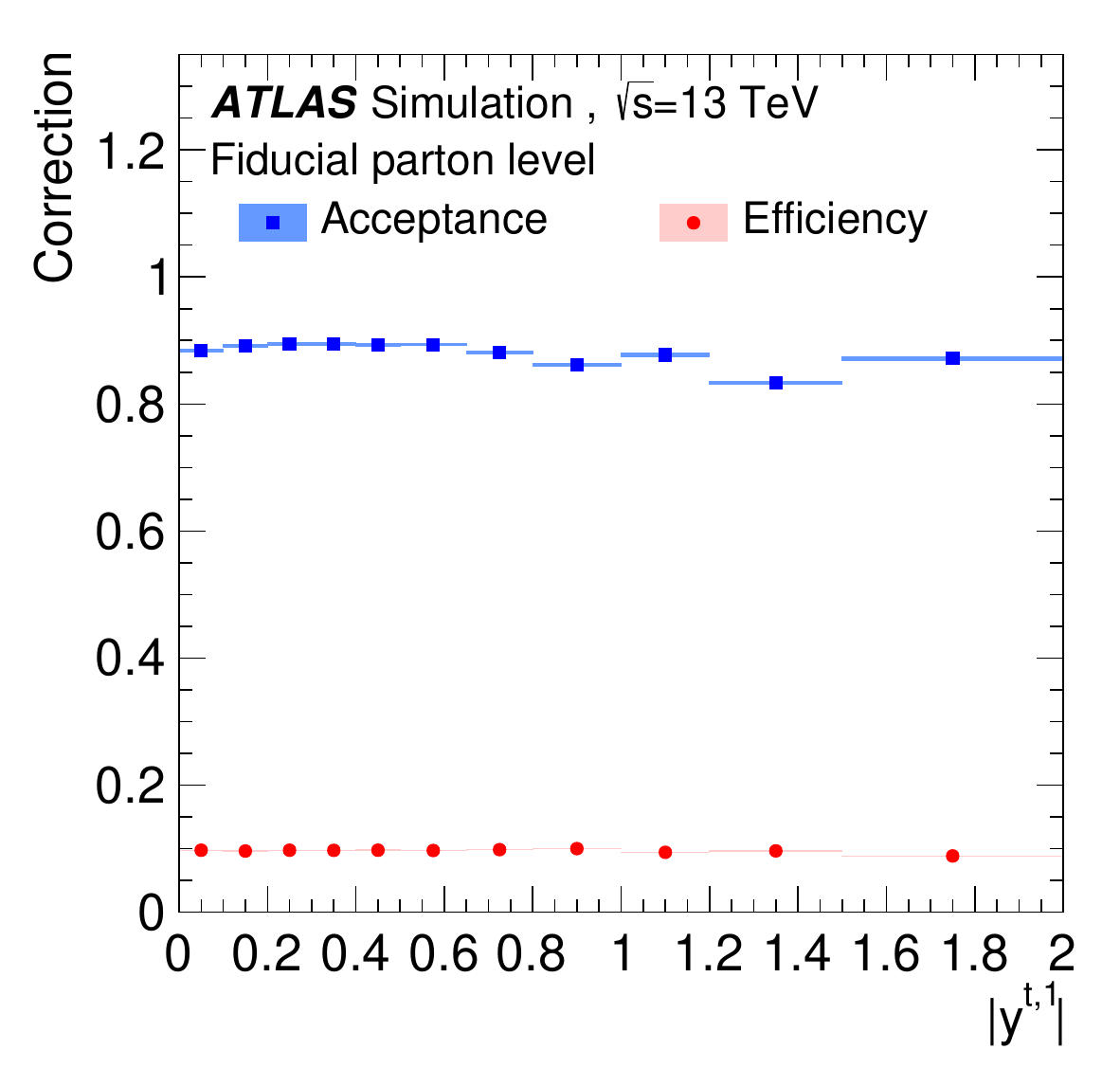}
\label{fig:corrections:parton:t1_y}}
\caption{Acceptance and efficiency corrections as a function of the leading top-quark-jet
\pT\ and $|y|$\
for the particle-level phase space are shown in
\subref{fig:corrections:particle:t1_pt}\ and \subref{fig:corrections:particle:t1_y}, respectively,
and for the parton-level fiducial phase space in
\subref{fig:corrections:parton:t1_pt}\ and \subref{fig:corrections:parton:t1_y}, respectively.
The observables on $x$-axes are at the truth level when used for the efficiency correction while they are at the detector level when used for the acceptance correction.
The \POWHEG{}+\Pythia[8] generator together with the \GEANT\ detector simulation framework is used to correct for detector effects.
The blue and red bars represent statistical uncertainties.}
\label{fig:corrections:particle:top_quarks}
\end{figure*}

The efficiency corrections $\epsilon_{\rm eff}^{i}$ correct for events
that are in the particle-level (parton-level) fiducial phase space but
are not reconstructed at detector level.
The acceptance corrections $f_{\rm acc}^j$\ account for events that are
generated outside the particle-level (parton-level) fiducial phase space but
pass the detector-level selection.
Figure~\ref{fig:corrections:particle:top_quarks}\ shows
the efficiency and acceptance corrections for the \pT\ and rapidity of
the leading jet.
The variations in acceptance as a function of rapidity arise from transitions from one
calorimeter system to another.
The corrections for the other observables show similar behaviour
except for observables sensitive to the
relative orientation of the two top-quark jets: \ptttbar, \absdeltaPhittbar, and
\absPoutttbar, which show modest decreases in acceptance.

The unfolding procedure for an observable $X$ at both particle- and parton-level is summarized by
the heuristic expression
\begin{equation}
\frac{{\dif}\sigma^{\rm fid}}{{\dif}X^i} \equiv \frac{1}{\int \!\mathcal{L}\,\dif\:\!t \cdot \Delta X^i} \cdot  \frac{1}{\epsilon_{\rm eff}^i} \cdot \sum_j \mathcal{M}_{ij}^{-1} \cdot  f_{\rm acc}^j \cdot \left(N_{\rm reco}^j - N_{\rm bg}^j\right)\hbox{,}
\label{eq:iterativeBayesian}
\end{equation}
where
$N_{\rm reco}^j$ and $N_{\rm bg}^j$ refer to the number of reconstructed signal and background
events in each detector-level bin, respectively;
the index $j$ runs over bins of $X$ at detector level while the index $i$\ labels bins at either particle or parton level;
$\Delta X^i$ is the bin width; and
$\int \!\mathcal{L}\,\dif\:\!t$ is the integrated luminosity.
The matrix $\mathcal{M}_{ij}^{-1}$\ denotes the unfolding procedure and, strictly speaking, is not
the inverse of the migration matrix defined earlier but is determined iteratively and
has the effect of inverting the smearing resulting from the measurement process.
 
This unfolding procedure, expressed in Eq.~(\ref{eq:iterativeBayesian}), is performed iteratively and
regularizes the smoothness of the unfolded distribution.
Studies of the performance of the algorithm using MC events show that four iterations provide
unbiased, high-precision unfolded distributions for all observables.
Other choices of the number of iterations between typically three to six provide similar results.
 
The inclusive cross-section, $\sigma^{\mathrm{fid}}$, for \ttbar\ events in the particle-level (parton-level)
phase space, obtained by integrating the differential cross-section,
is used to determine the normalized differential
cross-section $1/\sigma^{\rm fid}\cdot{\dif}\sigma^{\rm fid} / {\dif}X^i$.
The particle-level fiducial phase-space cross-section is not corrected for the all-hadronic \ttbar\ branching fraction
of 0.457 \cite{PDG}.
This branching fraction is used to correct the parton-level fiducial
phase-space cross-section measurement in order to facilitate a comparison with NNLO fixed-order predictions.
 
Tests are performed at both particle level and parton level to verify that the unfolding procedure recovers the generator-level distributions for detector-level distributions that vary from the
nominal predictions. These tests include linear reweighting of the input distributions
as a function of kinematic variables or reweighting based on possible
data/prediction discrepancies.
These stress tests show that the results of the unfolding procedure are unbiased as long as the variations in the input distributions are consistent with the measurement resolution of the
observable.
As part of the estimation of modelling systematic uncertainties,  other tests are performed where the underlying model is changed, e.g., the parton-shower model or the top-quark mass.
These are described in Section~\ref{sec:systematics}.


\section{Systematic and statistical uncertainties}
\label{sec:systematics}

Systematic uncertainties introduced by the particle and jet reconstruction and calibration,
the \ttbar\ modelling, and the background estimation are described below.
The propagation of systematic uncertainties through the unfolding procedure is described in Section~\ref{sec:prop_uncertainties}.
The treatment of the statistical uncertainties associated with the MC calculations is also discussed.
 
\subsection{Estimation of systematic uncertainties}
 
The systematic uncertainties of the measured distributions are estimated using simulation samples and
the data satisfying the final selection requirements.
 
A significant source of systematic uncertainty is the jet-energy scale (JES) for
the \largeR\ jets~\cite{JETM-2018-02}.
The \smallR\ jet JES~\cite{JETM-2018-05}\ does not contribute to the systematic uncertainties
as these jets are not used in the event selection nor the unfolding procedures.
Uncertainties in the jet energy resolution (JER) for \largeR\ jets is also
considered~\cite{JETM-2018-05,JETM-2018-02}.
The effect of correlations between the JES and JER systematic uncertainties is negligible in this analysis.
The JES uncertainty results in a cross-section uncertainty that is typically of 4\%--5\% but reaches
$12\%$ for rapidity-related observables at large rapidity values.
The JER uncertainty creates a cross-section uncertainty of 2\%--5\%.
 
The uncertainties in the \largeR\ jet mass scale (JMS) and resolution (JMR) are
derived from observations of the $W$~boson and top-quark masses in semileptonic \ttbar\
events~\cite{JETM-2018-02,ATLAS-CONF-2020-022}, and
by measuring the double ratio of data to MC simulation for calorimeter-only and
track-only quantities.
The effect of the JMS uncertainty is typically around 1\%--2\%,
while the effect of the JMR uncertainty is below $1$\%.
 
The efficiency to top-quark-tag \largeR\ jets is corrected in simulated
events by applying top-quark-tagging scale factors to account for
a residual difference between data and simulation samples~\cite{JETM-2018-03,ATL-PHYS-PUB-2020-017}.
The signal jets are required to be top-quark-tagged
while other jets are labelled as background jets.
Uncertainties in the rate of background jets were measured in two
phase-space regions enriched in multijet and $\gamma$\,+\,jet processes.
The signal-jet uncertainties were measured in boosted \ttbar\ lepton+jets events.
Additional uncertainties are assigned to cover signal-modelling effects and extrapolation beyond the fiducial phase-space regions.
The associated systematic uncertainties are computed by varying the
top-quark-tagging scale factors within their uncertainties and are found to
create differential cross-section uncertainties ranging from $7$\% to $10$\%.
 
The efficiency to tag \VRTrack\ jets containing $b$-hadrons is corrected in
simulated events by applying $b$-tagging scale factors, extracted from
\ttbar\ events, in order to account for residual tagging-efficiency differences
between data and simulation~\cite{FTAG-2018-01,ATL-PHYS-PUB-2017-013}.
An additional uncertainty is included for the
extrapolation of the measured uncertainties to the
high-\pt region of interest~\cite{ATL-PHYS-PUB-2021-003}.
Its estimation is improved for the jets passing the event selection of
this measurement by using the $b$-hadron \pt\ spectrum corresponding to these jets.
The systematic uncertainty is computed by varying the $b$-tagging scale factors within their uncertainties and is found to
be 3\%.
 
The cross-section uncertainties arising from the lepton energy scale, resolution, and identification efficiency
are below $0.1$\%~\cite{EGAM-2018-01,MUON-2018-03}.
 
For backgrounds estimated by MC simulation, the
uncertainties in the predicted production cross-sections are included.
An additional uncertainty of 50\% is assigned to the
$Wt$\ single-top-quark production cross-section to cover the large difference between the rates predicted
by the diagram-removal~\cite{Frixione:2008yi} and diagram-subtraction~\cite{Frixione:2008yi} schemes in the boosted
regime~\cite{ATL-PHYS-PUB-2016-004}. These schemes have different treatments of the overlap of
the $Wt$-channel with \ttbar\ production.
Systematic uncertainties affecting the multijet-background estimate
come from the subtraction of other background processes in the control regions
and from the uncertainties in the measured tagging correlations.
The detector-related uncertainties which affect the MC-based background
processes in the control regions used for the multijet-background estimates
are directly accounted for in the above-mentioned detector systematic uncertainty categories.
The remaining multijet-background uncertainties range
from 1\% to 6\%\ for leading \largeR\ jets with \pT\ from 500~\GeV\ to 2~\TeV, respectively.
 
Alternative MC generators are employed to assess modelling systematic uncertainties.
In these cases, the difference between the unfolded distribution from an
alternative model and its own particle-level or parton-level
distribution is used as the estimate of the corresponding systematic
uncertainty in the unfolded differential cross-section.
The matrices from the nominal MC simulation are used in the unfolding.
 
To assess the uncertainty related to the matrix-element calculation
and the parton-shower matching procedure, \AMCatNLO{}+\Pythia[8] events are
unfolded using the migration matrix and correction factors derived
from the \POWHEG{}+\Pythia[8] sample with the matrix-element correction turned off.
This uncertainty is typically a few percent, increasing to
5\%--10\%  at large \ptttbar ~and $|\Poutttbar|$, and
small \deltaPhittbar.
To assess the uncertainty associated with the choice of
parton-shower and hadronization model, a comparison is made between the
unfolded and generator-level distributions of simulated events created
with the \POWHEG{}+\HERWIG[7] generator but using the nominal corrections and migration matrices.
The resulting systematic uncertainties, taken
as the symmetrized difference, are found to be less than 5\%.
 
The uncertainty related to the modelling of initial- and final-state radiation
is determined by using \ttbar MC samples with modified ISR/FSR
settings~\cite{ATL-PHYS-PUB-2020-023}.
Four different upward/downward variations of MC simulation parameters are performed to assess the ISR uncertainty; these
have a significant effect on
initial-state radiation, while the effect on final-state radiation is small.
The upward (downward) variations are defined by scaling of each of $\muR$\ and $\muF$\ by a factor of 0.5 (2),
the setting of $h_{\text{damp}}$\ to $3m_{\text{top}}$ ($1.5m_{\text{top}}$),
and the variation of the A14 tuned set of parameters encoded by the Var3cUp (Var3cDown)
parameter~\cite{ATL-PHYS-PUB-2014-021,ATL-PHYS-PUB-2020-023}.
The effects caused by independent variations of individual parameters are summed in quadrature to define the ISR uncertainty.
For FSR, variations are defined by scaling $\muR$\ and $\muF$\ for FSR only.
The FSR-up variation  uses a scale factor of 0.5 while the FSR-down variation uses a factor of 2.
This uncertainty is found to be approximately 5\%\ or lower
depending on the observable considered.
 
The uncertainty arising from PDFs is assessed using the \POWHEG{}+\Pythia[8] \ttbar\ sample. The sample is reweighted to the nominal PDF4LHC15 PDF set and the uncertainty in the unfolded
distributions arising from the uncertainties in that PDF set is determined using the Hessian approach with 30 eigenvectors~\cite{Butterworth:2015oua}.
This uncertainty is found to be approximately $1$\%.
 
The effect of varying the top-quark mass by ${\pm}1$~\GeV\ had a negligible effect on
the unfolded results.
 
The uncertainty in the combined Run~2 integrated luminosity is
1.7\% \cite{ATLAS-CONF-2019-021}, obtained using the LUCID-2
detector \cite{LUCID2} for the primary luminosity measurements.
This uncertainty affects the rate of backgrounds estimated using MC calculations.
It also affects the overall normalization as seen in Eq.~(\ref{eq:iterativeBayesian}),
but has negligible effect on the normalized differential cross-section measurements.
The uncertainty arising from the size of the nominal MC sample is
approximately 1\%.
 
\subsection{Propagation of systematic uncertainties and treatment of correlations}
\label{sec:prop_uncertainties}
 
The statistical and systematic uncertainties are
propagated and combined in the same way for both the particle-level and parton-level
results, using pseudo-experiments created from the nominal and alternative MC samples.
 
To evaluate the impact of a systematic-uncertainty contribution to an unfolded distribution, a corresponding
distribution is obtained from simulations employing modified parameter settings reflecting this particular contribution.
This distribution is then unfolded using corrections obtained with the nominal \POWHEG{}+\Pythia[8] sample.
The resulting unfolded distribution is compared with the corresponding particle- or parton-level distribution
and the difference is taken as the uncertainty in the unfolded measurement.
For each systematic uncertainty, the correlation between the uncertainties in the
signal and background distributions
is taken into account.
All detector- and background-related systematic uncertainties are estimated using the
nominal \POWHEG{}+\Pythia[8] sample.
Residual hard-scattering, parton-shower and hadronization, ISR/FSR, and PDF uncertainties are
estimated from a
comparison between the unfolded cross-section and the corresponding particle- or parton-level
distribution produced using the corresponding MC generator.
This method is used to estimate the systematic uncertainties due to the choice of MC generator.
 
\begin{figure*}[htbp]
\centering
\subfigure[]{ \includegraphics[width=0.45\textwidth]{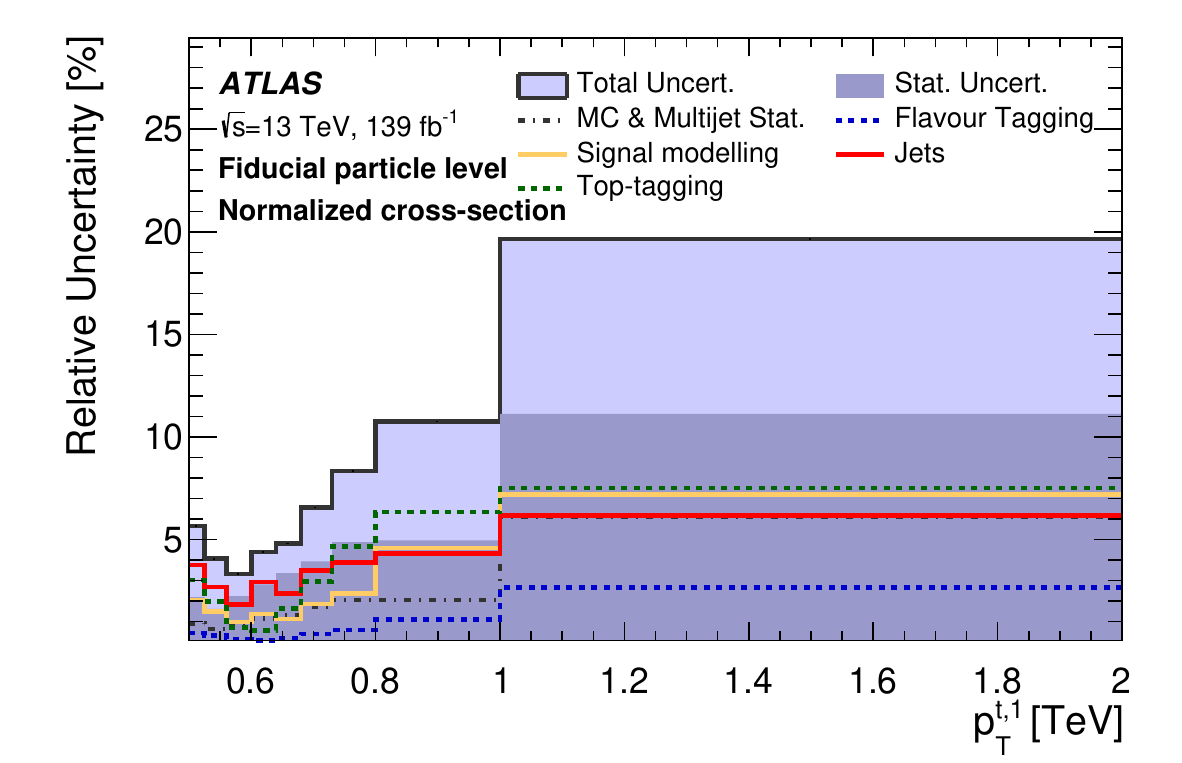}\label{fig:particle:t1_pt:rel:unc}}
\subfigure[]{ \includegraphics[width=0.45\textwidth]{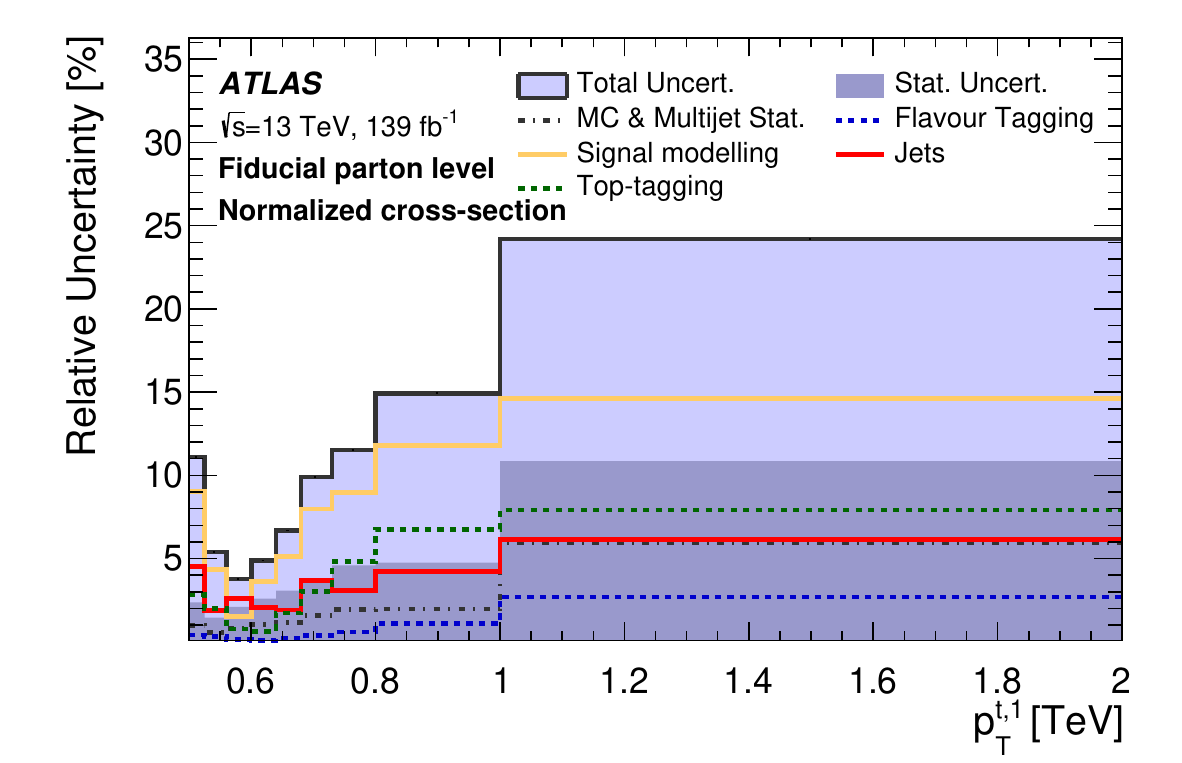}\label{fig:parton:t1_pt:rel:unc}}
\subfigure[]{ \includegraphics[width=0.45\textwidth]{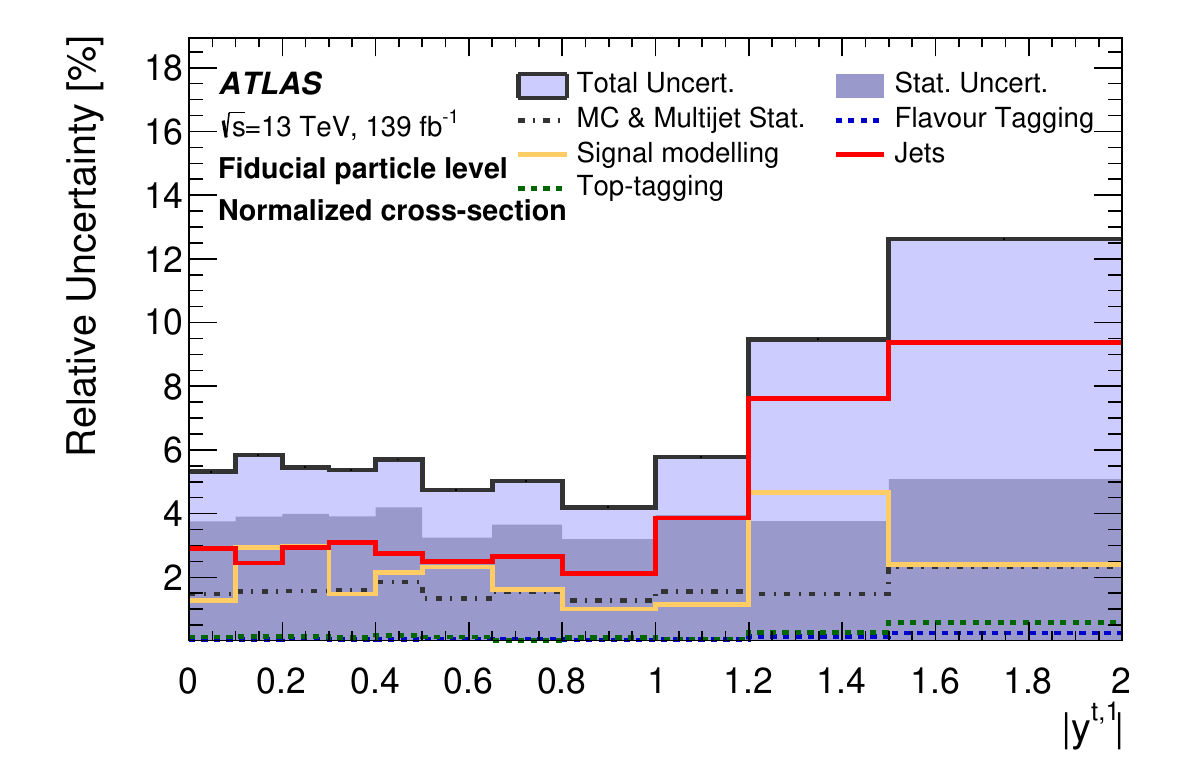}\label{fig:particle:t1_y:rel:unc}}
\subfigure[]{ \includegraphics[width=0.45\textwidth]{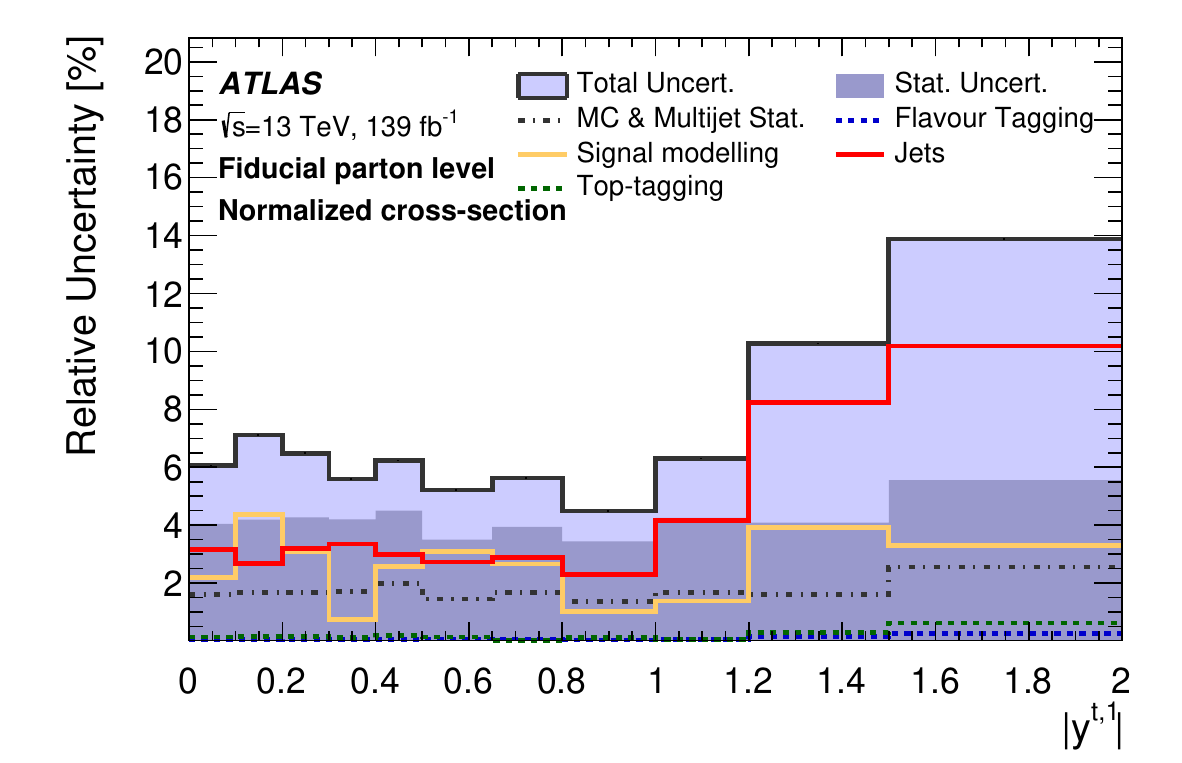}\label{fig:parton:t1_y:rel:unc}}
\caption{Relative uncertainties in the normalized differential cross-sections as a function
of the leading top-quark-jet \pT\ and rapidity at particle level (\subref{fig:particle:t1_pt:rel:unc}\ and
\subref{fig:particle:t1_y:rel:unc}) and parton level
(\subref{fig:parton:t1_pt:rel:unc}\ and \subref{fig:parton:t1_y:rel:unc}).
The light and dark blue areas represent the total and statistical uncertainty, respectively. The 'Jet' category includes JES, JER, JMS, and JMR systematic uncertainties. The 'Signal modelling' category includes alternative hard-scattering model, alternative parton-shower model, ISR/FSR, and PDF systematic uncertainties.
The \POWHEG{}+\Pythia[8] generator is used as the nominal prediction to correct for detector effects.
}
 
\label{fig:fractional_unc:t1:rel}
\end{figure*}

The systematic uncertainties for the particle-level fiducial phase-space cross-section measurement
described in Section~\ref{sec:measurements}\ are listed in Table~\ref{tab:systematics}.
Figure~\ref{fig:fractional_unc:t1:rel} shows a summary of the relative size of the systematic uncertainties for the normalized differential cross-sections as a function of the leading top-quark-jet \pT\ and rapidity at particle level and parton level.
For the second-leading jet, the uncertainty is ${\sim}8$\%\ at $\pT = 350$~\GeV\ and
${\sim}14$\%\ for $\pT > 800$~\GeV\ at particle level.
 
\begin{table}[t]
\begin{center}

\begin{tabular}{lc}
\hline
Source & Relative Uncertainty [\%] \\ \hline
JES $\oplus$ JER  &  $ \numRF{4.2227}{2}$ \\
JMS $\oplus$ JMR  &  $ \numRF{1.064}{2}$ \\
Top-tagging       &  $ \numRF{7.795}{2}$ \\
Flavour tagging    &  $ \numRF{2.932}{2}$ \\
Alternative hard-scattering model & $\numRF{0.90}{1}$  \\
Alternative parton-shower model   & $\numRF{4.25}{2}$  \\
ISR/FSR + scale               &  $ \numRF{4.858}{2}$  \\
PDF                           &  $ \numRF{0.81}{1}$ \\
Luminosity                    &  $ \numRF{1.66}{2}$ \\
MC \& Multijet sample statistics &  $ \numRF{0.39}{1}$  \\
\hline \hline
Total systematic uncertainty  &  $ \numRF{+11.79}{3}$~~ \\
Statistical uncertainty       &  $ \numRF{0.98}{1}$  \\
Total uncertainty             &  $ \numRF{+11.83}{3}$~~ \\
\hline
\end{tabular}


\vspace{.1cm}
\end{center}
\caption{
Summary of the largest systematic and statistical relative uncertainties for the total
particle-level fiducial phase-space cross-section measurement.
The uncertainties that are significantly less than 1\%\ are not listed.
}
\label{tab:systematics}
\end{table}
 
The effect of the statistical uncertainty of the data, the statistical uncertainty due to the size of MC samples, and the systematic uncertainties are incorporated into pseudo-experiments to determine the covariance matrix of the measurement, following the approach used in Refs.~\cite{TOPQ-2014-15,TOPQ-2016-01}.
The effect of the statistical uncertainty of the data is incorporated by independent Poisson fluctuations in each data bin.
The statistical uncertainty due to the size of the background MC samples is incorporated by adding independent Gaussian fluctuations in each bin of the signal region and control regions used for the multijet-background estimation.
The statistical uncertainty due to the size of the signal MC samples is incorporated by adding Gaussian fluctuations in the
unfolding corrections and migration matrices, and in each bin of the distributions in the control regions used in the
multijet-background estimation.
The effects of all other systematic uncertainties are incorporated into the
pseudo-experiments by including Gaussian fluctuations associated with each source of uncertainty.
A covariance matrix is constructed using these pseudo-experiments for each differential cross-section
in order to include the effect of all uncertainties and correlations on the bin-to-bin measurements to
allow quantitative comparisons with theoretical calculations.
 
The comparison between the measured differential cross-sections and a
variety of MC calculations is quantified by calculating $\chi^{2}$ values employing the covariance matrix and by calculating the
corresponding $p$-values from the $\chi^2$ and the number of degrees of freedom (NDF).
The $\chi^2$ values are obtained using
\begin{linenomath*}
\begin{equation*}
\chi_{N_{\textrm b}}^2 = V_{N_{\textrm b}}^{\textrm T} \cdot{\textrm{C}}_{N_{\textrm b}}^{-1} \cdot V_{N_{\textrm b}} \, ,
\end{equation*}
\end{linenomath*}
where $V_{N_{\textrm b}}$ is the vector of differences between the measured differential cross-section values and
calculations, $ {\textrm{C}}_{N_{\textrm b}}^{-1}$ is the inverse of the covariance matrix, and $N_{\textrm b}$ is the number of bins in the unfolded distribution. The theoretical uncertainties in the predictions are not included in the $\chi^2$ calculation.
 
The normalization constraint used to derive the normalized
differential cross-sections lowers the NDF to one less than the rank of the
$N_{\textrm b} \times N_{\textrm b}$ covariance matrix.
The $\chi^2$ for the normalized
differential cross-sections is
\begin{linenomath*}
\begin{equation*}
\chi_{N_{\textrm b}-1}^2 = {\hat V}_{N_{\textrm b}-1}^{\textrm T} \cdot{\hat{\textrm{C}}}_{N_{\textrm b}-1}^{-1} \cdot{\hat V}_{N_{\textrm b}-1} \, ,
\end{equation*}
\end{linenomath*}
where ${\hat V}_{N_{\textrm b}-1}$ is the vector of differences between measured normalized differential cross-section values
and calculations obtained by discarding one of the $N_{\textrm b}$
elements, and ${\hat{{\textrm{C}}}}_{N_{\textrm b}-1}$ is the
$(N_{\textrm b}-1) \times (N_{\textrm b}-1)$ sub-matrix derived from the
covariance matrix corresponding to normalized differential cross-section measurement
by discarding the corresponding row and column.
The $\chi^2$\ does not depend on the index of the discarded row and column.


\section{Measured differential cross-sections}
\label{sec:measurements}

All measurements are presented as single-differential, double-differential
or triple-differential cross-sections in the particle-level and
parton-level fiducial phase spaces.
The total cross-section measurements in these fiducial phase spaces also provide a test of the total cross-section calculations for different models.
Normalized differential cross-sections allow a comparison of their shapes between data and MC predictions while removing the effects of possible differences in the yields.
 
The particle-level and parton-level fiducial phase-space
cross-sections and the normalized fiducial phase-space differential cross-sections are presented below. The comparison of absolute differential cross-sections between data and MC
predictions is not performed given that the most significant discrepancies arises from
the ${\sim}20\%$ difference in total yields.
The one exception is Figure~\ref{fig:parton:NNLO:abs} where absolute differential cross-sections are presented together with the comparison to NNLO fixed-order predictions.
 
These measurements are compared with SM predictions using the nominal
\POWPY[8] MC samples,
the \POWHER[7] alternative parton-showering and hadronization calculations,
the \AMCatNLO+\PYTHIA[8] alternative matrix-element calculation,
and the \POWPY[8] samples using modified ISR and FSR settings.
The sample with less ISR and FSR (`less IFSR') has the factorization and renormalization scales increased by a factor of two compared to the nominal sample, and the A14 Var3c Down tune variation in the parton shower.
The sample with more ISR and FSR (`more IFSR') has $h_{\mathrm{damp}} = 3\mt$, the factorization and renormalization scales reduced by a factor of 0.5 compared to the nominal sample, and the A14 Var3c Up tune variation in the parton shower. The labels used in the plots for the above mentioned SM predictions are 'PWG+Py8', 'PWG+H7.1.3', 'MG5\_aMC@NLO+Py8', 'PWG+Py8 (more IFSR)', and 'PWG+Py8 (less IFSR)', respectively.
 
The discussion and the interpretation of these results is presented in Section~\ref{sec:comparisons}.
 
\subsection{Total particle-level cross-section in the fiducial phase space}
\label{sec:particle_level_total_cross_section}
The particle-level fiducial phase-space cross-section, multiplied by the \ttbar{} all-hadronic decay branching fraction, is used to
normalize the observed
particle-level fiducial phase-space differential cross-sections.
It is determined by taking the observed yield after background subtraction and
applying the correction factors to account for events that were produced
outside the fiducial phase-space region but passed the detector-level selection criteria, and the event-selection efficiency.
This procedure amounts to a single-bin unfolding.
All of the systematic uncertainties that affect the correction and acceptance are included
in this measurement.
 
\begin{figure*}[htbp]
\centering
\includegraphics[width=0.7\textwidth]{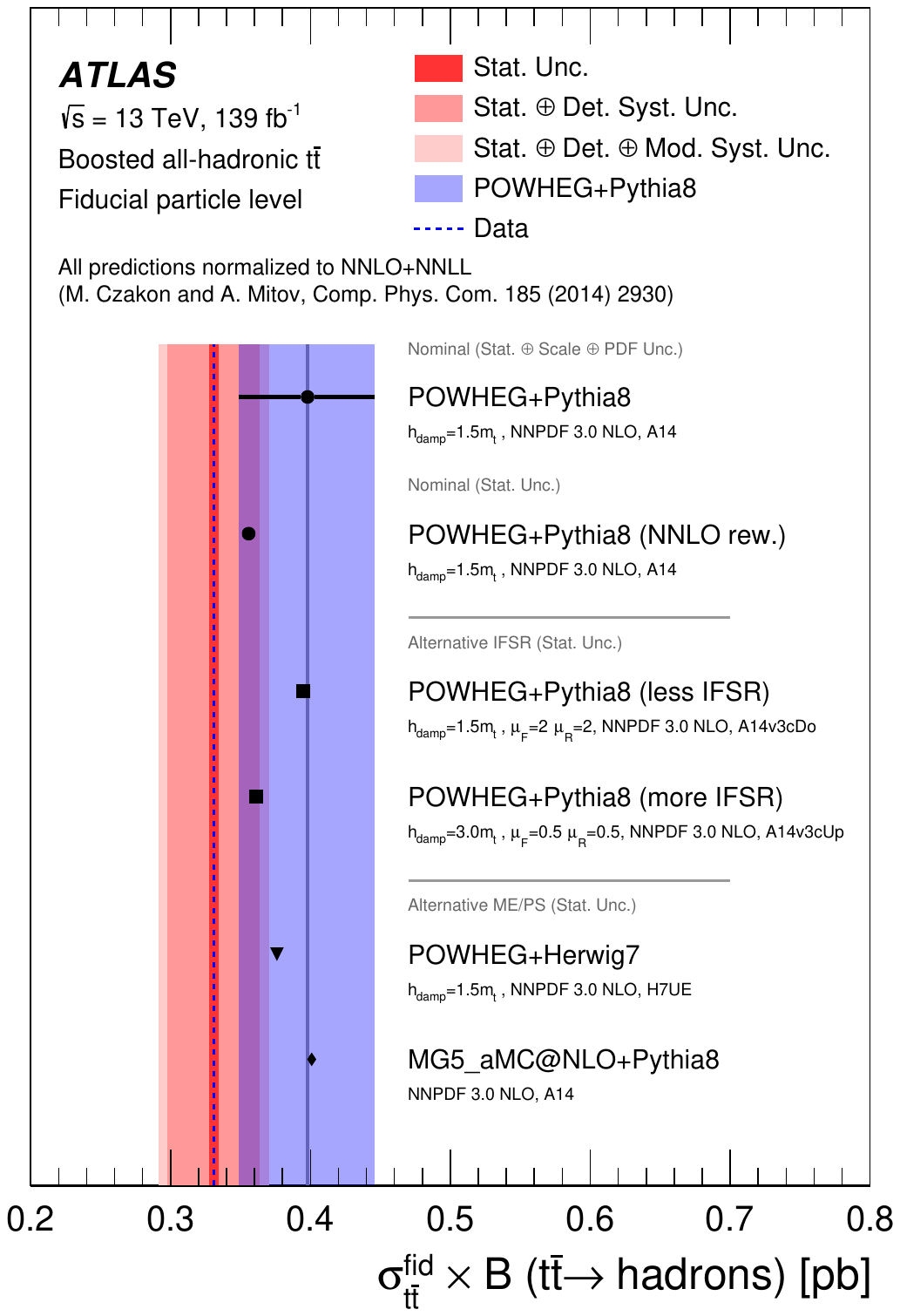}
\caption{
The particle-level cross-section in the fiducial phase space $\sigma_{\ttbar}^{\text{fid}}$ multiplied by the \ttbar{} all-hadronic decay branching fraction $B(\ttbar \rightarrow \text{hadrons})$.
The shaded (red) bands indicate the statistical, detector, and modelling uncertainties in the measurement.
The \POWPY[8] event generator is used as the nominal prediction to correct for detector effects.
The uncertainty associated with the nominal \POWPY[8] signal model (blue band) includes
the statistical, scale, PDF, and NNLO+NNLL total inclusive calculation uncertainty.
Other calculations show only the statistical uncertainty of the MC calculations, which is
negligible and not visible in the figure.
IFSR refers to both initial- and final-state radiation.
}
\label{fig:particle:inclusive:abs}
\end{figure*}

The particle-level fiducial phase-space cross-section is
\begin{equation}
\sigma^{\ttbar,\text{fid}}_\text{particle} \times B(\ttbar \rightarrow \text{hadrons})  = \inclXsecParticle\ {\rm fb}.  \nonumber
\end{equation} 
The measured fiducial phase-space cross-section times branching fraction can be compared with the cross-section
predicted by the \POWPY[8] particle-level calculation of
$\numRF{398.0}{3} \numpmRF{+48.16}{-49.00}{2}$~fb,
after normalizing its inclusive prediction to the NNLO+NNLL total \ttbar cross-section.
The associated uncertainty includes the statistical, scale, PDF, and NNLO+NNLL total inclusive
calculation uncertainty.
This measurement and the comparisons with predictions are shown
in Figure~\ref{fig:particle:inclusive:abs}.
 
\subsection{Particle-level fiducial phase-space differential cross-sections}
\label{sec:results:particle_differential}
 
\begin{figure*}[htbp]
\centering
\subfigure[]{ \includegraphics[width=0.49\textwidth]{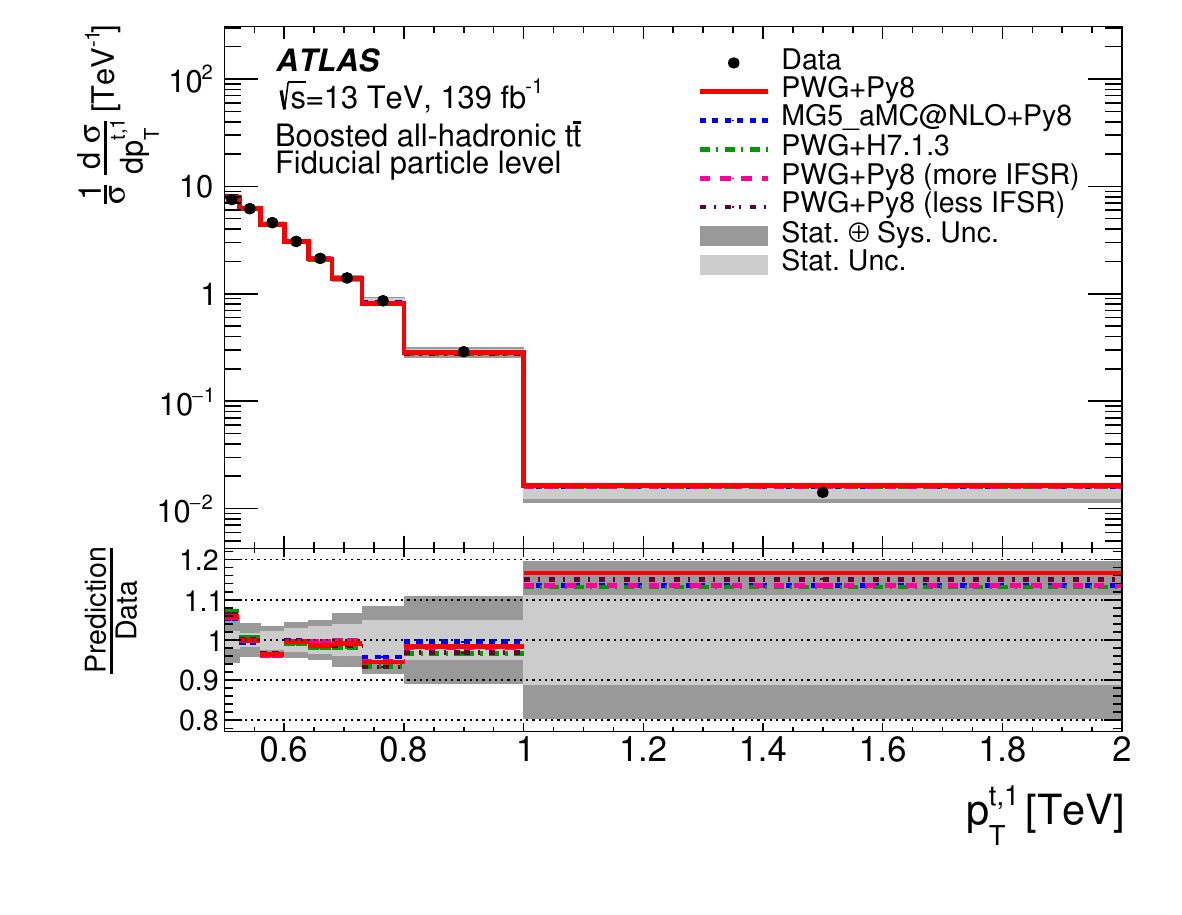}\label{fig:particle:t1_pt:rel}}
\subfigure[]{ \includegraphics[width=0.49\textwidth]{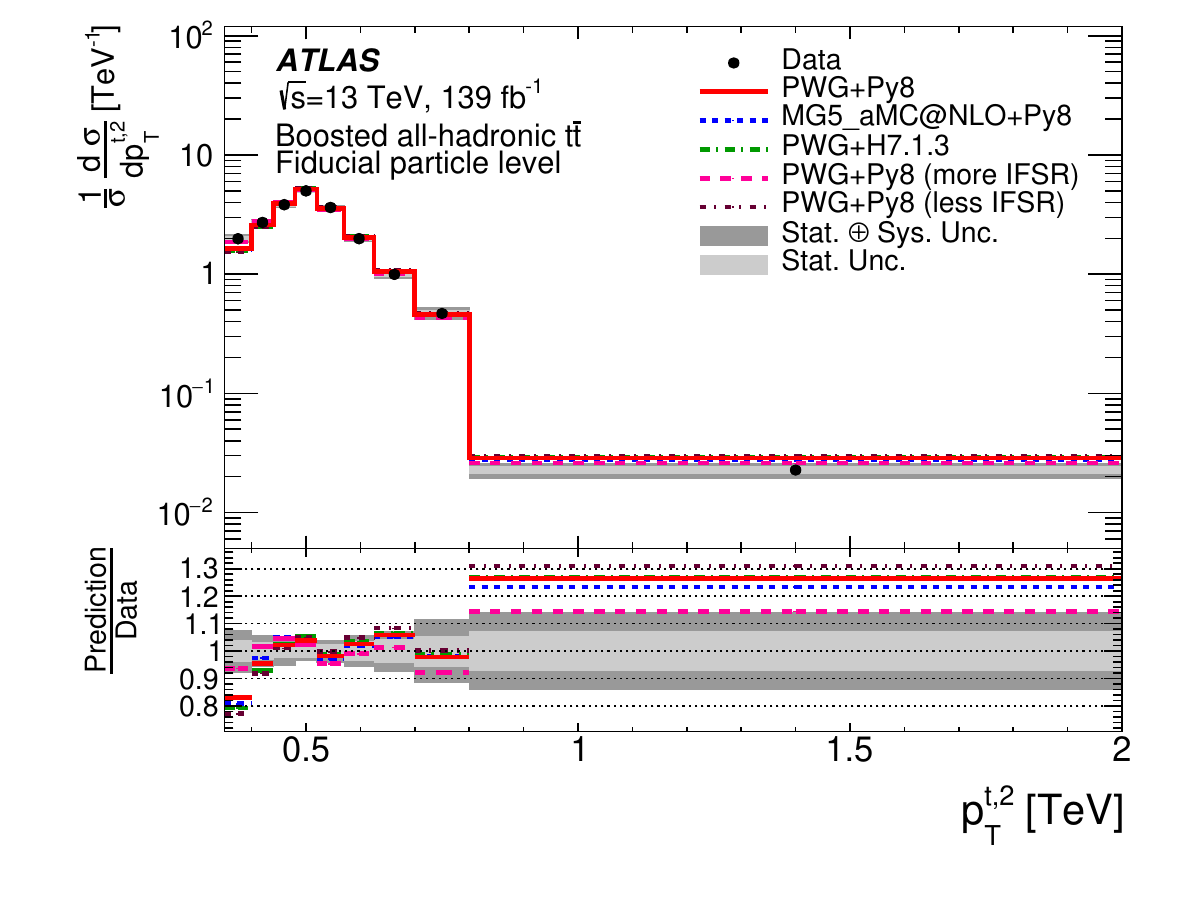}\label{fig:particle:t2_pt:rel}}
\subfigure[]{ \includegraphics[width=0.49\textwidth]{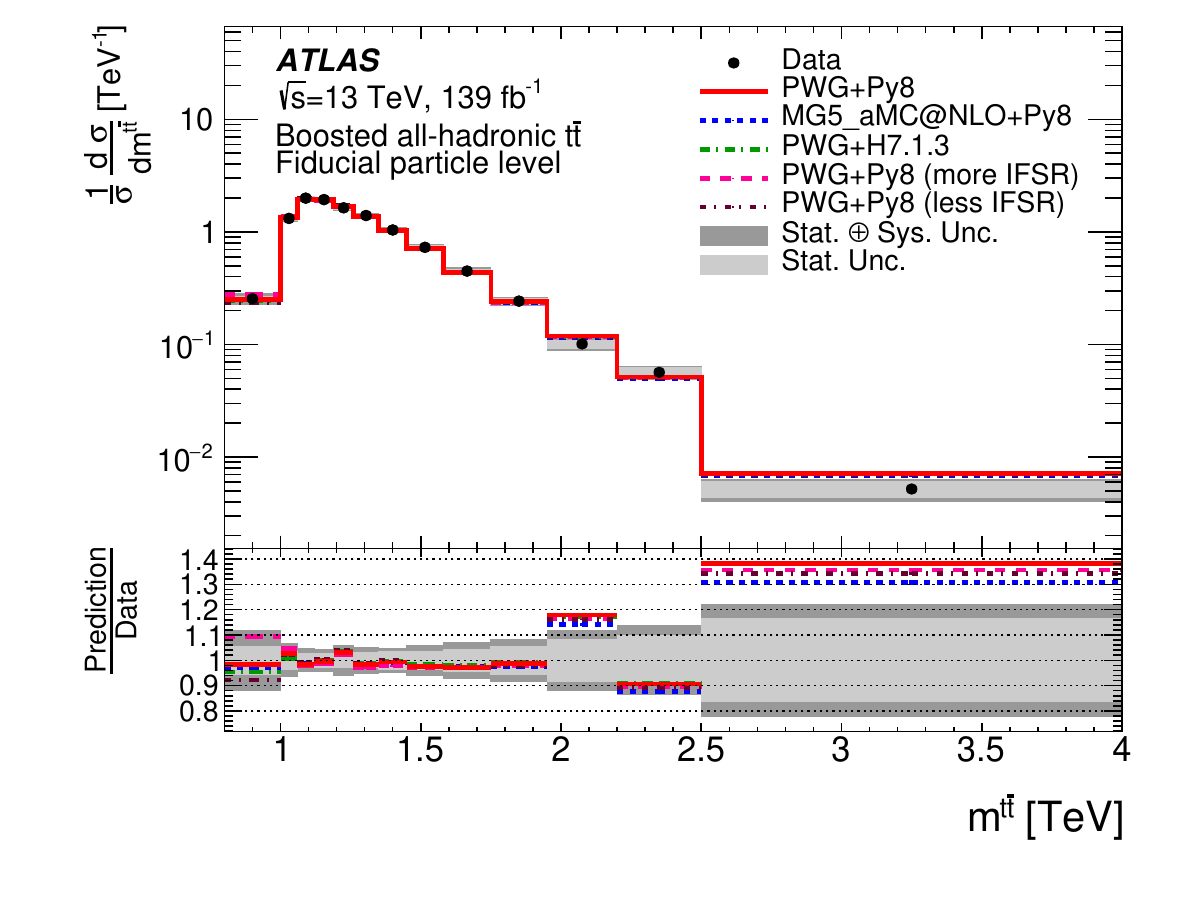}\label{fig:particle:tt_m:rel}}
\caption{
Normalized particle-level fiducial phase-space differential cross-sections as a function of
\subref{fig:particle:t1_pt:rel}~the \pT of the leading top-quark jet,
\subref{fig:particle:t2_pt:rel}~the \pT of the second-leading top-quark jet, and
\subref{fig:particle:tt_m:rel}~the invariant mass of the \ttbar{} system.
The dark and light grey bands indicate the total uncertainty and the statistical uncertainty, respectively, of the data in each bin.
Data points are placed at the centre of each bin.
The \POWPY[8] MC sample is used as the nominal prediction to correct the data to particle level.
}
\label{fig:particle:energy_observables:rel}
\end{figure*}

\begin{figure*}[htbp]
\centering
\subfigure[]{ \includegraphics[width=0.49\textwidth]{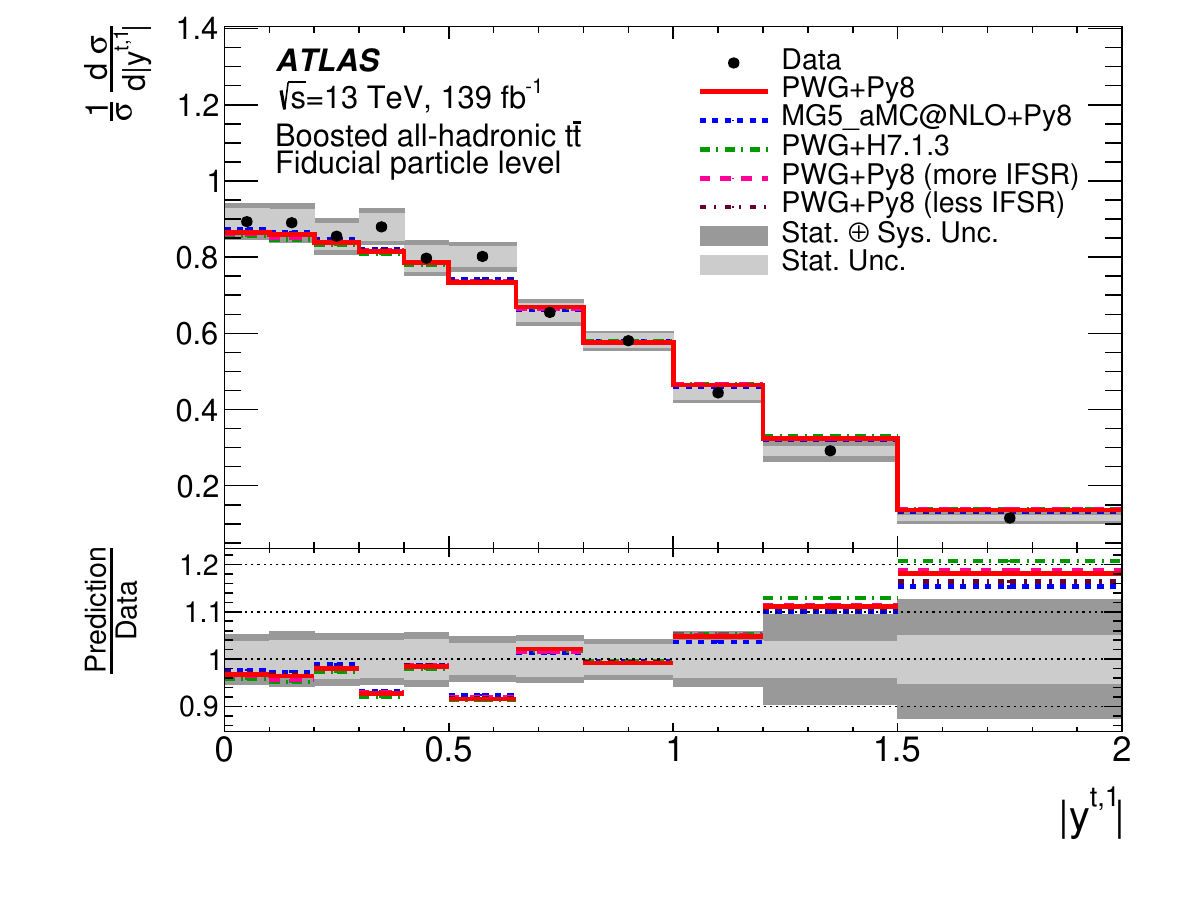}\label{fig:particle:t1_y:rel}}
\subfigure[]{ \includegraphics[width=0.49\textwidth]{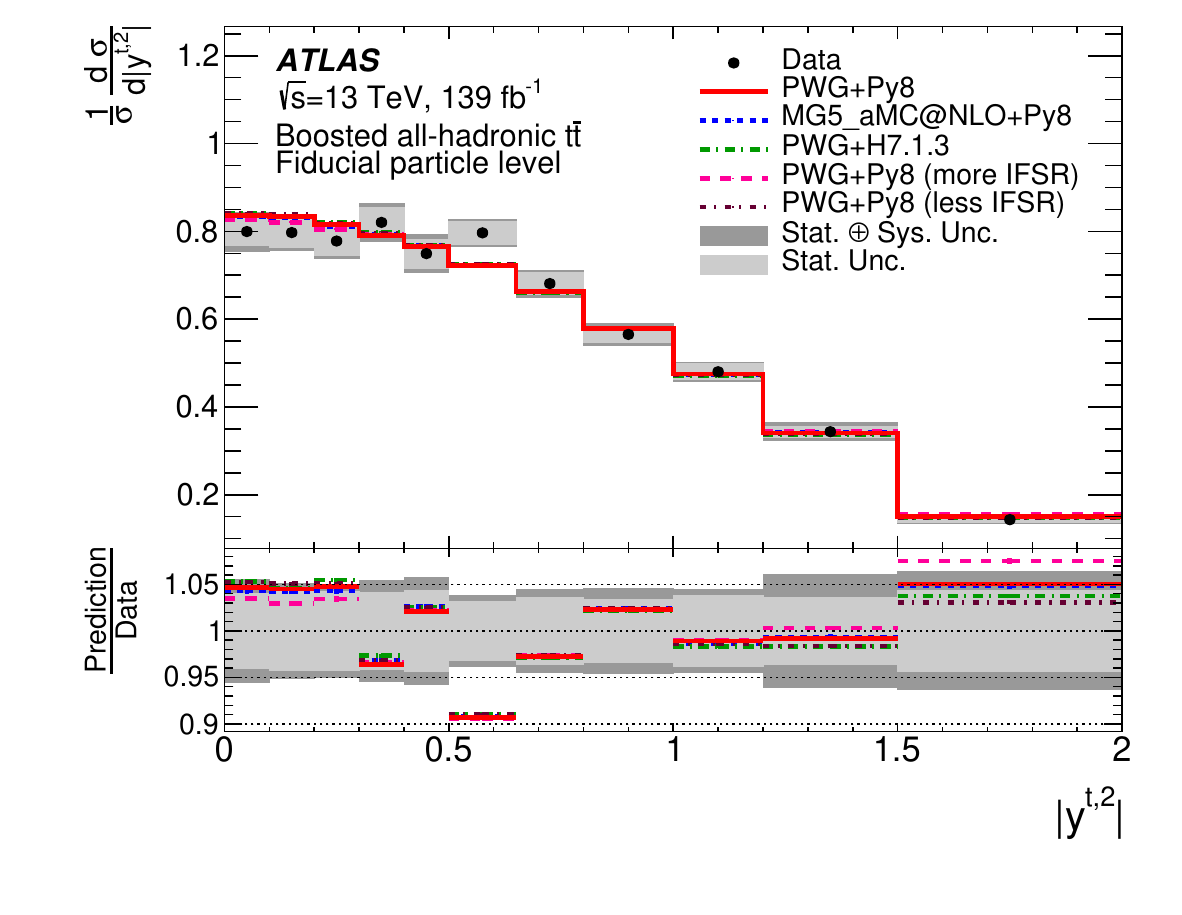}\label{fig:particle:t2_y:rel}}
\subfigure[]{ \includegraphics[width=0.49\textwidth]{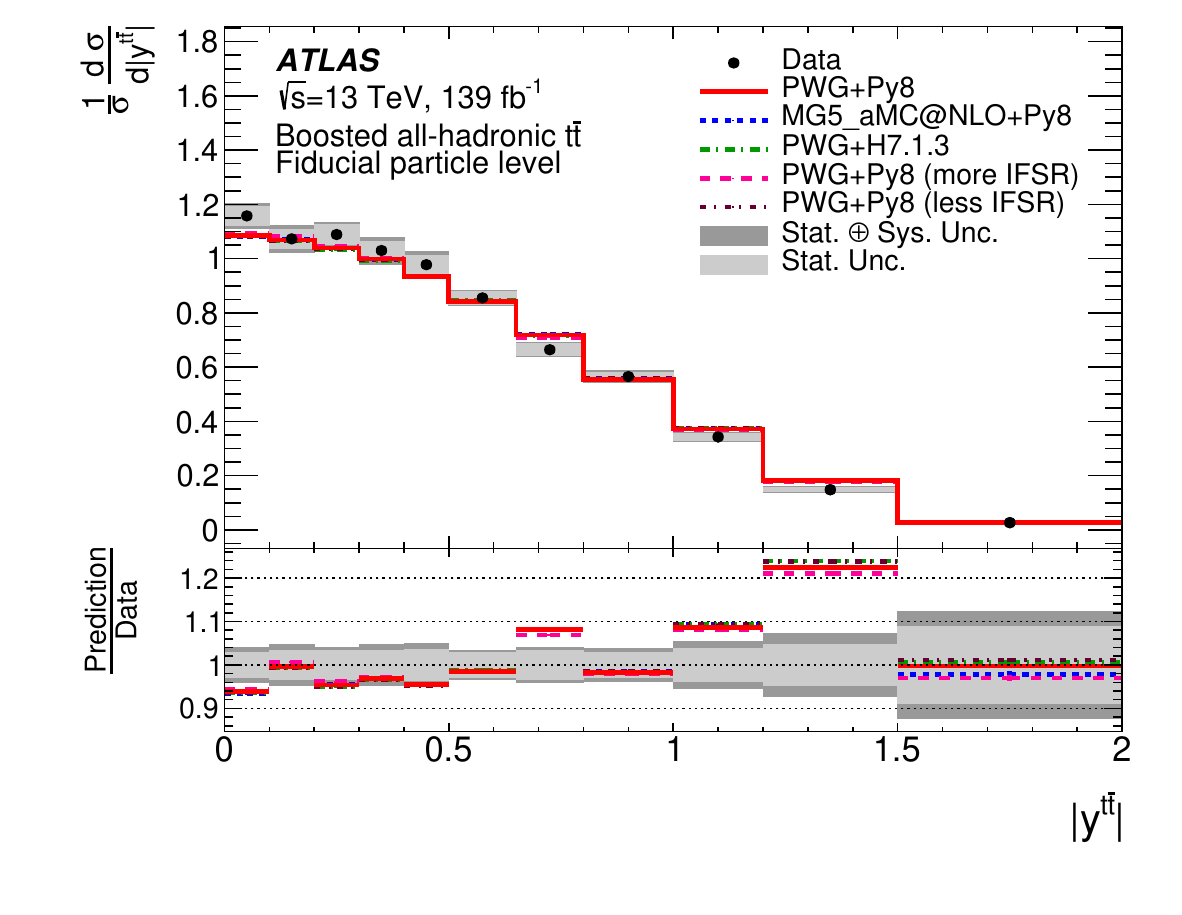}\label{fig:particle:tt_y:rel}}
\caption{
Normalized particle-level fiducial phase-space differential cross-sections as a function of the absolute value of the rapidity of
\subref{fig:particle:t1_y:rel}~the leading top-quark jet,
\subref{fig:particle:t2_y:rel}~the second-leading top-quark jet, and
\subref{fig:particle:tt_y:rel}~the \ttbar{}~system.
The dark and light grey bands indicate the total uncertainty and the statistical uncertainty, respectively, of the data in each bin.
Data points are placed at the centre of each bin.
The \POWPY[8] MC sample is used as the nominal prediction to correct the data to particle level.
}
\label{fig:particle:rapidity_observables:rel}
\end{figure*}

\begin{figure*}[htbp]
\centering
\subfigure[]{ \includegraphics[width=0.49\textwidth]{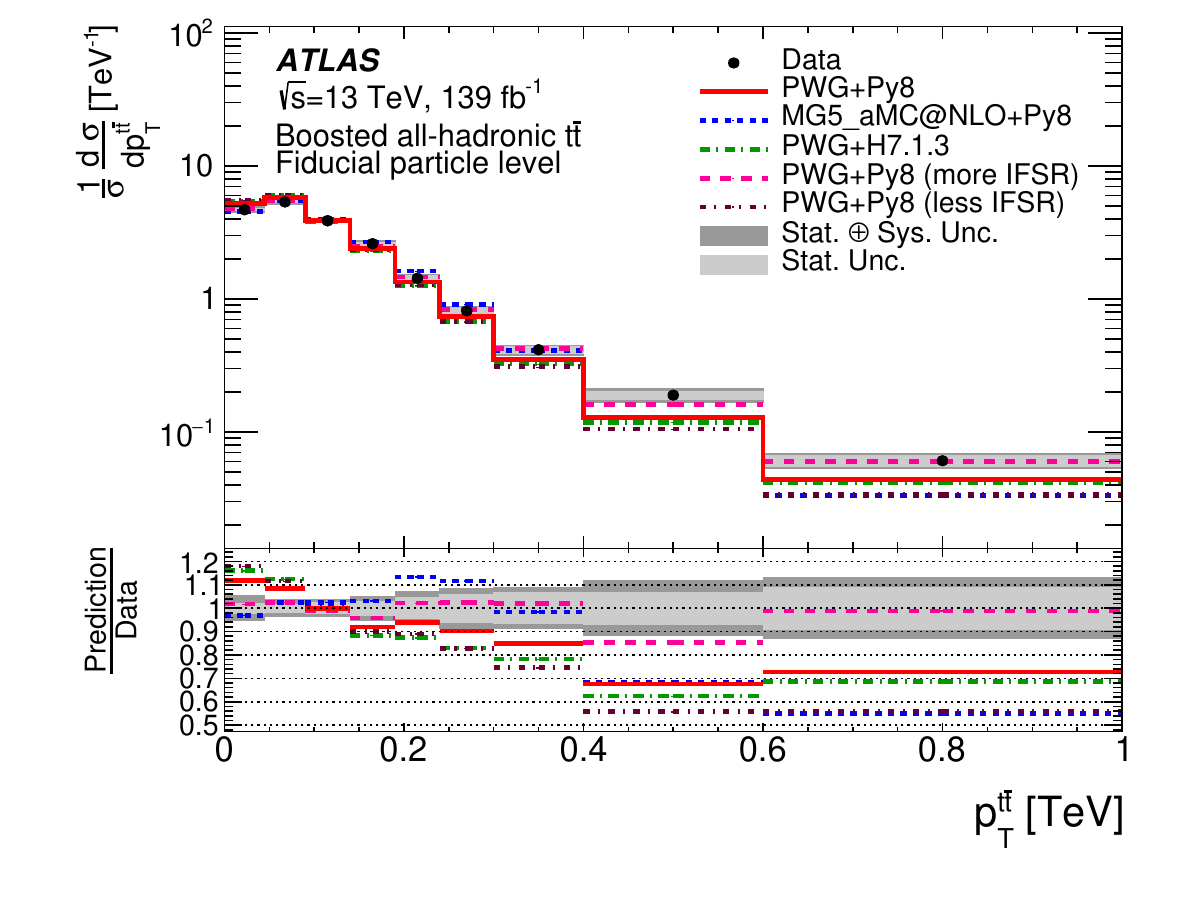}\label{fig:particle:tt_pt:rel}}
\subfigure[]{ \includegraphics[width=0.49\textwidth]{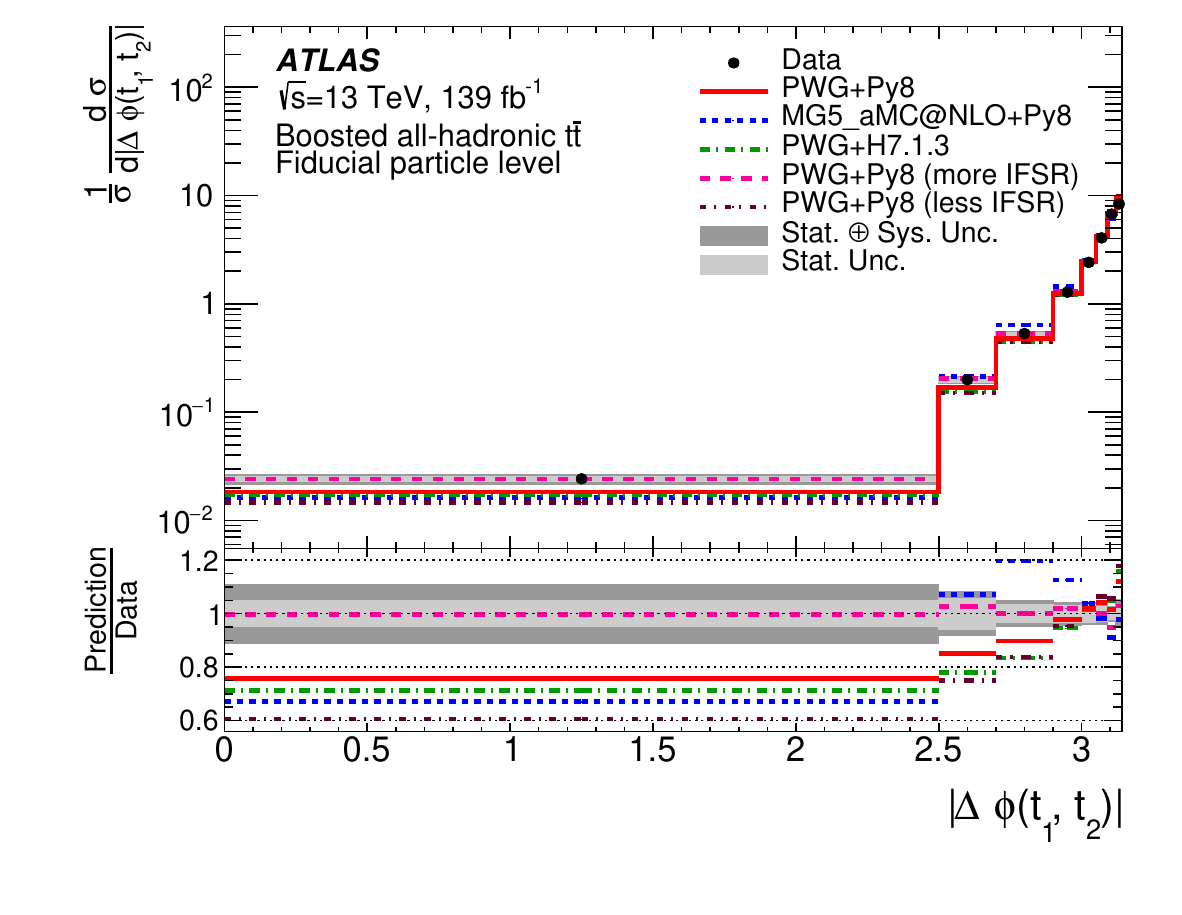}\label{fig:particle:tt_dPhittbar:rel}}
\subfigure[]{ \includegraphics[width=0.49\textwidth]{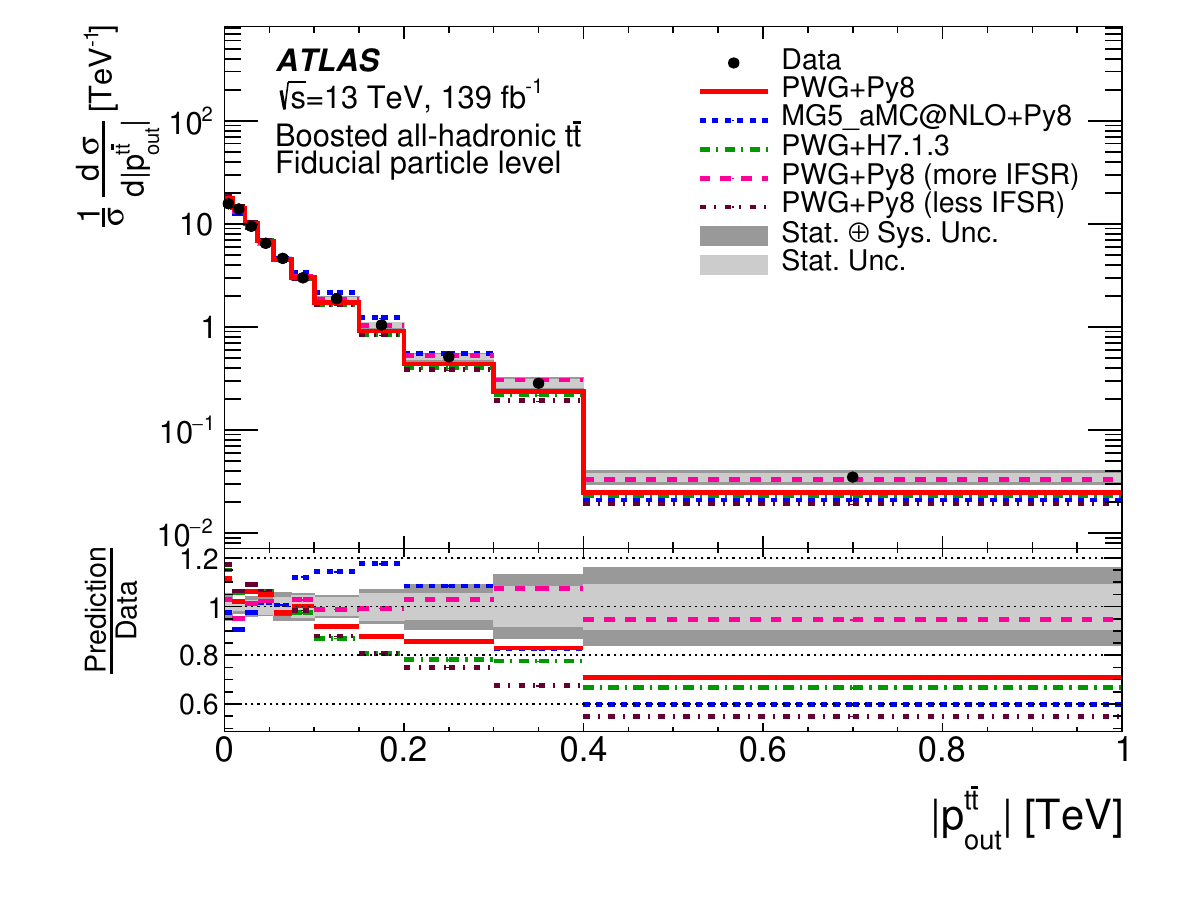}\label{fig:particle:tt_pout:rel}}
\caption{
Normalized particle-level fiducial phase-space differential cross-sections as a function of
\subref{fig:particle:tt_pt:rel}~the \pT of the \ttbar{} system, \ptttbar,
\subref{fig:particle:tt_dPhittbar:rel}~the azimuthal angle between the two top-quark jets, \deltaPhittbar, and
\subref{fig:particle:tt_pout:rel}~the absolute value of the out-of-plane momentum, \Poutttbar.
The dark and light grey bands indicate the total uncertainty and the statistical uncertainty, respectively, of the data in each bin.
Data points are placed at the centre of each bin.
The \POWPY[8] MC sample is used as the nominal prediction to correct the data to particle level.
}
\label{fig:particle:radiation_observables:rel}
\end{figure*}

The normalized particle-level fiducial phase-space
differential cross-sections for nine observables selected for comparison are presented
in Figures~\ref{fig:particle:energy_observables:rel}--\ref{fig:particle:radiation_observables:rel}.
Figure~\ref{fig:particle:energy_observables:rel} shows the differential cross-sections
for the \pT\ of the leading and second-leading top-quark jets,
and the invariant mass of the \ttbar\ system.
The differential cross-sections for the rapidity of the leading and second-leading top-quark jets,
and the rapidity of the \ttbar\ system are
shown in Figure~\ref{fig:particle:rapidity_observables:rel}.
Measured rapidity distributions have relatively large fluctuations between neighbouring bins.
These reflect fluctuations observed at the detector level and are consistent with statistical
uncertainties.
The differential cross-sections for observables sensitive to extra
radiation (the \pT of the \ttbar{} system, the out-of-plane momentum, and the
azimuthal separation of the top-quark jets) are shown in
Figure~\ref{fig:particle:radiation_observables:rel}.
The remaining distributions are presented in Appendix~\ref{sec:appendix:particle_level}.
 
\FloatBarrier
 
\begin{figure*}[htbp]
\centering
\subfigure[]{ \includegraphics[width=0.6\textwidth]{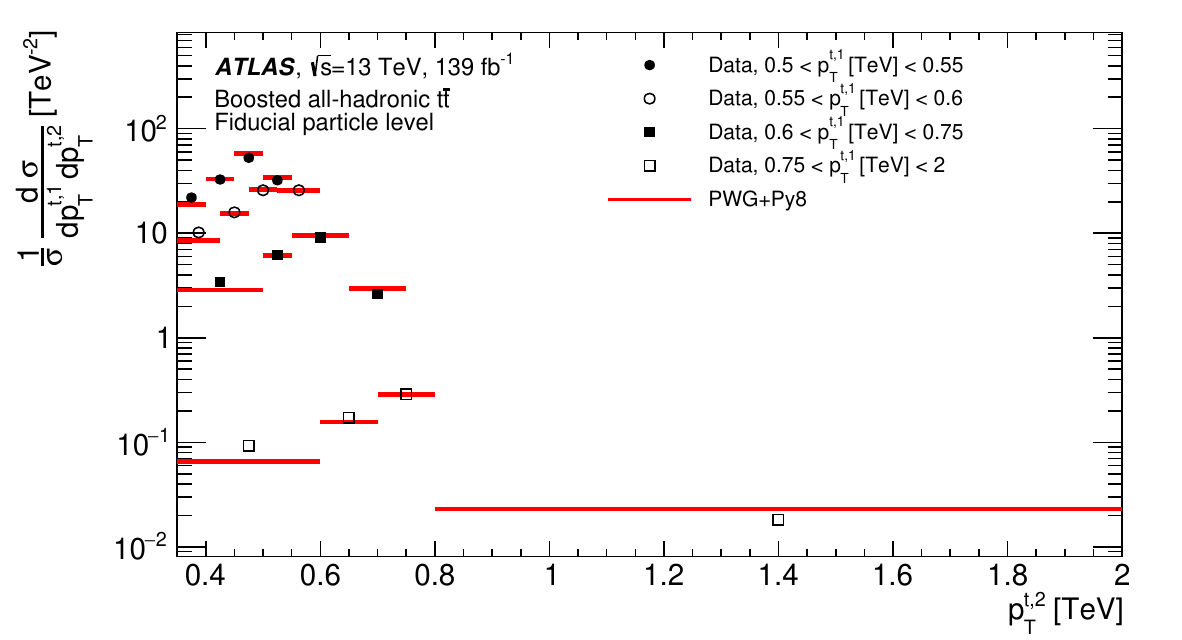}\label{fig:particle:t1_pt_vs_t2_pt:rel:shape}}
\subfigure[]{ \includegraphics[width=0.68\textwidth]{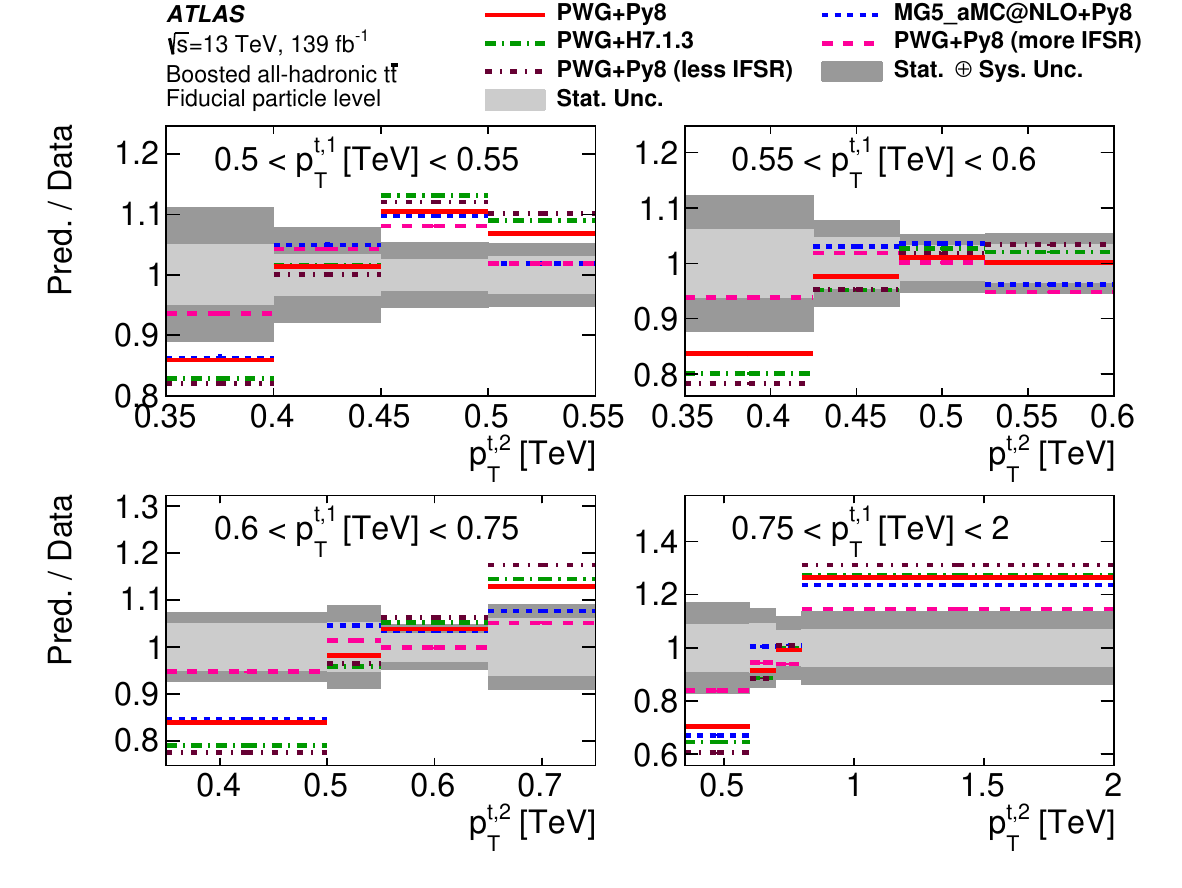}\label{fig:particle:t1_pt_vs_t2_pt:rel:ratio}}
\caption{
\subref{fig:particle:t1_pt_vs_t2_pt:rel:shape} Normalized particle-level fiducial phase-space double-differential cross-sections as a function of the transverse momenta of the leading and second-leading top-quark jets, compared with the \POWPY[8] calculation.
Data points are placed at the centre of each bin and the \POWPY[8] calculation is indicated by solid lines.
\subref{fig:particle:t1_pt_vs_t2_pt:rel:ratio}~The ratios of various MC calculations to the normalized particle-level fiducial phase-space differential cross-sections.
The dark and light grey bands indicate the total uncertainty and the statistical uncertainty, respectively, of the data in each bin.
}
\label{fig:particle:t1_pt_vs_t2_pt:rel}
\end{figure*}

\begin{figure*}[htbp]
\centering
\subfigure[]{ \includegraphics[width=0.6\textwidth]{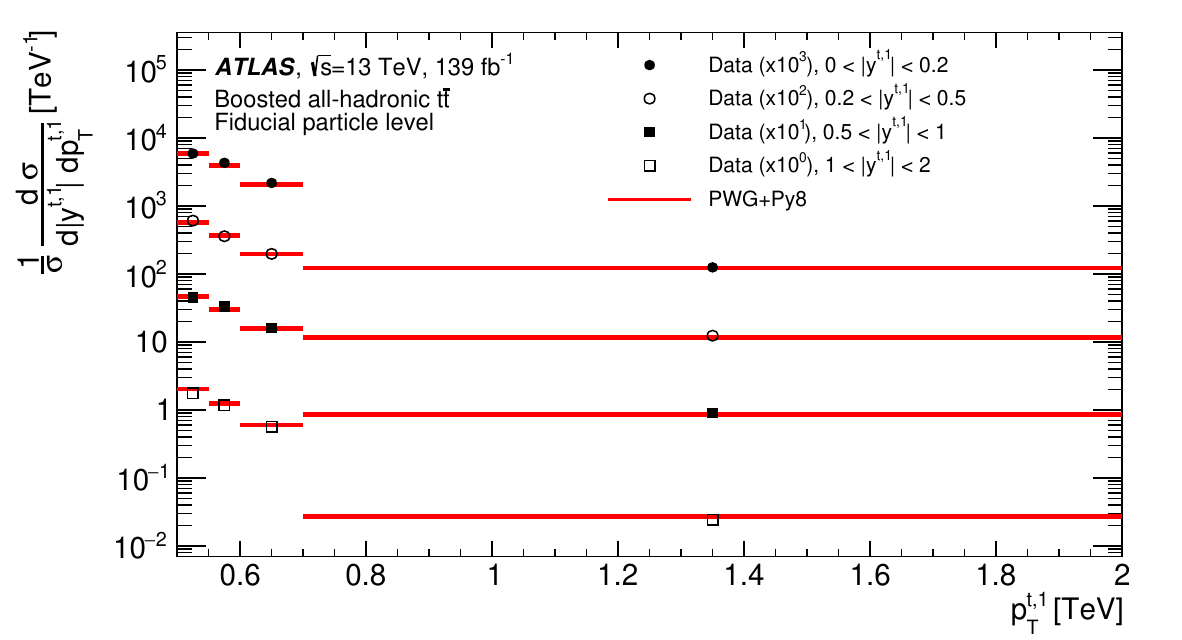}\label{fig:particle:t1_y_vs_t1_pt:rel:shape}}
\subfigure[]{ \includegraphics[width=0.68\textwidth]{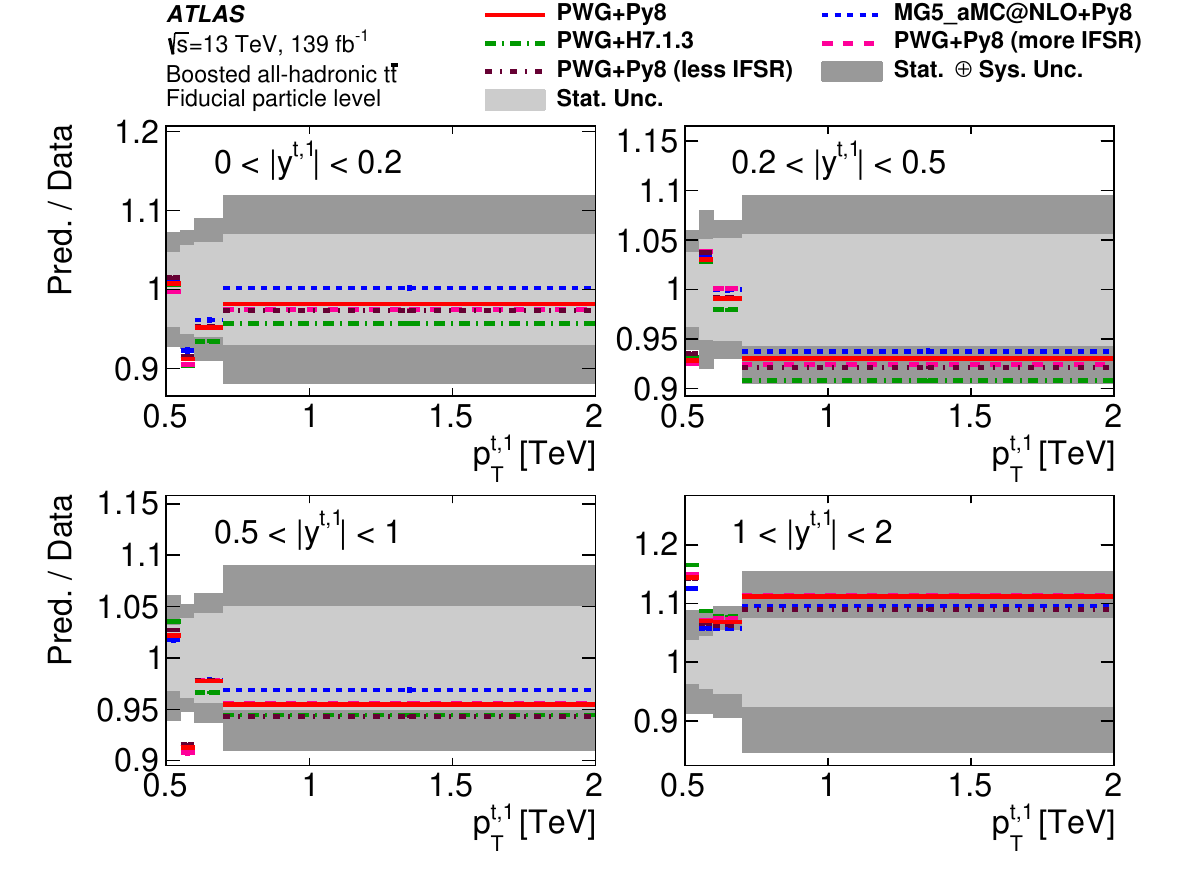}\label{fig:particle:t1_y_vs_t1_pt:rel:ratio}}
\caption{
\subref{fig:particle:t1_y_vs_t1_pt:rel:shape} Normalized particle-level fiducial phase-space double-differential cross-sections as a function of the absolute value of the rapidity and \pT of the leading top-quark jet, compared  with the \POWPY[8] calculation.
Data points are placed at the centre of each bin and the \POWPY[8] calculation is indicated by solid lines.
The measurement and the prediction are normalized by the factors shown in parentheses to aid visibility.
\subref{fig:particle:t1_y_vs_t1_pt:rel:ratio}~The ratios of various MC calculations to the normalized particle-level fiducial phase-space differential cross-sections.
The dark and light grey bands indicate the total uncertainty and the statistical uncertainty, respectively, of the data in each bin.
}
\label{fig:particle:t1_y_vs_t1_pt:rel}
\end{figure*}

\begin{figure*}[htbp]
\centering
\subfigure[]{ \includegraphics[width=0.6\textwidth]{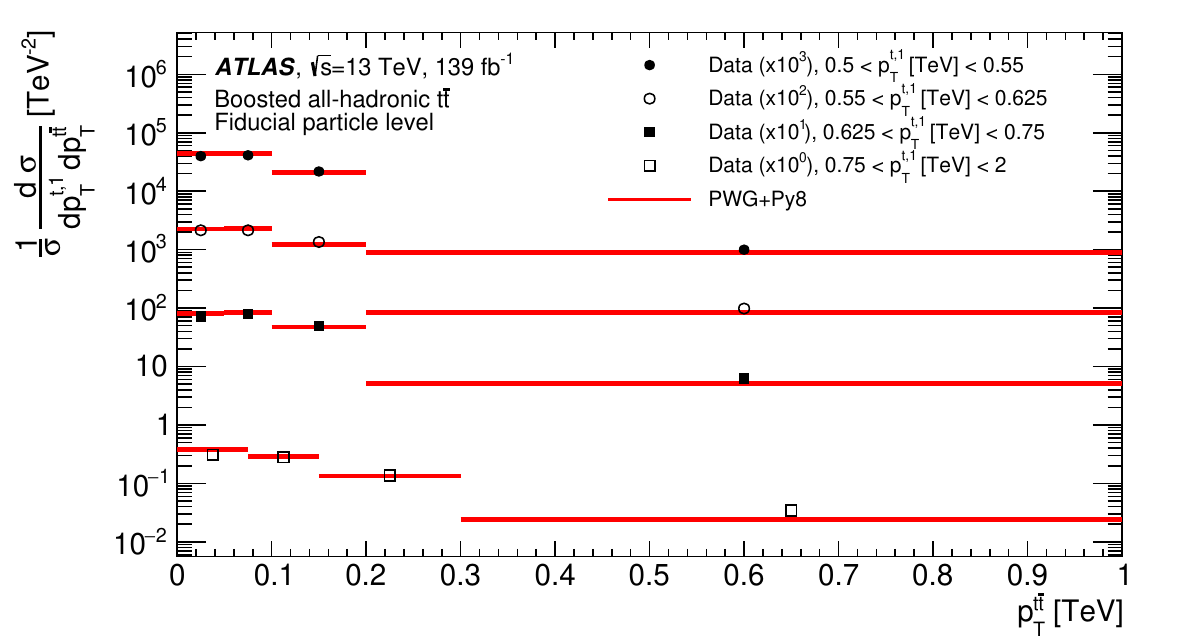}\label{fig:particle:t1_pt_vs_ttbar_pt:rel:shape}}
\subfigure[]{ \includegraphics[width=0.68\textwidth]{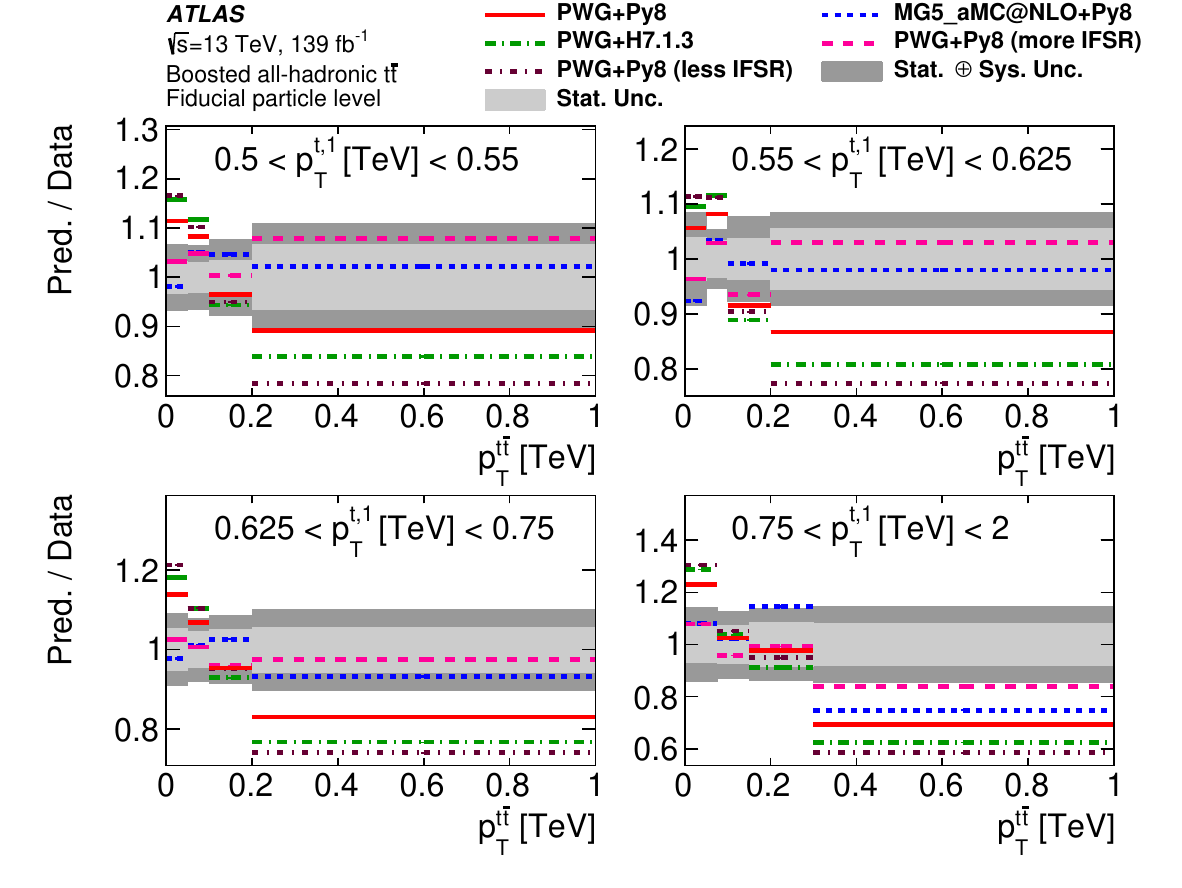}\label{fig:particle:t1_pt_vs_ttbar_pt:rel:ratio}}
\caption{
\subref{fig:particle:t1_pt_vs_ttbar_pt:rel:shape} Normalized particle-level fiducial phase-space double-differential cross-sections as a function of the \pT of the leading top-quark jet and the \pT of the \ttbar final state, \ptttbar,  compared  with the \POWPY[8] calculation.
Data points are placed at the centre of each bin and the \POWPY[8] calculation is indicated by solid lines.
The measurement and the prediction are normalized by the factors shown in parentheses to aid visibility.
\subref{fig:particle:t1_pt_vs_ttbar_pt:rel:ratio}~The ratios of various MC calculations to the normalized particle-level fiducial phase-space differential cross-sections.
The dark and light grey bands indicate the total uncertainty and the statistical uncertainty, respectively, of the data in each bin.
}
\label{fig:particle:t1_pt_vs_ttbar_pt:rel}
\end{figure*}

\begin{figure*}[htbp]
\centering
\subfigure[]{ \includegraphics[width=0.6\textwidth]{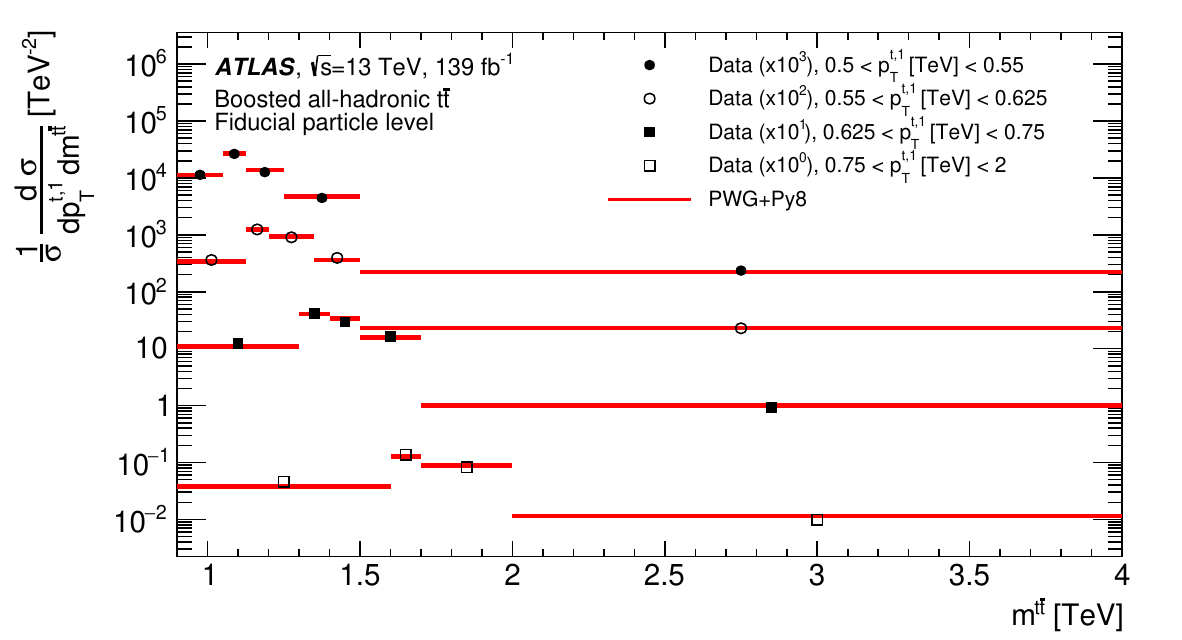}\label{fig:particle:t1_pt_vs_ttbar_mass:rel:shape}}
\subfigure[]{ \includegraphics[width=0.68\textwidth]{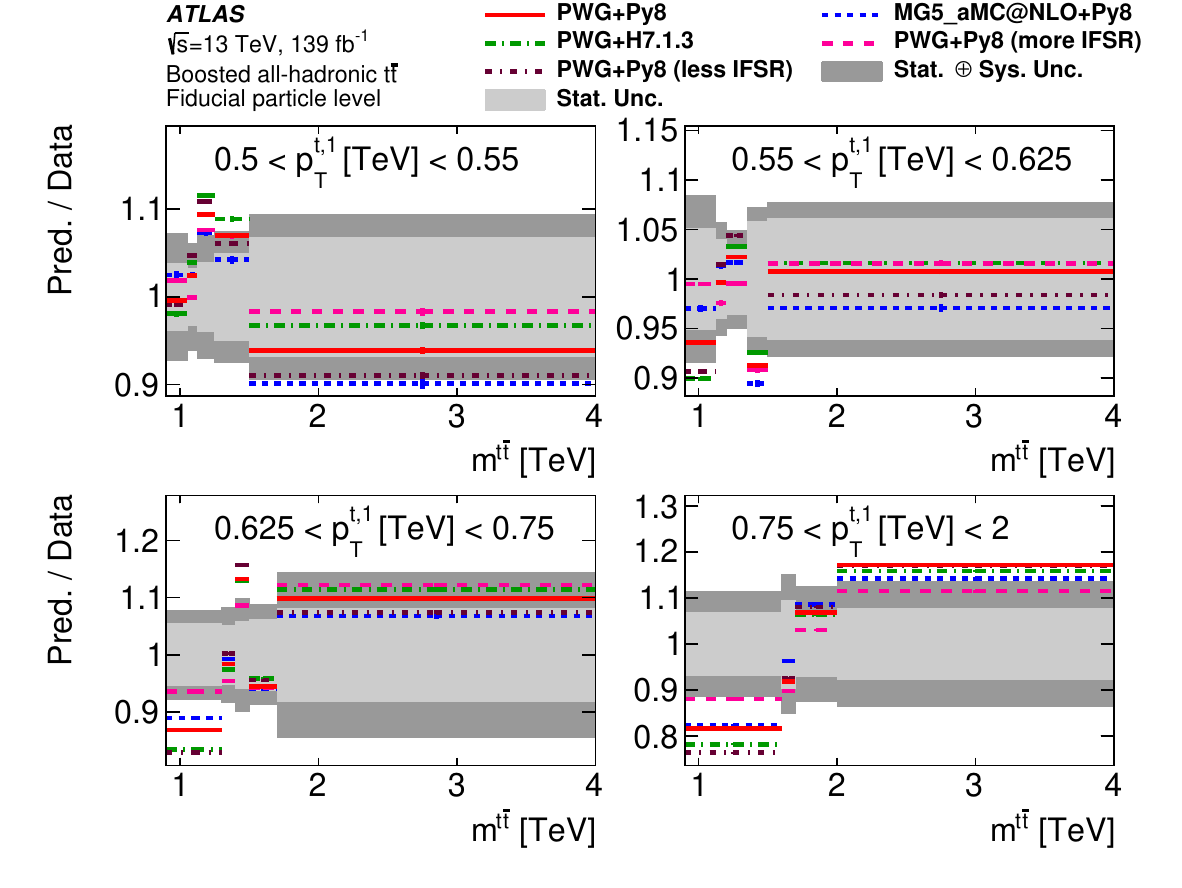}\label{fig:particle:t1_pt_vs_ttbar_mass:rel:ratio}}
\caption{
\subref{fig:particle:t1_pt_vs_ttbar_mass:rel:shape} Normalized particle-level fiducial phase-space double-differential cross-sections as a function of the \pT of the leading top-quark jet and the invariant mass of the \ttbar final state, \mttbar, compared  with the \POWPY[8] calculation.
Data points are placed at the centre of each bin and the \POWPY[8] calculation is indicated by solid lines.
The measurement and the prediction are normalized by the factors shown in parentheses to aid visibility.
\subref{fig:particle:t1_pt_vs_ttbar_mass:rel:ratio}~The ratios of various MC calculations to the normalized particle-level fiducial phase-space differential cross-sections.
The dark and light grey bands indicate the total uncertainty and the statistical uncertainty, respectively, of the data in each bin.
}
\label{fig:particle:t1_pt_vs_ttbar_mass:rel}
\end{figure*}

\begin{figure*}[htbp]
\centering
\subfigure[]{ \includegraphics[width=0.6\textwidth]{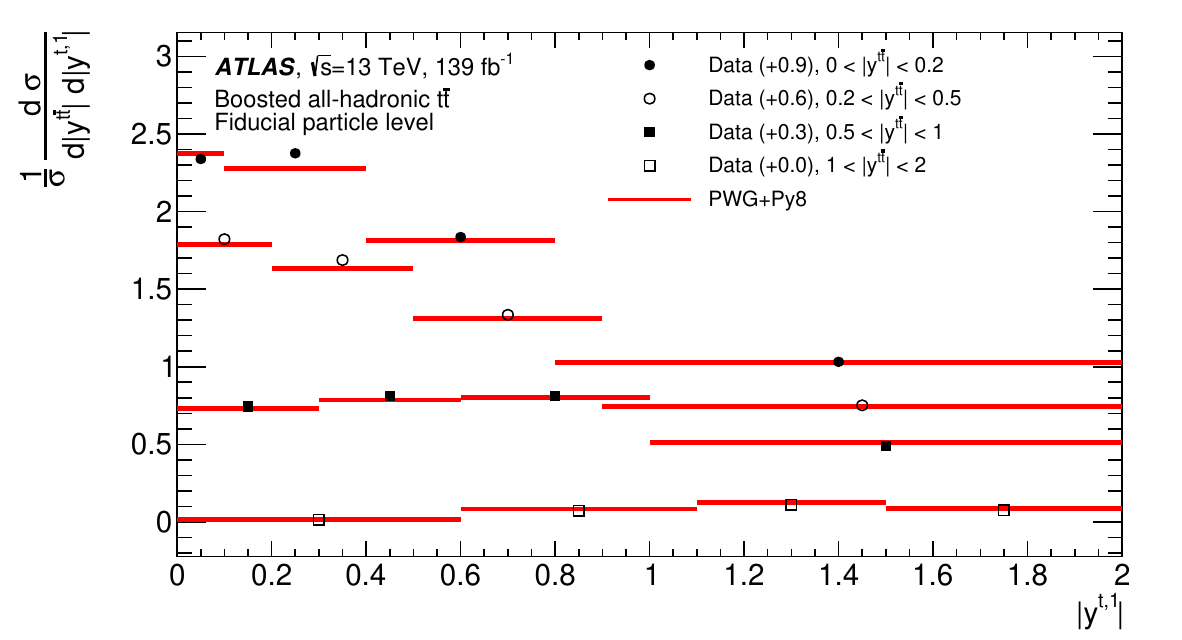}\label{fig:particle:ttbar_y_vs_t1_y:rel:shape}}
\subfigure[]{ \includegraphics[width=0.68\textwidth]{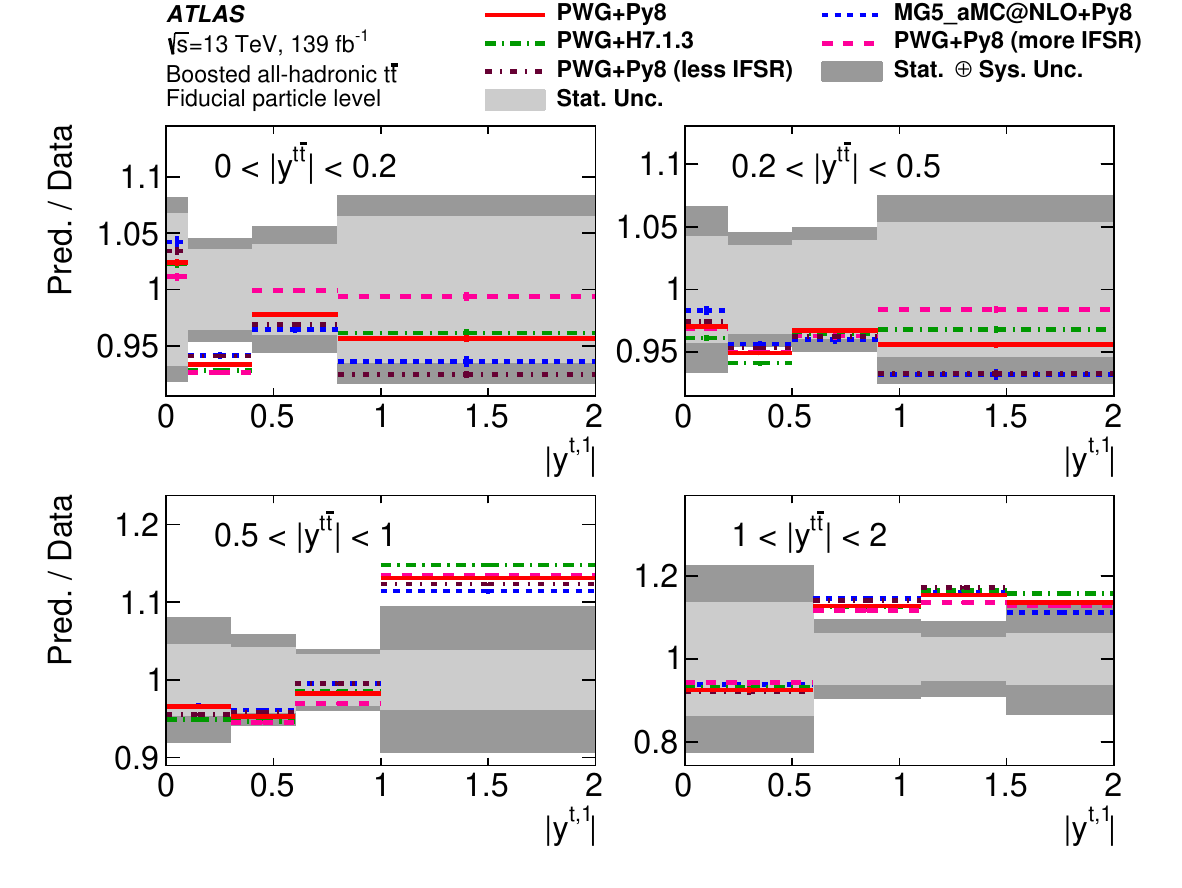}\label{fig:particle:ttbar_y_vs_t1_y:rel:ratio}}
\caption{
\subref{fig:particle:ttbar_y_vs_t1_y:rel:shape} Normalized particle-level fiducial phase-space double-differential cross-sections as a function of the absolute value of the rapidity of the \ttbar final state, \absyttbar, and the absolute value of the rapidity of the leading top-quark jet, compared  with the \POWPY[8] calculation.
Data points are placed at the centre of each bin and the \POWPY[8] calculation is indicated by solid lines.
The measurement and the prediction are shifted by the factors shown in parentheses to aid visibility.
\subref{fig:particle:ttbar_y_vs_t1_y:rel:ratio}~The ratios of various MC calculations to the normalized particle-level fiducial phase-space differential cross-sections.
The dark and light grey bands indicate the total uncertainty and the statistical uncertainty, respectively, of the data in each bin.
}
\label{fig:particle:ttbar_y_vs_t1_y:rel}
\end{figure*}

\begin{figure*}[htbp]
\centering
\subfigure[]{ \includegraphics[width=0.6\textwidth]{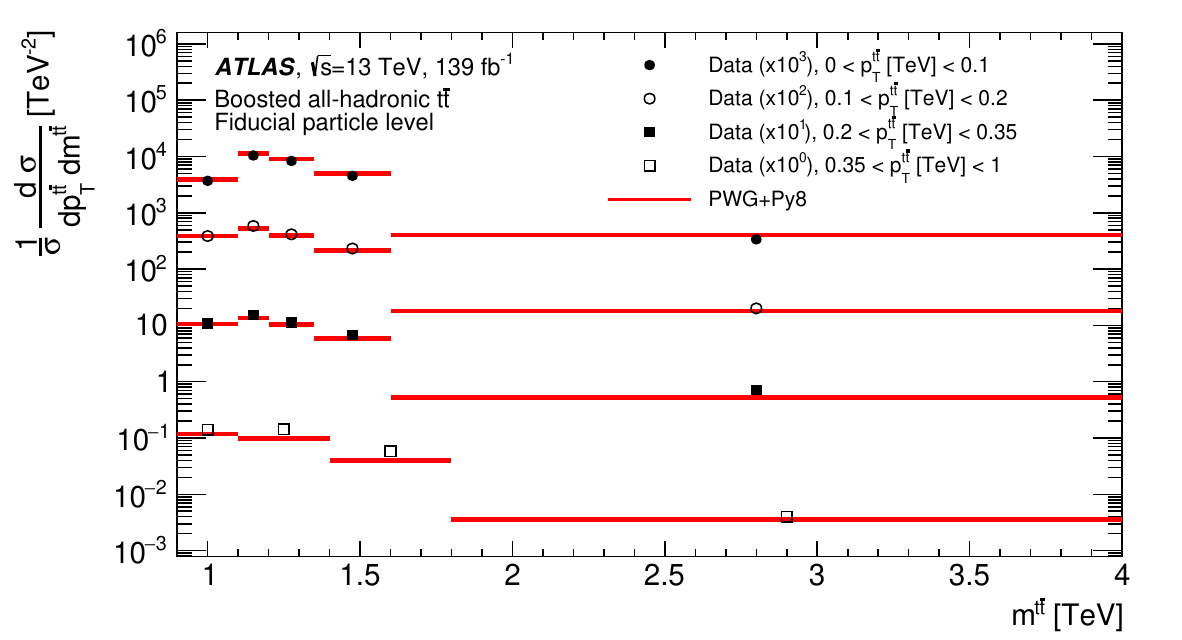}\label{fig:particle:ttbar_pt_vs_ttbar_mass:rel:shape}}
\subfigure[]{ \includegraphics[width=0.68\textwidth]{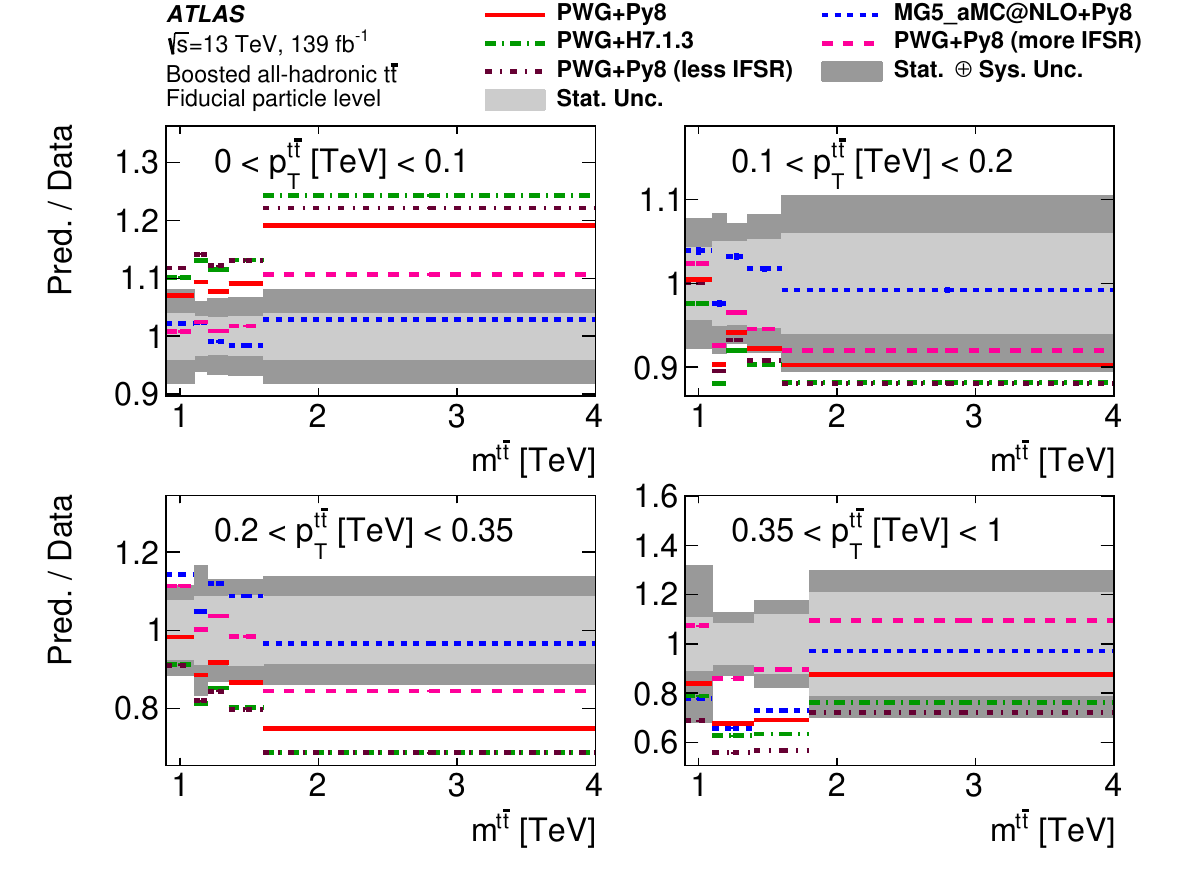}\label{fig:particle:ttbar_pt_vs_ttbar_mass:rel:ratio}}
\caption{
\subref{fig:particle:ttbar_pt_vs_ttbar_mass:rel:shape} Normalized particle-level fiducial phase-space double-differential cross-sections as a function of the \pT and the mass of the \ttbar\ final state, \ptttbar\ and \mttbar, compared  with the \POWPY[8] calculation.
Data points are placed at the centre of each bin and the \POWPY[8] calculation is indicated by solid lines.
The measurement and the prediction are normalized by the factors shown in parentheses to aid visibility.
\subref{fig:particle:ttbar_pt_vs_ttbar_mass:rel:ratio}~The ratios of various MC calculations to the normalized particle-level fiducial phase-space differential cross-sections.
The dark and light grey bands indicate the total uncertainty and the statistical uncertainty, respectively, of the data in each bin.
}
\label{fig:particle:ttbar_pt_vs_ttbar_mass:rel}
\end{figure*}

\begin{figure*}[htbp]
\centering
\subfigure[]{ \includegraphics[width=0.6\textwidth]{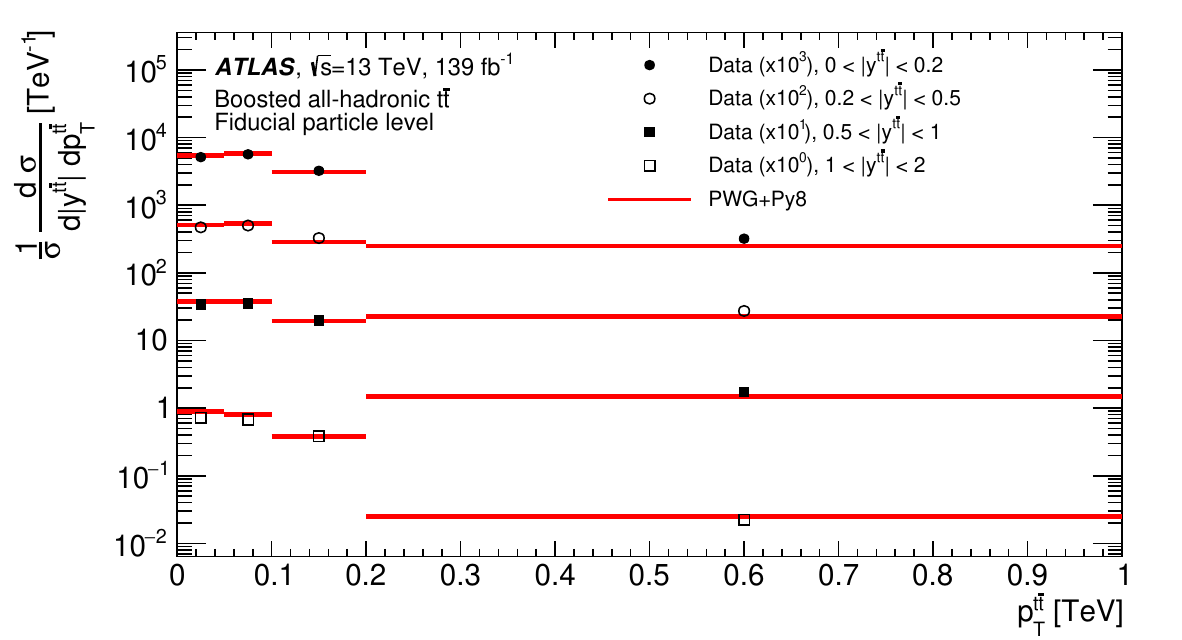}\label{fig:particle:ttbar_y_vs_ttbar_pt:rel:shape}}
\subfigure[]{ \includegraphics[width=0.68\textwidth]{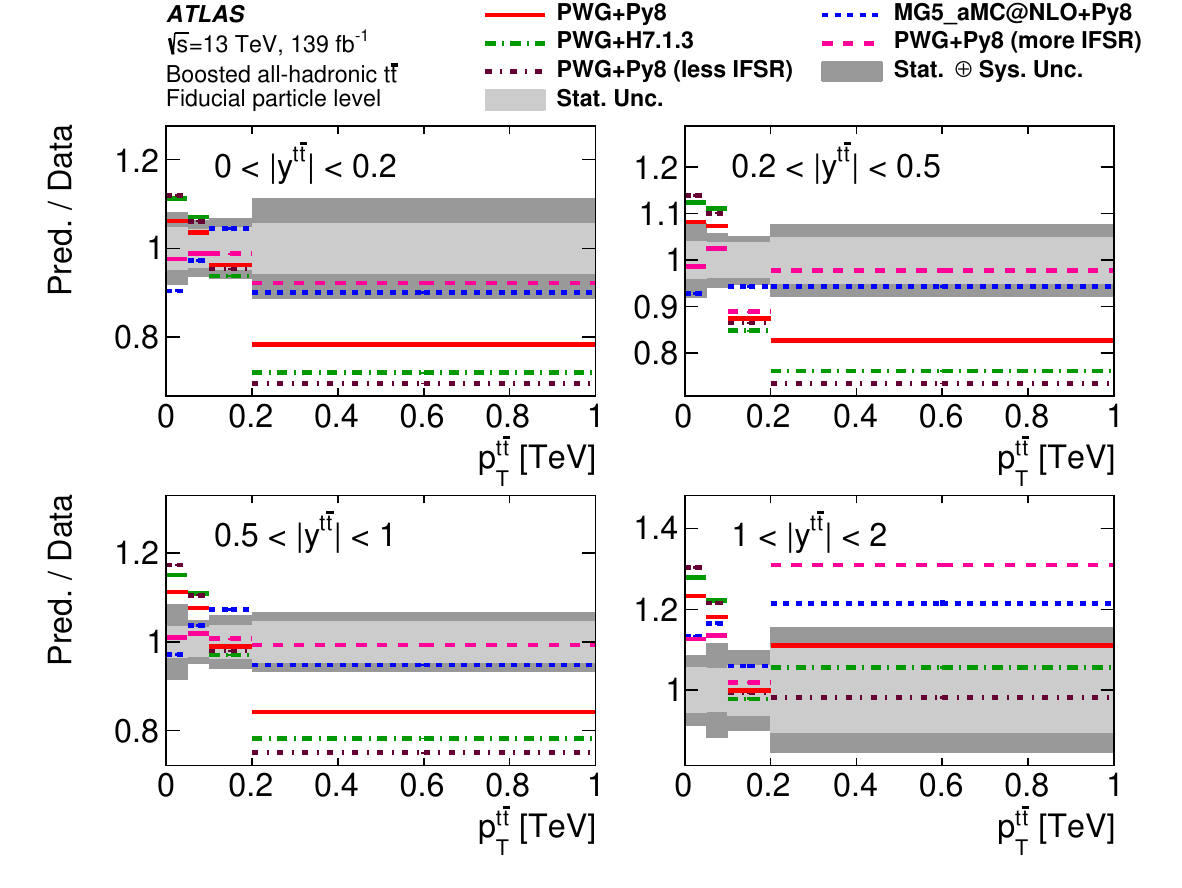}\label{fig:particle:ttbar_y_vs_ttbar_pt:rel:ratio}}
\caption{
\subref{fig:particle:ttbar_y_vs_ttbar_pt:rel:shape} Normalized particle-level fiducial phase-space double-differential cross-sections as a function of the absolute value of the rapidity and the \pT of the \ttbar\ final state, \absyttbar\ and \ptttbar, compared  with the \POWPY[8] calculation.
Data points are placed at the centre of each bin and the \POWPY[8] calculation is indicated by solid lines.
The measurement and the prediction are normalized by the factors shown in parentheses to aid visibility.
\subref{fig:particle:ttbar_y_vs_ttbar_pt:rel:ratio}~The ratios of various MC calculations to the normalized particle-level fiducial phase-space differential cross-sections.
The dark and light grey bands indicate the total uncertainty and the statistical uncertainty, respectively, of the data in each bin.
}
\label{fig:particle:ttbar_y_vs_ttbar_pt:rel}
\end{figure*}

\begin{figure*}[htbp]
\centering
\subfigure[]{ \includegraphics[width=0.45\textwidth]{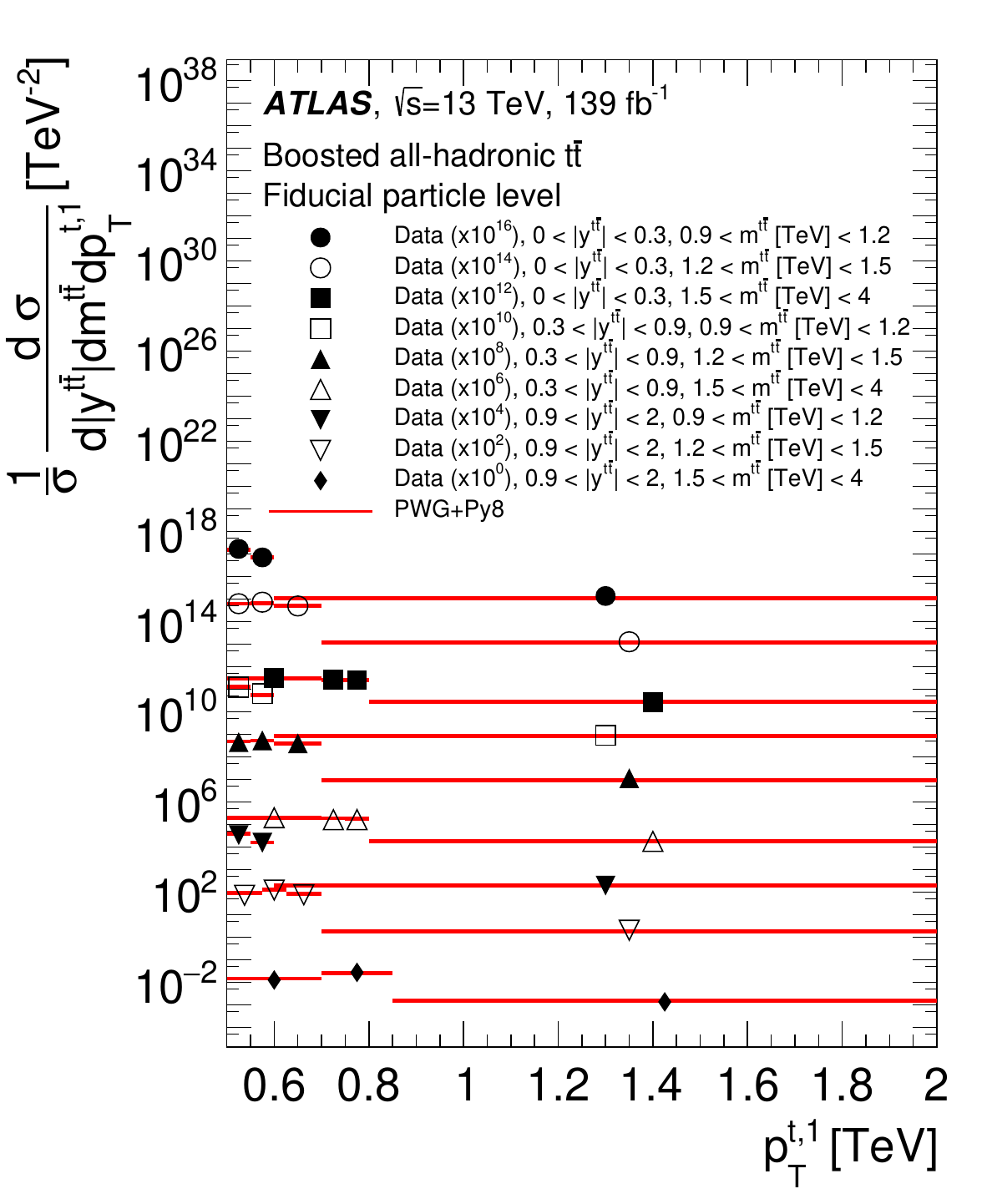}\label{fig:particle:ttbar_y_vs_ttbar_mass_vs_t1_pt:rel:shape}}
\subfigure[]{ \includegraphics[width=0.85\textwidth]{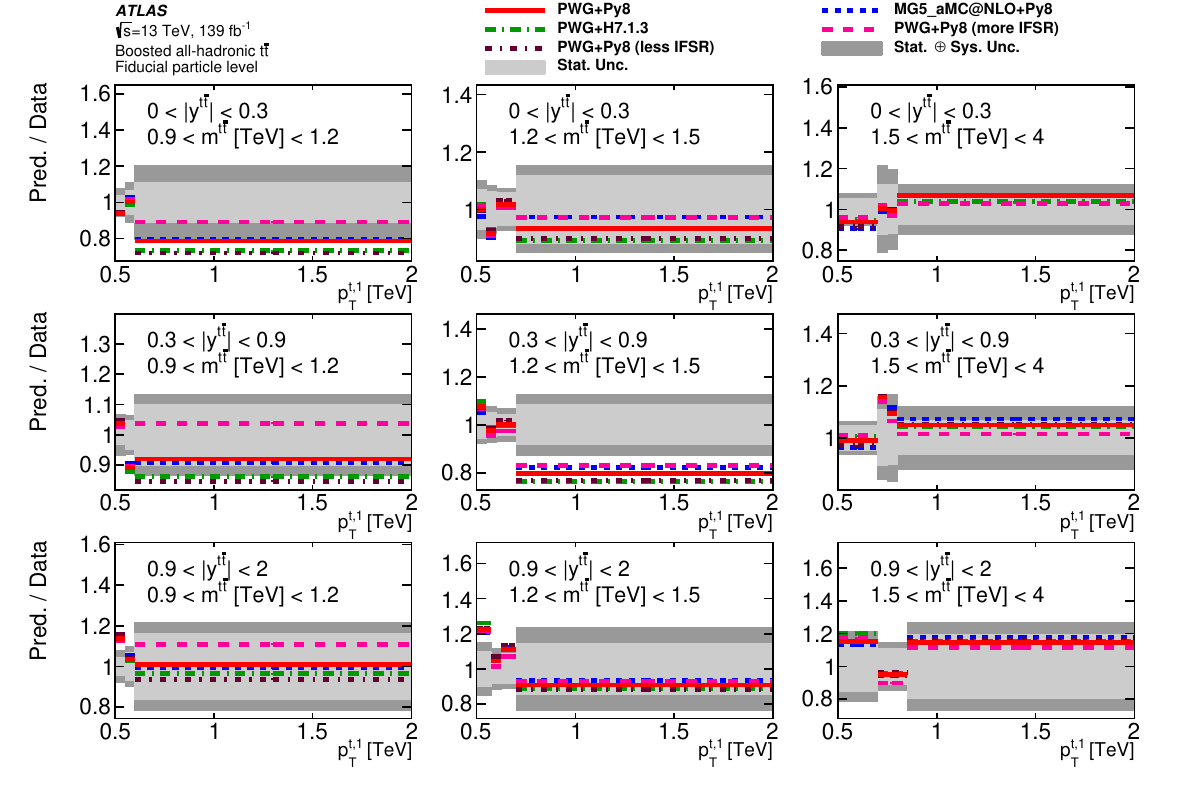}\label{fig:particle:ttbar_y_vs_ttbar_mass_vs_t1_pt:rel:ratio}}
\caption{
\subref{fig:particle:ttbar_y_vs_ttbar_mass_vs_t1_pt:rel:shape} Normalized particle-level fiducial phase-space triple-differential cross-sections as a function of the absolute value of the rapidity of the \ttbar\ final state, \absyttbar, the mass of the \ttbar\ final state, \mttbar, and the \pT\ of the leading top-quark jet, compared  with the \POWPY[8] calculation.
Data points are placed at the centre of each bin and the \POWPY[8] calculation is indicated by solid lines.
The measurement and the prediction are normalized by the factors shown in parentheses to aid visibility.
\subref{fig:particle:ttbar_y_vs_ttbar_mass_vs_t1_pt:rel:ratio}~The ratios of various MC calculations to the normalized particle-level fiducial phase-space differential cross-sections.
The dark and light grey bands indicate the total uncertainty and the statistical uncertainty, respectively, of the data in each bin.
}
\label{fig:particle:ttbar_y_vs_ttbar_mass_vs_t1_pt:rel}.
\end{figure*}

For a subset of the observables, pairs and triplets of variables are chosen to form double- and triple-differential
cross-sections.
These combinations of observables test specific aspects of the QCD predictions,
where particular combinations have correlations that potentially differentiate between models.
The selected set of fiducial phase-space double- and triple-differential cross-sections are
shown in Figures~\ref{fig:particle:t1_pt_vs_t2_pt:rel}--\ref{fig:particle:ttbar_y_vs_ttbar_mass_vs_t1_pt:rel}.
Additional double-differential cross-sections are presented in Appendix~\ref{sec:appendix:particle_level}.
 
\FloatBarrier
\subsection{Total parton-level cross-section in the fiducial phase-space}
\label{sec:parton_level_total_cross_section}
 
The measurement of the parton-level fiducial phase-space cross-section
is performed as a single-bin unfolding to the parton-level phase space.
This results in
\begin{equation}
\sigma^{\ttbar,\text{fid}}_\text{parton} = \inclXsecParton\ {\rm pb},   \nonumber
\end{equation}
where a correction has been made for the \ttbar\ branching fraction to the all-hadronic
final state.
 
\begin{figure*}[htbp]
\centering
\includegraphics[width=0.7\textwidth]{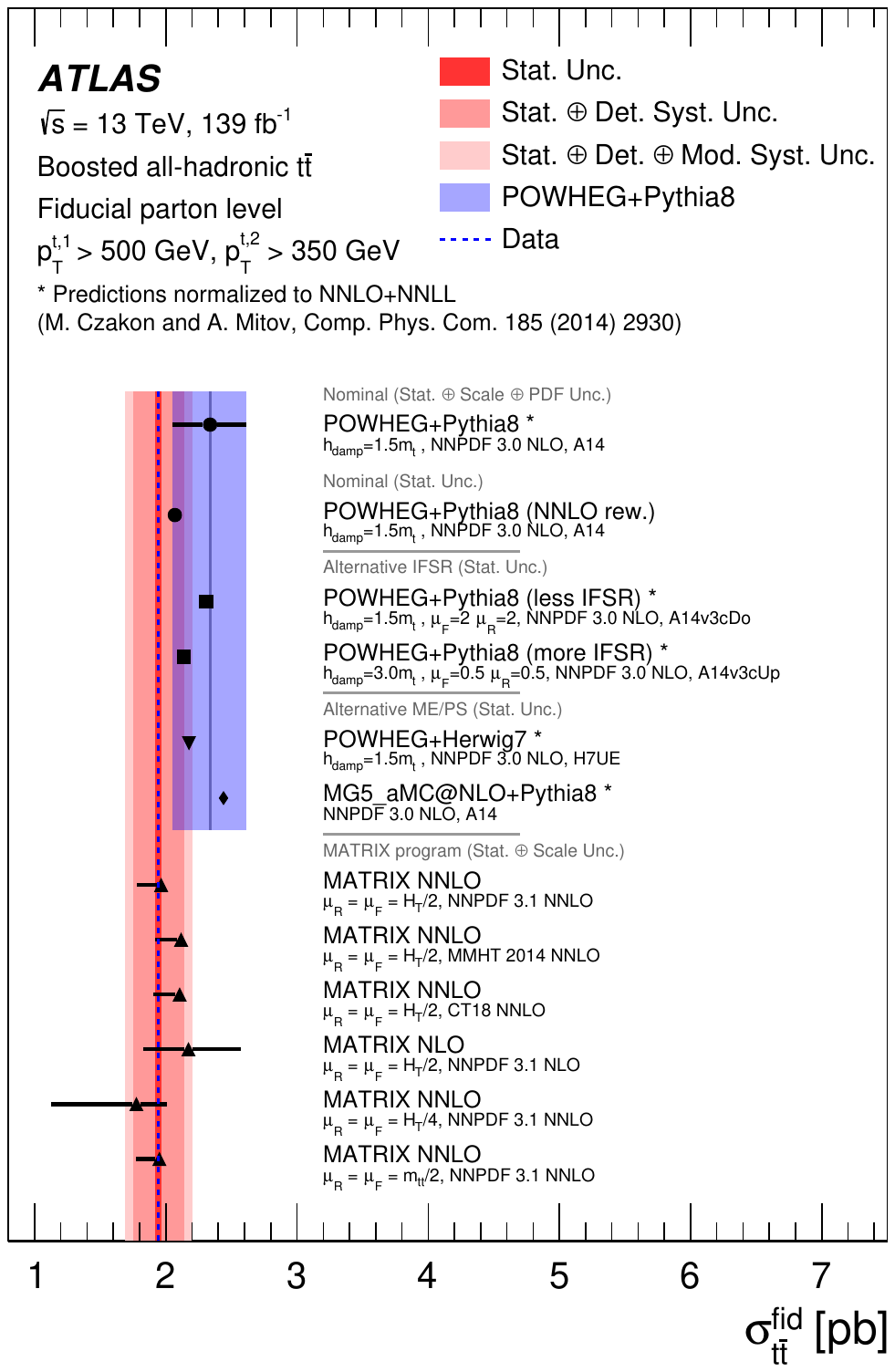}
\caption{
Comparison of the parton-level fiducial phase-space cross-section with the calculations from several MC
generators and the fixed-order NNLO prediction with various PDF sets and dynamical scales obtained
using the \MATRIX program.
A correction has been made for the \ttbar\ branching fraction to the all-hadronic final state.
The shaded (red) bands indicate the statistical, detector, and modelling uncertainties in the measurement.
The uncertainty associated with the \POWPY[8] signal model (blue band) includes the statistical, scale, PDF, and NNLO+NNLL total inclusive prediction uncertainty.
Other MC calculations show only the statistical uncertainty of the MC calculation, which is negligible and not visible in the figure.
The fixed-order NNLO calculations include the scale uncertainties, which are asymmetric, and the statistical uncertainties.}
\label{fig:parton:inclusive:abs}
\end{figure*}

The measured cross-section can be compared with the cross-section
calculation of $\numRF{2.338}{3} \pm \numRF{0.282}{2}$~pb by
the \POWPY[8] calculation after normalizing
its full phase-space calculation to the NNLO+NNLL \ttbar cross-section.
It can be also compared with the nominal fixed-order NNLO cross-section calculation
of $\numRF{1.9645}{3} \numpmRP{+0.0188}{-0.174}{2}$~pb obtained using the \MATRIX program, described in Section~\ref{sec:datasets}.
The  \POWPY[8] associated uncertainty includes the statistical, scale, PDF, and NNLO+NNLL total inclusive calculation uncertainty, while the NNLO calculation includes the scale uncertainties, which are asymmetric, and the statistical uncertainties.
Figure~\ref{fig:parton:inclusive:abs}\ compares the measured
parton-level cross-section with various MC NLO calculations and also with the fixed-order NNLO calculation
for various PDF sets and dynamical scales.
 
The difference between the particle-level and parton-level total cross-sections stems mainly from
correcting the parton-level cross-section for the \ttbar\ branching fraction to the all-hadronic final state,
the particle-level requirements on the leading and second-leading \largeR\ jet masses, and
the matching of $b$-hadrons to \largeR\ jets.
 
\FloatBarrier
\subsection{Parton-level differential cross-sections}
\label{sec:results:parton}
 
\begin{figure*}[htbp]
\centering
\subfigure[]{ \includegraphics[width=0.49\textwidth]{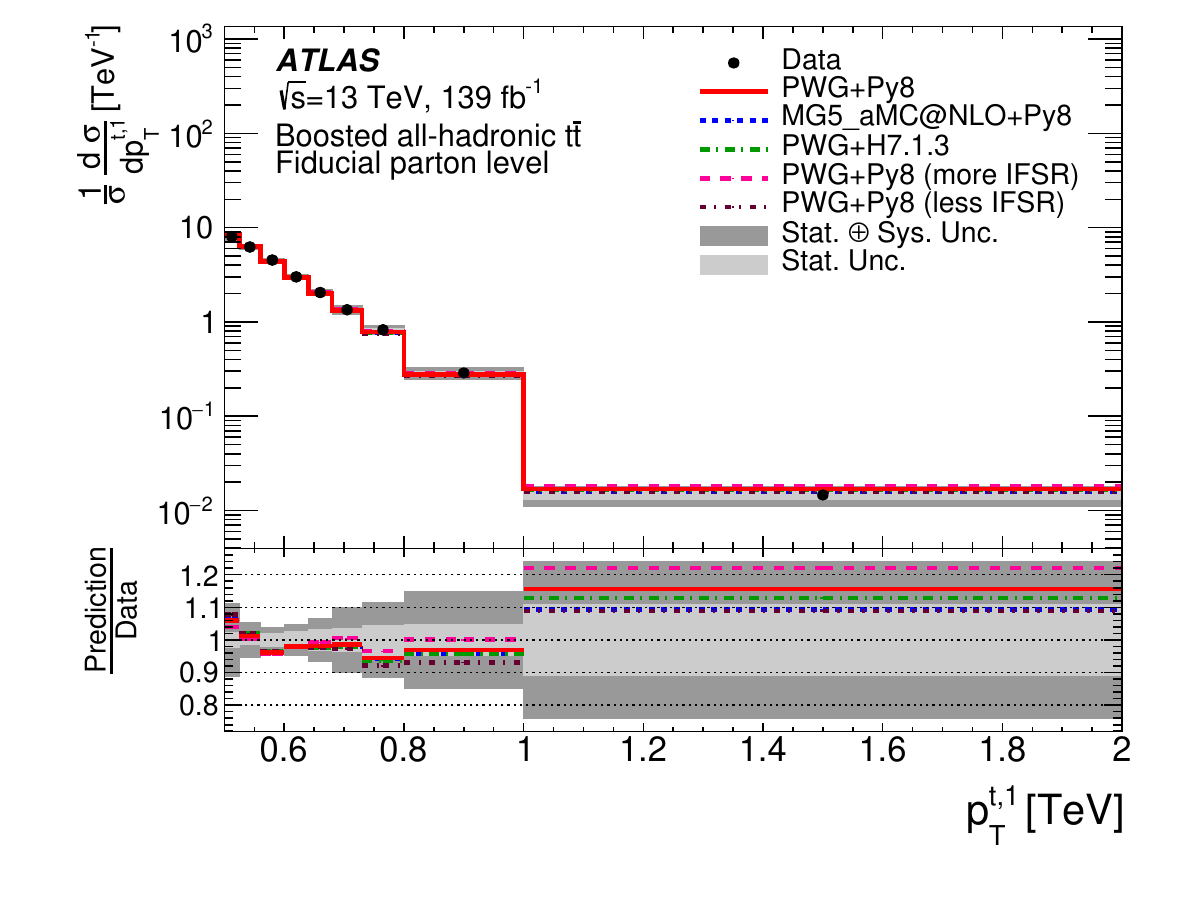}\label{fig:parton:t1_pt:rel}}
\subfigure[]{ \includegraphics[width=0.49\textwidth]{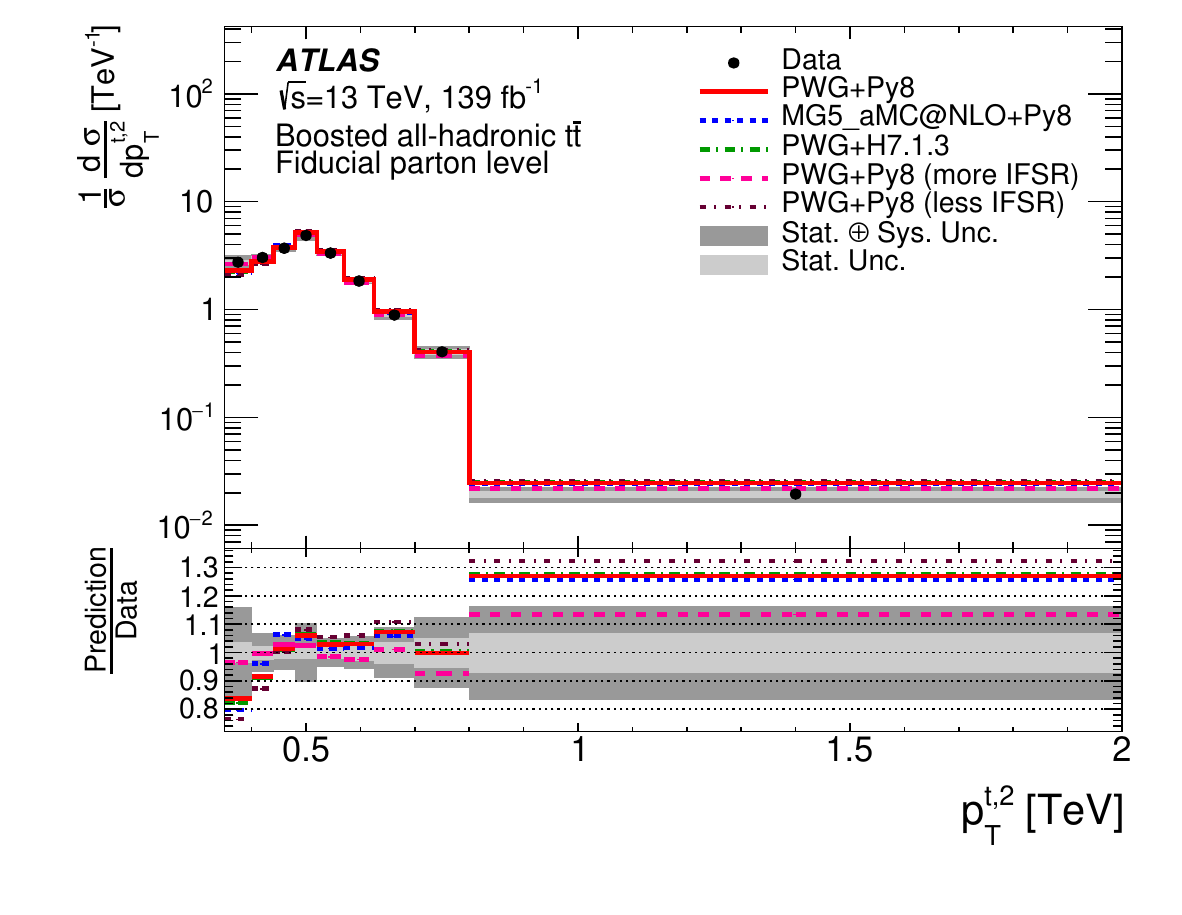}\label{fig:parton:t2_pt:rel}}
\subfigure[]{ \includegraphics[width=0.49\textwidth]{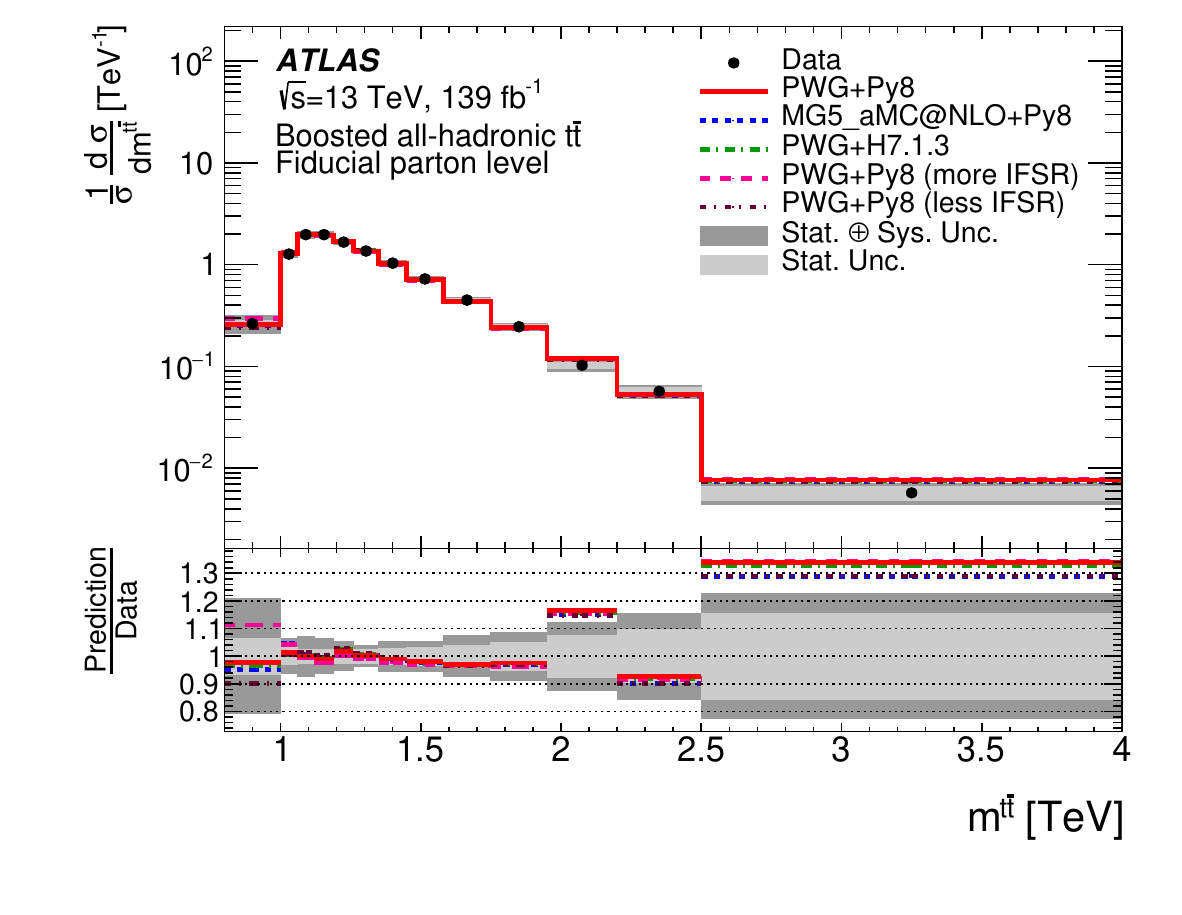}\label{fig:parton:tt_m:rel}}
\caption{
Normalized parton-level fiducial phase-space differential cross-sections as a function of
\subref{fig:parton:t1_pt:rel}~the \pT of the leading top quark,
\subref{fig:parton:t2_pt:rel}~the \pT of the second-leading top quark,
and \subref{fig:parton:tt_m:rel}~the invariant mass of the \ttbar{} system.
The dark and light grey bands indicate the total uncertainty and the statistical uncertainty, respectively, of the data in each bin.
Data points are placed at the centre of each bin.
The \POWPY[8] MC sample is used as the nominal prediction to correct the data to parton level.}
\label{fig:parton:energy_observables:rel}
\end{figure*}

\begin{figure*}[htbp]
\centering
\subfigure[]{ \includegraphics[width=0.49\textwidth]{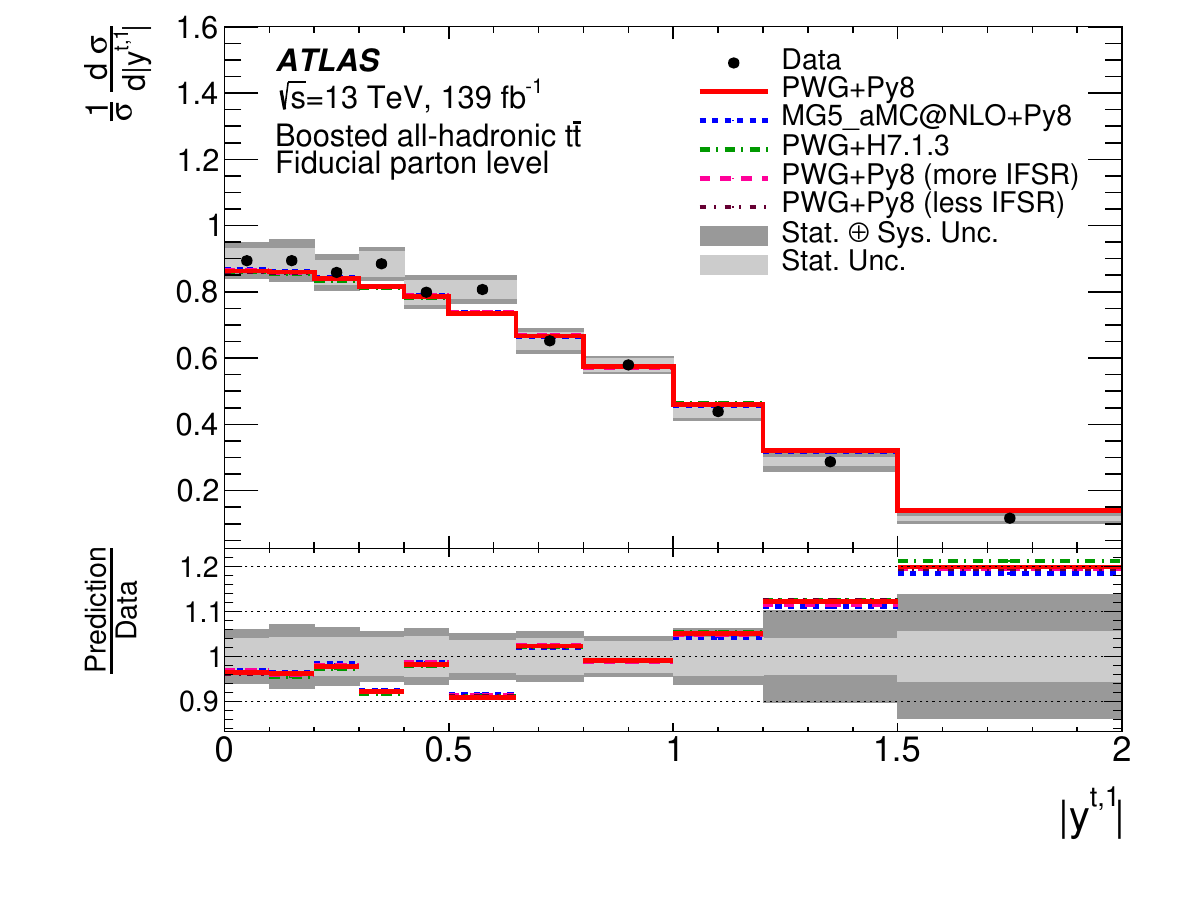}\label{fig:parton:t1_y:rel}}
\subfigure[]{ \includegraphics[width=0.49\textwidth]{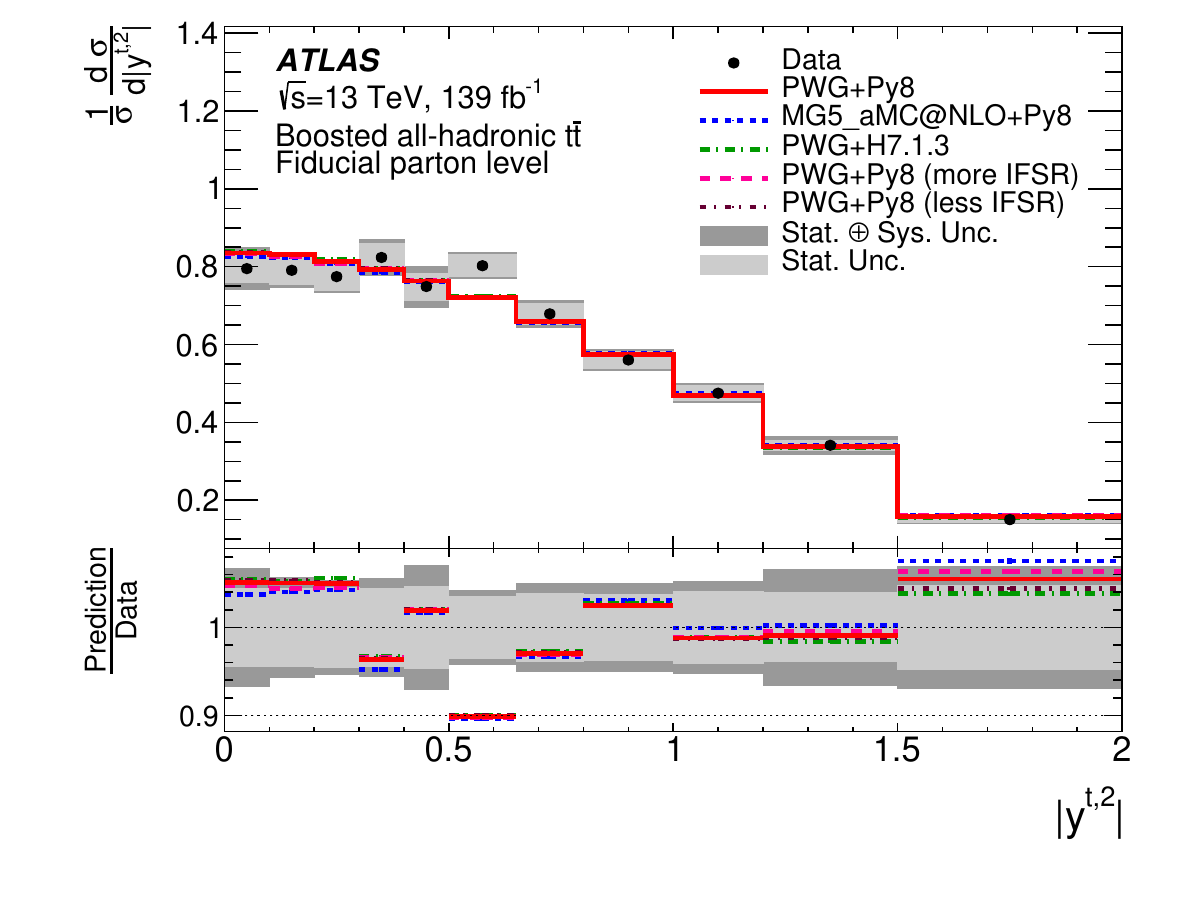}\label{fig:parton:t2_y:rel}}
\subfigure[]{ \includegraphics[width=0.49\textwidth]{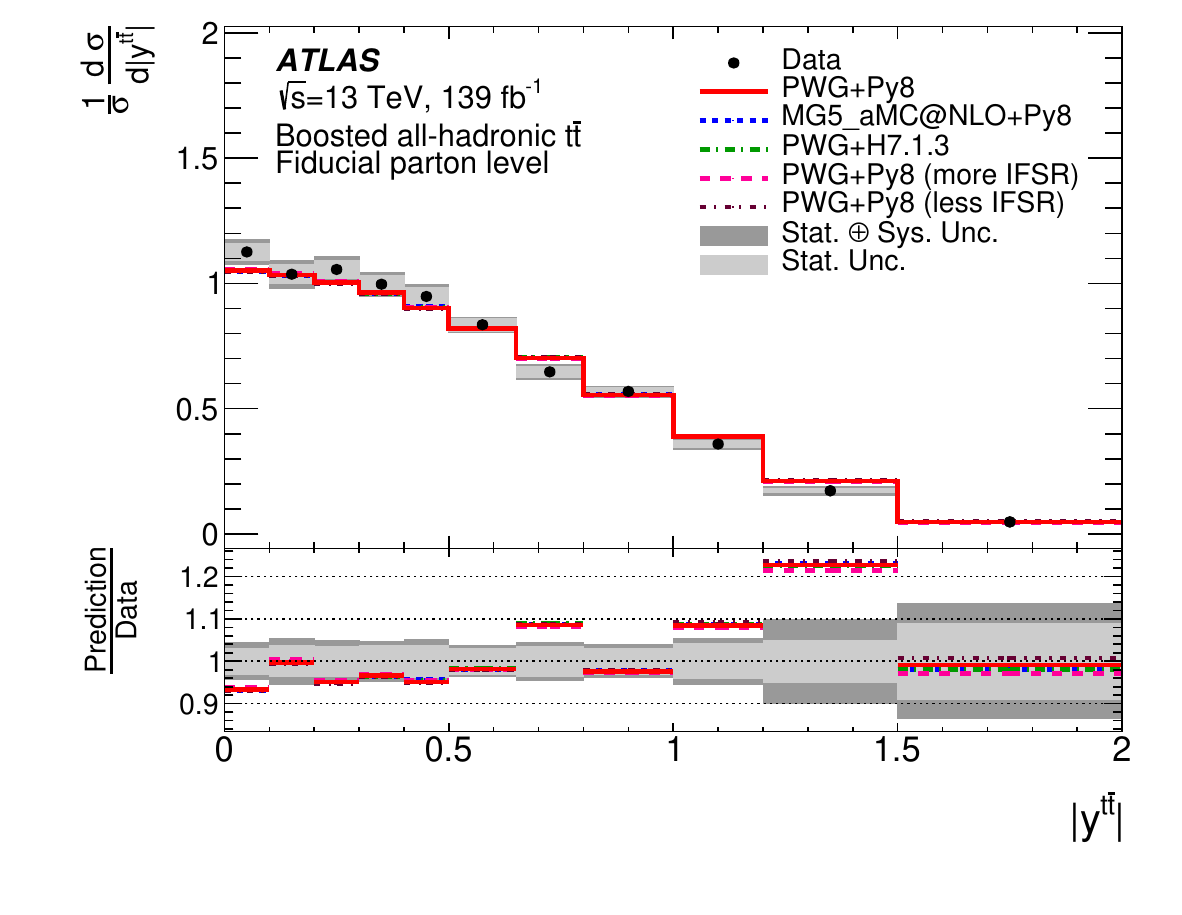}\label{fig:parton:tt_y:rel}}
\caption{
Normalized parton-level fiducial phase-space differential cross-sections as a function of the absolute value of the rapidity of
\subref{fig:parton:t1_y:rel}~the leading top quark,
\subref{fig:parton:t2_y:rel}~the second-leading top quark, and
\subref{fig:parton:tt_y:rel}~the \ttbar{}~system, \absyttbar.
The dark and light grey bands indicate the total uncertainty and the statistical uncertainty, respectively, of the data in each bin.
Data points are placed at the centre of each bin.
The \POWPY[8] MC sample is used as the nominal prediction to correct the data to parton level.
}
\label{fig:parton:rapidity_observables:rel}
\end{figure*}

\begin{figure*}[htbp]
\centering
\subfigure[]{ \includegraphics[width=0.49\textwidth]{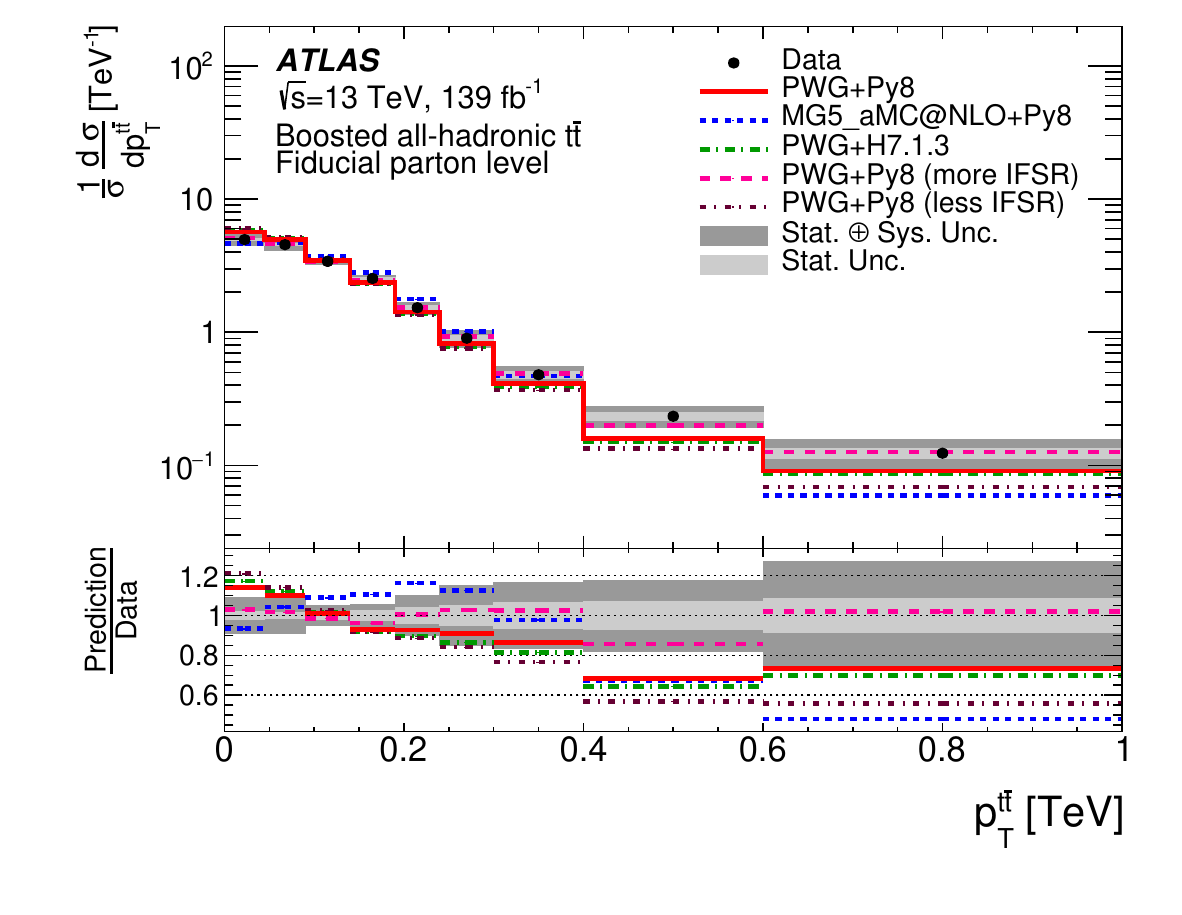}\label{fig:parton:tt_pt:rel_dupl}}
\subfigure[]{ \includegraphics[width=0.49\textwidth]{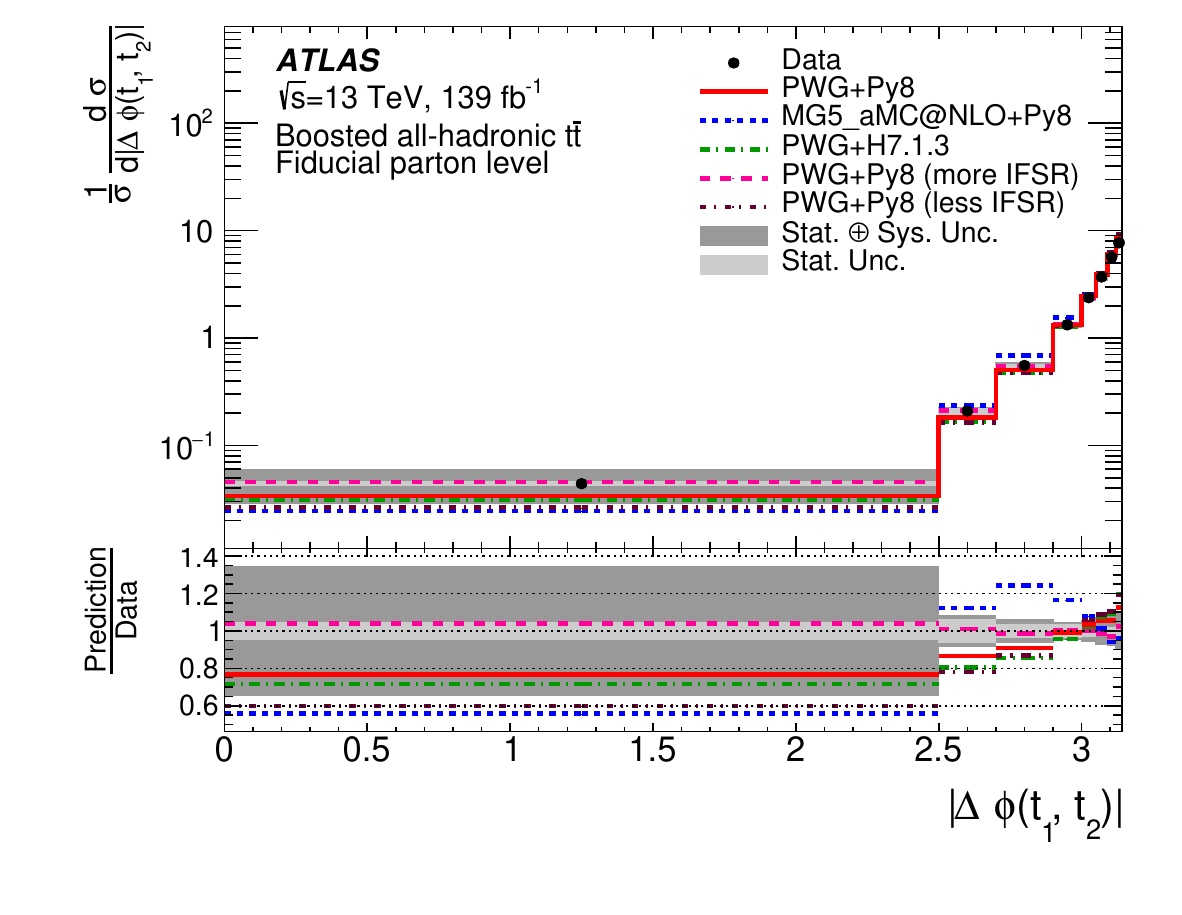}\label{fig:parton:tt_dPhittbar:rel}}
\subfigure[]{ \includegraphics[width=0.49\textwidth]{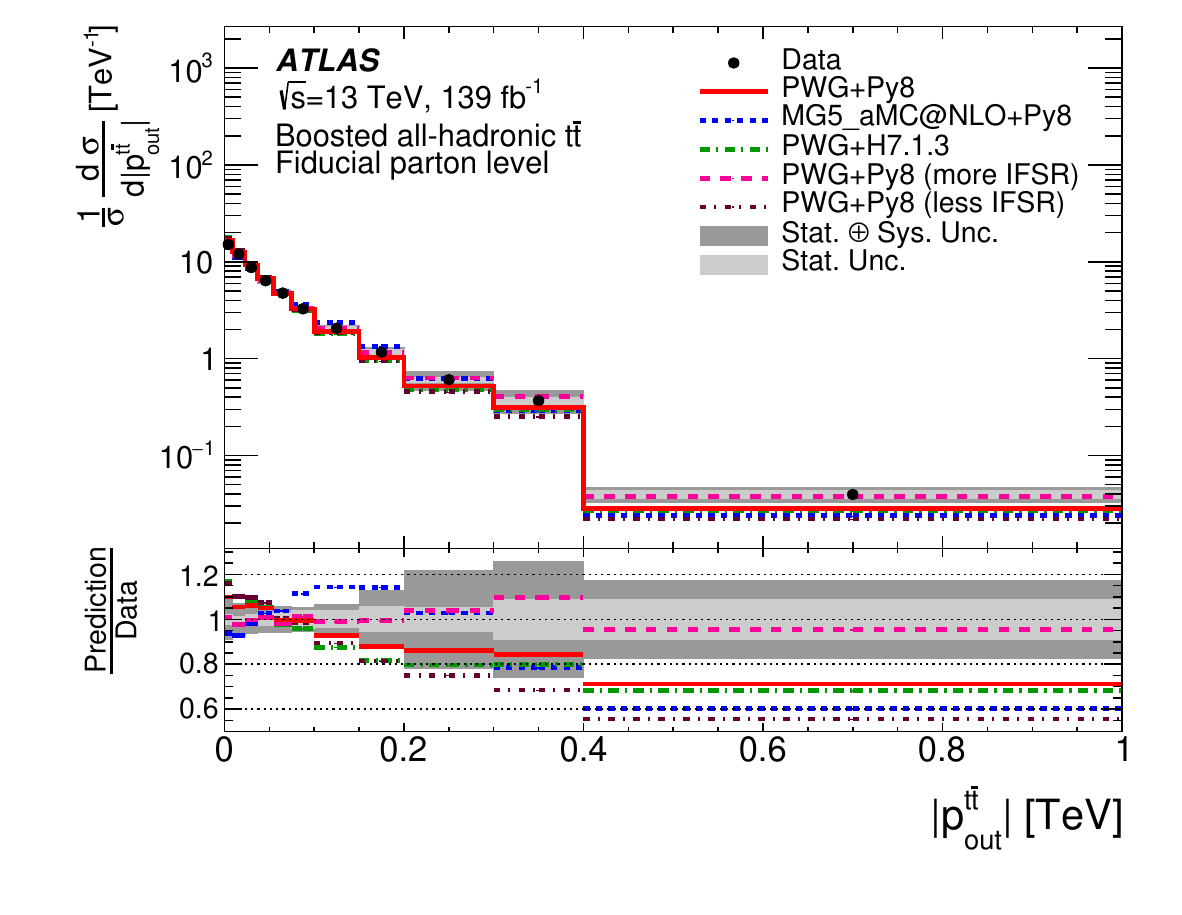}\label{fig:parton:tt_pout:rel}}
\caption{
Normalized parton-level differential cross-sections as a function of
\subref{fig:parton:tt_pt:rel_dupl}~the \pT of the \ttbar{} system, \ptttbar,
\subref{fig:parton:tt_dPhittbar:rel}~the azimuthal angle between the two top quarks, \deltaPhittbar, and
\subref{fig:parton:tt_pout:rel}~the absolute value of the out-of-plane momentum, \Poutttbar.
The dark and light grey bands indicate the total uncertainty and the statistical uncertainty, respectively, of the data in each bin.
Data points are placed at the centre of each bin.
The \POWPY[8] MC sample is used as the nominal prediction to correct the data to parton level.
}
\label{fig:parton:radiation_observables:rel}
\end{figure*}

The normalized parton-level fiducial phase-space differential cross-section distributions are compared
with SM predictions in
Figures~\ref{fig:parton:energy_observables:rel}--\ref{fig:parton:radiation_observables:rel}:
the differential cross-sections that probe the \pT of the top quarks and the invariant mass of the \ttbar{} system
(Figure~\ref{fig:parton:energy_observables:rel}),
the rapidity of top quarks and of the \ttbar{} system (Figure~\ref{fig:parton:rapidity_observables:rel}),
and the extra radiation from the \ttbar{} system (Figure~\ref{fig:parton:radiation_observables:rel}).
The remaining distributions are presented in Appendix~\ref{sec:appendix:parton_level}.
 
\begin{figure*}[htbp]
\centering
\subfigure[]{ \includegraphics[width=0.6\textwidth]{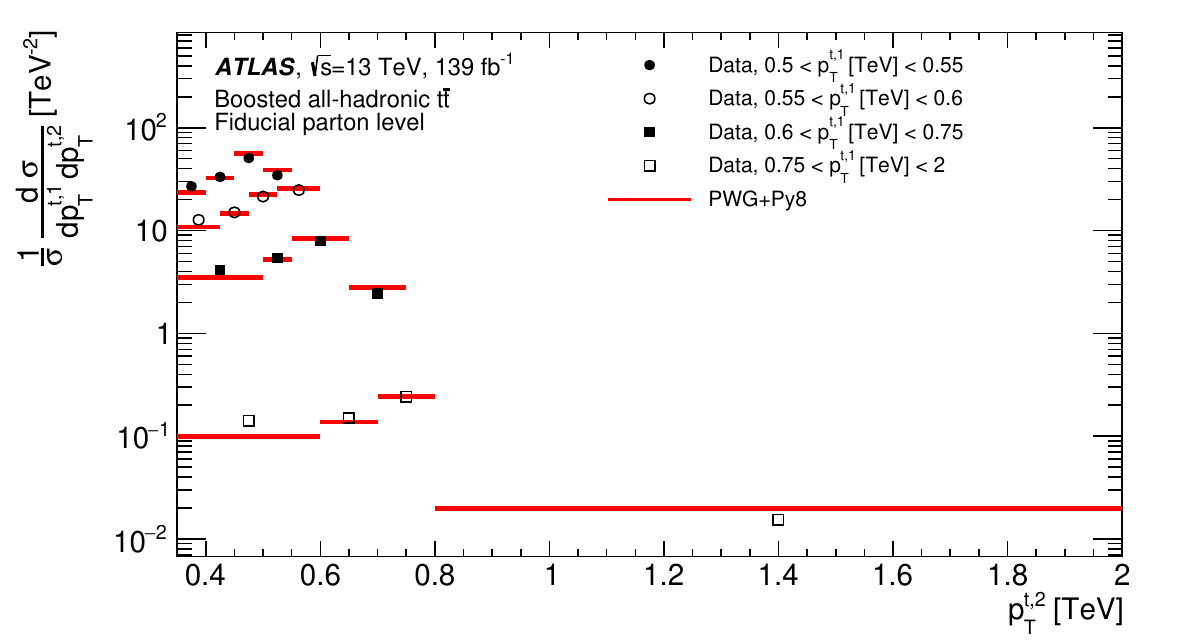}\label{fig:parton:t1_pt_vs_t2_pt:rel:shape}}
\subfigure[]{ \includegraphics[width=0.68\textwidth]{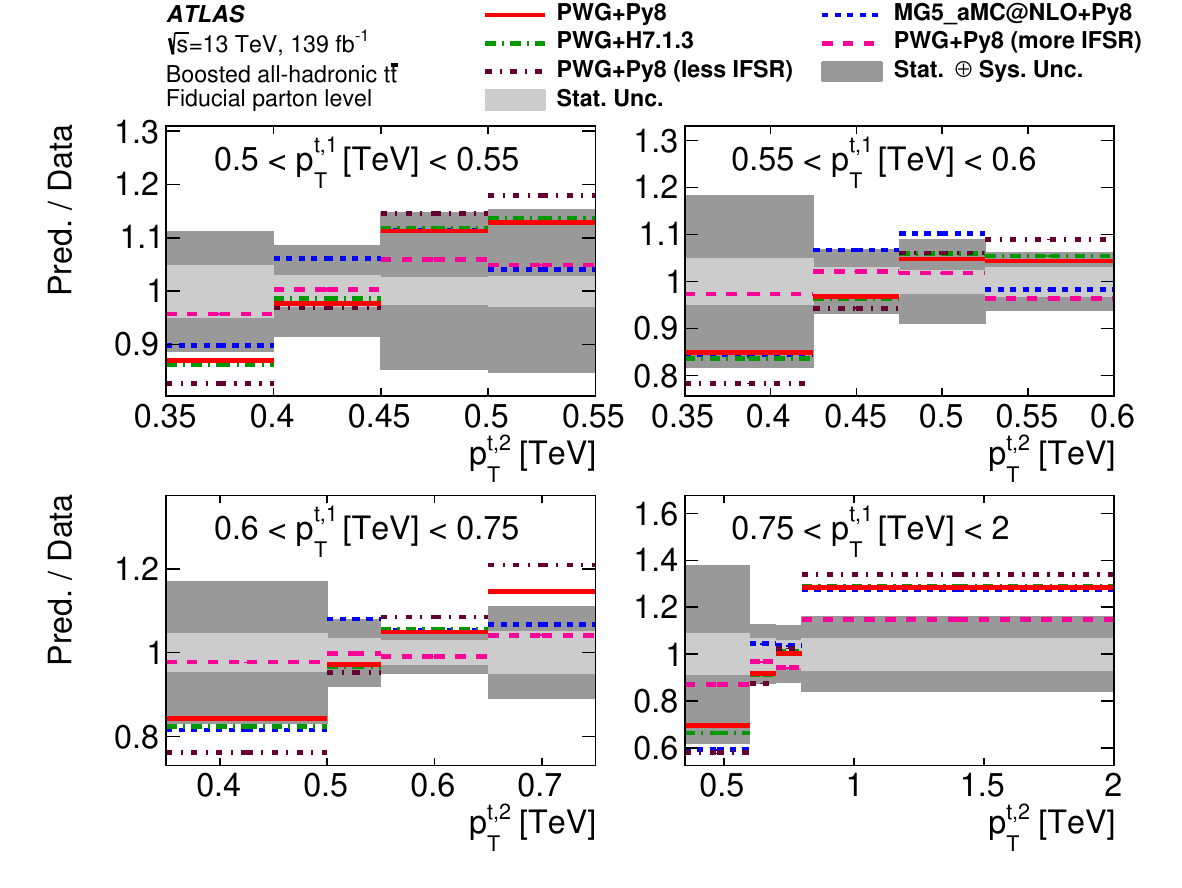}\label{fig:parton:t1_pt_vs_t2_pt:rel:ratio}}
\caption{
\subref{fig:parton:t1_pt_vs_t2_pt:rel:shape} Normalized parton-level fiducial phase-space double-differential cross-sections as a function of the transverse momenta of the leading and second-leading top quarks, compared with the \POWPY[8] calculation.
Data points are placed at the centre of each bin and the \POWPY[8] calculation is indicated by solid lines.
\subref{fig:parton:t1_pt_vs_t2_pt:rel:ratio}~The ratios of various MC calculations to the normalized parton-level fiducial phase-space differential cross-sections.
The dark and light grey bands indicate the total uncertainty and the statistical uncertainty, respectively, of the data in each bin.
}
\label{fig:parton:t1_pt_vs_t2_pt:rel}
\end{figure*}

\begin{figure*}[htbp]
\centering
\subfigure[]{ \includegraphics[width=0.6\textwidth]{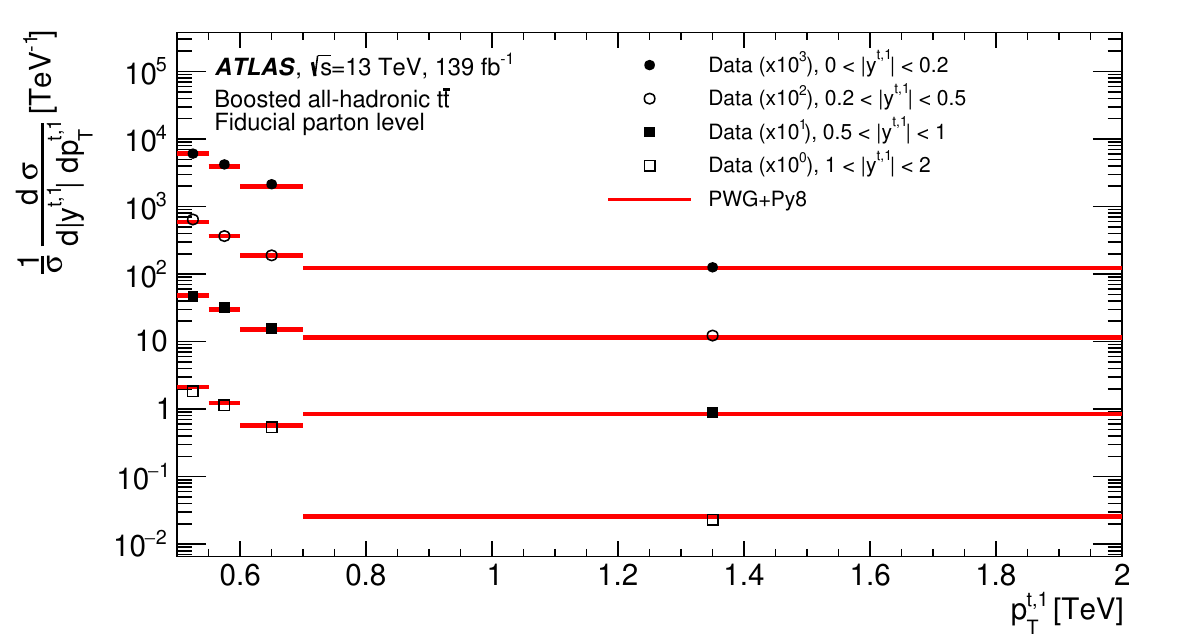}\label{fig:parton:t1_y_vs_t1_pt:rel:shape}}
\subfigure[]{ \includegraphics[width=0.68\textwidth]{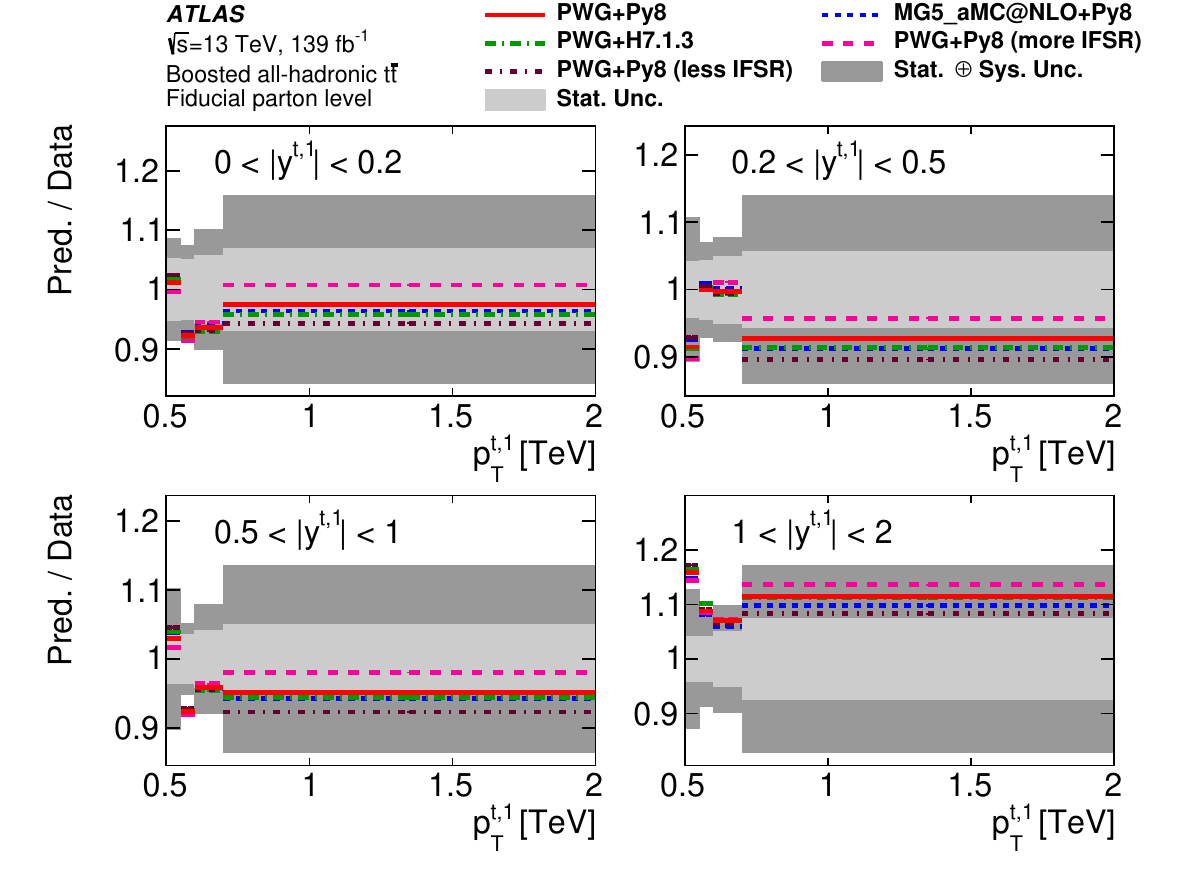}\label{fig:parton:t1_y_vs_t1_pt:rel:ratio}}
\caption{
\subref{fig:parton:t1_y_vs_t1_pt:rel:shape} Normalized parton-level fiducial phase-space double-differential cross-sections as a function of the absolute value of the leading top-quark rapidity and \pT, compared with the \POWPY[8] calculation.
Data points are placed at the centre of each bin and the \POWPY[8] calculation is indicated by solid lines.
The measurement and the prediction are normalized by the factors shown in parentheses to aid visibility.
\subref{fig:parton:t1_y_vs_t1_pt:rel:ratio}~The ratios of various MC calculations to the normalized parton-level fiducial phase-space differential cross-sections.
The dark and light grey bands indicate the total uncertainty and the statistical uncertainty, respectively, of the data in each bin.
}
\label{fig:parton:t1_y_vs_t1_pt:rel}
\end{figure*}

\begin{figure*}[htbp]
\centering
\subfigure[]{ \includegraphics[width=0.6\textwidth]{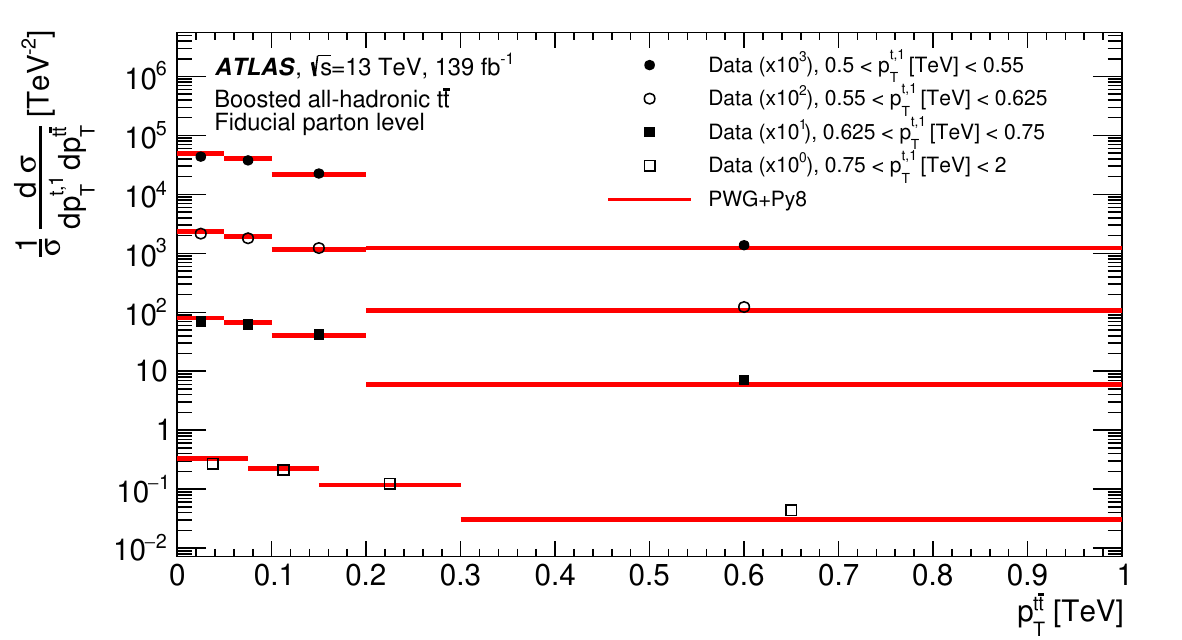}\label{fig:parton:t1_pt_vs_ttbar_pt:rel:shape}}
\subfigure[]{ \includegraphics[width=0.68\textwidth]{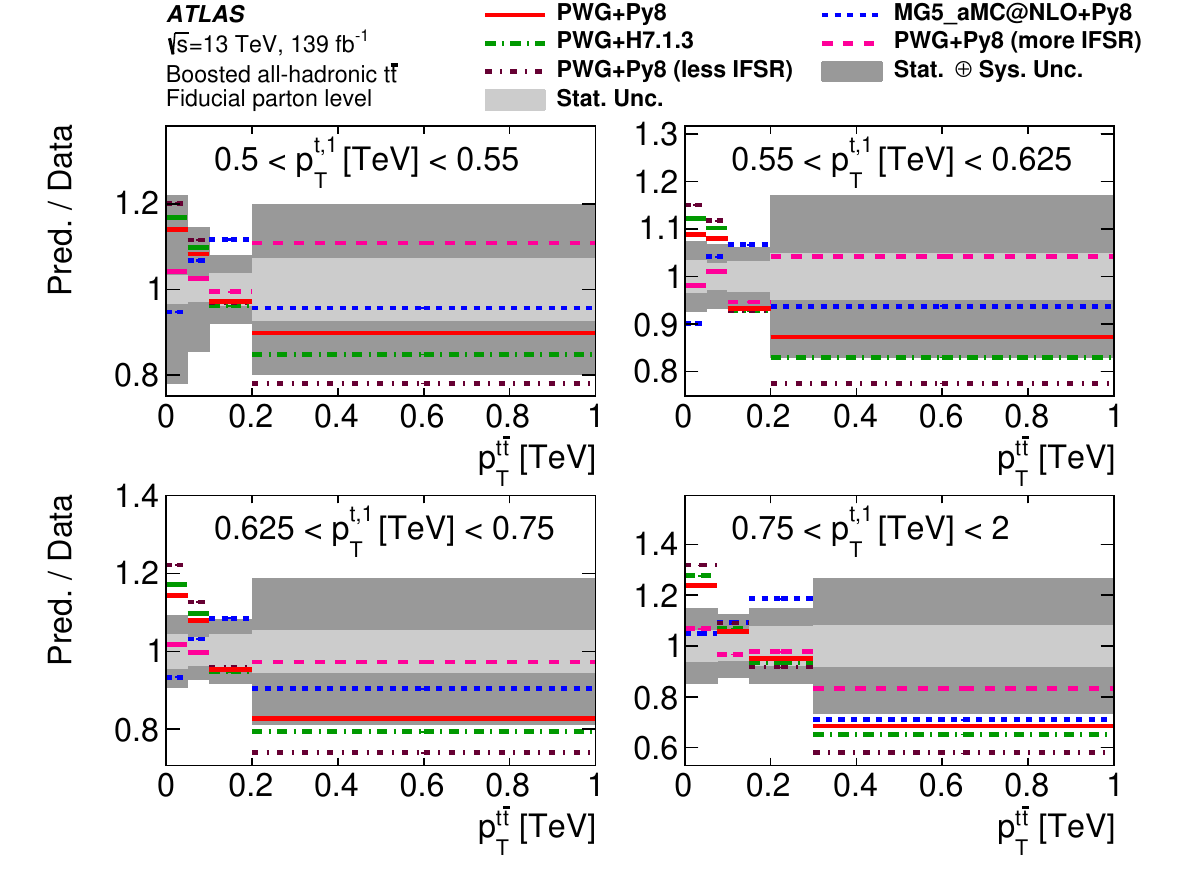}\label{fig:parton:t1_pt_vs_ttbar_pt:rel:ratio}}
\caption{
\subref{fig:parton:t1_pt_vs_ttbar_pt:rel:shape} Normalized parton-level fiducial phase-space double-differential cross-sections as a function of the \pT of the leading top quark and the \pT of the \ttbar system, \ptttbar, compared with the \POWPY[8] calculation.
Data points are placed at the centre of each bin and the \POWPY[8] calculation is indicated by solid lines.
The measurement and the prediction are normalized by the factors shown in parentheses to aid visibility.
\subref{fig:parton:t1_pt_vs_ttbar_pt:rel:ratio}~The ratios of various MC calculations to the normalized parton-level fiducial phase-space differential cross-sections.
The dark and light grey bands indicate the total uncertainty and the statistical uncertainty, respectively, of the data in each bin.
}
\label{fig:parton:t1_pt_vs_ttbar_pt:rel}
\end{figure*}

\begin{figure*}[htbp]
\centering
\subfigure[]{ \includegraphics[width=0.6\textwidth]{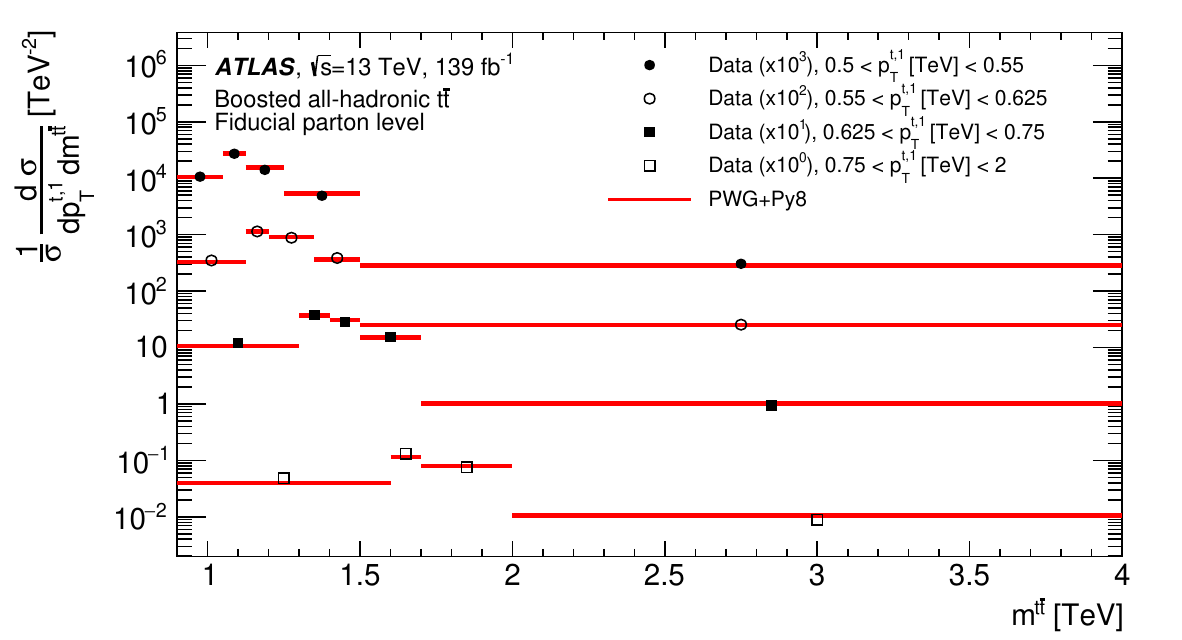}\label{fig:parton:t1_pt_vs_ttbar_mass:rel:shape}}
\subfigure[]{ \includegraphics[width=0.68\textwidth]{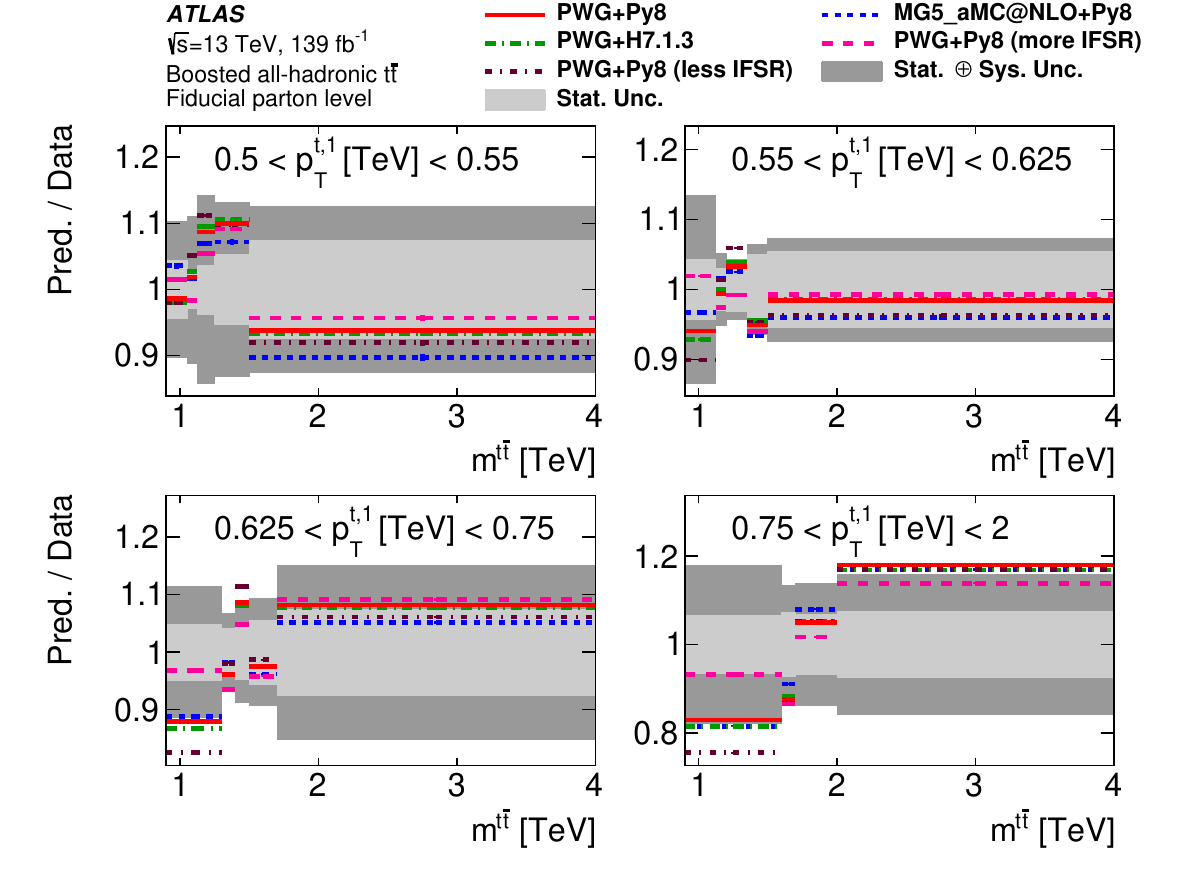}\label{fig:parton:t1_pt_vs_ttbar_mass:rel:ratio}}
\caption{
\subref{fig:parton:t1_pt_vs_ttbar_mass:rel:shape} Normalized parton-level fiducial phase-space double-differential cross-sections as a function of the \pT of the leading top quark and the invariant mass of the \ttbar system, \mttbar, compared with the \POWPY[8] calculation.
Data points are placed at the centre of each bin and the \POWPY[8] calculation is indicated by solid lines.
The measurement and the prediction are normalized by the factors shown in parentheses to aid visibility.
\subref{fig:parton:t1_pt_vs_ttbar_mass:rel:ratio}~The ratios of various MC calculations to the normalized parton-level fiducial phase-space differential cross-sections.
The dark and light grey bands indicate the total uncertainty and the statistical uncertainty, respectively, of the data in each bin.
}
\label{fig:parton:t1_pt_vs_ttbar_mass:rel}
\end{figure*}

\begin{figure*}[htbp]
\centering
\subfigure[]{ \includegraphics[width=0.6\textwidth]{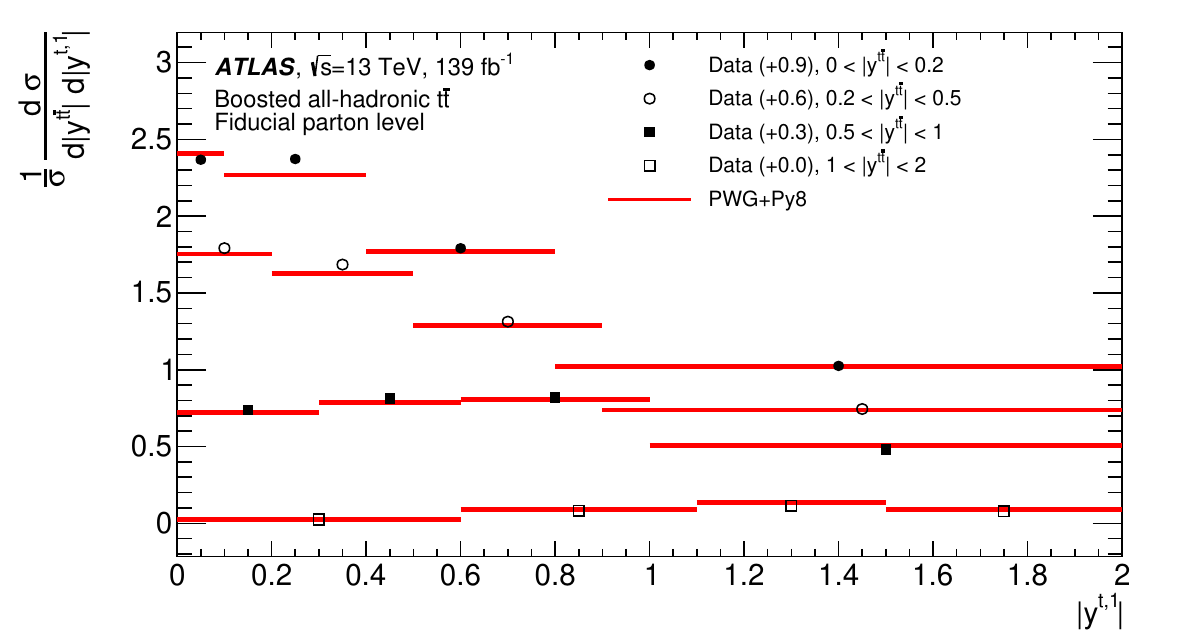}\label{fig:parton:ttbar_y_vs_t1_y:rel:shape}}
\subfigure[]{ \includegraphics[width=0.68\textwidth]{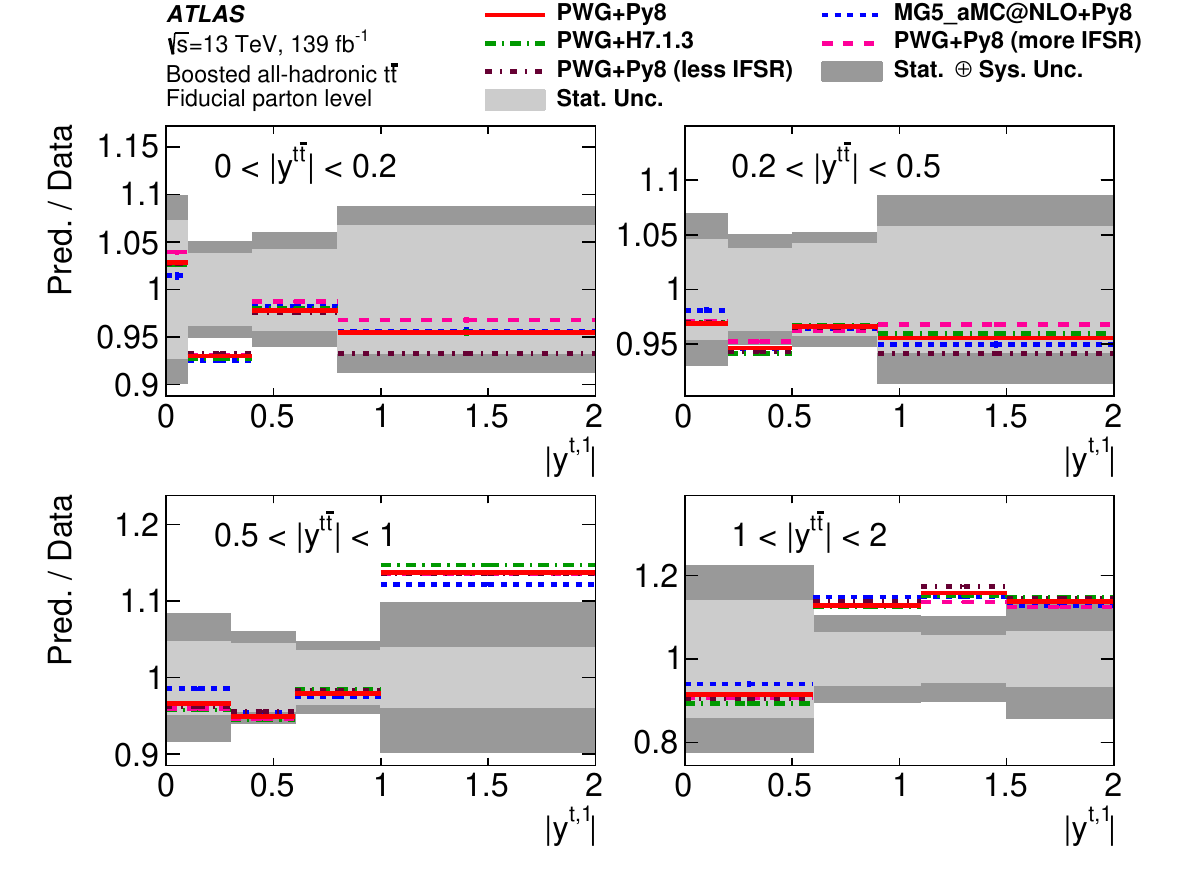}\label{fig:parton:ttbar_y_vs_t1_y:rel:ratio}}
\caption{
\subref{fig:parton:ttbar_y_vs_t1_y:rel:shape} Normalized parton-level fiducial phase-space double-differential cross-sections as a function of the absolute value of the rapidity of the \ttbar system, \absyttbar, and the absolute value of the rapidity of the leading top quark, compared with the \POWPY[8] calculation.
Data points are placed at the centre of each bin and the \POWPY[8] calculation is indicated by solid lines.
The measurement and the prediction are shifted by the factors shown in parentheses to aid visibility.
\subref{fig:parton:ttbar_y_vs_t1_y:rel:ratio}~The ratios of various MC calculations to the normalized parton-level fiducial phase-space differential cross-sections.
The dark and light grey bands indicate the total uncertainty and the statistical uncertainty, respectively, of the data in each bin.
}
\label{fig:parton:ttbar_y_vs_t1_y:rel}
\end{figure*}

\begin{figure*}[htbp]
\centering
\subfigure[]{ \includegraphics[width=0.6\textwidth]{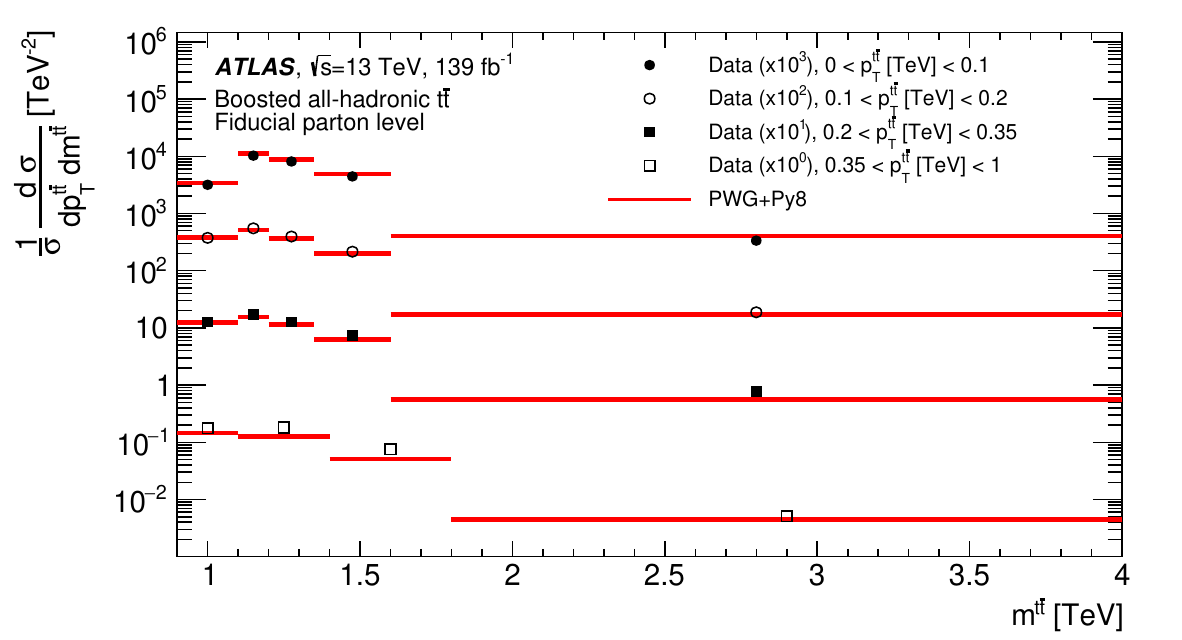}\label{fig:parton:ttbar_pt_vs_ttbar_mass:rel:shape}}
\subfigure[]{ \includegraphics[width=0.68\textwidth]{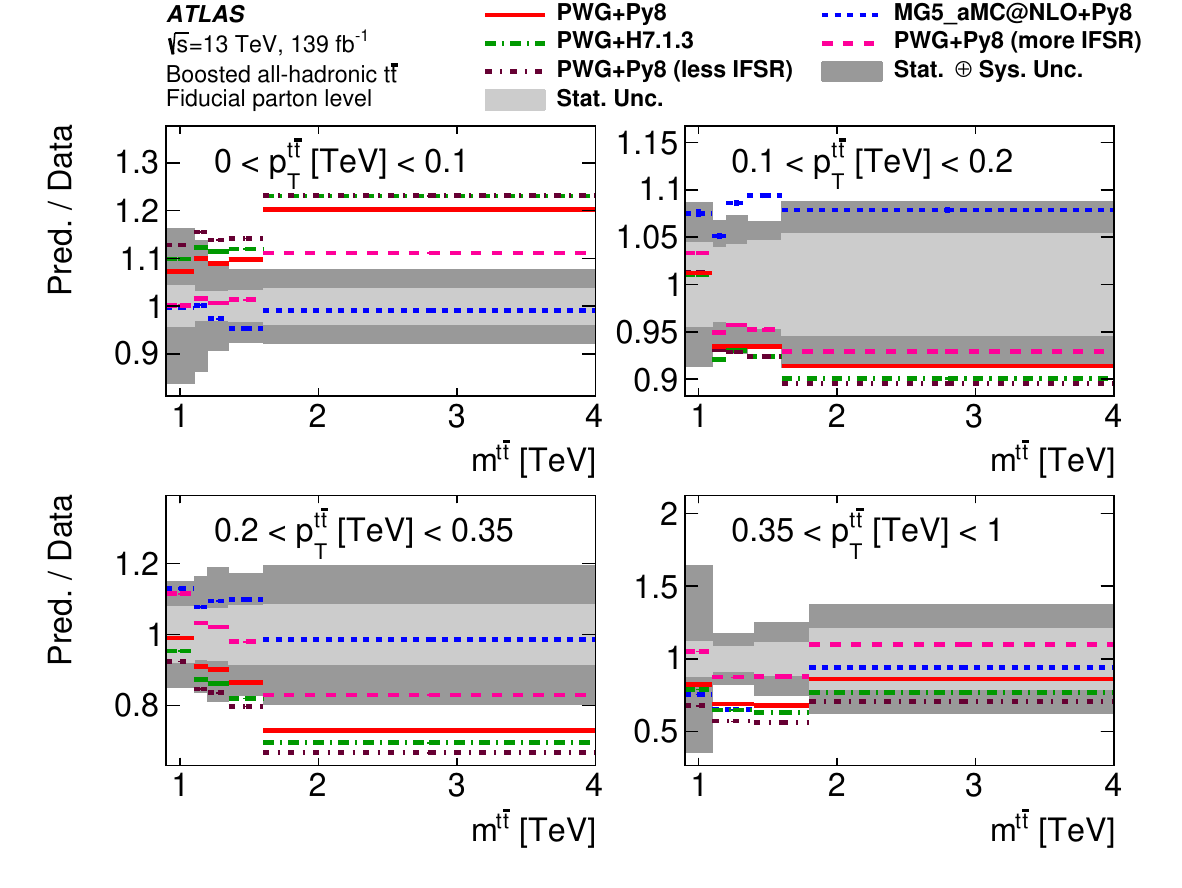}\label{fig:parton:ttbar_pt_vs_ttbar_mass:rel:ratio}}
\caption{
\subref{fig:parton:ttbar_pt_vs_ttbar_mass:rel:shape} Normalized parton-level fiducial phase-space double-differential cross-sections as a function of the \pT and the mass of the \ttbar system, \ptttbar\ and \mttbar, compared with the \POWPY[8] calculation.
Data points are placed at the centre of each bin and the \POWPY[8] calculation is indicated by solid lines.
The measurement and the prediction are normalized by the factors shown in parentheses to aid visibility.
\subref{fig:parton:ttbar_pt_vs_ttbar_mass:rel:ratio}~The ratios of various MC calculations to the normalized parton-level fiducial phase-space differential cross-sections.
The dark and light grey bands indicate the total uncertainty and the statistical uncertainty, respectively, of the data in each bin.
}
\label{fig:parton:ttbar_pt_vs_ttbar_mass:rel}
\end{figure*}

\begin{figure*}[htbp]
\centering
\subfigure[]{ \includegraphics[width=0.6\textwidth]{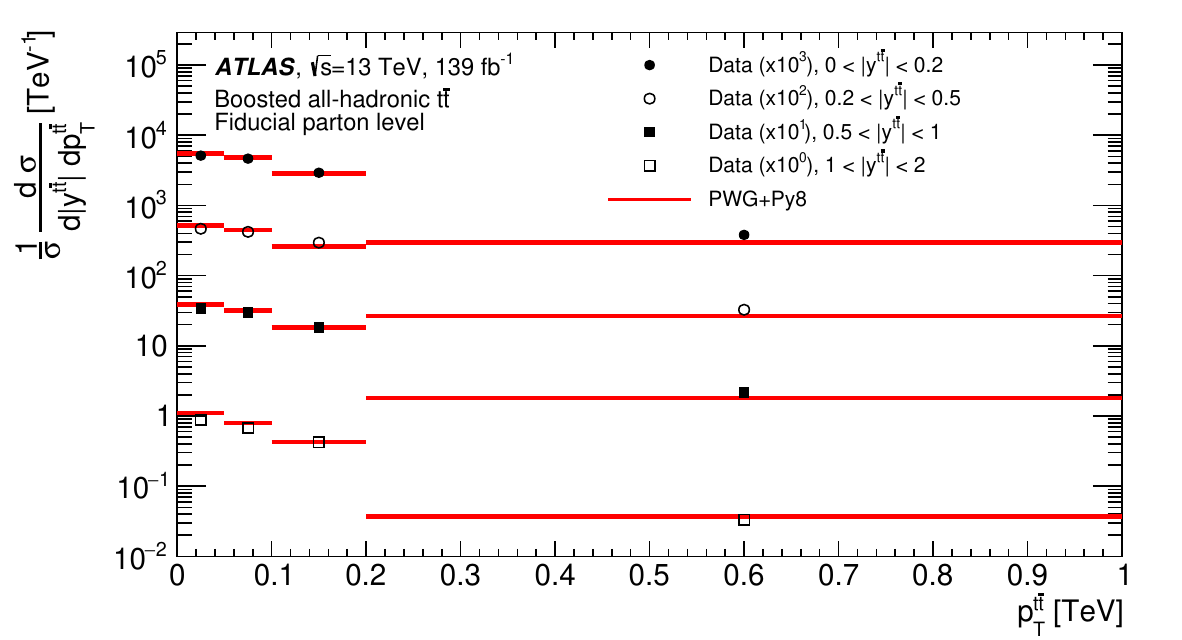}\label{fig:parton:ttbar_y_vs_ttbar_pt:rel:shape}}
\subfigure[]{ \includegraphics[width=0.68\textwidth]{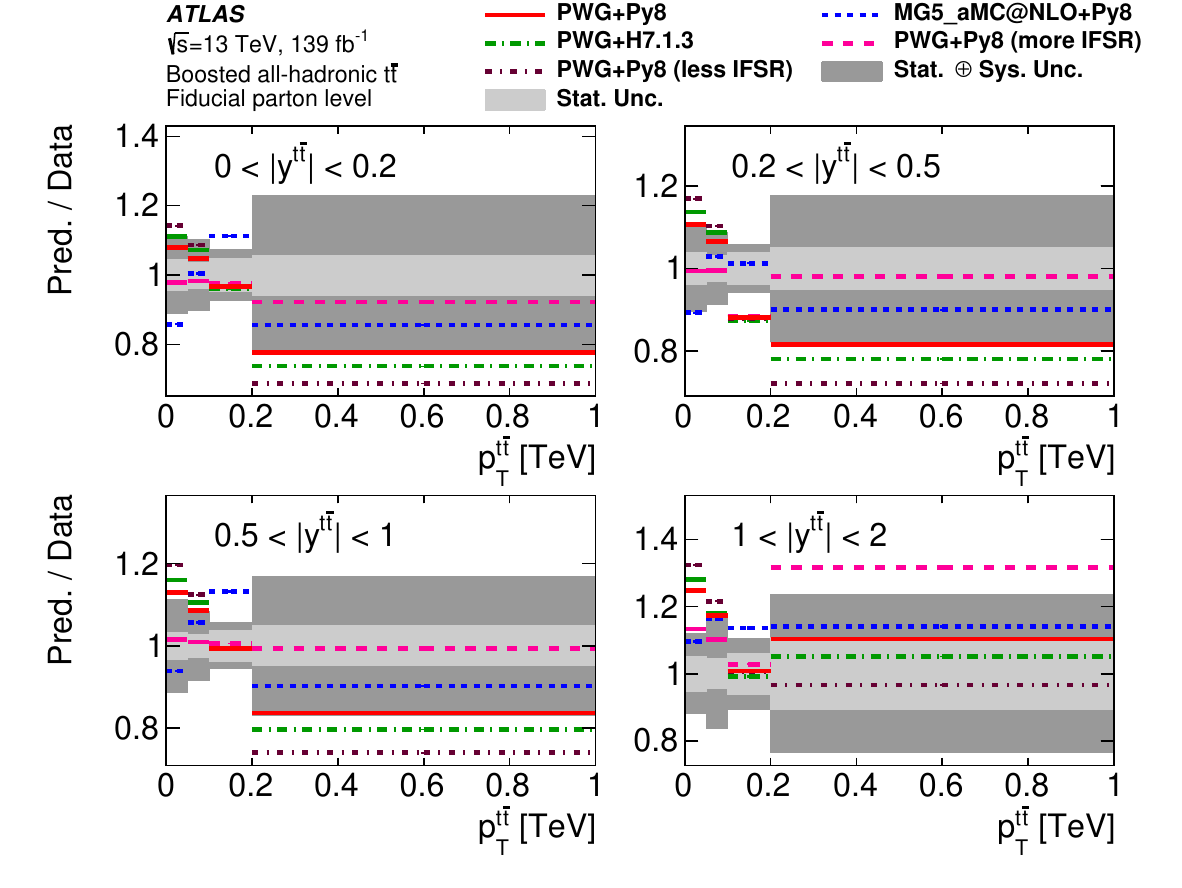}\label{fig:parton:ttbar_y_vs_ttbar_pt:rel:ratio}}
\caption{
\subref{fig:parton:ttbar_y_vs_ttbar_pt:rel:shape} Normalized parton-level fiducial phase-space double-differential cross-sections as a function of the absolute value of the rapidity and the \pT of the \ttbar system, \absyttbar\ and \ptttbar, compared with the \POWPY[8] calculation.
Data points are placed at the centre of each bin and the \POWPY[8] calculation is indicated by solid lines.
The measurement and the prediction are normalized by the factors shown in parentheses to aid visibility.
\subref{fig:parton:ttbar_y_vs_ttbar_pt:rel:ratio}~The ratios of various MC calculations to the normalized parton-level fiducial phase-space differential cross-sections.
The dark and light grey bands indicate the total uncertainty and the statistical uncertainty, respectively, of the data in each bin.
}
\label{fig:parton:ttbar_y_vs_ttbar_pt:rel}
\end{figure*}

\begin{figure*}[htbp]
\centering
\subfigure[]{ \includegraphics[width=0.45\textwidth]{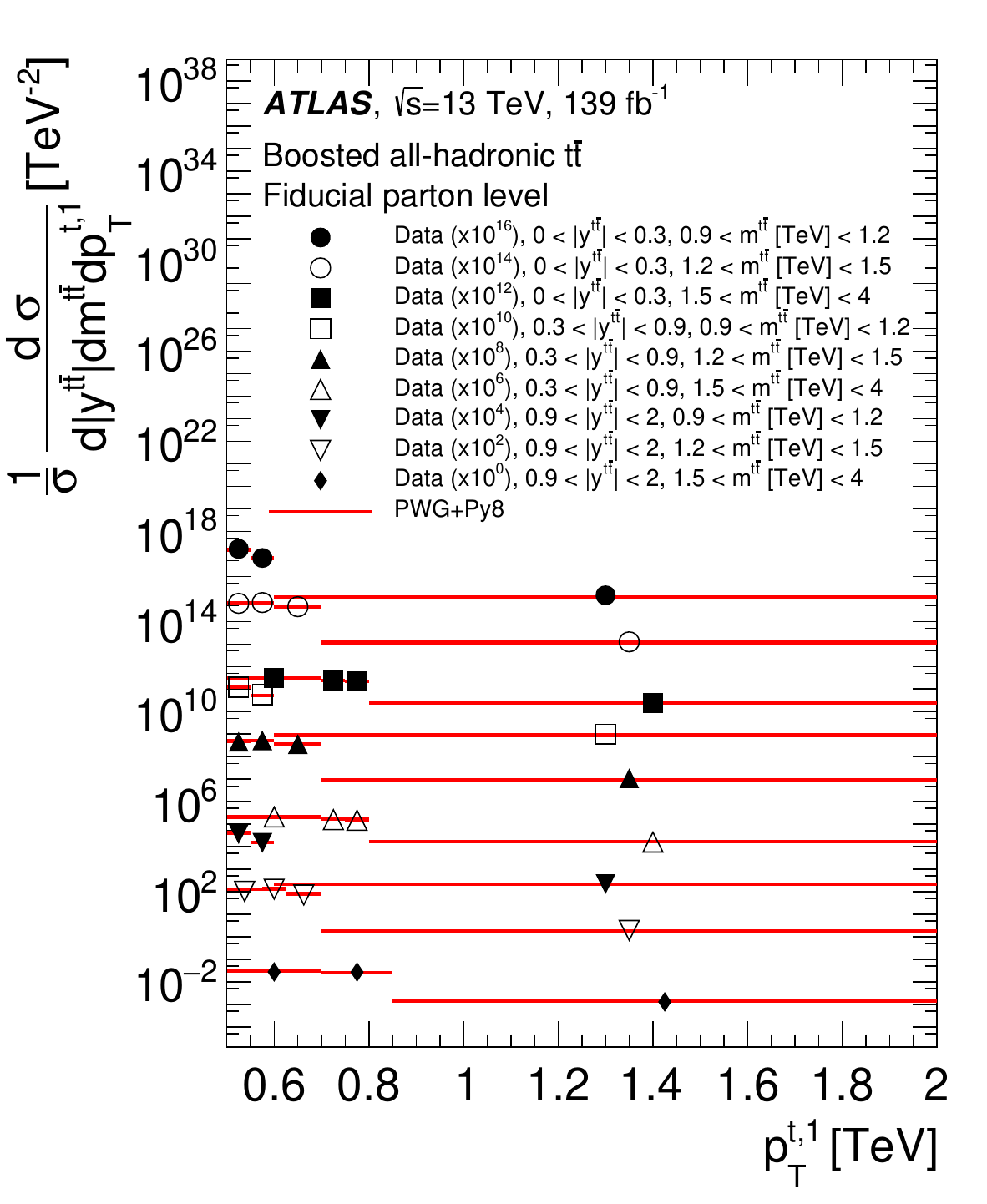}\label{fig:parton:ttbar_y_vs_ttbar_mass_vs_t1_pt:rel:shape}}
\subfigure[]{ \includegraphics[width=0.85\textwidth]{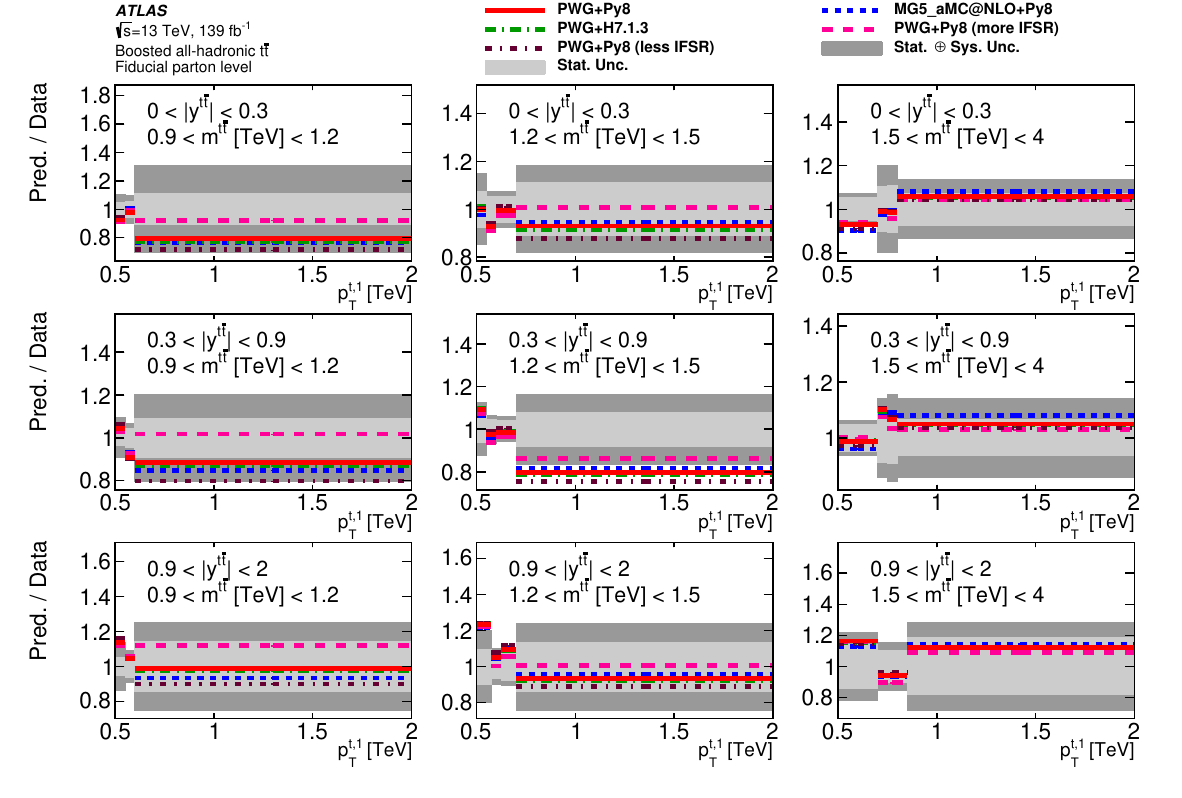}\label{fig:parton:ttbar_y_vs_ttbar_mass_vs_t1_pt:rel:ratio}}
\caption{
\subref{fig:parton:ttbar_y_vs_ttbar_mass_vs_t1_pt:rel:shape} Normalized parton-level fiducial phase-space triple-differential cross-sections as a function of the absolute value of the rapidity of the \ttbar system, \absyttbar, the mass of the \ttbar system, \mttbar, and the \pT of the leading top quark, compared with the \POWPY[8] calculation.
Data points are placed at the centre of each bin and the \POWPY[8] calculation is indicated by solid lines.
The measurement and the prediction are normalized by the factors shown in parentheses to aid visibility.
\subref{fig:parton:ttbar_y_vs_ttbar_mass_vs_t1_pt:rel:ratio}~The ratios of various MC calculations to the normalized parton-level fiducial phase-space differential cross-sections.
The dark and light grey bands indicate the total uncertainty and the statistical uncertainty, respectively, of the data in each bin.
}
\label{fig:parton:ttbar_y_vs_ttbar_mass_vs_t1_pt:rel}
\end{figure*}

A selection of the fiducial phase-space double- and triple-differential cross-sections are shown in
Figures~\ref{fig:parton:t1_pt_vs_t2_pt:rel}--\ref{fig:parton:ttbar_y_vs_ttbar_mass_vs_t1_pt:rel}.
Other double- and triple-differential cross-sections are presented in Appendix~\ref{sec:appendix:parton_level}.
 
\begin{figure*}[htbp]
\centering
\subfigure[]{ \includegraphics[width=0.49\textwidth]{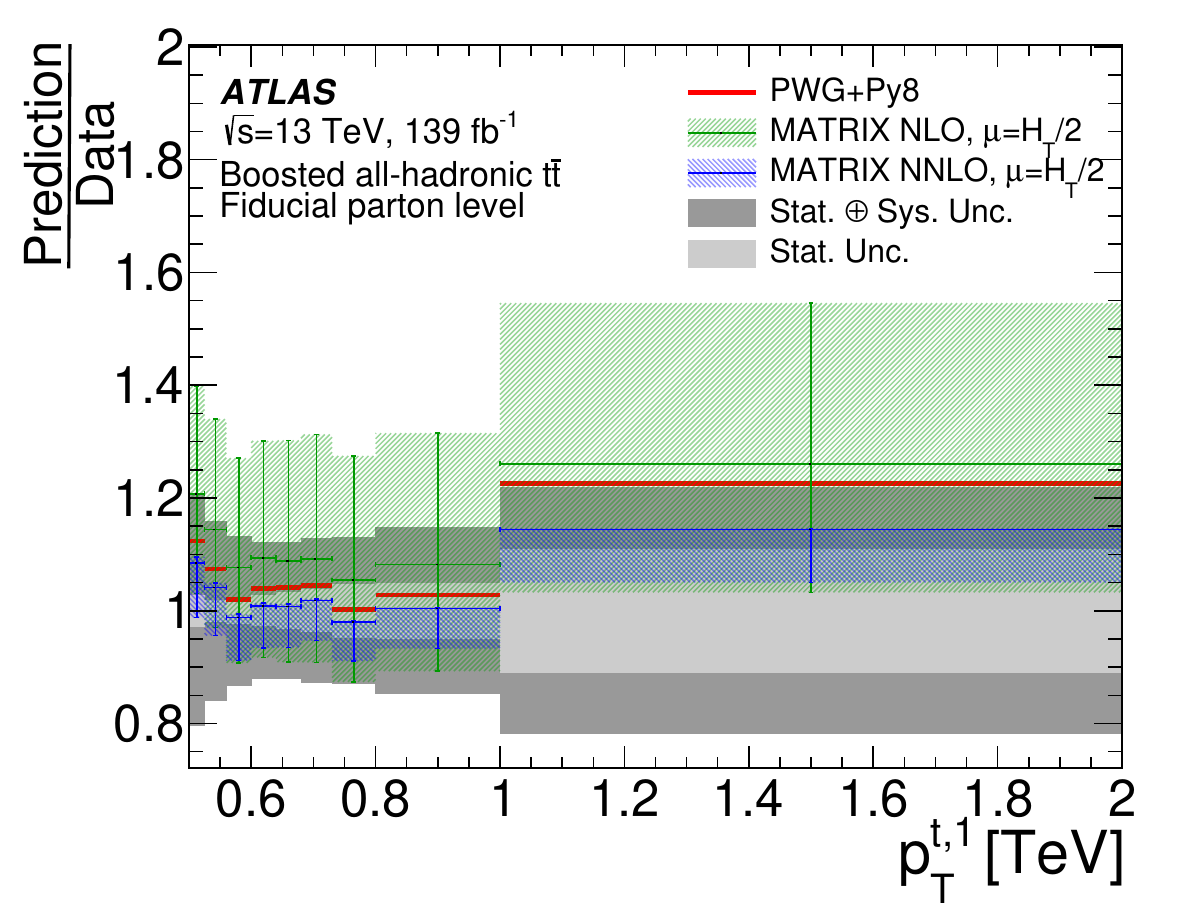}\label{fig:parton:NNLO:t1_pt:abs}}
\subfigure[]{ \includegraphics[width=0.49\textwidth]{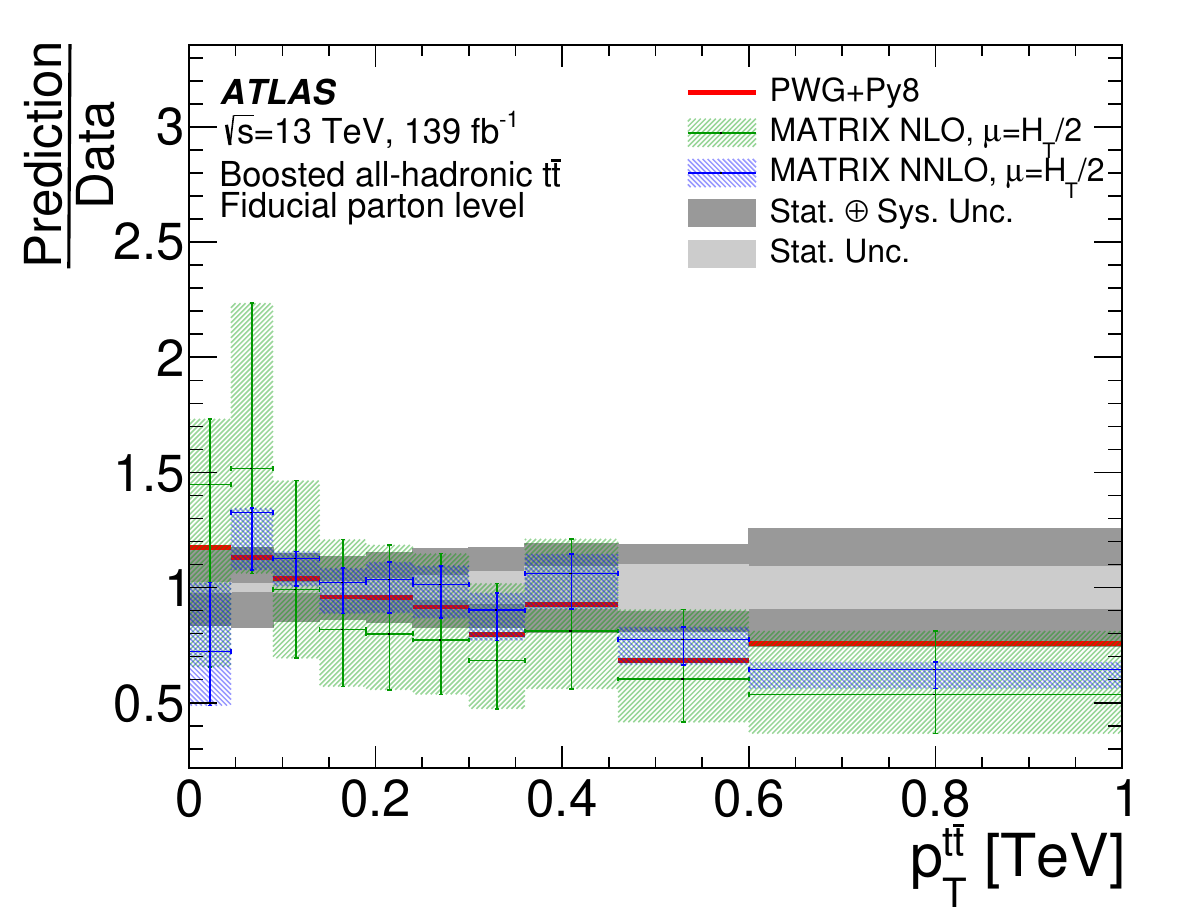}\label{fig:parton:NNLO:tt_pt:abs}}
\caption{Comparison of the NLO, NNLO, and \POWPY[8] calculations with measured parton-level
fiducial phase-space absolute differential cross-sections for \subref{fig:parton:NNLO:t1_pt:abs}~the \pT of the
leading top quark, and \subref{fig:parton:NNLO:tt_pt:abs}~~the \pT of the \ttbar{}~system, \ptttbar.
The dark and light grey bands indicate the total uncertainty and the statistical uncertainty, respectively,
of the data in each bin.
The fixed-order calculation bands correspond to the scale uncertainty. The \POWPY[8] calculation is not normalized to the NNLO total inclusive cross-section.}
\label{fig:parton:NNLO:abs}
\end{figure*}

The examples of comparisons of absolute differential cross-sections are also presented. Figure~\ref{fig:parton:NNLO:abs} shows the ratio of calculation to data for absolute differential cross-sections in \pTtone\ and \ptttbar. The leading jet \pT\ has the largest sensitivity in the subsequent EFT analysis in Section~\ref{sec:eft} and there was a discrepancy in the \pT\ of the \ttbar\ system in the comparison with NLO MC predictions.
Ratios are shown for various fixed-order calculations at NLO and NNLO together with the nominal \POWPY[8] calculation.
The fixed-order predictions are plotted with the scale uncertainties, which are asymmetric.
 
\FloatBarrier


\section{Comparisons with QCD calculations}
\label{sec:comparisons}

 
The nominal \POWPY[8] particle-level and
parton-level cross-sections for top-quark pair production in
their corresponding fiducial phase-space regions are $20\%$ larger than the
observed values as shown in Figure~\ref{fig:particle:inclusive:abs} and Figure~\ref{fig:parton:inclusive:abs},
respectively.
The significance of this difference is ${\sim}1.1$\ standard deviations  when taking into account both the uncertainties
of the measurements and the corresponding uncertainties in the \POWPY[8] predictions.
At both the particle level and parton level, there is better agreement with the \POWHEG{}+\HERWIG[7.1.3]\
calculations and the
predictions of \POWPY[8] with increased initial- and final-state radiation,
where the differences correspond to ${\sim}0.5$\ standard deviations.
Agreement with the nominal \POWPY[8] calculation improves
after reweighting those to the NNLO calculation.
At parton level, even better agreement with the NNLO \MATRIX\ calculation is observed for various definitions of the
renormalization and factorization scales, as well as for different PDF sets.
This is consistent with previous measurements that have observed
that the top-quark \pT\ spectra are softer than in various NLO+PS predictions at
high top-quark \pt~\cite{TOPQ-2012-08,TOPQ-2013-07,TOPQ-2014-15,TOPQ-2015-06,TOPQ-2016-04,TOPQ-2016-01,TOPQ-2016-09}.
 
The particle-level and parton-level fiducial phase-space differential cross-sections are compared with several SM predictions. In this comparison, there are no uncertainties associated with the
predictions.
The information provided by the shapes of the differential cross-section measurements is
compared with the calculations using the $\chi^2$ test described in Section~\ref{sec:prop_uncertainties},
which takes into account the correlations between the measured quantities while the uncertainty in
the prediction is not included.
The largest correlations at the detector level arise from sources of uncertainty that affect all bins equally.
The most sensitive comparison uses the normalized differential cross-sections because many of the common detector-level
uncertainties largely cancel out.
The $\chi^2$ values and associated $p$-values are shown in Table~\ref{tab:chi2:particle:relMain} and Table~\ref{tab:chi2:parton:relMain} for
the normalized particle-level and parton-level fiducial phase-space differential cross-sections, respectively.
The multidimensional distributions are labeled such that the innermost observable is listed as the last one.
No comparison with the fixed-order NNLO calculation is shown because numerical instabilities were observed in certain phase-space regions
(e.g.\ around $\ptttbar \sim 0$).
 
\begin{table*} [ht!]
\footnotesize
\centering
\scalebox{0.75}{

\begin{tabular}{|c|cc|cc|cc|cc|cc|} \hline
Observable & \multicolumn{2}{|c|}{PWG+Py8} & \multicolumn{2}{|c|}{MG5\_aMC@NLO+Py8} & \multicolumn{2}{|c|}{PWG+H7.1.3} & \multicolumn{2}{|c|}{PWG+Py8 (more IFSR)} & \multicolumn{2}{|c|}{PWG+Py8 (less IFSR)} \\
& \multicolumn{2}{|c|}{NNPDF30 A14} & \multicolumn{2}{|c|}{NNPDF30 UE-EE-5} & \multicolumn{2}{|c|}{NNPDF30 A14} & \multicolumn{2}{|c|}{NNPDF30 A14} & \multicolumn{2}{|c|}{NNPDF30 A14} \\
& \multicolumn{1}{|c|}{$\chi^{2}$/NDF} & \multicolumn{1}{|c|}{$p$-value} & \multicolumn{1}{|c|}{$\chi^{2}$/NDF} & \multicolumn{1}{|c|}{$p$-value} & \multicolumn{1}{|c|}{$\chi^{2}$/NDF} & \multicolumn{1}{|c|}{$p$-value} & \multicolumn{1}{|c|}{$\chi^{2}$/NDF} & \multicolumn{1}{|c|}{$p$-value} & \multicolumn{1}{|c|}{$\chi^{2}$/NDF} & \multicolumn{1}{|c|}{$p$-value} \\ \hline \hline
$p_\text{T}^{t}$  & 3.9/9 &  0.92  & 3.1/9 &  0.96  & 6.2/9 &  0.72  & 1.2/9 &  1.00  & 7.7/9 &  0.57  \\
$|y^{t}|$  & ~~6.8/10 &  0.75  & ~~5.8/10 &  0.83  & ~~6.8/10 &  0.74  & ~~7.5/10 &  0.68  & ~~5.9/10 &  0.83  \\
$p_\text{T}^{t,1}$  & 5.1/8 &  0.75  & 3.9/8 &  0.86  & 5.3/8 &  0.72  & 4.3/8 &  0.83  & 5.3/8 &  0.72  \\
$|y^{t,1}|$  & ~~6.1/10 &  0.81  & ~~4.7/10 &  0.91  & ~~6.7/10 &  0.76  & ~~5.7/10 &  0.84  & ~~5.6/10 &  0.84  \\
$p_\text{T}^{t,2}$  & 9.9/8 &  0.27  & 10.2/8~~ &  0.25  & 13.9/8~~ &  0.08  & 4.4/8 &  0.82  & 16.0/8~~ &  0.04  \\
$|y^{t,2}|$  & ~~9.4/10 &  0.49  & ~~9.0/10 &  0.53  & ~~9.4/10 &  0.50  & ~~8.9/10 &  0.54  & ~~9.3/10 &  0.50  \\
$m^{t\bar{t}}$  & ~~8.1/12 &  0.78  & ~~6.9/12 &  0.87  & ~~7.4/12 &  0.83  & ~~8.9/12 &  0.71  & ~~7.9/12 &  0.79  \\
$p_\text{T}^{t\bar{t}}$  & 14.3/8~~ &  0.07  & 35.2/8~~ &  $<\ $0.01~~~~  & 24.5/8~~ &  $<\ $0.01~~~~  & 2.7/8 &  0.95  & 33.5/8~~ &  $<\ $0.01~~~~  \\
$|y^{t\bar{t}}|$  & 16.7/10 &  0.08  & 17.3/10 &  0.07  & 18.1/10 &  0.05  & 14.8/10 &  0.14  & 17.9/10 &  0.06  \\
$\chi^{t\bar{t}}$  & ~~8.0/11 &  0.71  & 10.0/11 &  0.53  & ~~8.1/11 &  0.71  & ~~9.5/11 &  0.57  & 12.4/11 &  0.34  \\
$|y_\text{B}^{t\bar{t}}|$  & 15.3/10 &  0.12  & 15.7/10 &  0.11  & 16.6/10 &  0.08  & 14.1/10 &  0.17  & 16.6/10 &  0.08  \\
$|p_{\mathrm{out}}^{t\bar{t}}|$  & 17.1/10 &  0.07  & 53.6/10 &  $<\ $0.01~~~~  & 30.9/10 &  $<\ $0.01~~~~  & ~~8.6/10 &  0.57  & 32.7/10 &  $<\ $0.01~~~~  \\
$H_\text{T}^{t\bar{t}}$  & 5.4/9 &  0.80  & 5.0/9 &  0.83  & 6.4/9 &  0.70  & 3.6/9 &  0.94  & 6.8/9 &  0.66  \\
$|\Delta \phi(t_{1}, t_{2})|$  & 12.2/7~~ &  0.09  & 73.4/7~~ &  $<\ $0.01~~~~  & 23.6/7~~ &  $<\ $0.01~~~~  & 5.3/7 &  0.63  & 28.5/7~~ &  $<\ $0.01~~~~  \\
$|\cosThetaStar|$  & ~~7.0/10 &  0.72  & ~~9.8/10 &  0.46  & ~~6.8/10 &  0.74  & ~~7.4/10 &  0.69  & 10.5/10 &  0.39  \\
$p_\text{T}^{t,1}\otimes p_\text{T}^{t,2}$  & 27.1/15 &  0.03  & 27.0/15 &  0.03  & 36.7/15 &  $<\ $0.01~~~~  & 12.0/15 &  0.68  & 41.0/15 &  $<\ $0.01~~~~  \\
$|y^{t,1}|\otimes |y^{t,2}|$  & 11.6/19 &  0.90  & ~~9.8/19 &  0.96  & 12.0/19 &  0.88  & 14.3/19 &  0.77  & ~~9.7/19 &  0.96  \\
$|y^{t,1}|\otimes p_\text{T}^{t,1}$  & ~~8.5/15 &  0.90  & ~~7.6/15 &  0.94  & ~~9.4/15 &  0.85  & ~~9.5/15 &  0.85  & ~~8.4/15 &  0.91  \\
$|y^{t,2}|\otimes p_\text{T}^{t,2}$  & 15.9/20 &  0.72  & 17.1/20 &  0.65  & 19.5/20 &  0.49  & 10.8/20 &  0.95  & 20.7/20 &  0.41  \\
$p_\text{T}^{t,1}\otimes p_\text{T}^{t\bar{t}}$  & 16.1/15 &  0.37  & 12.6/15 &  0.63  & 26.7/15 &  0.03  & ~~7.3/15 &  0.95  & 30.7/15 &  $<\ $0.01~~~~  \\
$p_\text{T}^{t,1}\otimes m^{t\bar{t}}$  & 23.1/18 &  0.19  & 21.9/18 &  0.24  & 26.7/18 &  0.08  & 13.8/18 &  0.74  & 30.5/18 &  0.03  \\
$|y^{t\bar{t}}|\otimes p_\text{T}^{t,1}$  & 14.4/15 &  0.50  & 14.5/15 &  0.49  & 15.0/15 &  0.45  & 12.8/15 &  0.62  & 15.6/15 &  0.41  \\
$|y^{t\bar{t}}|\otimes |y^{t,1}|$  & 14.7/15 &  0.47  & 18.0/15 &  0.26  & 15.6/15 &  0.41  & 11.6/15 &  0.71  & 19.1/15 &  0.21  \\
$|y^{t,1}|\otimes m^{t\bar{t}}$  & 20.0/19 &  0.40  & 20.1/19 &  0.39  & 20.0/19 &  0.39  & 19.5/19 &  0.42  & 20.3/19 &  0.38  \\
$|y^{t\bar{t}}|\otimes m^{t\bar{t}}$  & 12.5/18 &  0.82  & 12.1/18 &  0.84  & 13.2/18 &  0.78  & 12.5/18 &  0.82  & 12.9/18 &  0.80  \\
$p_\text{T}^{t\bar{t}}\otimes m^{t\bar{t}}$  & 20.2/18 &  0.32  & 17.9/18 &  0.46  & 30.9/18 &  0.03  & ~~9.4/18 &  0.95  & 35.2/18 &  $<\ $0.01~~~~  \\
$|y^{t\bar{t}}|\otimes p_\text{T}^{t\bar{t}}$  & 19.1/15 &  0.21  & 14.5/15 &  0.49  & 29.4/15 &  0.01  & 12.2/15 &  0.66  & 33.4/15 &  $<\ $0.01~~~~  \\
$|y^{t\bar{t}}|\otimes m^{t\bar{t}}\otimes p_\text{T}^{t,1}$  & 21.9/31 &  0.88  & 24.1/31 &  0.81  & 24.6/31 &  0.79  & 18.0/31 &  0.97  & 26.9/31 &  0.68  \\
\hline
\end{tabular}


}
\vspace{0.2cm}
\caption{
Comparison between the measured normalized particle-level fiducial phase-space differential cross-sections and
the predictions of several MC event generators.
For each observed and predicted differential cross-section, a $\chi^2$ and a $p$-value are calculated using the
covariance matrix described in the text,
which includes all sources of uncertainty in the measurement.
The uncertainty in the prediction is not included.
The number of degrees of freedom (NDF) is equal to $N_{\rm b}-1$, where $N_{\rm b}$ is the number of measured values
in the distribution.
}
\label{tab:chi2:particle:relMain}
\end{table*}
 
\begin{table*} [ht!]
\footnotesize
\centering
\scalebox{0.75}{

\begin{tabular}{|c|cc|cc|cc|cc|cc|} \hline
Observable & \multicolumn{2}{|c|}{PWG+Py8} & \multicolumn{2}{|c|}{MG5\_aMC@NLO+Py8} & \multicolumn{2}{|c|}{PWG+H7.1.3} & \multicolumn{2}{|c|}{PWG+Py8 (more IFSR)} & \multicolumn{2}{|c|}{PWG+Py8 (less IFSR)} \\
& \multicolumn{2}{|c|}{NNPDF30 A14} & \multicolumn{2}{|c|}{NNPDF30 UE-EE-5} & \multicolumn{2}{|c|}{NNPDF30 A14} & \multicolumn{2}{|c|}{NNPDF30 A14} & \multicolumn{2}{|c|}{NNPDF30 A14} \\
& \multicolumn{1}{|c|}{$\chi^{2}$/NDF} & \multicolumn{1}{|c|}{$p$-value} & \multicolumn{1}{|c|}{$\chi^{2}$/NDF} & \multicolumn{1}{|c|}{$p$-value} & \multicolumn{1}{|c|}{$\chi^{2}$/NDF} & \multicolumn{1}{|c|}{$p$-value} & \multicolumn{1}{|c|}{$\chi^{2}$/NDF} & \multicolumn{1}{|c|}{$p$-value} & \multicolumn{1}{|c|}{$\chi^{2}$/NDF} & \multicolumn{1}{|c|}{$p$-value} \\ \hline \hline
$p_\text{T}^{t}$  & 3.1/9 &  0.96  & 3.7/9 &  0.93  & 4.3/9 &  0.89  & 1.4/9 &  1.00  & 6.2/9 &  0.72  \\
$|y^{t}|$  & ~~6.2/10 &  0.80  & ~~6.1/10 &  0.81  & ~~6.0/10 &  0.82  & ~~6.1/10 &  0.80  & ~~5.8/10 &  0.83  \\
$p_\text{T}^{t,1}$  & 3.2/8 &  0.92  & 2.6/8 &  0.96  & 3.6/8 &  0.89  & 4.0/8 &  0.86  & 3.1/8 &  0.93  \\
$|y^{t,1}|$  & ~~5.7/10 &  0.84  & ~~5.0/10 &  0.89  & ~~5.9/10 &  0.82  & ~~5.5/10 &  0.86  & ~~5.5/10 &  0.86  \\
$p_\text{T}^{t,2}$  & 5.4/8 &  0.71  & 9.6/8 &  0.30  & 5.9/8 &  0.66  & 3.2/8 &  0.92  & 8.3/8 &  0.41  \\
$|y^{t,2}|$  & ~~9.3/10 &  0.51  & ~~9.6/10 &  0.48  & ~~9.2/10 &  0.51  & ~~9.1/10 &  0.52  & ~~9.2/10 &  0.52  \\
$m^{t\bar{t}}$  & ~~7.4/12 &  0.83  & ~~8.6/12 &  0.73  & ~~7.4/12 &  0.83  & ~~7.6/12 &  0.81  & ~~7.1/12 &  0.85  \\
$p_\text{T}^{t\bar{t}}$  & 7.2/8 &  0.51  & 23.5/8~~ &  $<\ $0.01~~~~  & 8.6/8 &  0.38  & 3.1/8 &  0.93  & 13.0/8~~ &  0.11  \\
$|y^{t\bar{t}}|$  & 13.1/10 &  0.22  & 13.5/10 &  0.20  & 13.6/10 &  0.19  & 12.1/10 &  0.28  & 13.9/10 &  0.18  \\
$\chi^{t\bar{t}}$  & ~~7.6/11 &  0.74  & ~~8.0/11 &  0.71  & ~~8.3/11 &  0.69  & ~~7.4/11 &  0.77  & ~~9.9/11 &  0.54  \\
$|y_\text{B}^{t\bar{t}}|$  & 11.7/10 &  0.31  & 12.0/10 &  0.29  & 11.7/10 &  0.31  & 11.1/10 &  0.35  & 12.5/10 &  0.26  \\
$|p_{\mathrm{out}}^{t\bar{t}}|$  & ~~7.1/10 &  0.72  & 44.9/10 &  $<\ $0.01~~~~  & 12.5/10 &  0.25  & ~~4.6/10 &  0.92  & 11.2/10 &  0.34  \\
$H_\text{T}^{t\bar{t}}$  & 3.4/9 &  0.95  & 3.3/9 &  0.95  & 3.8/9 &  0.93  & 3.3/9 &  0.95  & 3.7/9 &  0.93  \\
$|\Delta \phi(t_{1}, t_{2})|$  & 10.5/7~~ &  0.16  & 81.1/7~~ &  $<\ $0.01~~~~  & 25.9/7~~ &  $<\ $0.01~~~~  & 4.2/7 &  0.76  & 19.2/7~~ &  $<\ $0.01~~~~  \\
$|\cosThetaStar|$  & ~~7.1/10 &  0.72  & ~~7.8/10 &  0.65  & ~~7.5/10 &  0.67  & ~~6.6/10 &  0.76  & ~~8.6/10 &  0.57  \\
$p_\text{T}^{t,1}\otimes p_\text{T}^{t,2}$  & 13.7/15 &  0.55  & 23.2/15 &  0.08  & 16.5/15 &  0.35  & ~~5.8/15 &  0.98  & 22.5/15 &  0.10  \\
$|y^{t,1}|\otimes |y^{t,2}|$  & ~~9.8/15 &  0.83  & ~~9.6/15 &  0.85  & ~~9.5/15 &  0.85  & 10.3/15 &  0.80  & ~~9.2/15 &  0.86  \\
$|y^{t,1}|\otimes p_\text{T}^{t,1}$  & ~~8.0/15 &  0.92  & ~~7.5/15 &  0.94  & ~~8.6/15 &  0.90  & ~~8.8/15 &  0.89  & ~~8.1/15 &  0.92  \\
$|y^{t,2}|\otimes p_\text{T}^{t,2}$  & 13.5/20 &  0.86  & 15.7/20 &  0.74  & 13.5/20 &  0.86  & 11.3/20 &  0.94  & 16.5/20 &  0.68  \\
$p_\text{T}^{t,1}\otimes p_\text{T}^{t\bar{t}}$  & 11.9/15 &  0.69  & 21.5/15 &  0.12  & 15.4/15 &  0.42  & ~~6.9/15 &  0.96  & 22.2/15 &  0.10  \\
$p_\text{T}^{t,1}\otimes m^{t\bar{t}}$  & 17.8/18 &  0.47  & 19.5/18 &  0.36  & 17.6/18 &  0.48  & 12.9/18 &  0.80  & 23.8/18 &  0.16  \\
$|y^{t\bar{t}}|\otimes p_\text{T}^{t,1}$  & 12.0/15 &  0.68  & 11.6/15 &  0.71  & 11.4/15 &  0.72  & 11.5/15 &  0.71  & 12.7/15 &  0.63  \\
$|y^{t\bar{t}}|\otimes |y^{t,1}|$  & 14.2/15 &  0.51  & 14.7/15 &  0.47  & 14.1/15 &  0.52  & 12.2/15 &  0.67  & 17.2/15 &  0.31  \\
$|y^{t,1}|\otimes m^{t\bar{t}}$  & 19.0/19 &  0.46  & 18.6/19 &  0.49  & 19.3/19 &  0.44  & 19.0/19 &  0.46  & 19.2/19 &  0.44  \\
$|y^{t\bar{t}}|\otimes m^{t\bar{t}}$  & 12.3/18 &  0.83  & 12.1/18 &  0.84  & 12.2/18 &  0.84  & 13.6/18 &  0.75  & 11.8/18 &  0.86  \\
$p_\text{T}^{t\bar{t}}\otimes m^{t\bar{t}}$  & 25.9/18 &  0.10  & 22.0/18 &  0.23  & 32.0/18 &  0.02  & 13.8/18 &  0.74  & 35.2/18 &  $<\ $0.01~~~~  \\
$|y^{t\bar{t}}|\otimes p_\text{T}^{t\bar{t}}$  & 13.5/15 &  0.56  & 18.9/15 &  0.22  & 15.6/15 &  0.41  & 12.7/15 &  0.63  & 16.3/15 &  0.36  \\
$|y^{t\bar{t}}|\otimes m^{t\bar{t}}\otimes p_\text{T}^{t,1}$  & 15.5/31 &  0.99  & 17.9/31 &  0.97  & 15.1/31 &  0.99  & 15.5/31 &  0.99  & 17.7/31 &  0.97  \\
\hline
\end{tabular}


}
\vspace{0.2cm}
\caption{
Comparison between the measured normalized parton-level differential cross-sections and the
predictions from several MC event generators.
For each observable and calculation, a $\chi^2$ and a $p$-value are calculated using the covariance matrix described in the text, which includes all sources of uncertainty in the measurement. The uncertainty in the calculation is not included.  The number of degrees of freedom (NDF) is equal to $N_{\rm b}-1$, where $N_{\rm b}$ is the number of bins in the distribution.
}
\label{tab:chi2:parton:relMain}
\end{table*}
 
In the case of the normalized particle-level fiducial phase-space differential cross-sections,
good agreement is generally observed.
The one-dimensional distributions that are sensitive to extra radiation
(i.e.\ the \pT\ of the \ttbar\ system, \ptttbar, the out-of-plane momentum, \absPoutttbar, and the absolute value
of the azimuthal separation of the
top-quark jets, \absdeltaPhittbar) yield $p$-values below 1\%\ for all MC predictions
except for the nominal prediction
of \POWPY[8] and those including more initial- and final-state radiation.
These distributions indicate a deficit of radiation in the MC predictions, i.e.\ \ptttbar\ (Figure~\ref{fig:particle:tt_pt:rel}) and \Poutttbar\ (Figure~\ref{fig:particle:tt_pout:rel}) are softer, while \absdeltaPhittbar\ (Figure~\ref{fig:particle:tt_dPhittbar:rel}) is closer to $\pi$ for the predictions.
Moreover, the \AMCatNLO{}+\Pythia[8] calculations for these observables differ significantly from
the predictions of the other MC generators.
It is notable that these discrepancies are not evident in the parton-level comparisons.
 
The \pTttwo\ distribution (Figure~\ref{fig:particle:t2_pt:rel}), and consequently \HTttbar\ to a lesser
extent (Figure~\ref{fig:particle:tt_HTttbar:rel}), indicates that the MC particle-level predictions have a
harder \ptttbar\ distribution
than is observed in data, except for the prediction of \POWPY[8] with more initial- and final-state radiation.
Low $p$-values are seen for the comparison of the multi-dimensional distributions for
$\pTtone \otimes \pTttwo$ (Figure~\ref{fig:particle:t1_pt_vs_t2_pt:rel}) for all MC calculations except for the prediction of \POWPY[8]
with more initial- and final-state radiation.
The largest slope in the calculation/data ratio is observed for the largest values of \pTtone.
The \POWHEG{}+\Herwig[7.1.3] calculations and the calculations of \POWPY[8] with less initial- and
final-state radiation give low $p$-values for the
$\pTtone \otimes \ptttbar$ (Figure~\ref{fig:particle:t1_pt_vs_ttbar_pt:rel}), $\ptttbar \otimes \mttbar$ (Figure~\ref{fig:particle:ttbar_pt_vs_ttbar_mass:rel}) and $\absyttbar \otimes \ptttbar$ (Figure~\ref{fig:particle:ttbar_y_vs_ttbar_pt:rel}) distributions.
There are large slopes in the calculation/data ratios for all
$\pTtone \otimes \ptttbar$ (Figure~\ref{fig:particle:t1_pt_vs_ttbar_pt:rel}) and $\absyttbar \otimes \ptttbar$ (Figure~\ref{fig:particle:ttbar_y_vs_ttbar_pt:rel})
distributions, which confirm the trends observed in the
$\ptttbar$ differential cross-section, while different trends are
observed in different $\ptttbar \otimes \mttbar$\ differential cross-sections
(Figure~\ref{fig:particle:ttbar_pt_vs_ttbar_mass:rel}).
A steep gradient in the calculation/data ratio can also be observed in the
$\absyttbar \otimes |\ytone|$ (Figure~\ref{fig:particle:ttbar_y_vs_t1_y:rel}) and $\pTtone \otimes \mttbar$ (Figure~\ref{fig:particle:t1_pt_vs_ttbar_mass:rel})
differential cross-sections for large values of \absyttbar and \pTtone,
respectively, for all MC calculations, except for the calculations of \POWPY[8] with more initial- and final-state radiation.
 
The level of agreement of the normalized parton-level fiducial phase-space differential cross-section calculations
with the measurements is generally better, as evidenced by the differential cross-section comparisons and confirmed
by the $p$-values in Table~\ref{tab:chi2:parton:relMain}.
The better agreement at the parton level, especially in the differential cross-sections as a function of \ptttbar,
\absPoutttbar, and \absdeltaPhittbar,  suggests that the
poorer descriptions at the particle level are introduced by parton-showering and hadronization models,
and/or ISR/FSR modelling.


\section{EFT interpretation}
\label{sec:eft}

 
The SMEFT~\cite{Buchmuller:1985jz} provides a theoretically
elegant way to encode the modifications of the top-quark
properties induced by a wide class of BSM theories that
reduce to the SM at low energies.
Within the mathematical language of the SMEFT relevant to top-quark
physics, the effects of BSM dynamics are characterized by an energy scale $\Lambda$\ at which BSM effects become apparent and which is
well above the typical scale for top-quark processes given by \mtop.
These BSM effects can be parameterized at low energies, $E \ll \Lambda$, in terms of higher-dimensional operators built from the SM fields while respecting symmetries of the SM such as gauge invariance using the Lagrangian
\begin{equation}
\mathcal{L}_{\textrm{SMEFT}} = \mathcal{L}_{\textrm{SM}} + \sum_{i} {\frac{C_i}{\Lambda^2} \mathcal{O}_i^{(6)}} +
\sum_{j} {\frac{B_j}{\Lambda^4} \mathcal{O}_j^{(8)}} + ...,
\label{eq:SMEFT}
\end{equation}
where $\mathcal{L}_{\textrm{SM}}$ is the SM Lagrangian, $O_i^{(6)}$ and
$O_j^{(8)}$ represent a complete set of operators of mass-dimensions
$d=6$ and $d=8$, and $C_i$ and $B_j$ are the corresponding complex-valued
Wilson coefficients that determine the strength of the
operators. Operators with $d=5$ and $d=7$ violate lepton
and/or baryon number conservation and are not relevant for top-quark
physics. The effective-theory expansion in Eq.~(\ref{eq:SMEFT}) is robust, fully
general, and can be systematically matched to explicit
ultraviolet-complete BSM scenarios. Contributions from operators of
mass-dimension $d=8$ or higher are not considered in this analysis.
 
Measurements of top-quark differential cross-sections can
place constraints on SMEFT Wilson coefficients. For any cross-section $\sigma(C_i)$,
the corresponding expression
including SM and SMEFT operators up to dimension-6 becomes
\begin{eqnarray}
\sigma(C_i)
& = &
\sigma_{\textrm{SM}} +
\sigma_{\textrm{SM--EFT}}   +
\sigma_{\textrm{EFT--EFT}}  \nonumber \\
& = &
\sigma_{\textrm{SM}}
+ \frac{1}{\Lambda^2} \sum_{i} \alpha_i C_{i}
+ \frac{1}{\Lambda^4} \sum_{i} \beta_i C_{i}^2
+ \frac{1}{\Lambda^4}\sum_{i,j,i<j} \tilde{\beta}_{ij} C_{i}C_{j}, \label{eq:EFT:crossTerms}
\end{eqnarray}
where $\sigma_{\textrm{SM}}$ is the SM cross-section for the given process and
$\sigma_{\textrm{SM--EFT}}$ is an interference term between SM and BSM
operators, which depends linearly on the Wilson coefficients $C_i$.
The last term, $\sigma_{\textrm{EFT--EFT}}$, includes products of BSM operators,
including possible interference between SMEFT operators, and depends
quadratically on the Wilson coefficients $C_i$.
The constants $\alpha_i$, $\beta_i$, and $\tilde{\beta}_{ij}$\ are used to parameterize
the dependence of the cross-section on each Wilson coefficient.
Their determination is described below.
 
For the presented results, the `dim6top'
model~\cite{AguilarSaavedra:2018nen,www:dim6top} is used
to implement SMEFT at leading order, using the Warsaw basis for the
operators~\cite{Grzadkowski:2010es}. Signal events for this study were
generated with the \AMCatNLO\ MC generator and include contributions from the SM term,
SM--EFT interference, and the EFT--EFT interference term.
Within the dim6top model, there are numerous Wilson coefficients describing non-SM top-quark interactions.
A total of 43 coefficients are systematically explored to identify those
that could be constrained by the differential cross-section measurements.
 
Three different sets of operators are identified among these
coefficients: 2-liqht-quark and 2-heavy-quark (2LQ2HQ) operators,
4-heavy-quark (4HQ) operators, and 2-heavy-quark plus boson (2HQV)
operators.
Measurements are made for a subset of coefficients, chosen with regard to
sensitivity, stability of results for linear and linear+quadratic
terms, and competitiveness with the results reported in global
EFT fits~\cite{Hartland:2019bjb,Ethier:2021bye}.
 
These sensitivity considerations lead to individual measurements of seven Wilson
coefficients, all corresponding to 2LQ2HQ operators:
$C_{Qq}^{3,8}$, $C_{Qq}^{1,8}$, $C_{Qu}^{8}$, $C_{Qd}^{8}$, $C_{tq}^{8}$, $C_{tu}^{8}$, and $C_{td}^{8}$.
The operators feature different chiral and colour structures indicated by lower and upper indexes~\cite{AguilarSaavedra:2018nen}.
All these coefficients are purely real with no imaginary part.
No individual limits are placed on 2HQV operators since the only sensitive coefficient, $C_{tG}$,
does not provide limits competitive with the best one-dimensional limits available.
However, the real part of $C_{tG}$\ is measured in combination with $C_{Qq}^{3,8}$\ while
no limit is placed on the imaginary part of $C_{tG}$.
No limits on 4HQ operators are presented since their sensitivity largely
originates from terms in $\sigma_{\textrm{EFT--EFT}}$ suppressed by
$\Lambda^{-4}$.
As interference effects from dimension-8 operators with SM
operators, for which no calculations are available, contribute to the
cross-section at the same power of $\Lambda$, the interpretation of such
limits is difficult.
Simultaneous measurements of pairs of Wilson coefficients are made for
three combinations that serve as an example for other
combinations from the same set of operators.
In particular, the measurement of  $C_{tG}$ vs $C_{Qq}^{3,8}$ is an example of the 2HQV vs 2LQ2HQ
combinations and $C_{Qq}^{3,8}$ vs $C_{Qq}^{1,8}$ and $C_{Qq}^{1,8}$ vs $C_{tq}^{8}$ are examples of
2LQ2HQ vs 2LQ2HQ combinations.
 
The absolute differential cross-section at parton level as a function of \pTtone\ is found to be
more sensitive to EFT effects than the differential cross-sections for other observables
(|\ytone|, \HTttbar, \mttbar, \absyttbar) for which it is possible to make NNLO calculations
using the \MATRIX program.
EFT constraints are therefore set using the absolute differential cross-section as a function of \pTtone.
The unfolding procedure recovers within 1\% the
generator-level distribution for an input \pTtone\ distribution that
includes the EFT contributions for both \ttbar\ signal and \ttbar\ non-all-hadronic background where the Wilson coefficients values are set to the expected upper limits of this measurement.
 
For each bin of the $\pt^{t,1}$ distribution, a parameterization using the quadratic dependence of the
differential cross-section as a function of a given Wilson coefficient is developed according to
Eq.~(\ref{eq:EFT:crossTerms}), i.e.\ the $\alpha_i,\beta_i$\ constants for each Wilson coefficient and for each
bin of the distribution are determined.
The $\tilde{\beta}_{ij}$\ constants are then estimated by fitting the above formula to samples having
two $C_i$ non-zero and using the previously determined values of
$\alpha_i$ and $\beta_i$.
The $\sigma_{\textrm{SM}}$ parameter determined in the above parameterization gives the LO SM \ttbar\ calculation and is
not used.
 
The nominal \MATRIX NNLO calculation with the symmetrized scale uncertainty is employed as the SM prediction for the full model used to
interpret the differential cross-section. The
parameterization of LO EFT effects obtained above includes
linear, quadratic, and cross-terms and is also used in the fit.
The model includes all systematic and statistical uncertainties of the
measurement, as well as their correlations as described by the covariance
matrix of the differential measurement. In the model, the NNLO QCD
scale uncertainty is used as a theoretical uncertainty and it is taken as fully correlated between the measurements.
It was verified that a given scale variation always provides the
maximum (or minimum) value for all the predictions. It was also verified that
changing the correlation from 100\%\ to 50\%\ has little impact on the limits.
The full fit is implemented in the Bayesian inference
tool \EFTfitter~\cite{Castro:2016jjv}.
To make the dependence of the Wilson coefficients on the energy scale of the new physics explicit, the results are
presented as the product $C_i(\TeV/\Lambda)^2$.
This also facilitates comparisons with other results where $\Lambda = 1$~\TeV.
 
\begin{figure*}[!thbp]
\centering
\includegraphics[width=0.7\textwidth]{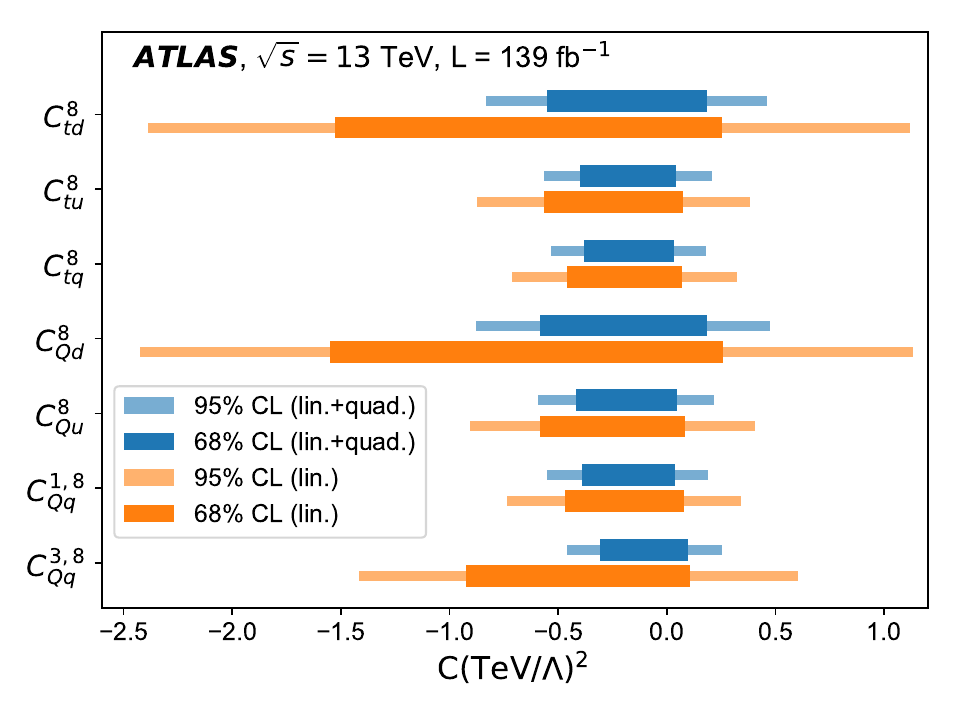}
\caption{A summary of one-dimensional limits on selected Wilson coefficients corresponding to 2-light-quark
2-heavy-quark operators.
The limits are provided for cases where both linear and quadratic terms are included
and for cases where only linear terms are included.}
\label{fig:EFTfit:1D}
\end{figure*}

\begin{figure*}[!thbp]
\centering
\includegraphics[width=0.7\textwidth]{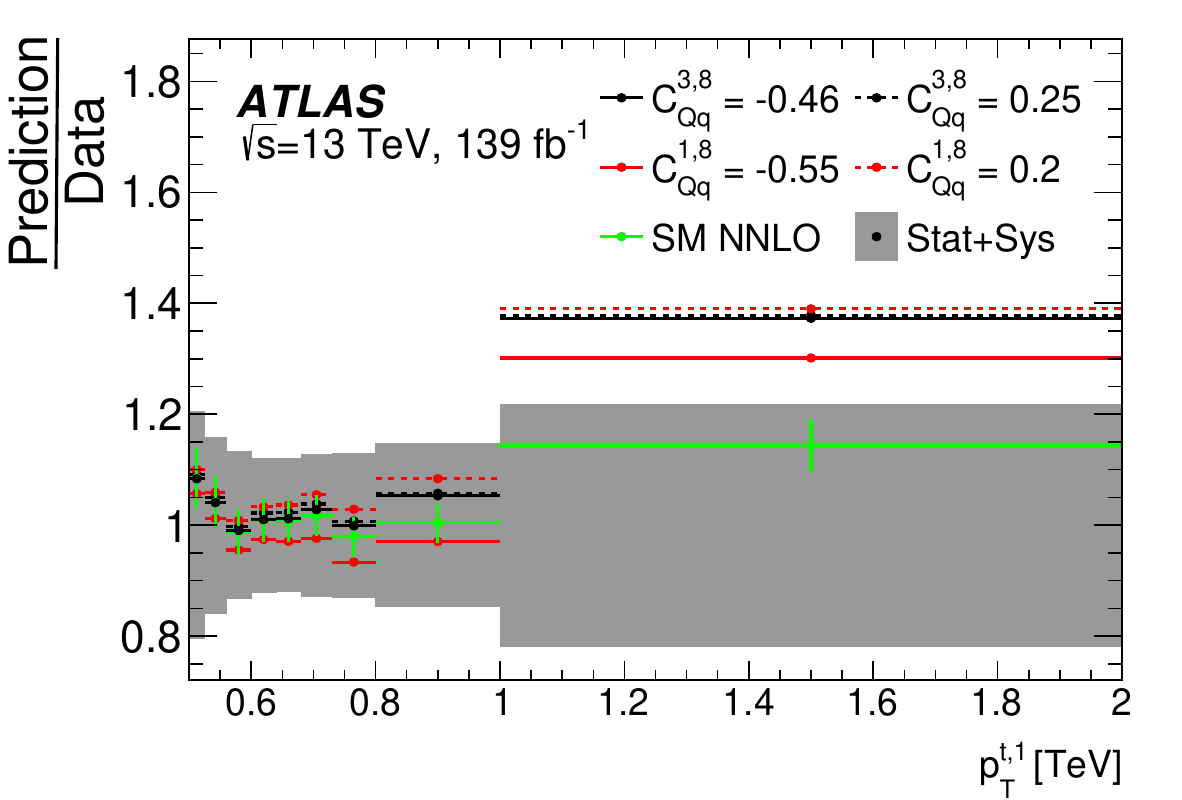}
\caption{The ratio of various SMEFT predictions that include non-zero Wilson coefficients to the data for the
leading top-quark \pTtone\  distribution. No uncertainties are included for these calculations.
The values of non-zero Wilson coefficients correspond approximately to the 95\% CL limits obtained by this measurement.
The SM NNLO calculations as obtained by the nominal \MATRIX\ settings is also shown. It includes the symmetrized scale uncertainty.
}
\label{fig:EFT:pt1}
\end{figure*}

\begin{figure*}[!thbp] \centering
\subfigure[]{
\includegraphics[width=0.49\textwidth]{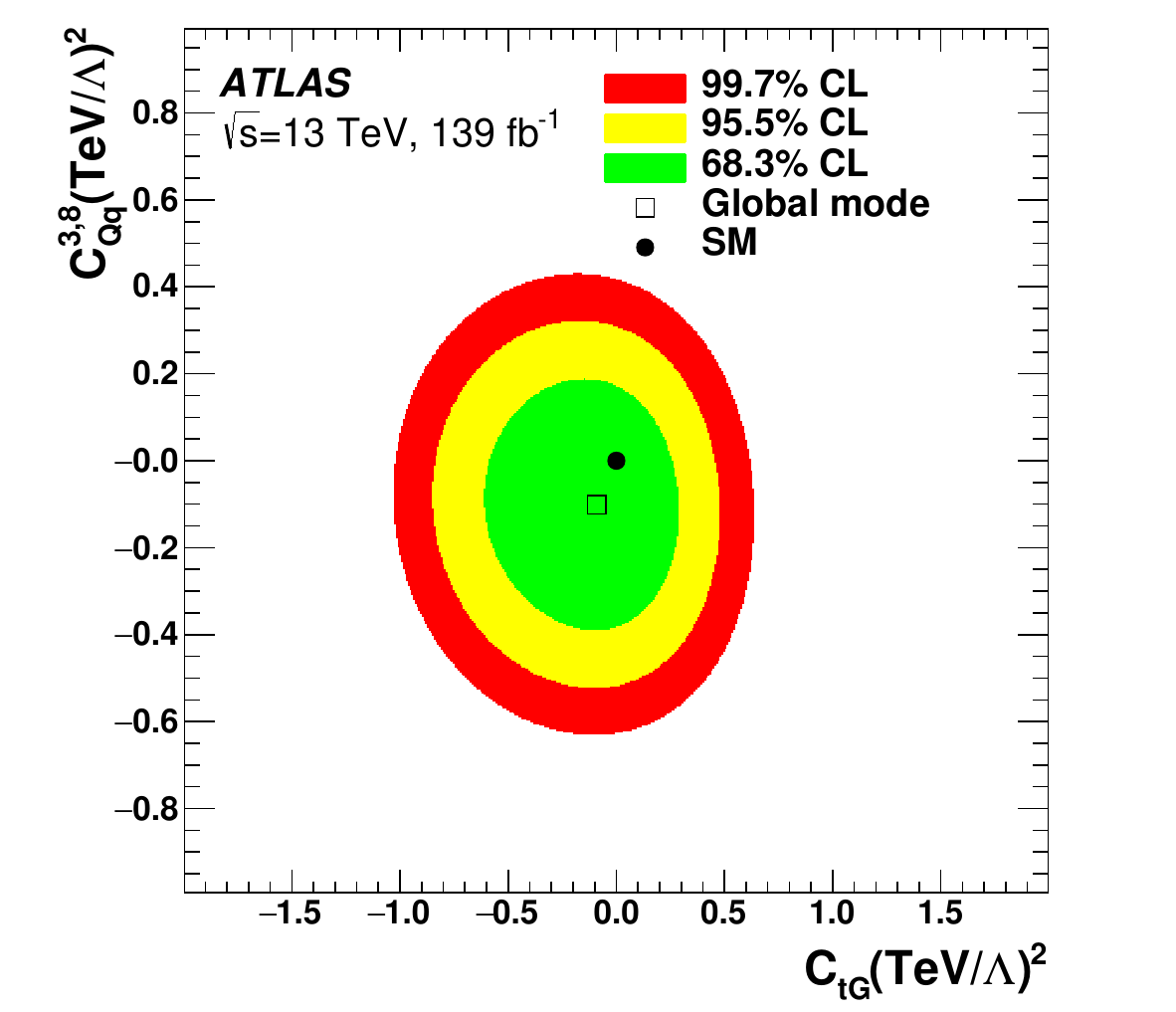}\label{fig:EFTfit:2D:ID16_51}}
\subfigure[]{
\includegraphics[width=0.49\textwidth]{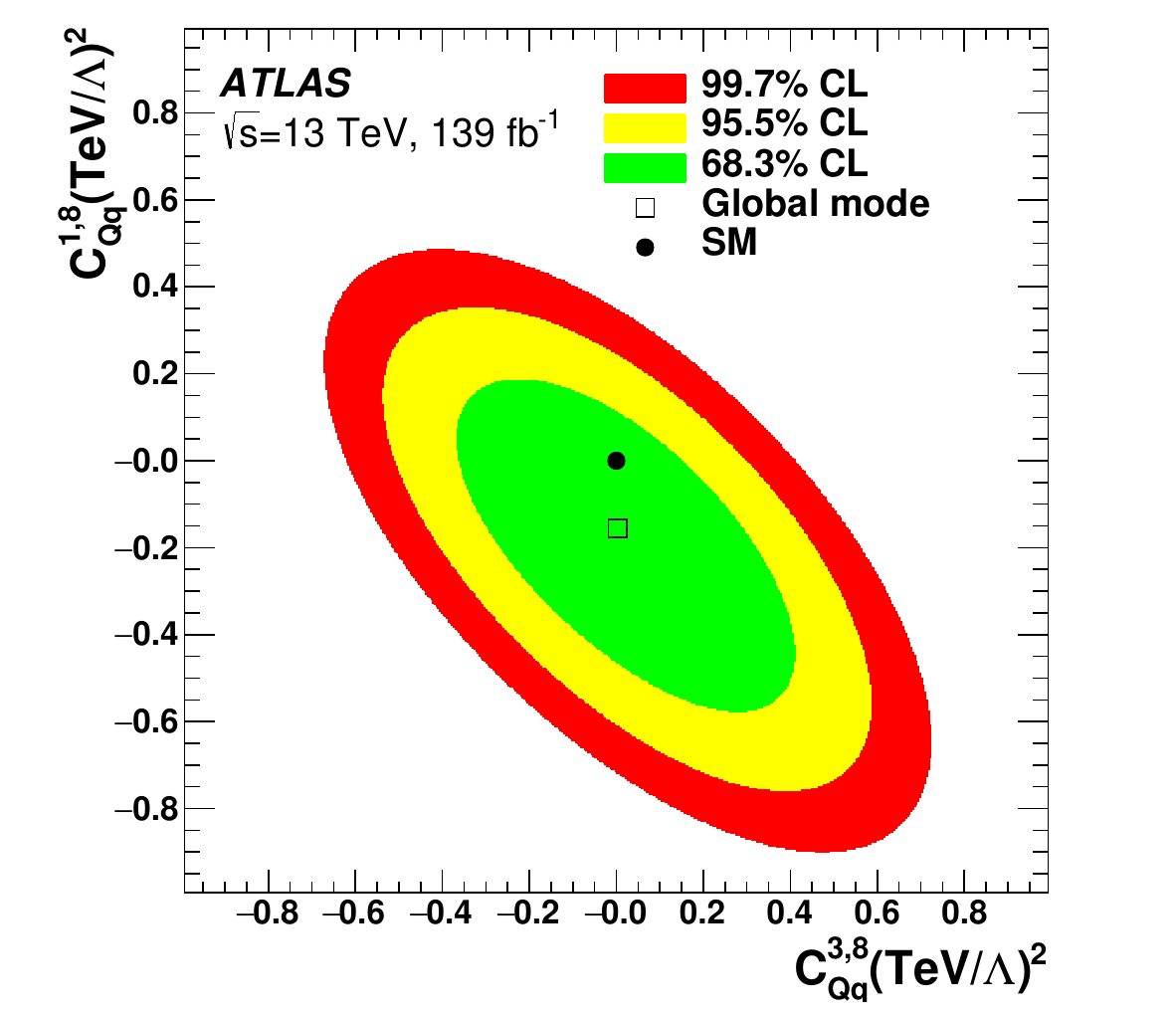}\label{fig:EFTfit:2D:ID51_52}}
\subfigure[]{
\includegraphics[width=0.49\textwidth]{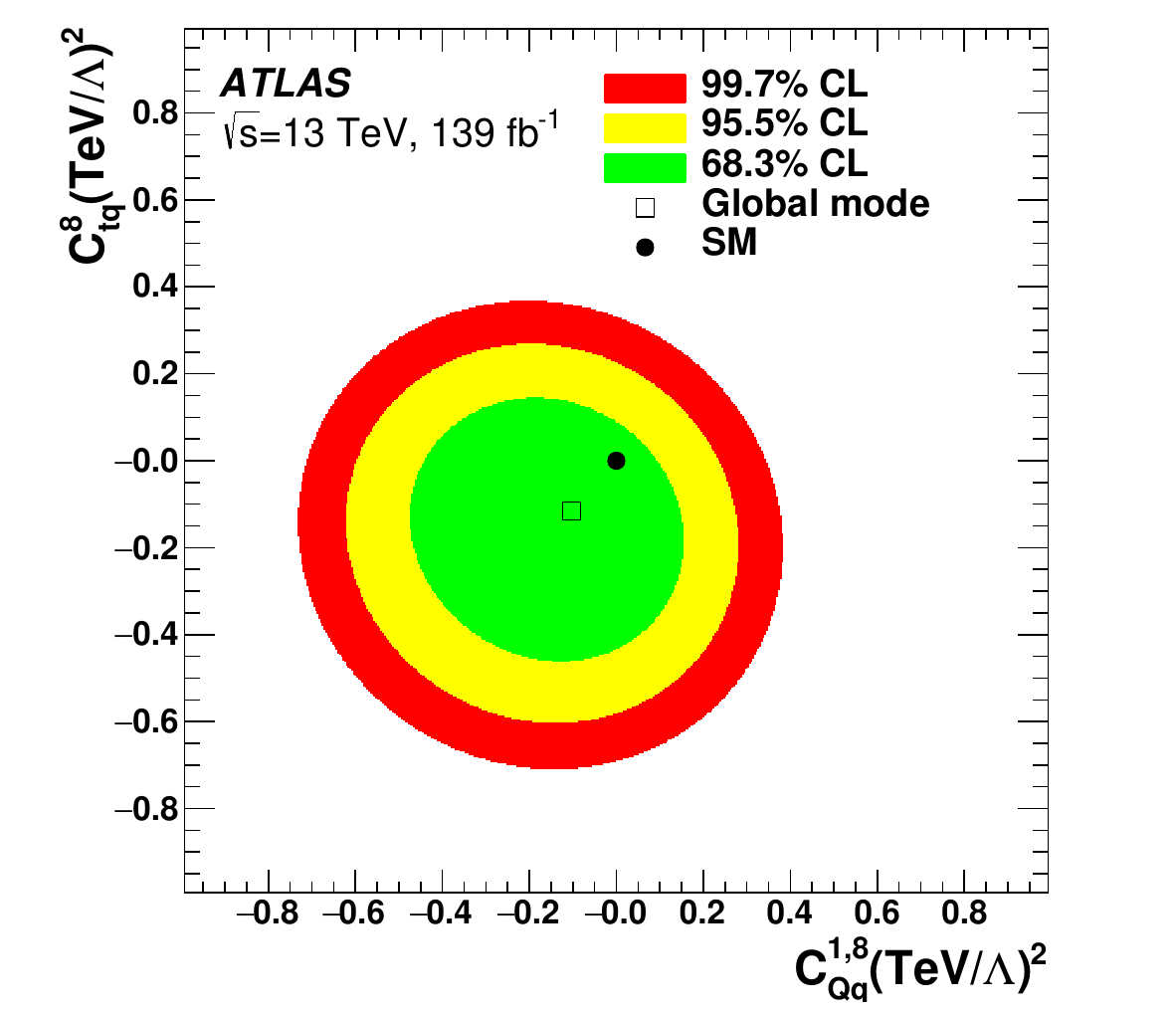}\label{fig:EFTfit:2D:ID52_55}}
\caption{The two-dimensional limits on
\subref{fig:EFTfit:2D:ID16_51}\  $C_{tG}$\ vs $C_{Qq}^{3,8}$,
\subref{fig:EFTfit:2D:ID51_52}\ $C_{Qq}^{3,8}$\ vs $C_{Qq}^{1,8}$, and
\subref{fig:EFTfit:2D:ID52_55}\ $C_{Qq}^{1,8}$\ vs $C_{tq}^{8}$ coefficients. Here, $C_{tG}$ is the real part of the
Wilson coefficient. The global mode corresponds to the most probable value of the posterior probability distribution.}
\label{fig:EFTfit:2D}
\end{figure*}

Fits of seven individual Wilson coefficients are performed with all other coefficients set to zero.
A summary of the 95\% confidence level (CL) and 68\% CL limits is shown in Figure~\ref{fig:EFTfit:1D}.
The limits are provided for cases where both linear and quadratic terms are included and for cases
where only linear terms are included.
The 95\% CL limits on $C_i(\TeV/\Lambda)^2$ are within the range of ($-0.9,+0.5$) for cases where both linear and quadratic terms are included.
The inclusion of the quadratic terms leads to tighter bounds by 30\%--60\% for all Wilson coefficients.
The ratios of various SMEFT predictions that  include non-zero Wilson coefficients to the data for the leading
top-quark \pTtone\  distribution are shown in Figure~\ref{fig:EFT:pt1}.
The effect of EFT contributions is seen mainly in the highest \pTtone\ bin.
Limits for selected pairs of Wilson coefficients are shown in Figure~\ref{fig:EFTfit:2D}.
 
The 95\% CL limit intervals for individual Wilson coefficients
$C_{Qq}^{1,8}$, $C_{tq}^{8}$, and $C_{tu}^{8}$ for cases where both
linear and quadratic terms are included are about a factor of 4--5
smaller than those obtained from the measurement of the \ttbar energy
asymmetry~\cite{TOPQ-2019-28}.
Also, the 95\% CL limits on individual Wilson coefficients are typically
10\%--50\% more restrictive than the currently available
individual limits from the global fits~\cite{Hartland:2019bjb,Ethier:2021bye}.
 
The EFT analysis presented here shows that boosted \ttbar\ differential cross-section measurements
provide significant constraints on the SMEFT Wilson coefficients and can be used in
global fits that include these top-quark Wilson coefficients.
 
\FloatBarrier
 
\section{Conclusion}
\label{sec:conclusion}

 
Kinematic distributions of top quarks and the \ttbar\ system are measured
by selecting boosted top-quark jets and unfolding the observed distributions
to a particle-level fiducial phase space and a parton-level fiducial phase space.
The fiducial phase-space cross-sections and differential cross-sections
are compared with several NLO calculations with and without parton showering and hadronization,
and with a parton-level NNLO calculation.
The \ttbar\ events were produced in 13~\TeV{} $pp$\ collisions
and recorded by the ATLAS detector at the LHC.
The data set corresponds to an integrated luminosity of \lumitot.
 
The observed particle-level fiducial phase-space and parton-level fiducial phase-space
cross-sections are
\begin{eqnarray}
\sigma^{\ttbar,\text{fid}}_\text{particle} \times B(\ttbar \rightarrow \text{hadrons}) & = & \inclXsecParticle\ {\rm fb},\ {\rm and} \nonumber \\
\sigma^{\ttbar,\text{fid}}_\text{parton} & = & \inclXsecParton\ {\rm pb}. \nonumber
\end{eqnarray} 
Both are approximately 20\%\ lower than the \POWPY[8] NLO+PS
predictions scaled to the NNLO total cross-section of $\numRF{398.0}{3} \numpmRF{+48.16}{-49.00}{2}$~fb and $\numRF{2.338}{3} \pm \numRF{0.282}{2}$~pb for particle and parton level, respectively, but still compatible within the uncertainties.
The result at parton level is in excellent agreement with the NNLO prediction $\numRF{1.9645}{3} \numpmRP{+0.0188}{-0.174}{2}$~pb.
 
Normalized particle-level fiducial phase-space differential cross-sections
are measured as a function of the transverse momentum and rapidity of
the leading and second-leading top-quark jets, and a top-quark jet chosen
at random from each event.
Also, normalized differential cross-sections are measured as a function of the
mass, \pT, and rapidity of the \ttbar{} system.
In addition, a set of observables describing the hard-scattering interaction
(\cosThetaStar, \chittbar, and \boostttbar) and sensitive to the emission of radiation along with
the \ttbar{} final state (\absdeltaPhittbar, \absPoutttbar, and \HTttbar) are presented.
Normalized parton-level fiducial phase-space differential cross-sections are also shown for the same set of observables.
Furthermore, several normalized double- and triple-differential distributions as a function of a subset of the  observables
described above are presented.
 
The normalized differential cross-sections exhibit a precision of 10\%--20\% and are
in agreement with several NLO+PS predictions for most of the observables measured.
These result in the most precise differential cross-sections measured in the
boosted \ttbar\ all-hadronic final state,
with uncertainties being a factor of two smaller than in
previous ATLAS measurements overall,
and up to a factor of four smaller in the region with top-quark $\pt > 1~\TeV$.
A notable disagreement between the measurement and the NLO+PS prediction is observed in the second-leading
top-quark \pt\ distribution, where the data are softer than
predicted, as seen in several previous ATLAS
studies.
Also, observables sensitive to gluon radiation are not well described by most NLO+PS MC calculations.
Agreement with the NNLO predictions at the parton level is in general better than with the NLO+PS calculations.
These observations point to the need for NNLO+PS MC calculations, as well as a better
understanding of initial- and final-state radiation.
 
An interpretation of the measurements of these boosted \ttbar\ events
within the SMEFT framework is performed for the first time.
Using the measurement of the parton-level leading top-quark \pt differential cross-section,
limits are set on individual coefficients of several four-fermion operators.
These are competitive with, and typically more stringent than, existing
limits from global EFT fits.
Moreover, two-dimensional limits are also set on several pairs of coefficients.
This SMEFT interpretation shows that boosted \ttbar cross-section measurements are
well-suited to constraining several four-fermion operators and will be
useful in future global EFT analyses.


\section*{Acknowledgements}


We thank CERN for the very successful operation of the LHC, as well as the
support staff from our institutions without whom ATLAS could not be
operated efficiently.
 
We acknowledge the support of
ANPCyT, Argentina;
YerPhI, Armenia;
ARC, Australia;
BMWFW and FWF, Austria;
ANAS, Azerbaijan;
CNPq and FAPESP, Brazil;
NSERC, NRC and CFI, Canada;
CERN;
ANID, Chile;
CAS, MOST and NSFC, China;
Minciencias, Colombia;
MEYS CR, Czech Republic;
DNRF and DNSRC, Denmark;
IN2P3-CNRS and CEA-DRF/IRFU, France;
SRNSFG, Georgia;
BMBF, HGF and MPG, Germany;
GSRI, Greece;
RGC and Hong Kong SAR, China;
ISF and Benoziyo Center, Israel;
INFN, Italy;
MEXT and JSPS, Japan;
CNRST, Morocco;
NWO, Netherlands;
RCN, Norway;
MEiN, Poland;
FCT, Portugal;
MNE/IFA, Romania;
MESTD, Serbia;
MSSR, Slovakia;
ARRS and MIZ\v{S}, Slovenia;
DSI/NRF, South Africa;
MICINN, Spain;
SRC and Wallenberg Foundation, Sweden;
SERI, SNSF and Cantons of Bern and Geneva, Switzerland;
MOST, Taiwan;
TENMAK, T\"urkiye;
STFC, United Kingdom;
DOE and NSF, United States of America.
In addition, individual groups and members have received support from
BCKDF, CANARIE, Compute Canada and CRC, Canada;
PRIMUS 21/SCI/017 and UNCE SCI/013, Czech Republic;
COST, ERC, ERDF, Horizon 2020 and Marie Sk{\l}odowska-Curie Actions, European Union;
Investissements d'Avenir Labex, Investissements d'Avenir Idex and ANR, France;
DFG and AvH Foundation, Germany;
Herakleitos, Thales and Aristeia programmes co-financed by EU-ESF and the Greek NSRF, Greece;
BSF-NSF and MINERVA, Israel;
Norwegian Financial Mechanism 2014-2021, Norway;
NCN and NAWA, Poland;
La Caixa Banking Foundation, CERCA Programme Generalitat de Catalunya and PROMETEO and GenT Programmes Generalitat Valenciana, Spain;
G\"{o}ran Gustafssons Stiftelse, Sweden;
The Royal Society and Leverhulme Trust, United Kingdom.
 
The crucial computing support from all WLCG partners is acknowledged gratefully, in particular from CERN, the ATLAS Tier-1 facilities at TRIUMF (Canada), NDGF (Denmark, Norway, Sweden), CC-IN2P3 (France), KIT/GridKA (Germany), INFN-CNAF (Italy), NL-T1 (Netherlands), PIC (Spain), ASGC (Taiwan), RAL (UK) and BNL (USA), the Tier-2 facilities worldwide and large non-WLCG resource providers. Major contributors of computing resources are listed in Ref.~\cite{ATL-SOFT-PUB-2021-003}.


\clearpage
\appendix
\part*{Appendix}
\addcontentsline{toc}{part}{Appendix}
 
\section{Additional particle-level fiducial phase-space differential cross-sections}
\label{sec:appendix:particle_level}

\begin{figure*}[htbp]
\centering
\subfigure[]{\includegraphics[width=0.49\textwidth]{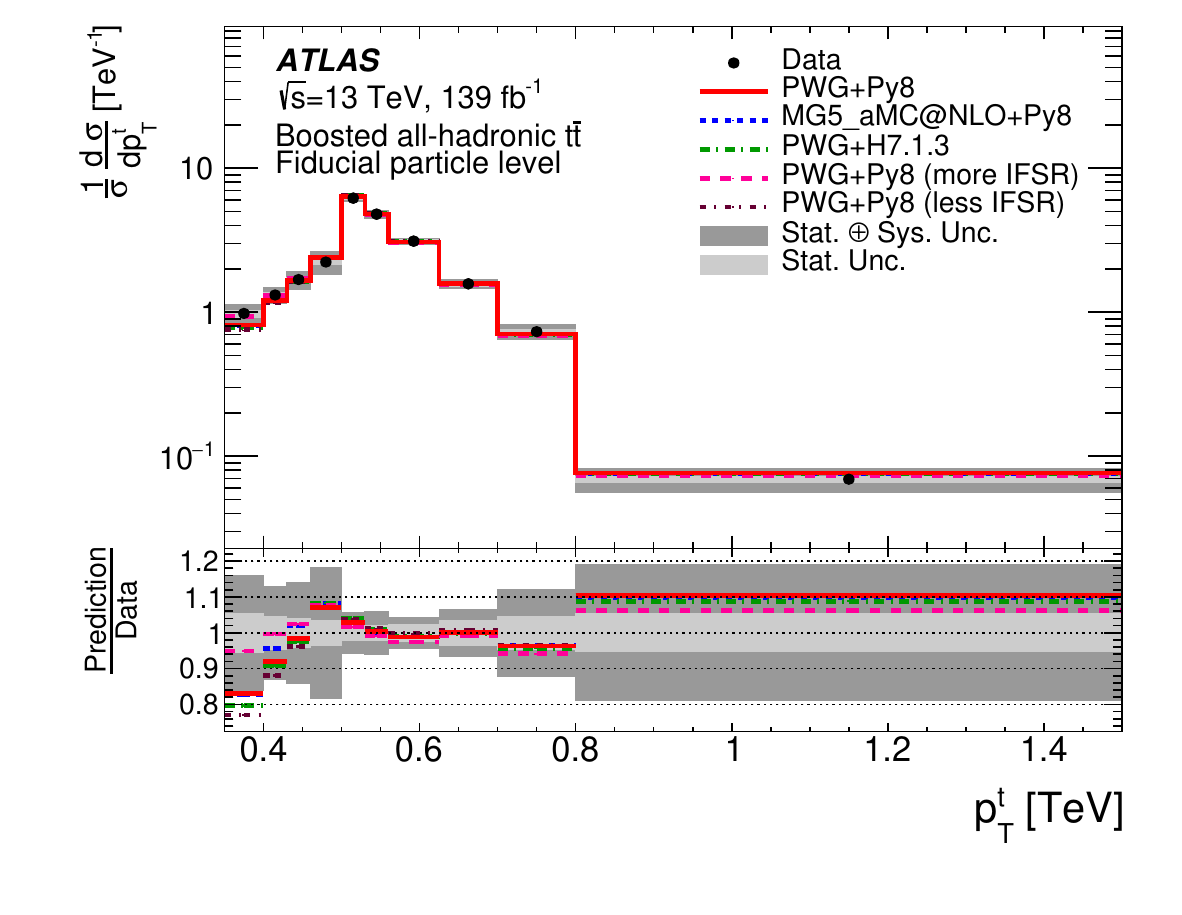}\label{fig:results:particle:top_pt_rel}}
\subfigure[]{\includegraphics[width=0.49\textwidth]{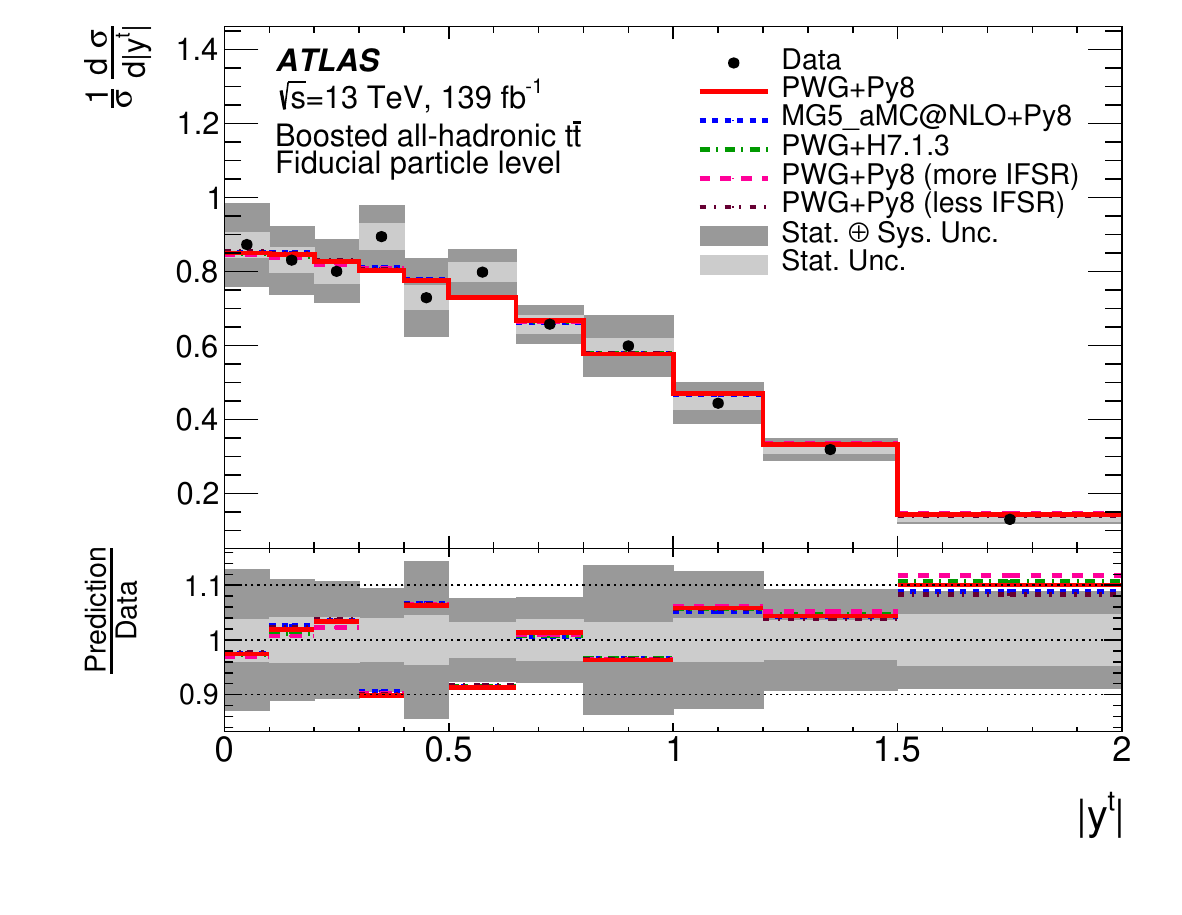}\label{fig:results:particle:randomTop_y_rel}}
\caption{
Normalized particle-level fiducial phase-space differential cross-sections as a function of
\subref{fig:results:particle:top_pt_rel}~the transverse momentum and
\subref{fig:results:particle:randomTop_y_rel}~the absolute value of the rapidity of the randomly chosen top-quark jet.
The dark and light grey bands indicate the total uncertainty and the statistical uncertainty, respectively, of the data in each bin.
Data points are placed at the centre of each bin.
The \POWPY[8] MC sample is used as the nominal prediction to correct the data to particle level.
}
\label{fig:particle:random_top:rel}
\end{figure*}

\begin{figure*}[htbp]
\centering
\subfigure[]{ \includegraphics[width=0.49\textwidth]{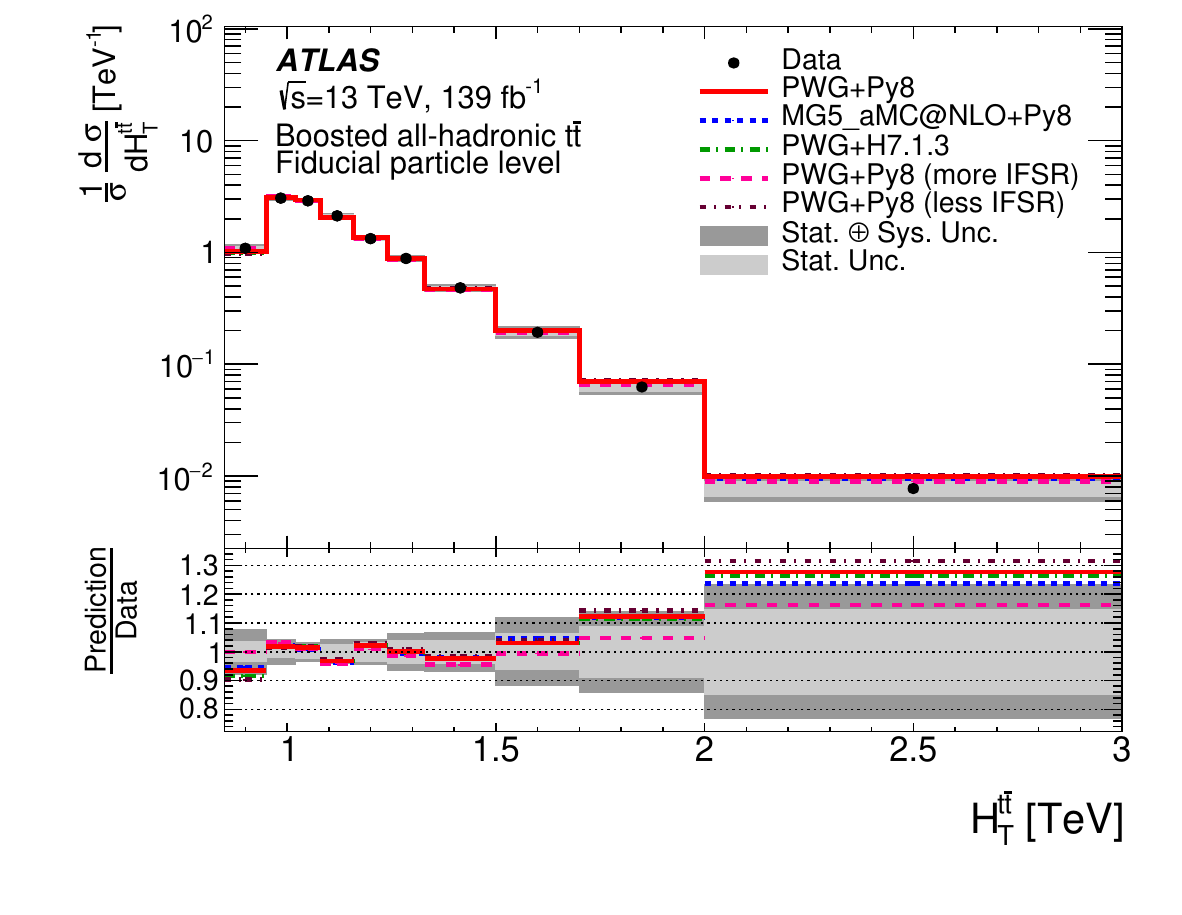}\label{fig:particle:tt_HTttbar:rel}}
\subfigure[]{ \includegraphics[width=0.49\textwidth]{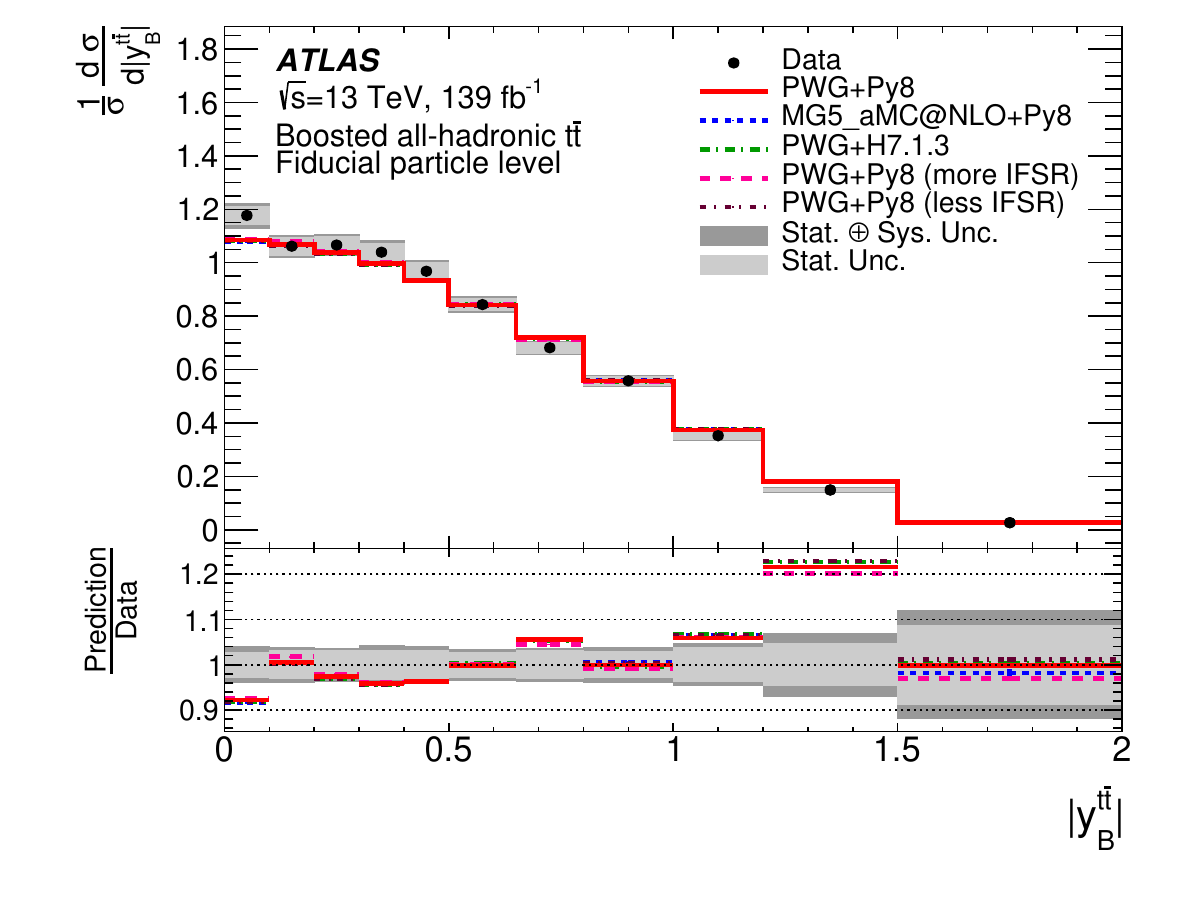}\label{fig:particle:tt_yboost:rel}}
\subfigure[]{ \includegraphics[width=0.49\textwidth]{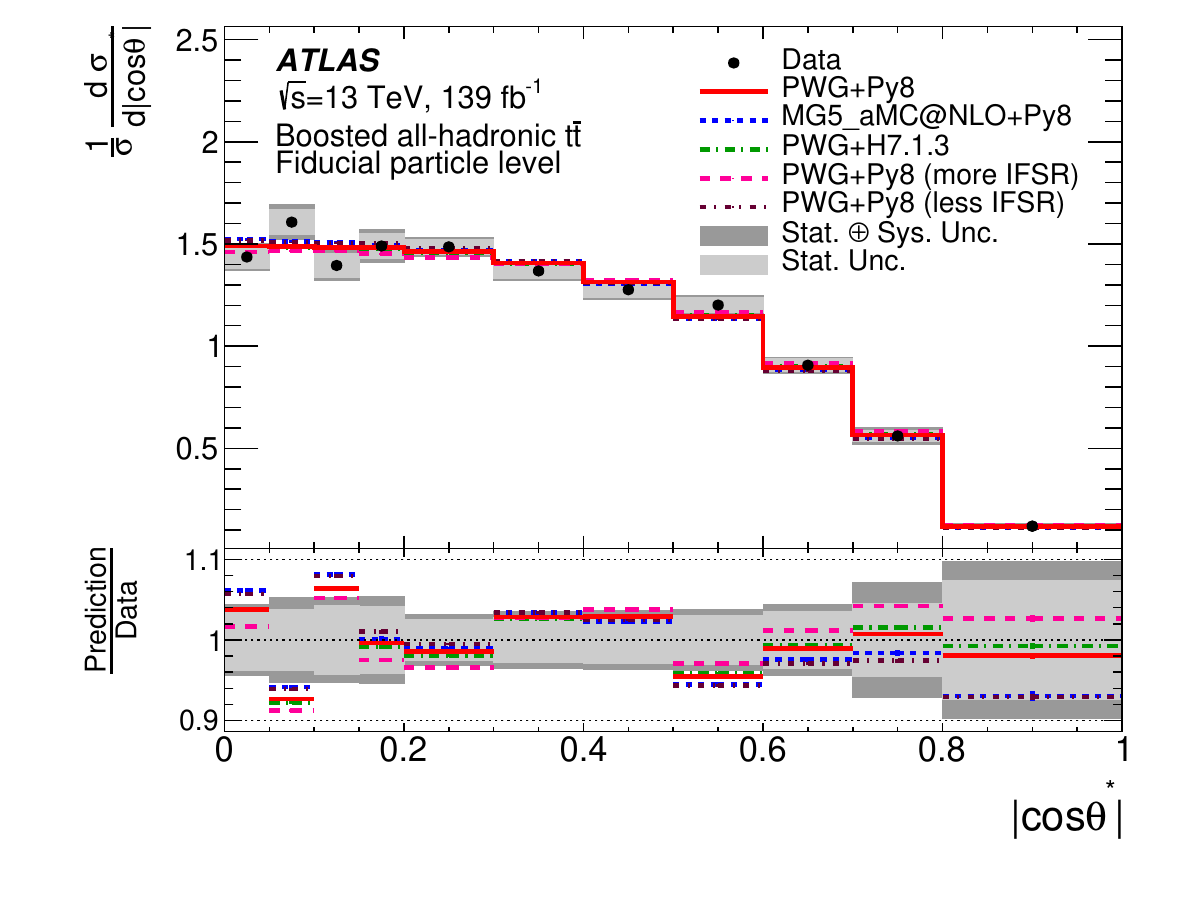}\label{fig:particle:tt_cosThetaStar:rel}}
\subfigure[]{ \includegraphics[width=0.49\textwidth]{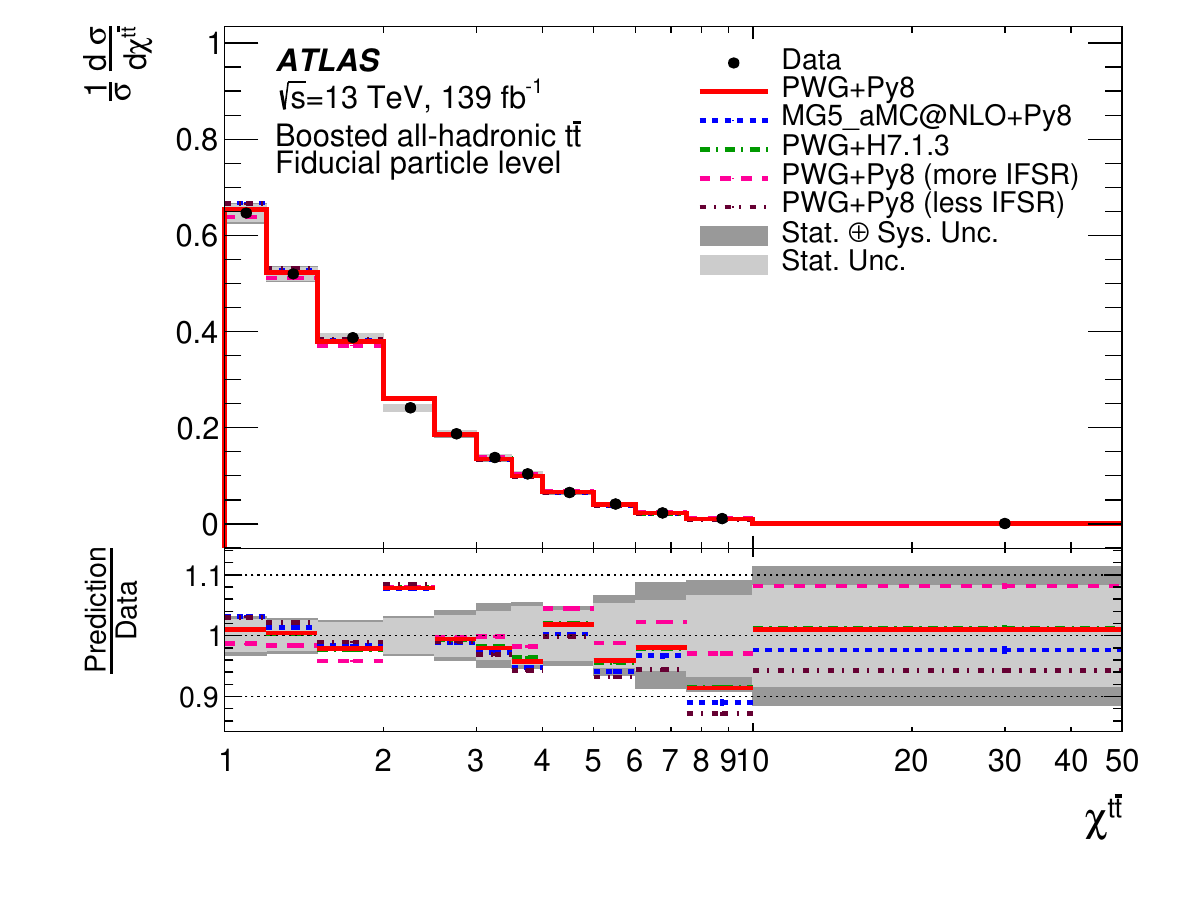}\label{fig:particle:tt_chittbar:rel}}
\caption{
Normalized particle-level fiducial phase-space differential cross-sections as a function of
\subref{fig:particle:tt_HTttbar:rel}~the scalar sum of the transverse momenta of the top-quark jets,
\subref{fig:particle:tt_yboost:rel}~the \ttbar\ rapidity boost,
\subref{fig:particle:tt_cosThetaStar:rel}~the production angle in the Collins--Soper reference frame, and
\subref{fig:particle:tt_chittbar:rel}~the production angle \chittbar.
The dark and light grey bands indicate the total uncertainty and the statistical uncertainty, respectively, of the data in each bin.
Data points are placed at the centre of each bin.
The \POWPY[8] MC sample is used as the nominal prediction to correct the data to particle level.
}
\label{fig:particle:others:rel}
\end{figure*}

 
The normalized particle-level fiducial phase-space
differential cross-sections for six additional observables selected for comparison are presented in Figure~\ref{fig:particle:random_top:rel} and Figure~\ref{fig:particle:others:rel}.
Figure~\ref{fig:particle:random_top:rel}
shows differential cross-sections for the \pT\
and rapidity of the top-quark jet, where the top-quark jet is chosen at random on an
event-by-event basis. These distributions are equivalent to the average of the top-quark and
top-antiquark distributions.
Figure~\ref{fig:particle:others:rel} shows the differential cross-sections for
the scalar sum of the \pT\ of the top-quark jets, $H_\text{T}^{\ttbar}$,
the rapidity boost of \ttbar{} system,
the cosine of the production angle in the Collins--Soper reference frame, and $\chi^{\ttbar{}}$, which
measures the production angle with respect to the beam direction.
 
\begin{figure*}[htbp]
\centering
\subfigure[]{ \includegraphics[width=0.6\textwidth]{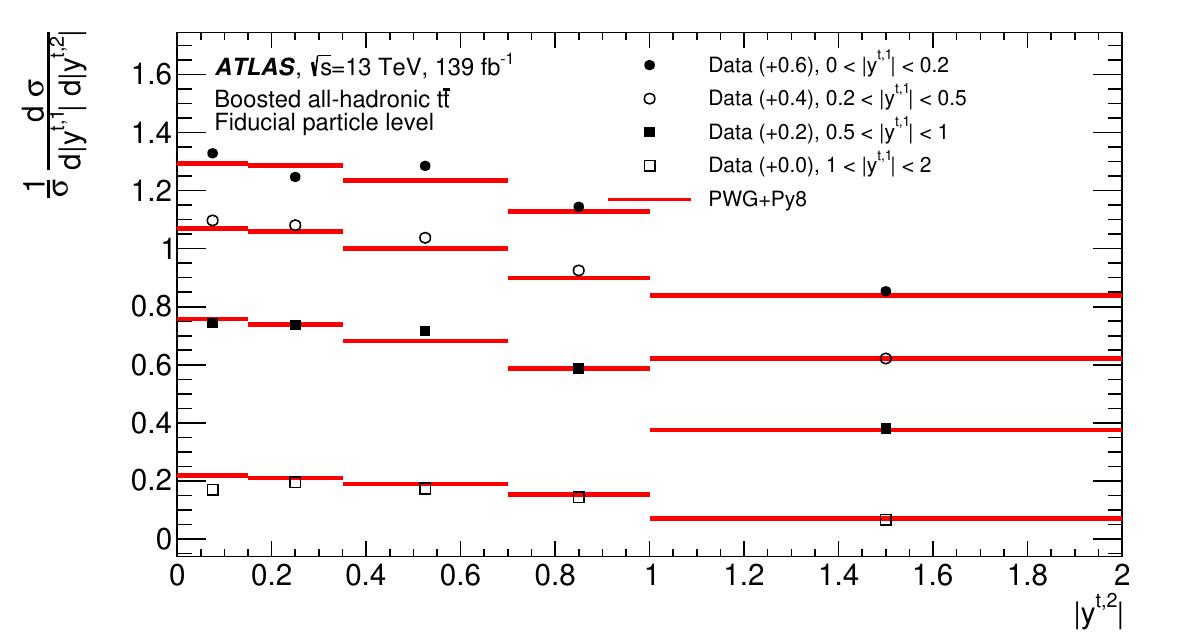}\label{fig:particle:t1_y_vs_t2_y:rel:shape}}
\subfigure[]{ \includegraphics[width=0.68\textwidth]{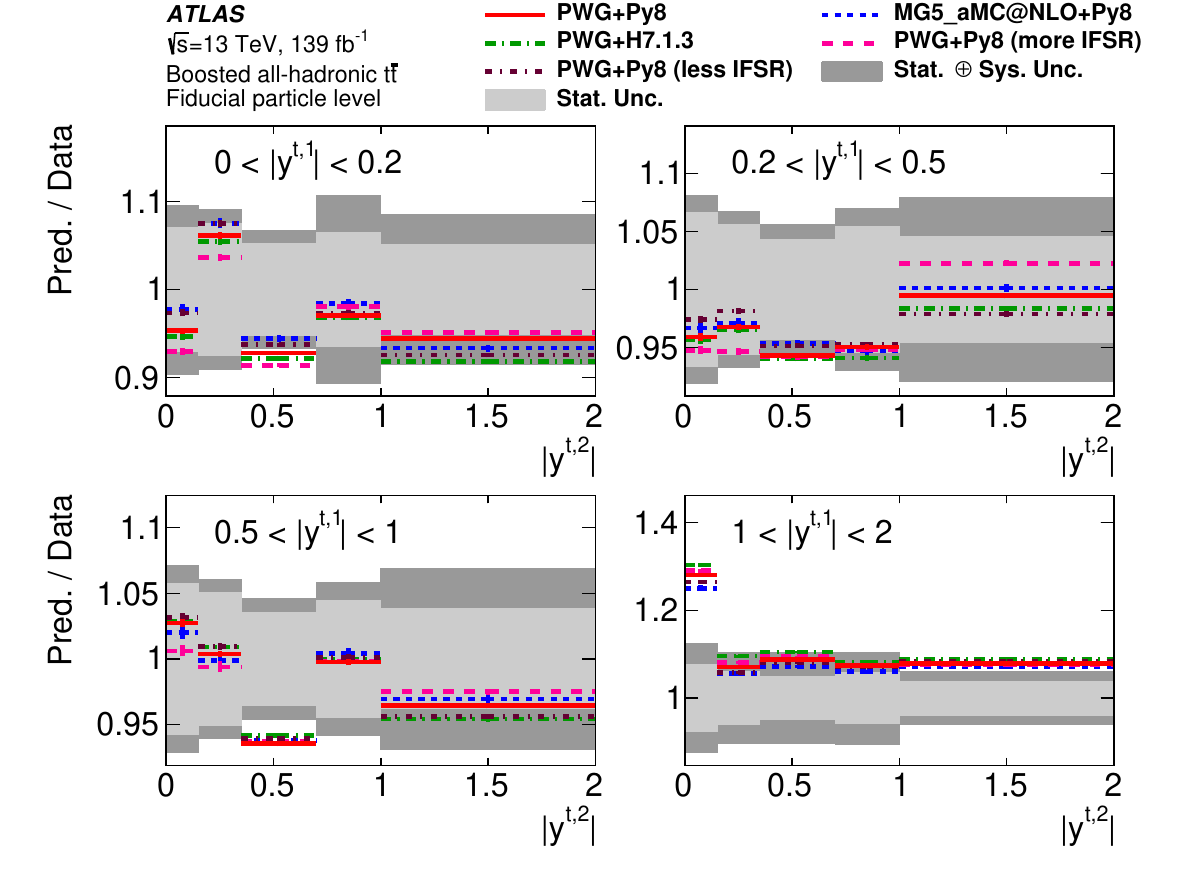}\label{fig:particle:t1_y_vs_t2_y:rel:ratio}}
\caption{
\subref{fig:particle:t1_y_vs_t2_y:rel:shape} Normalized particle-level fiducial phase-space double-differential cross-sections as a function of the absolute value of the rapidity of the leading and the second-leading top-quark jet, compared with the \POWPY[8] calculation.
Data points are placed at the centre of each bin and the \POWPY[8] calculation is indicated by solid lines.
The measurement and the prediction are shifted by the factors shown in parentheses to aid visibility.
\subref{fig:particle:t1_y_vs_t2_y:rel:ratio}~The ratios of various MC calculations to the normalized particle-level fiducial phase-space differential cross-sections.
The dark and light grey bands indicate the total uncertainty and the statistical uncertainty, respectively, of the data in each bin.
}
\label{fig:particle:t1_y_vs_t2_y:rel}
\end{figure*}

\begin{figure*}[htbp]
\centering
\subfigure[]{ \includegraphics[width=0.6\textwidth]{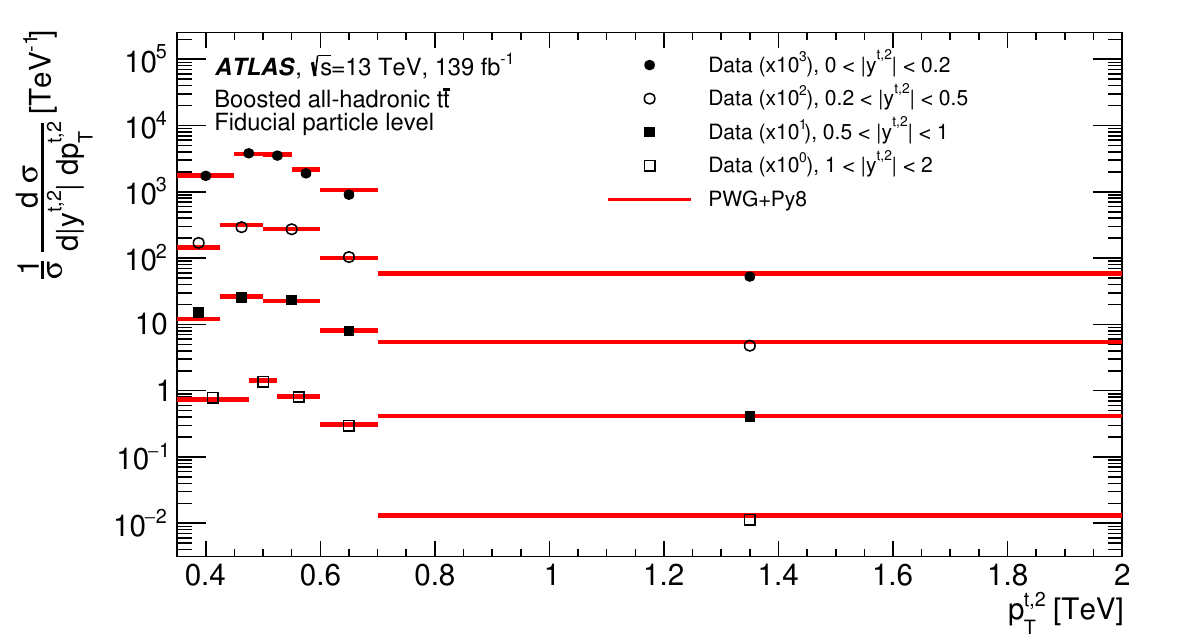}\label{fig:particle:t2_y_vs_t2_pt:rel:shape}}
\subfigure[]{ \includegraphics[width=0.68\textwidth]{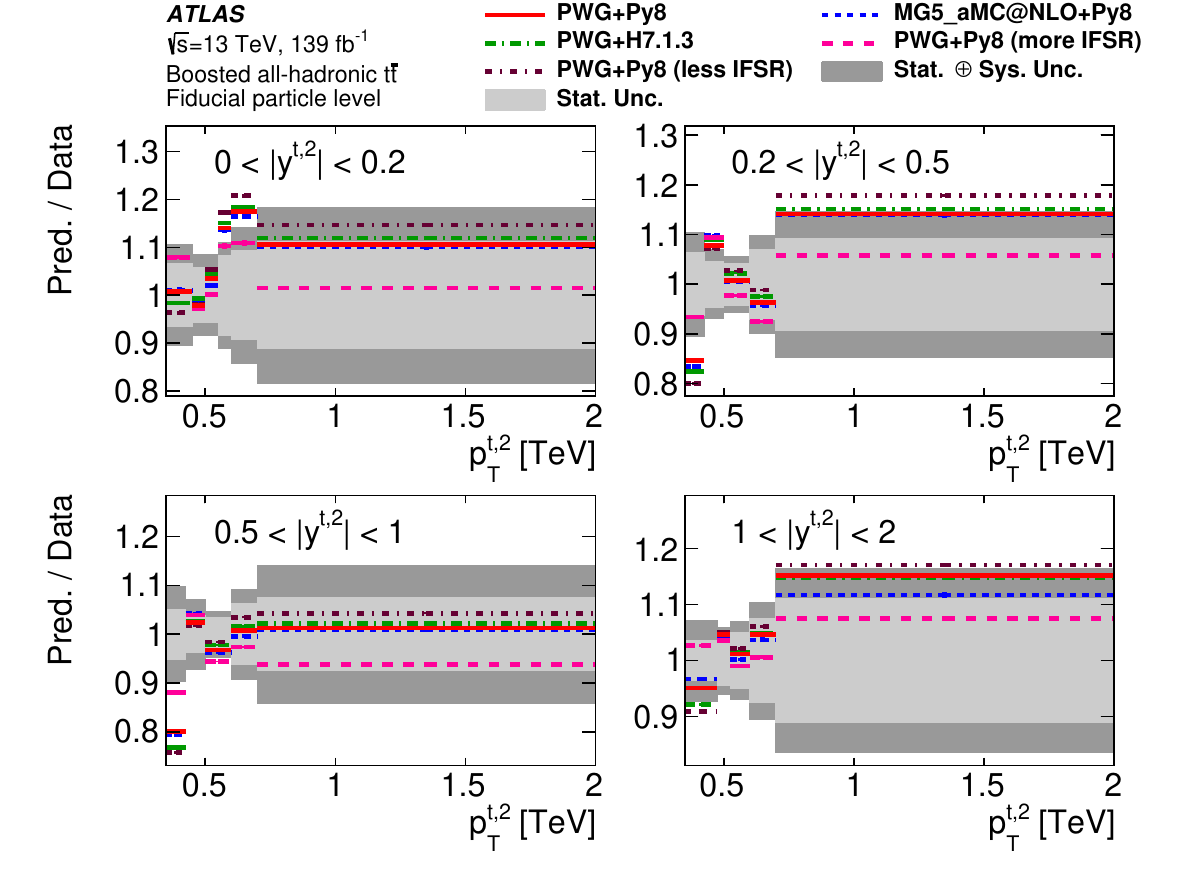}\label{fig:particle:t2_y_vs_t2_pt:rel:ratio}}
\caption{
\subref{fig:particle:t2_y_vs_t2_pt:rel:shape} Normalized particle-level fiducial phase-space double-differential cross-sections as a function of the absolute value of the rapidity and transverse momentum of the second-leading top-quark jet, compared with the \POWPY[8] calculation.
Data points are placed at the centre of each bin and the \POWPY[8] calculation is indicated by solid lines.
The measurement and the prediction are normalized by the factors shown in parentheses to aid visibility.
\subref{fig:particle:t2_y_vs_t2_pt:rel:ratio}~The ratios of various MC calculations to the normalized particle-level fiducial phase-space differential cross-sections.
The dark and light grey bands indicate the total uncertainty and the statistical uncertainty, respectively, of the data in each bin.
}
\label{fig:particle:t2_y_vs_t2_pt:rel}
\end{figure*}

\begin{figure*}[htbp]
\centering
\subfigure[]{ \includegraphics[width=0.6\textwidth]{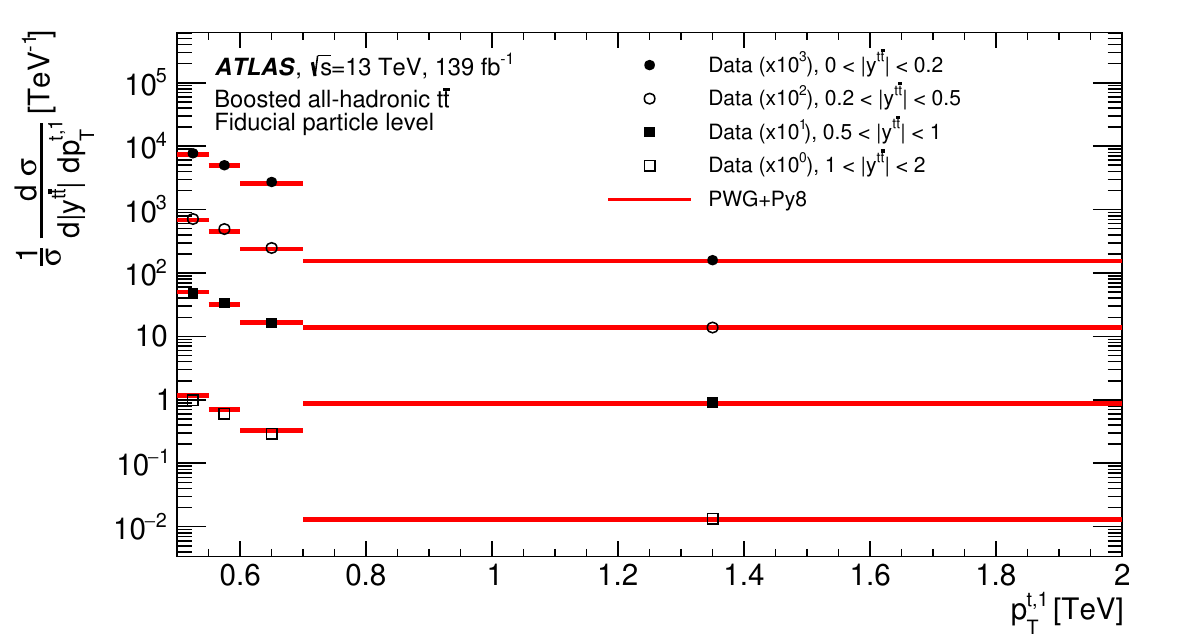}\label{fig:particle:ttbar_y_vs_t1_pt:rel:shape}}
\subfigure[]{ \includegraphics[width=0.68\textwidth]{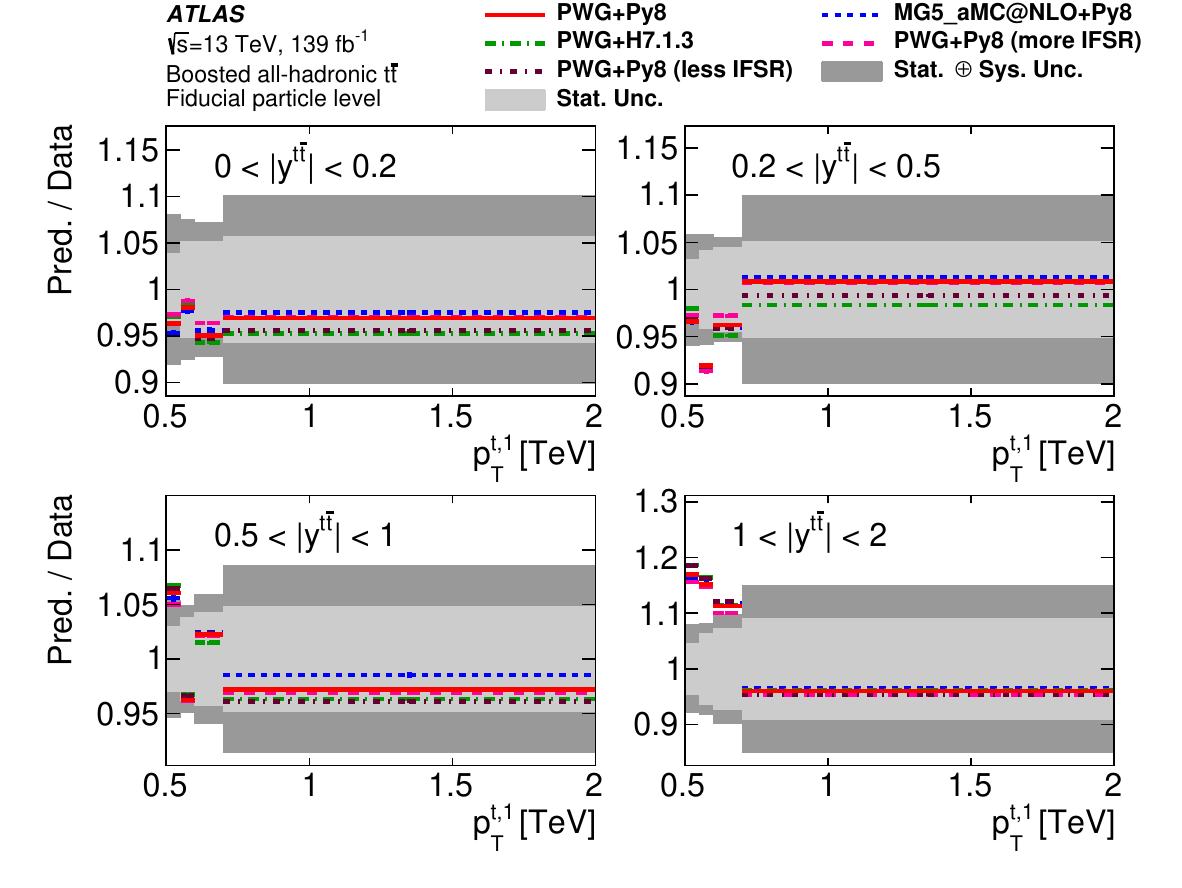}\label{fig:particle:ttbar_y_vs_t1_pt:rel:ratio}}
\caption{
\subref{fig:particle:ttbar_y_vs_t1_pt:rel:shape} Normalized particle-level fiducial phase-space double-differential cross-sections as a function of the absolute value of the rapidity of the \ttbar final state and the transverse momentum of the leading top-quark jet, compared with the \POWPY[8] calculation.
Data points are placed at the centre of each bin and the \POWPY[8] calculation is indicated by solid lines.
The measurement and the prediction are normalized by the factors shown in parentheses to aid visibility.
\subref{fig:particle:ttbar_y_vs_t1_pt:rel:ratio}~The ratios of various MC calculations to the normalized particle-level fiducial phase-space differential cross-sections.
The dark and light grey bands indicate the total uncertainty and the statistical uncertainty, respectively, of the data in each bin.
}
\label{fig:particle:ttbar_y_vs_t1_pt:rel}
\end{figure*}

\begin{figure*}[htbp]
\centering
\subfigure[]{ \includegraphics[width=0.6\textwidth]{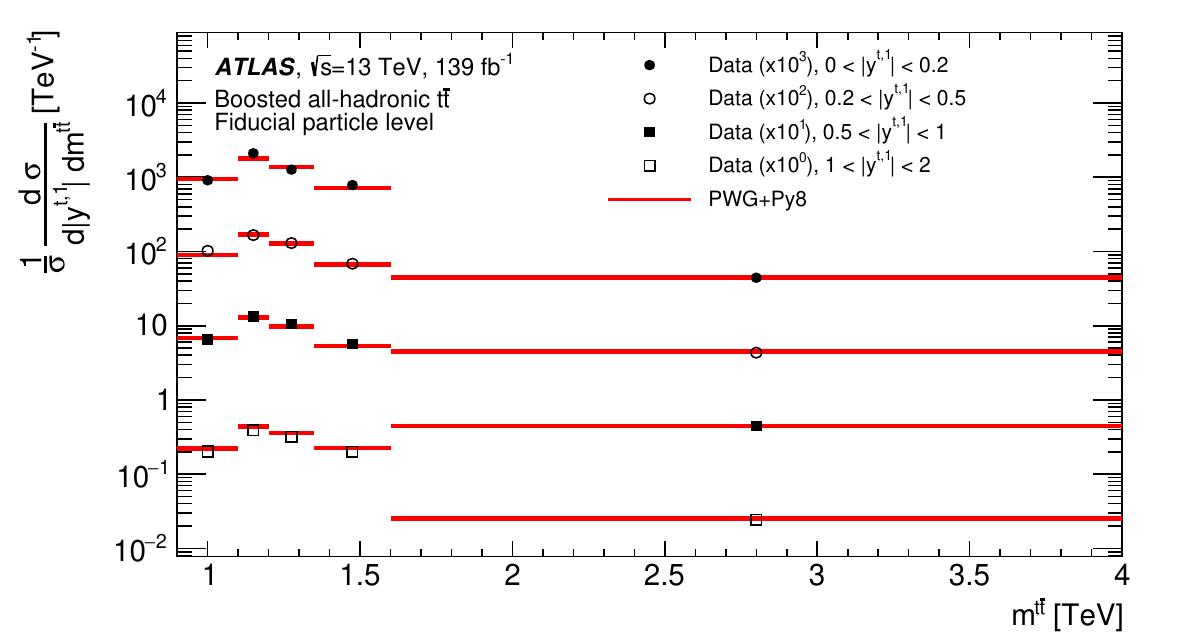}\label{fig:particle:t1_y_vs_ttbar_mass:rel:shape}}
\subfigure[]{ \includegraphics[width=0.68\textwidth]{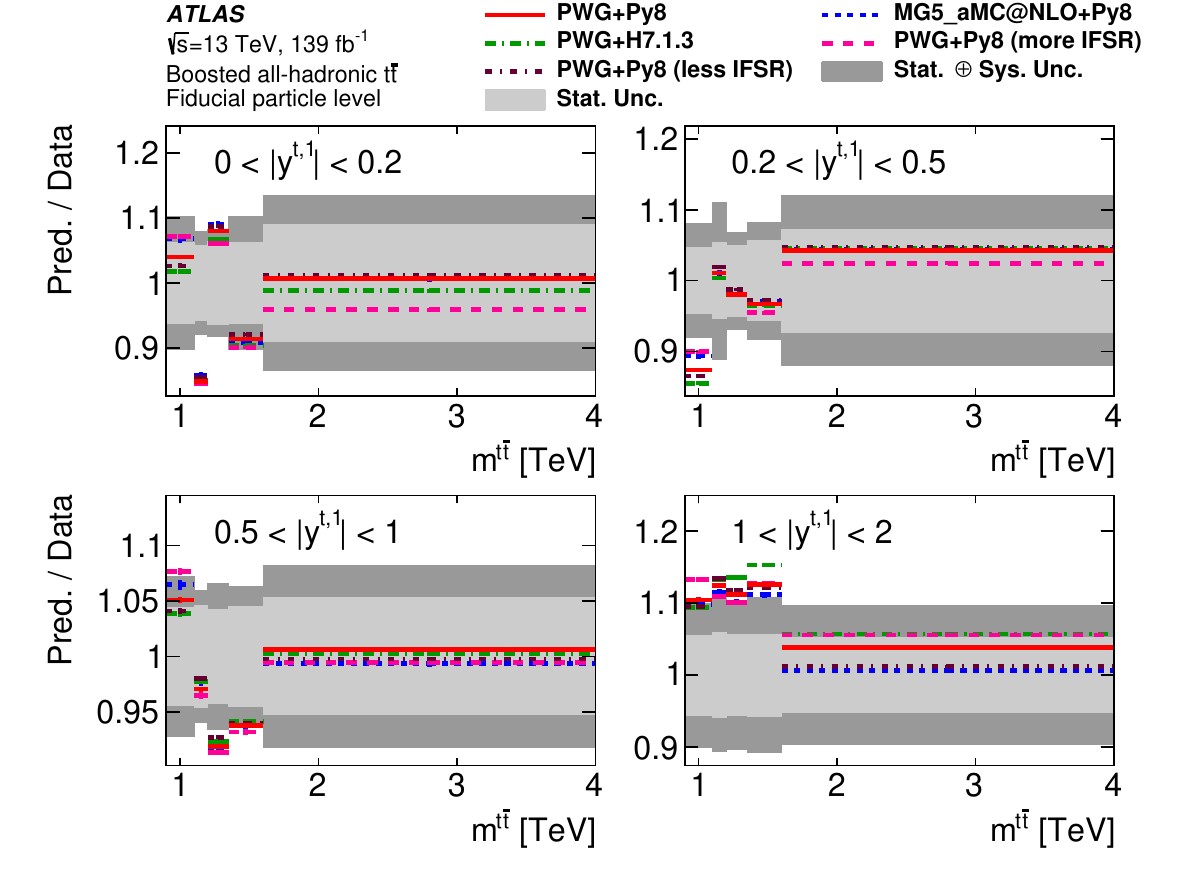}\label{fig:particle:t1_y_vs_ttbar_mass:rel:ratio}}
\caption{
\subref{fig:particle:t1_y_vs_ttbar_mass:rel:shape} Normalized particle-level fiducial phase-space double-differential cross-sections as a function of the absolute value of the rapidity of the leading top-quark jet and the mass of the \ttbar\ final state, compared with the \POWPY[8] calculation.
Data points are placed at the centre of each bin and the \POWPY[8] calculation is indicated by solid lines.
The measurement and the prediction are normalized by the factors shown in parentheses to aid visibility.
\subref{fig:particle:t1_y_vs_ttbar_mass:rel:ratio}~The ratios of various MC calculations to the normalized particle-level fiducial phase-space differential cross-sections.
The dark and light grey bands indicate the total uncertainty and the statistical uncertainty, respectively, of the data in each bin.
}
\label{fig:particle:t1_y_vs_ttbar_mass:rel}
\end{figure*}

\begin{figure*}[htbp]
\centering
\subfigure[]{ \includegraphics[width=0.6\textwidth]{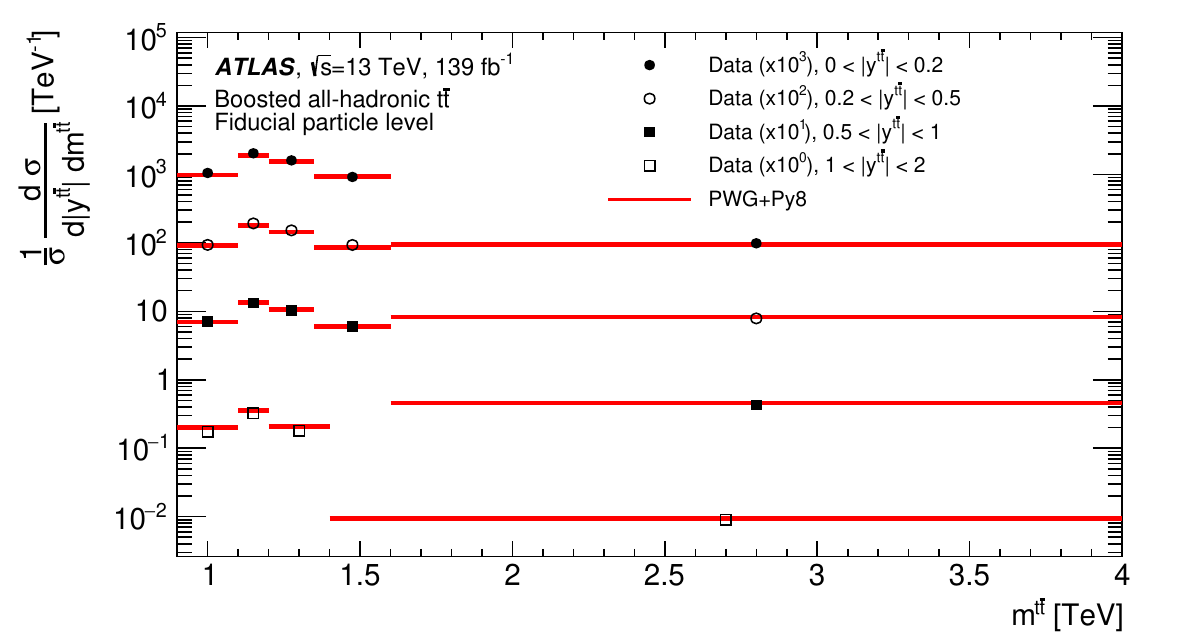}\label{fig:particle:ttbar_y_vs_ttbar_mass:rel:shape}}
\subfigure[]{ \includegraphics[width=0.68\textwidth]{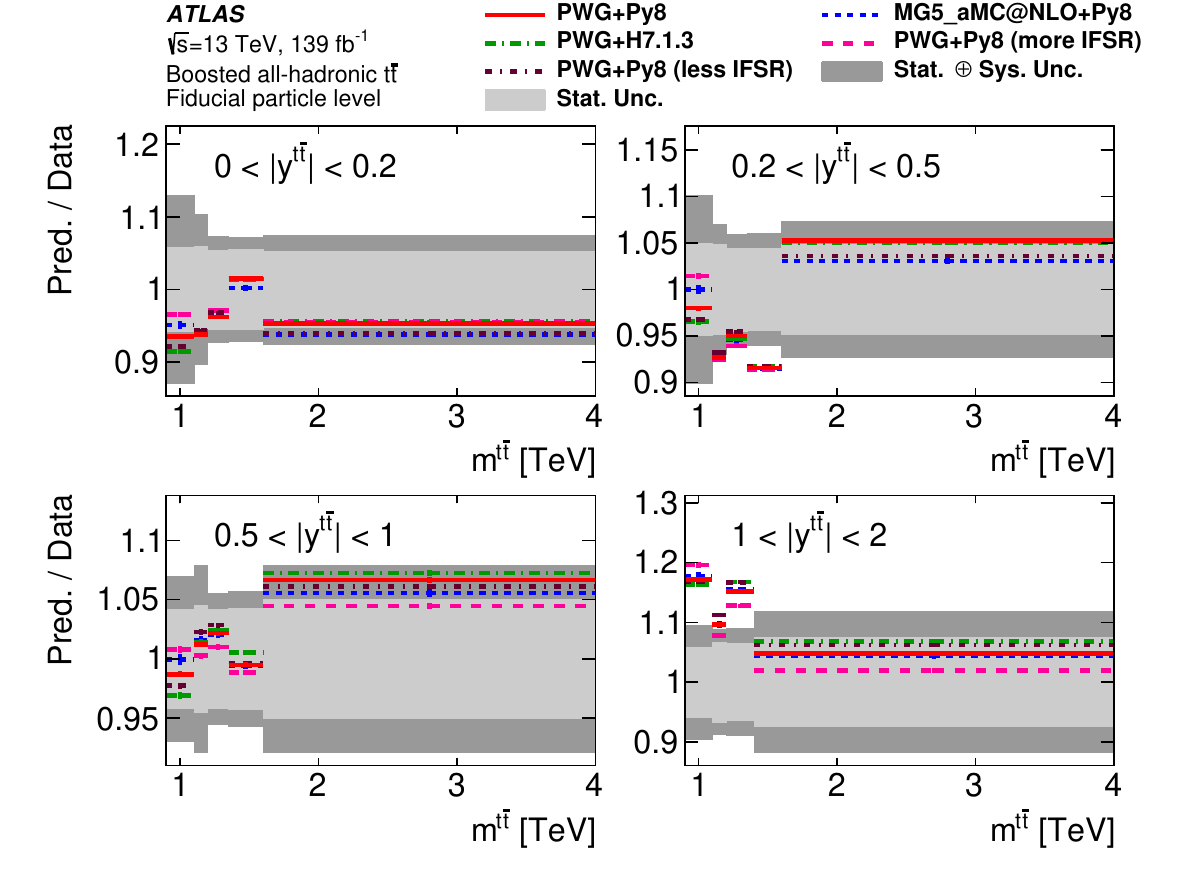}\label{fig:particle:ttbar_y_vs_ttbar_mass:rel:ratio}}
\caption{
\subref{fig:particle:ttbar_y_vs_ttbar_mass:rel:shape} Normalized particle-level fiducial phase-space double-differential cross-sections as a function of the absolute value of the rapidity and the mass of the \ttbar\ final state, compared with the \POWPY[8] calculation.
Data points are placed at the centre of each bin and the \POWPY[8] calculation is indicated by solid lines.
The measurement and the prediction are normalized by the factors shown in parentheses to aid visibility.
\subref{fig:particle:ttbar_y_vs_ttbar_mass:rel:ratio}~The ratios of various MC calculations to the normalized particle-level fiducial phase-space differential cross-sections.
The dark and light grey bands indicate the total uncertainty and the statistical uncertainty, respectively, of the data in each bin.
}
\label{fig:particle:ttbar_y_vs_ttbar_mass:rel}
\end{figure*}

Additional double-differential cross-sections are presented in Figures~\ref{fig:particle:t1_y_vs_t2_y:rel}--\ref{fig:particle:ttbar_y_vs_ttbar_mass:rel}.
 
\clearpage
 
\section{Additional parton-level fiducial phase-space differential cross-sections}
\label{sec:appendix:parton_level}

\begin{figure*}[htbp]
\centering
\subfigure[]{\includegraphics[width=0.49\textwidth]{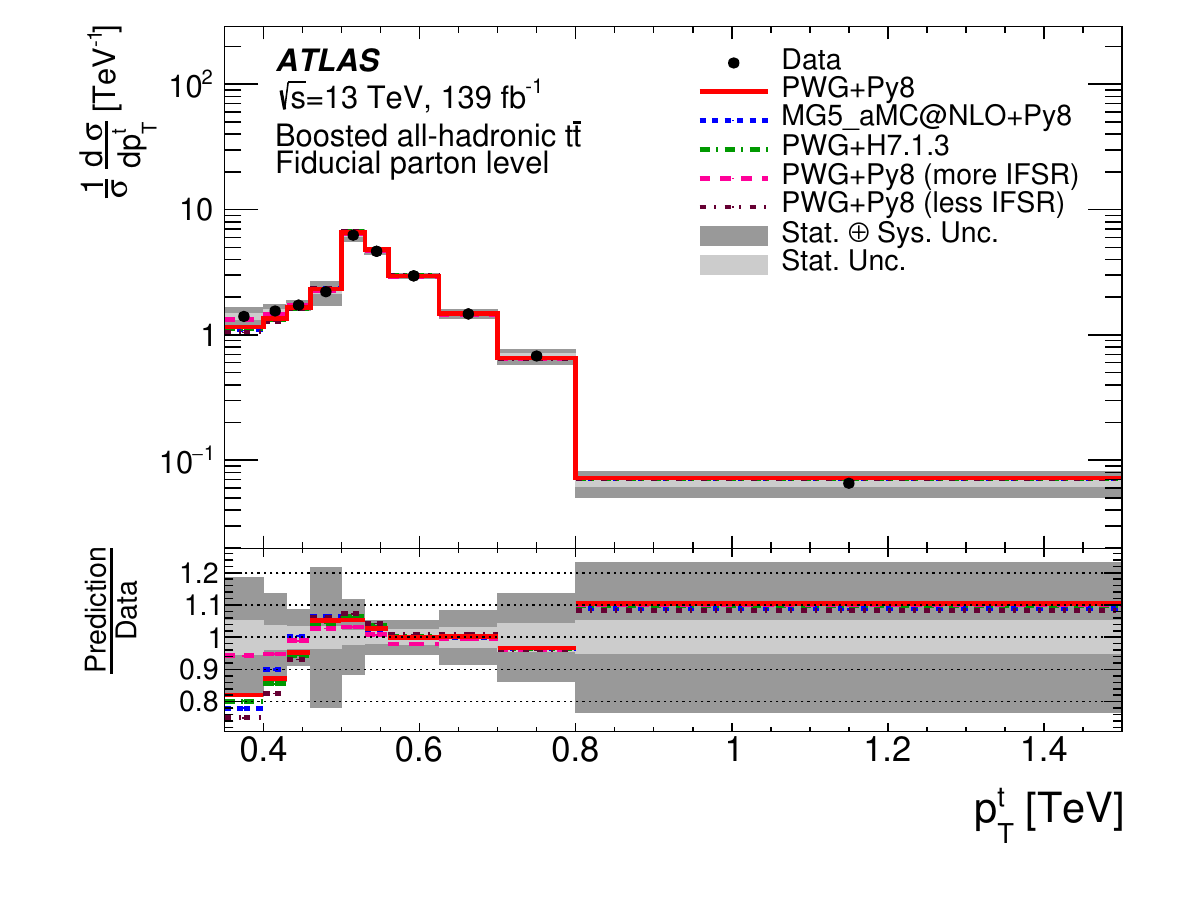}\label{fig:results:parton:top_pt_rel}}
\subfigure[]{\includegraphics[width=0.49\textwidth]{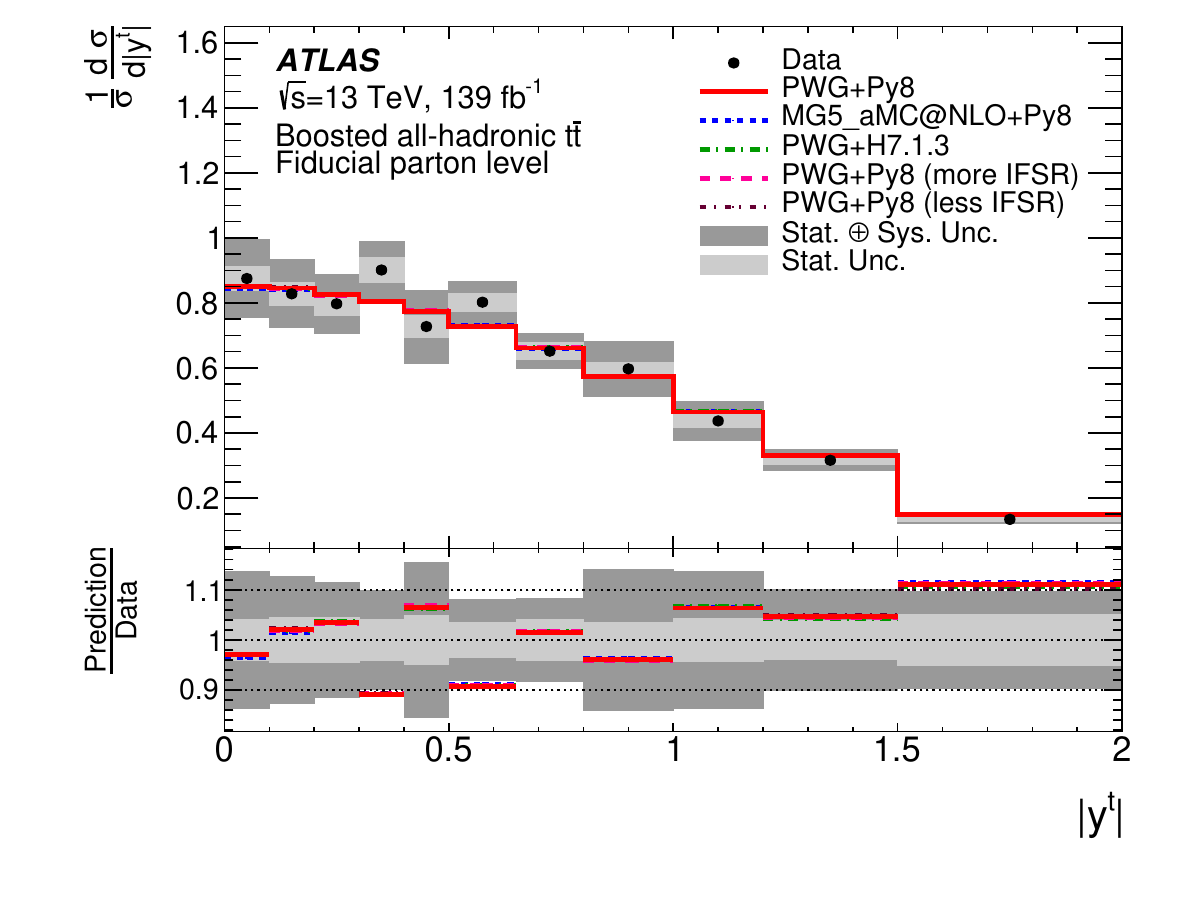}\label{fig:results:parton:randomTop_y_rel}}
\caption{
Normalized parton-level fiducial phase-space differential cross-sections as a function of
\subref{fig:results:parton:top_pt_rel}~the transverse momentum and
\subref{fig:results:parton:randomTop_y_rel}~the absolute value of the rapidity of the randomly chosen top quark.
The dark and light grey bands indicate the total uncertainty and the statistical uncertainty, respectively, of the data in each bin.
Data points are placed at the centre of each bin.
The \POWPY[8] MC sample is used as the nominal prediction to correct the data to parton level.
}
\label{fig:parton:random_top:rel}
\end{figure*}

\begin{figure*}[htbp]
\centering
\subfigure[]{ \includegraphics[width=0.49\textwidth]{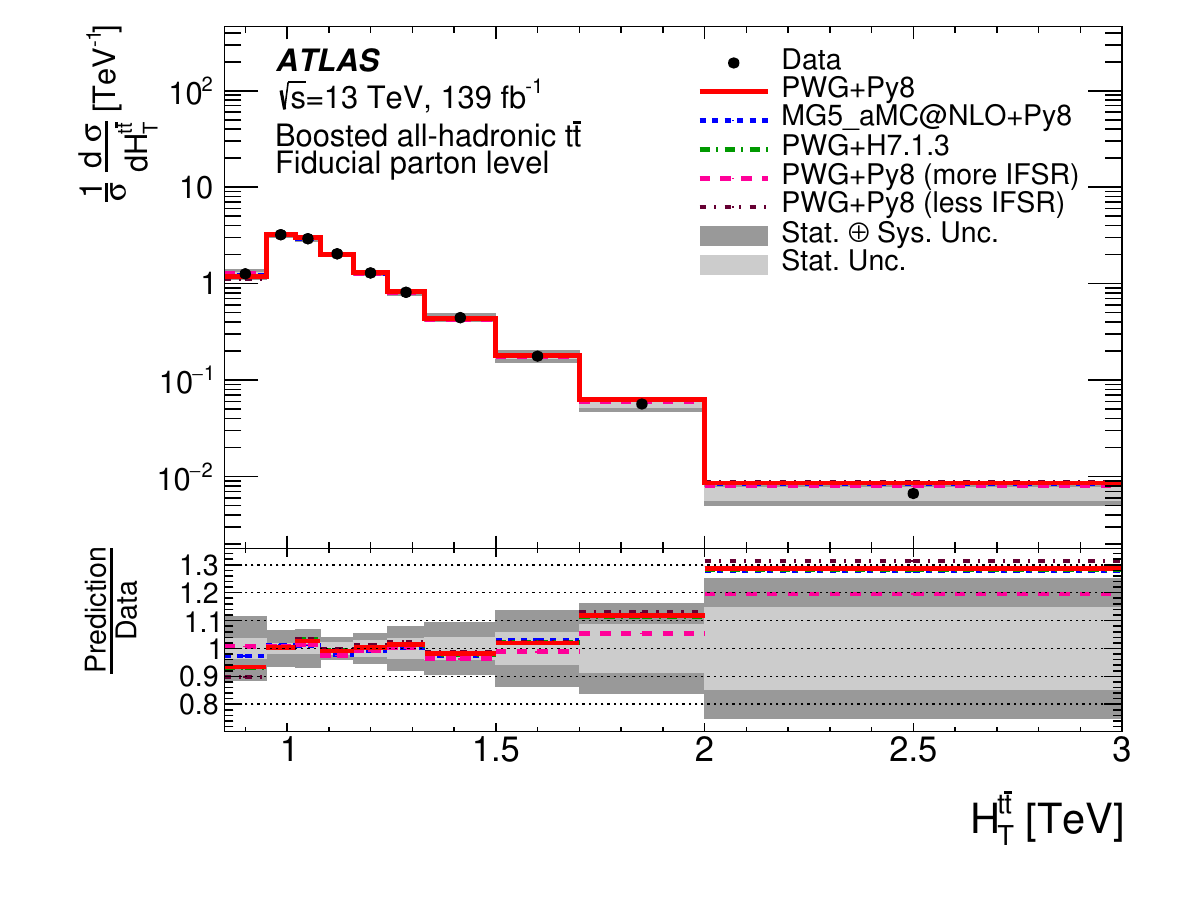}\label{fig:parton:tt_HT:rel}}
\subfigure[]{ \includegraphics[width=0.49\textwidth]{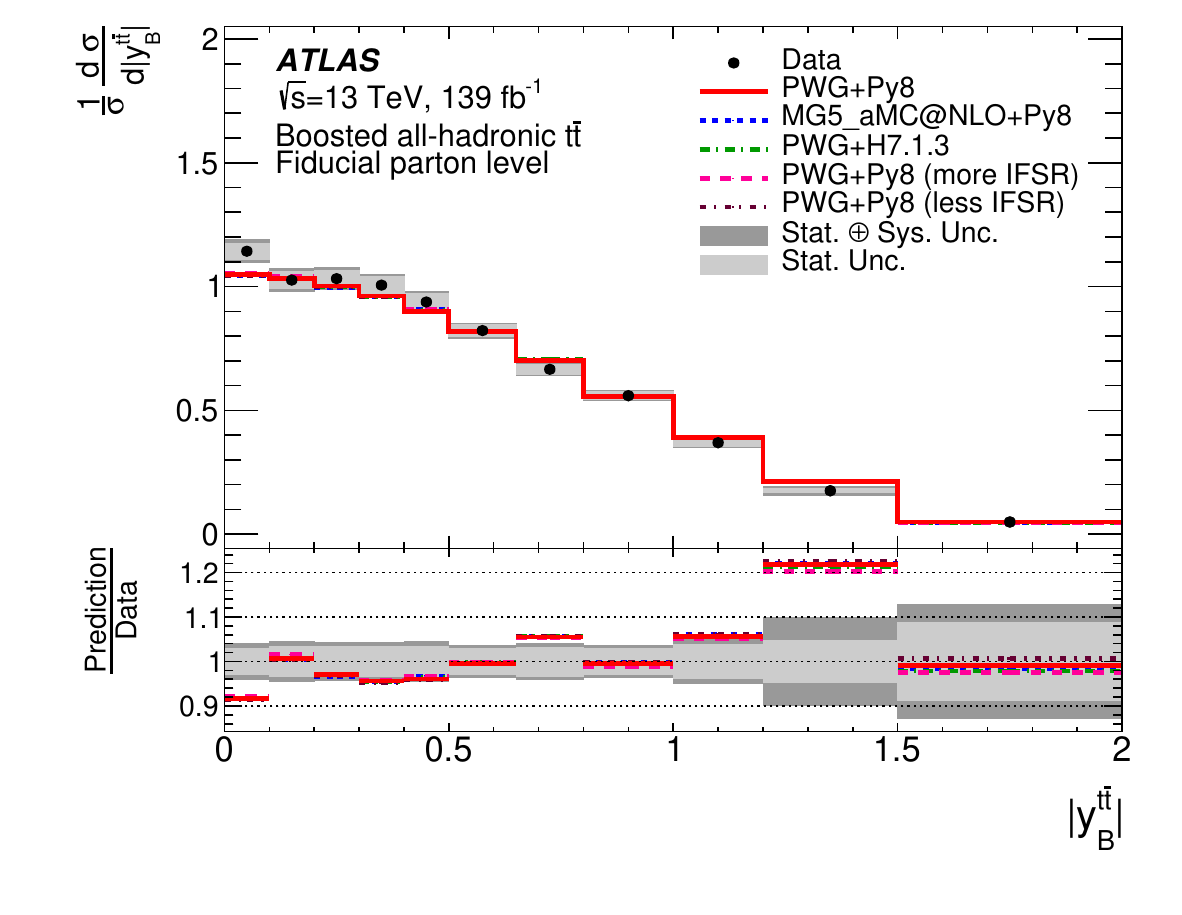}\label{fig:parton:tt_yboost:rel}}
\subfigure[]{ \includegraphics[width=0.49\textwidth]{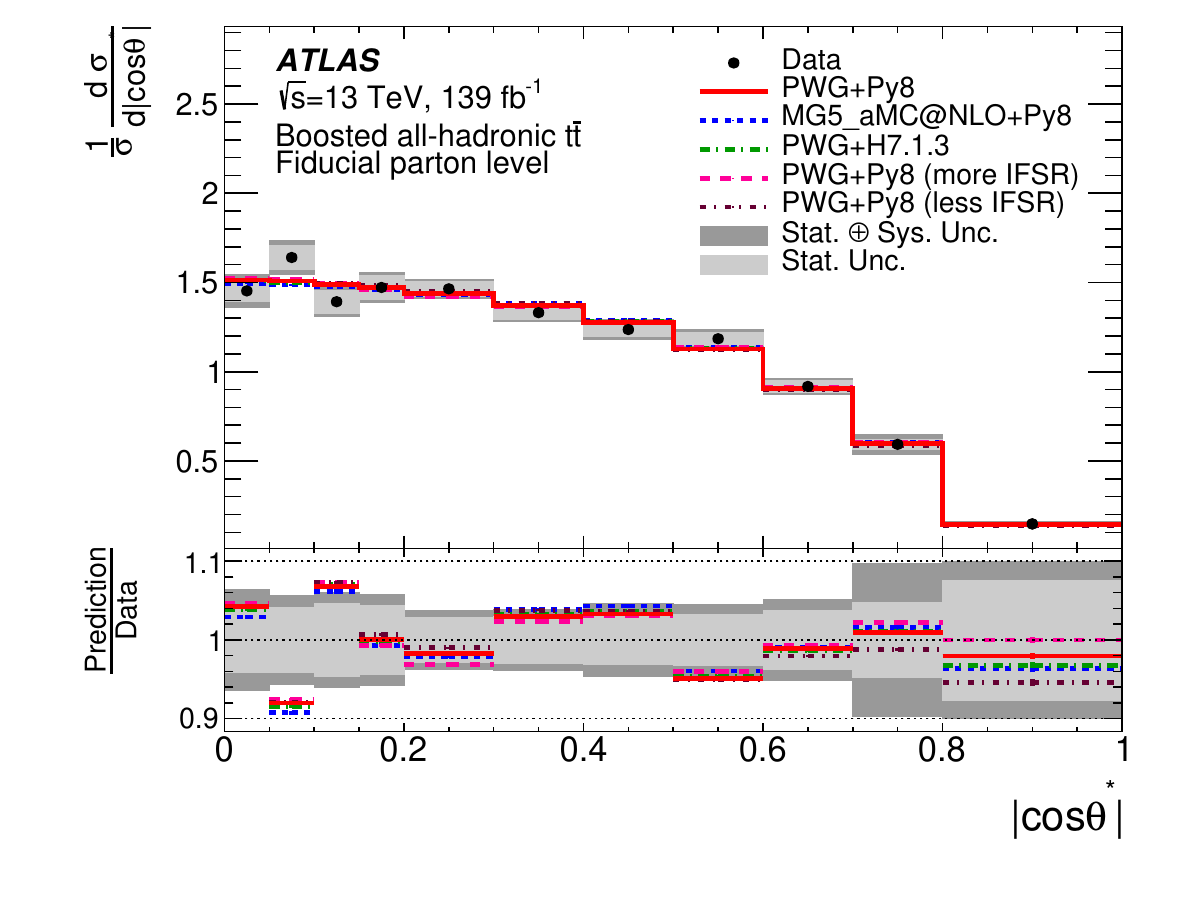}\label{fig:parton:tt_cosThetaStar:rel}}
\subfigure[]{ \includegraphics[width=0.49\textwidth]{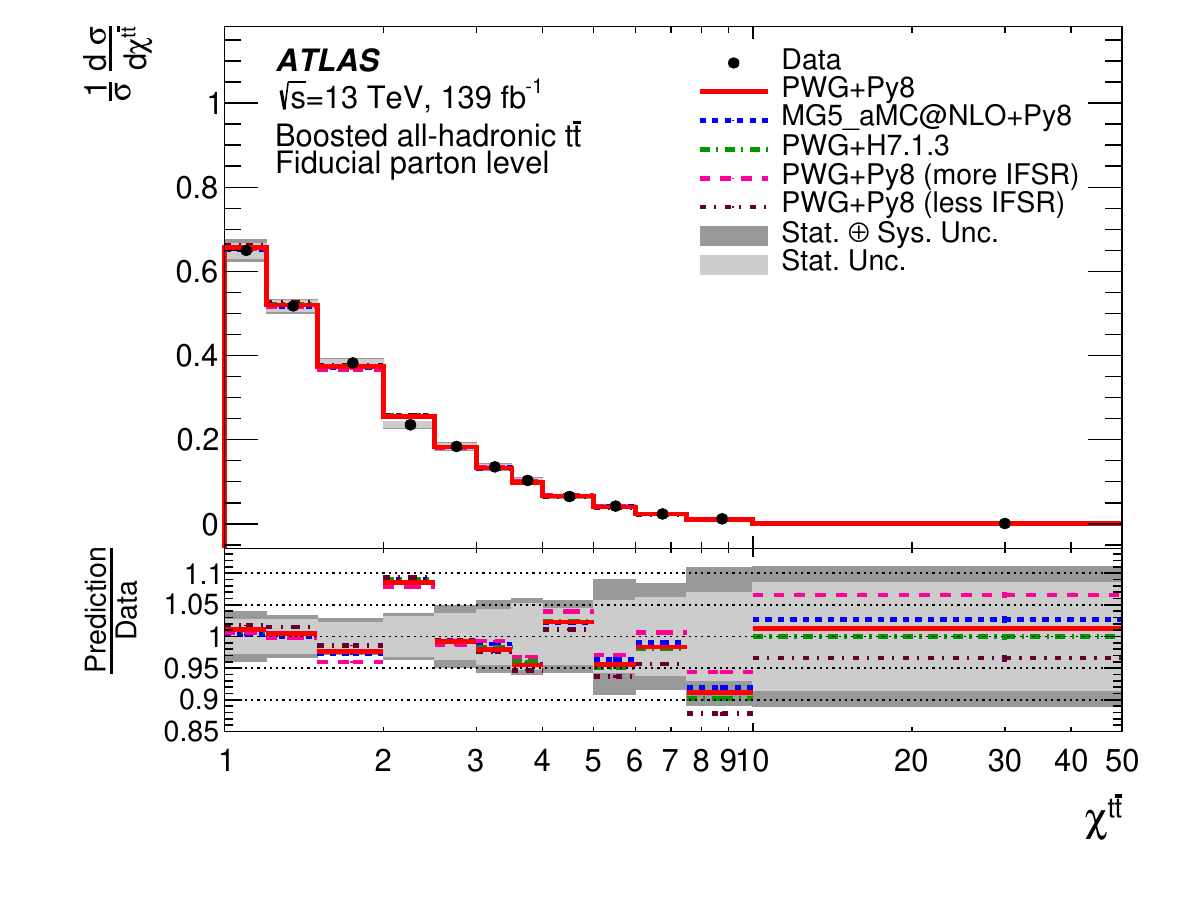}\label{fig:parton:tt_chittbar:rel}}
\caption{
Normalized parton-level fiducial phase-space differential cross-sections as a function of
\subref{fig:parton:tt_HT:rel}~the scalar sum of the transverse momenta of the top quarks,
\subref{fig:parton:tt_yboost:rel}~the rapidity boost,
\subref{fig:parton:tt_cosThetaStar:rel}~the production angle in the Collins--Soper reference frame, and
\subref{fig:parton:tt_chittbar:rel}~the production angle \chittbar.
The dark and light grey bands indicate the total uncertainty and the statistical uncertainty, respectively, of the data in each bin.
Data points are placed at the centre of each bin.
The \POWPY[8] MC sample is used as the nominal prediction to correct the data to parton level.
}
\label{fig:parton:others:rel}
\end{figure*}

 
The normalized parton-level differential cross-section distributions are compared
with Standard Model calculations in Figure~\ref{fig:parton:random_top:rel} and Figure~\ref{fig:parton:others:rel}.
Figure~\ref{fig:parton:random_top:rel}
shows differential cross-sections for the \pT\
and rapidity of the top quark, where the top quark is chosen at random on an
event-by-event basis.
These distributions are equivalent to the average of the top-quark and
top-antiquark distributions.
Figure~\ref{fig:parton:others:rel} shows the measurements of four observables:
the scalar sum of the transverse momenta of the top quarks, $H_\text{T}^{\ttbar}$,
the rapidity boost, the production angle in the Collins--Soper reference frame, and
the production angle \chittbar.
 
\begin{figure*}[htbp]
\centering
\subfigure[]{ \includegraphics[width=0.6\textwidth]{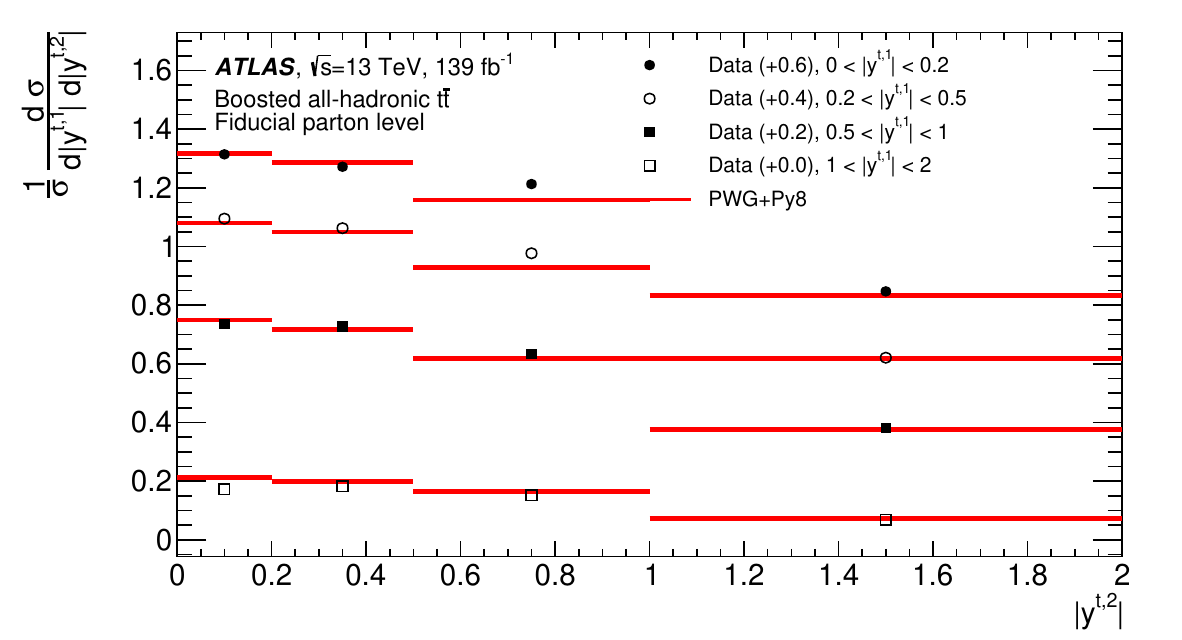}\label{fig:parton:t1_y_vs_t2_y:rel:shape}}
\subfigure[]{ \includegraphics[width=0.68\textwidth]{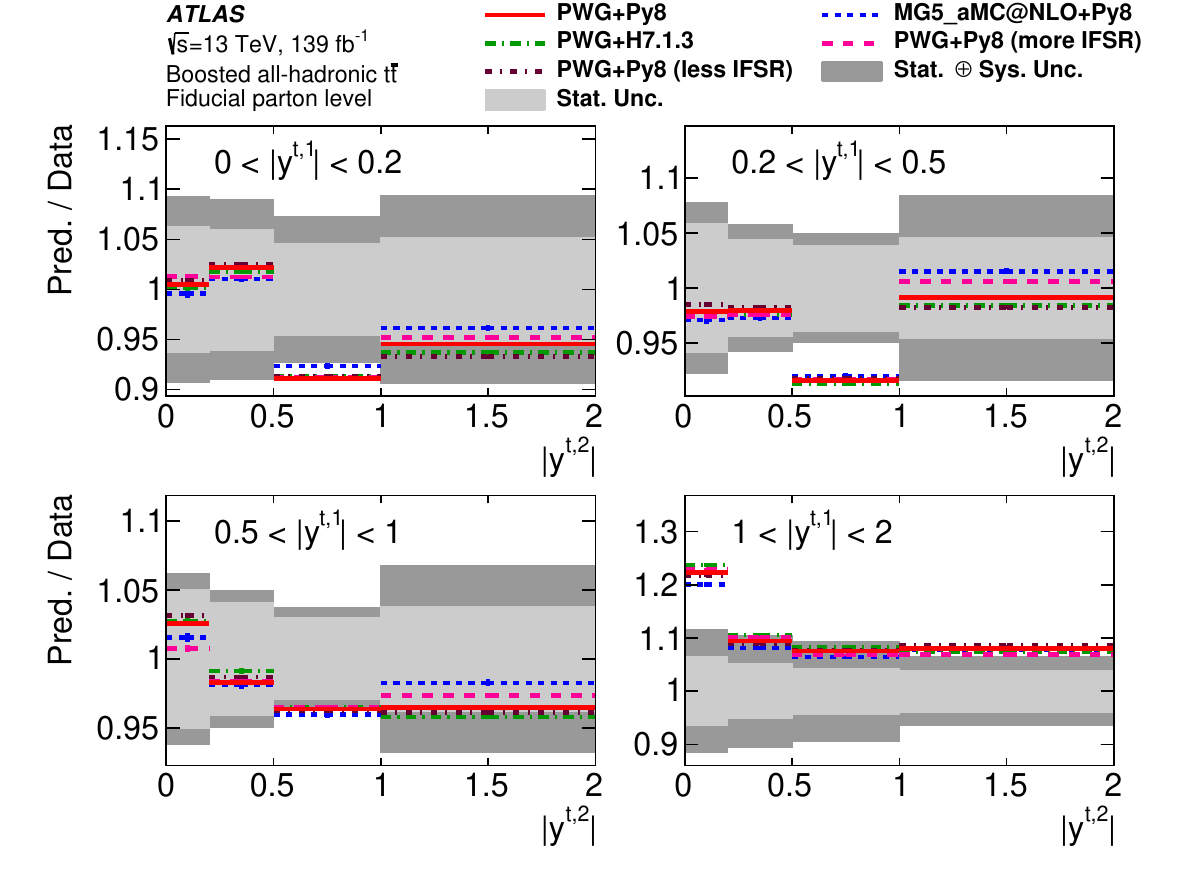}\label{fig:parton:t1_y_vs_t2_y:rel:ratio}}
\caption{
\subref{fig:parton:t1_y_vs_t2_y:rel:shape} Normalized parton-level fiducial phase-space double-differential cross-sections as a function of the absolute value of the rapidity of the leading and the second-leading top-quark, compared with the \POWPY[8] calculation.
Data points are placed at the centre of each bin and the \POWPY[8] calculation is indicated by solid lines.
The measurement and the prediction are shifted by the factors shown in parentheses to aid visibility.
\subref{fig:parton:t1_y_vs_t2_y:rel:ratio}~The ratios of various MC calculations to the normalized parton-level fiducial phase-space differential cross-sections.
The dark and light grey bands indicate the total uncertainty and the statistical uncertainty, respectively, of the data in each bin.
}
\label{fig:parton:t1_y_vs_t2_y:rel}
\end{figure*}

\begin{figure*}[htbp]
\centering
\subfigure[]{ \includegraphics[width=0.6\textwidth]{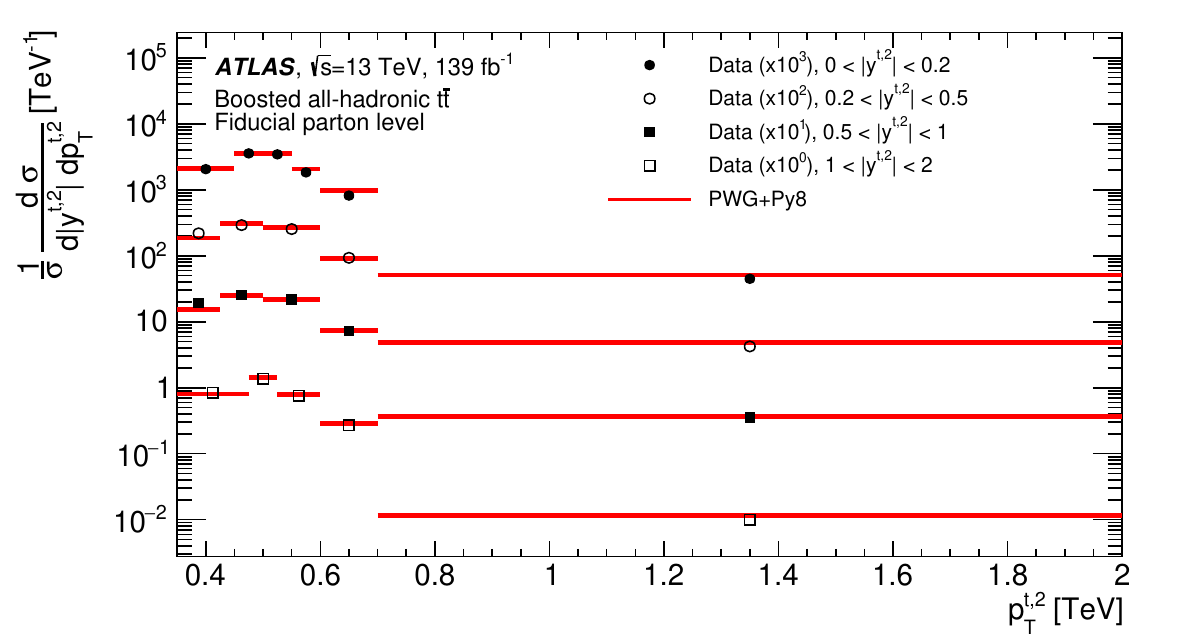}\label{fig:parton:t2_y_vs_t2_pt:rel:shape}}
\subfigure[]{ \includegraphics[width=0.68\textwidth]{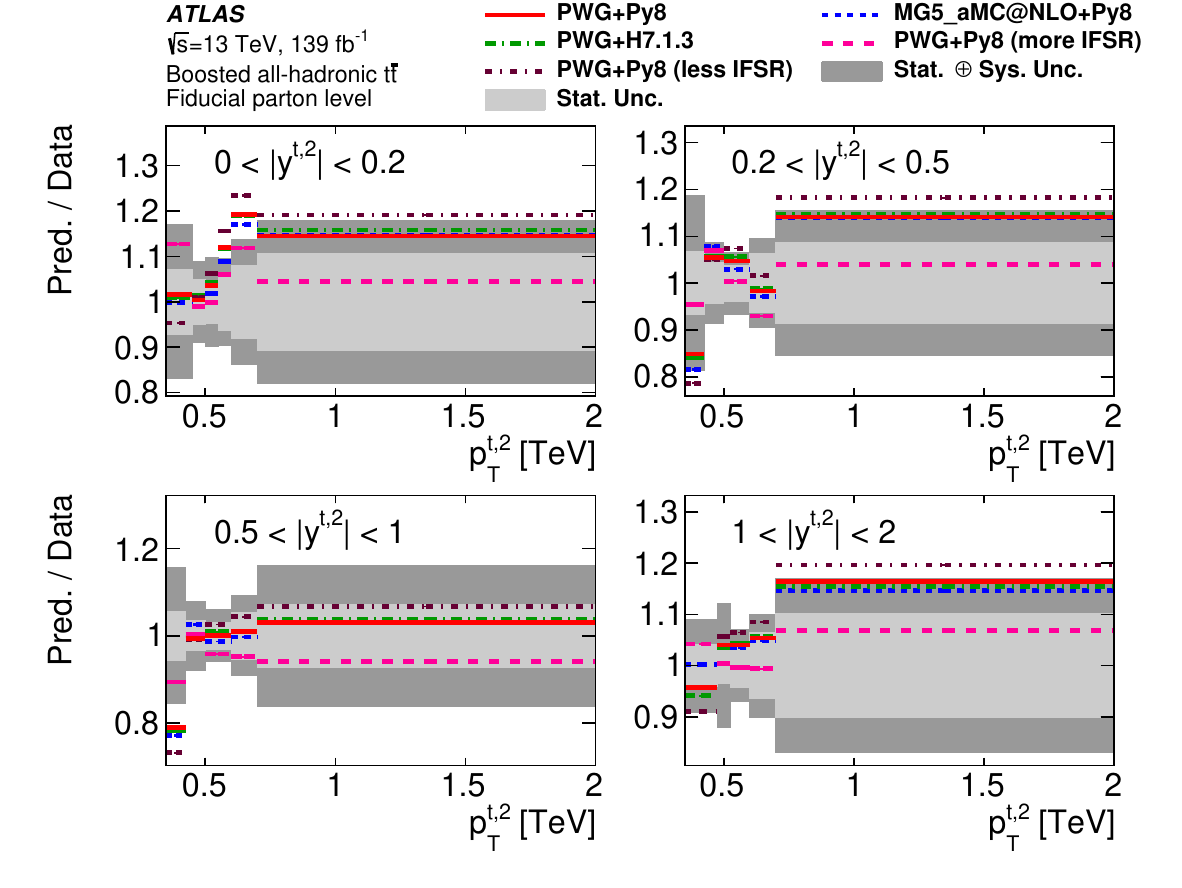}\label{fig:parton:t2_y_vs_t2_pt:rel:ratio}}
\caption{
\subref{fig:parton:t2_y_vs_t2_pt:rel:shape} Normalized parton-level fiducial phase-space double-differential cross-sections as a function of the absolute value of the rapidity and transverse momentum of the second-leading top-quark, compared with the \POWPY[8] calculation.
Data points are placed at the centre of each bin and the \POWPY[8] calculation is indicated by solid lines.
The measurement and the prediction are normalized by the factors shown in parentheses to aid visibility.
\subref{fig:parton:t2_y_vs_t2_pt:rel:ratio}~The ratios of various MC calculations to the normalized parton-level fiducial phase-space differential cross-sections.
The dark and light grey bands indicate the total uncertainty and the statistical uncertainty, respectively, of the data in each bin.
}
\label{fig:parton:t2_y_vs_t2_pt:rel}
\end{figure*}

\begin{figure*}[htbp]
\centering
\subfigure[]{ \includegraphics[width=0.6\textwidth]{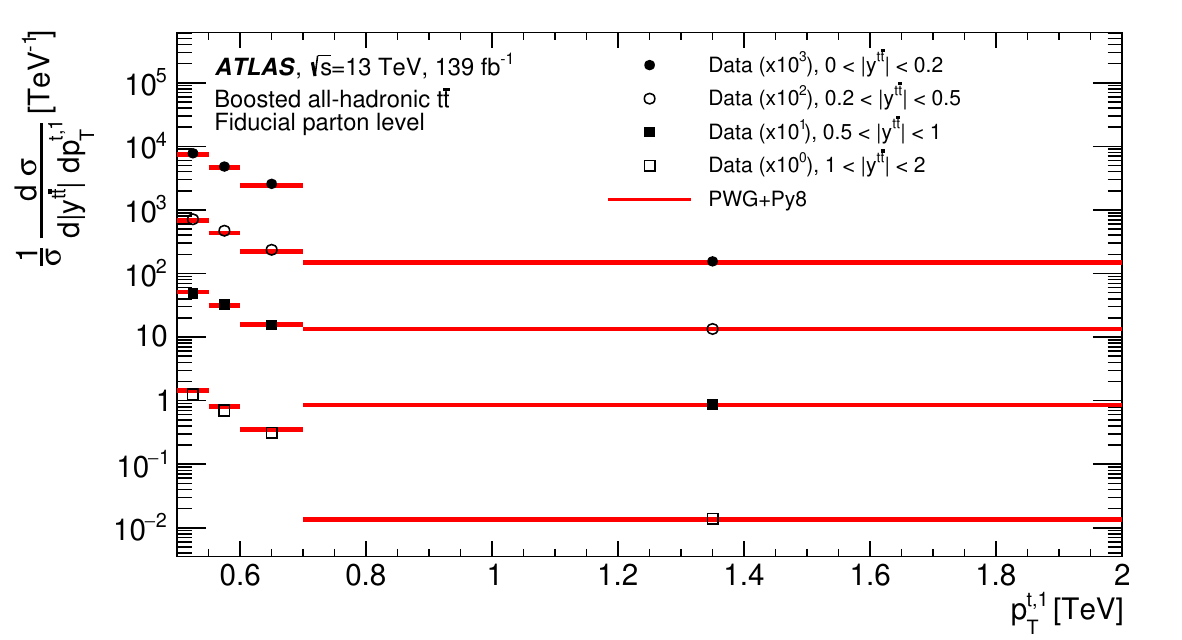}\label{fig:parton:ttbar_y_vs_t1_pt:rel:shape}}
\subfigure[]{ \includegraphics[width=0.68\textwidth]{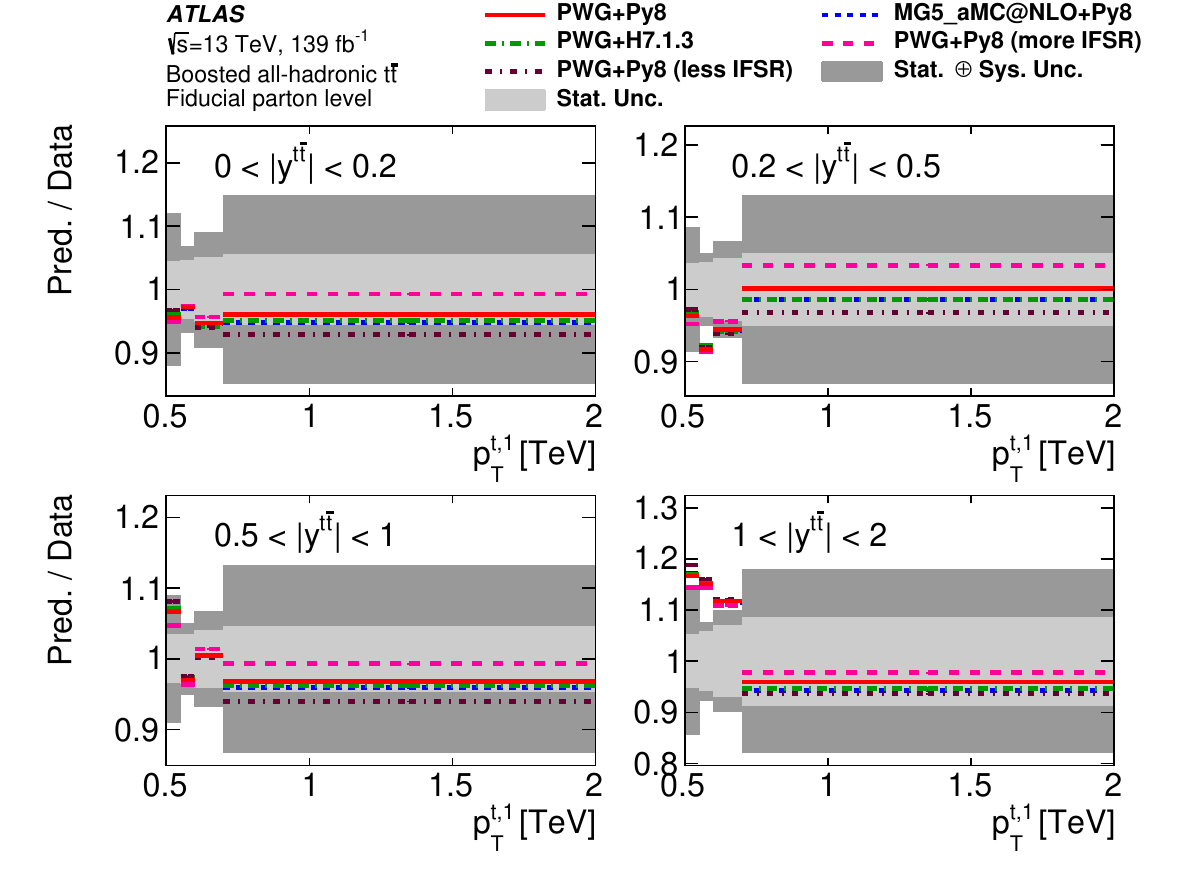}\label{fig:parton:ttbar_y_vs_t1_pt:rel:ratio}}
\caption{
\subref{fig:parton:ttbar_y_vs_t1_pt:rel:shape} Normalized parton-level fiducial phase-space double-differential cross-sections as a function of the absolute value of the rapidity of the \ttbar system and the transverse momentum of the leading top quark, compared with the \POWPY[8] calculation.
Data points are placed at the centre of each bin and the \POWPY[8] calculation is indicated by solid lines.
The measurement and the prediction are normalized by the factors shown in parentheses to aid visibility.
\subref{fig:parton:ttbar_y_vs_t1_pt:rel:ratio}~The ratios of various MC calculations to the normalized parton-level fiducial phase-space differential cross-sections.
The dark and light grey bands indicate the total uncertainty and the statistical uncertainty, respectively, of the data in each bin.
}
\label{fig:parton:ttbar_y_vs_t1_pt:rel}
\end{figure*}

\begin{figure*}[htbp]
\centering
\subfigure[]{ \includegraphics[width=0.6\textwidth]{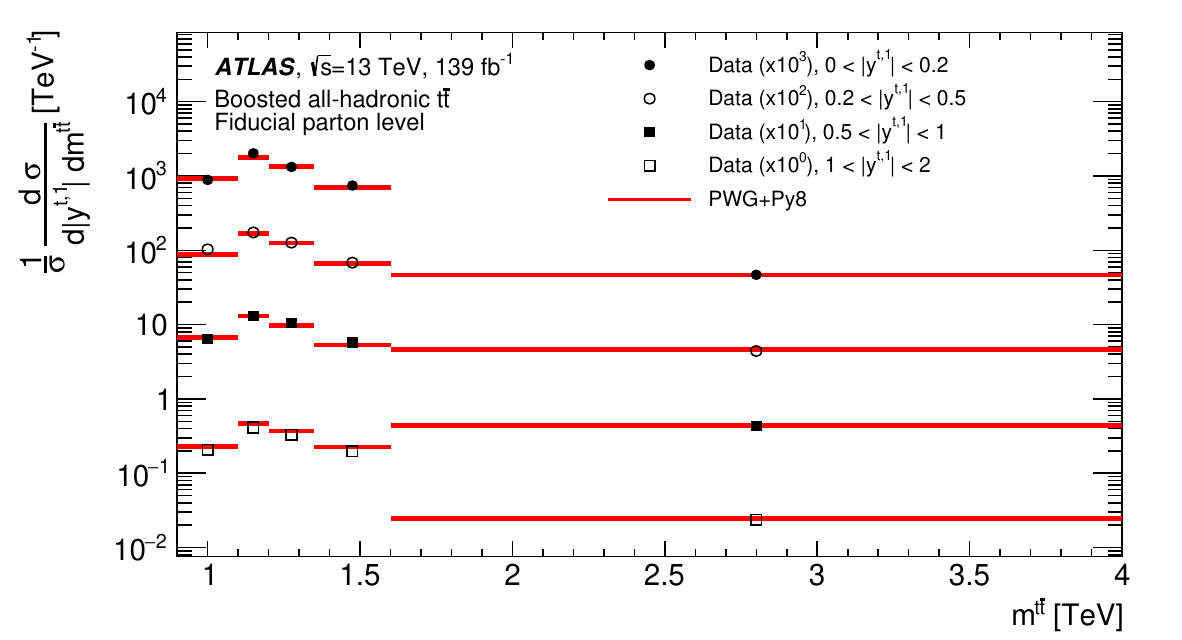}\label{fig:parton:t1_y_vs_ttbar_mass:rel:shape}}
\subfigure[]{ \includegraphics[width=0.68\textwidth]{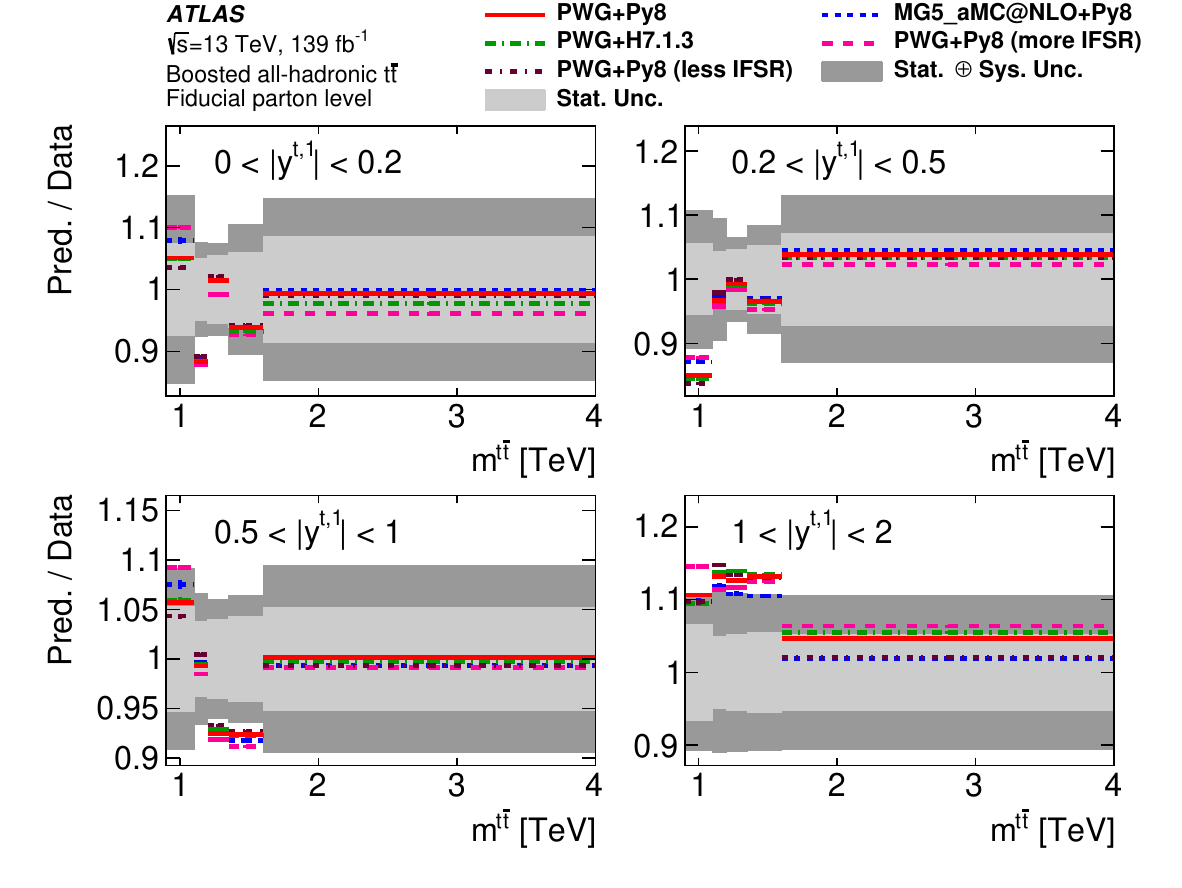}\label{fig:parton:t1_y_vs_ttbar_mass:rel:ratio}}
\caption{
\subref{fig:parton:t1_y_vs_ttbar_mass:rel:shape} Normalized parton-level fiducial phase-space double-differential cross-sections as a function of the absolute value of the rapidity of the leading top-quark and the mass of the \ttbar system, compared with the \POWPY[8] calculation.
Data points are placed at the centre of each bin and the \POWPY[8] calculation is indicated by solid lines.
The measurement and the prediction are normalized by the factors shown in parentheses to aid visibility.
\subref{fig:parton:t1_y_vs_ttbar_mass:rel:ratio}~The ratios of various MC calculations to the normalized parton-level fiducial phase-space differential cross-sections.
The dark and light grey bands indicate the total uncertainty and the statistical uncertainty, respectively, of the data in each bin.
}
\label{fig:parton:t1_y_vs_ttbar_mass:rel}
\end{figure*}

\begin{figure*}[htbp]
\centering
\subfigure[]{ \includegraphics[width=0.6\textwidth]{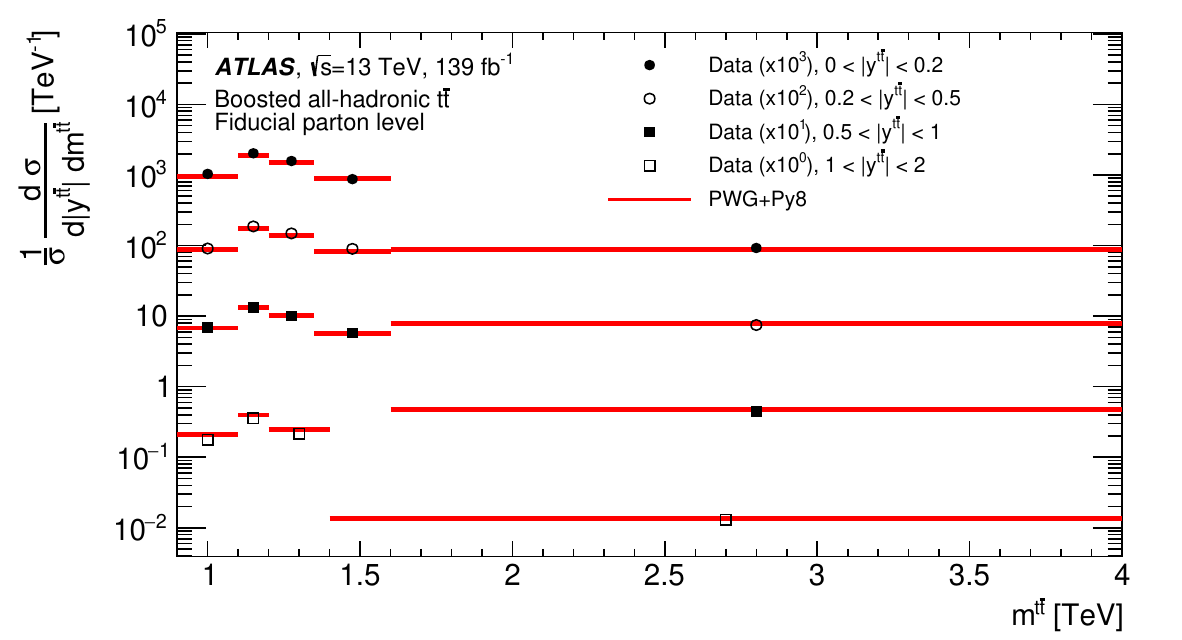}\label{fig:parton:ttbar_y_vs_ttbar_mass:rel:shape}}
\subfigure[]{ \includegraphics[width=0.68\textwidth]{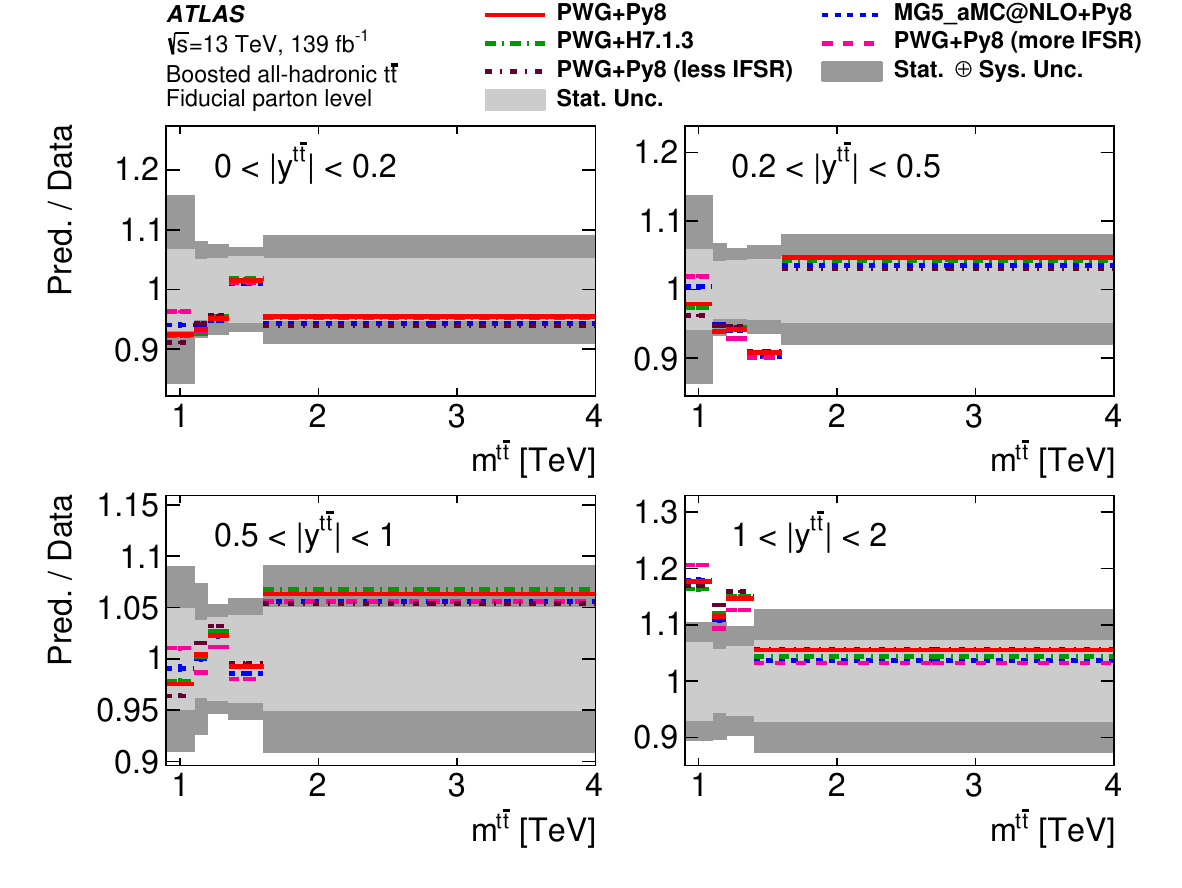}\label{fig:parton:ttbar_y_vs_ttbar_mass:rel:ratio}}
\caption{
\subref{fig:parton:ttbar_y_vs_ttbar_mass:rel:shape} Normalized parton-level fiducial phase-space double-differential cross-sections as a function of the absolute value of the rapidity and the mass of the \ttbar system, compared with the \POWPY[8] calculation.
Data points are placed at the centre of each bin and the \POWPY[8] calculation is indicated by solid lines.
The measurement and the prediction are normalized by the factors shown in parentheses to aid visibility.
\subref{fig:parton:ttbar_y_vs_ttbar_mass:rel:ratio}~The ratios of various MC calculations to the normalized parton-level fiducial phase-space differential cross-sections.
The dark and light grey bands indicate the total uncertainty and the statistical uncertainty, respectively, of the data in each bin.
}
\label{fig:parton:ttbar_y_vs_ttbar_mass:rel}
\end{figure*}

The additional  double-differential cross-section distributions are presented in Figures~\ref{fig:parton:t1_y_vs_t2_y:rel}--\ref{fig:parton:ttbar_y_vs_ttbar_mass:rel}.
 
\clearpage

\printbibliography

@article{Ball:2017nwa,
    author = "Ball, Richard D. and others",
    collaboration = "NNPDF",
    title = "{Parton distributions from high-precision collider data}",
    eprint = "1706.00428",
    archivePrefix = "arXiv",
    primaryClass = "hep-ph",
    reportNumber = "CAVENDISH-HEP-17-06, CERN-TH-2017-077, EDINBURGH-2017-08, NIKHEF-2017-006, OUTP-17-04P, TIF-UNIMI-2017-3",
    doi = "10.1140/epjc/s10052-017-5199-5",
    journal = "Eur. Phys. J. C",
    volume = "77",
    number = "10",
    pages = "663",
    year = "2017"
}

@article{Hou:2019qau,
	author = "Hou, Tie-Jiun and others",
	title = "{Progress in the CTEQ-TEA NNLO global QCD analysis}",
	eprint = "1908.11394",
	archivePrefix = "arXiv",
	primaryClass = "hep-ph",
	reportNumber = "MSUHEP-19-020, PITT-PACC-1905",
	month = "8",
	year = "2019"
}

@article{Grazzini:2017mhc,
    author = "Grazzini, Massimiliano and Kallweit, Stefan and Wiesemann, Marius",
    title = "{Fully differential NNLO computations with MATRIX}",
    eprint = "1711.06631",
    archivePrefix = "arXiv",
    primaryClass = "hep-ph",
    reportNumber = "ZU-TH-30-17, CERN-TH-2017-232, ZU-TH 30/17",
    doi = "10.1140/epjc/s10052-018-5771-7",
    journal = "Eur. Phys. J. C",
    volume = "78",
    number = "7",
    pages = "537",
    year = "2018"
}

@article{Catani:2019iny,
    author = "Catani, Stefano and Devoto, Simone and Grazzini, Massimiliano and Kallweit, Stefan and Mazzitelli, Javier and Sargsyan, Hayk",
    title = "{Top-quark pair hadroproduction at next-to-next-to-leading order in QCD}",
    eprint = "1901.04005",
    archivePrefix = "arXiv",
    primaryClass = "hep-ph",
    reportNumber = "ZU-TH 02/19",
    doi = "10.1103/PhysRevD.99.051501",
    journal = "Phys. Rev. D",
    volume = "99",
    number = "5",
    pages = "051501",
    year = "2019"
}

@article{Catani:2019hip,
    author = "Catani, Stefano and Devoto, Simone and Grazzini, Massimiliano and Kallweit, Stefan and Mazzitelli, Javier",
    title = "{Top-quark pair production at the LHC: Fully differential QCD predictions at NNLO}",
    eprint = "1906.06535",
    archivePrefix = "arXiv",
    primaryClass = "hep-ph",
    reportNumber = "ZU-TH 31/19",
    doi = "10.1007/JHEP07(2019)100",
    journal = "JHEP",
    volume = "07",
    pages = "100",
    year = "2019"
}

@inproceedings{Serkin:2021bbn,
	author = "Serkin, Leonid",
	collaboration = "ATLAS, CMS",
	title = "{Treatment of top-quark backgrounds in extreme phase spaces: the ''top $p_{T}$ reweighting'' and novel data-driven estimations in ATLAS and CMS}",
	booktitle = "{13th International Workshop on Top Quark Physics}",
	eprint = "2105.03977",
	archivePrefix = "arXiv",
	primaryClass = "hep-ex",
	reportNumber = "ATL-PHYS-PROC-2021-016",
	month = "5",
	year = "2021"
}

@article{LUCID2,
  author={G. Avoni and others},
  title={The new LUCID-2 detector for luminosity measurement and monitoring in ATLAS},
  journal={JINST},
  volume={13},
  number={07},
  pages={P07017},
  doi="10.1088/1748-0221/13/07/P07017",
  year={2018},
}

@article{VRjets,
    author = "Krohn, David and Thaler, Jesse and Wang, Lian-Tao",
    title = "{Jets with Variable R}",
    eprint = "0903.0392",
    archivePrefix = "arXiv",
    primaryClass = "hep-ph",
    doi = "10.1088/1126-6708/2009/06/059",
    journal = "JHEP",
    volume = "06",
    pages = "059",
    year = "2009"
}

@article{ghostmatch,
    author = "Cacciari, Matteo and Salam, Gavin P.",
    title = "{Pileup subtraction using jet areas}",
    eprint = "0707.1378",
    archivePrefix = "arXiv",
    primaryClass = "hep-ph",
    reportNumber = "LPTHE-07-01",
    doi = "10.1016/j.physletb.2007.09.077",
    journal = "Phys. Lett. B",
    volume = "659",
    pages = "119--126",
    year = "2008"
}

@article{Grzadkowski:2010es,
    author = "Grzadkowski, B. and Iskrzynski, M. and Misiak, M. and Rosiek, J.",
    title = "{Dimension-Six Terms in the Standard Model Lagrangian}",
    eprint = "1008.4884",
    archivePrefix = "arXiv",
    primaryClass = "hep-ph",
    reportNumber = "IFT-9-2010, TTP10-35",
    doi = "10.1007/JHEP10(2010)085",
    journal = "JHEP",
    volume = "10",
    pages = "085",
    year = "2010"
}

@article{AguilarSaavedra:2018nen,
    author = "Barducci, D. and others",
    editor = "Aguilar-Saavedra, Juan Antonio and Degrande, C. and Durieux, G. and Maltoni, F. and Vryonidou, E. and Zhang, C.",
    title = "{Interpreting top-quark LHC measurements in the standard-model effective field theory}",
    eprint = "1802.07237",
    archivePrefix = "arXiv",
    primaryClass = "hep-ph",
    reportNumber = "CERN-LPCC-2018-01",
    month = "2",
    year = "2018"
}

@article{www:dim6top,
      title         = "{dim6top model wiki page}",
      author        = "",
      year          = "2021",
      url           = "https://feynrules.irmp.ucl.ac.be/wiki/dim6top",
}

@article{Hartland:2019bjb,
    author = "Hartland, Nathan P. and Maltoni, Fabio and Nocera, Emanuele R. and Rojo, Juan and Slade, Emma and Vryonidou, Eleni and Zhang, Cen",
    title = "{A Monte Carlo global analysis of the Standard Model Effective Field Theory: the top quark sector}",
    eprint = "1901.05965",
    archivePrefix = "arXiv",
    primaryClass = "hep-ph",
    reportNumber = "OUTP-18-07P, Nikhef-2018-058, CP3-19-02, CERN-TH-2018-274",
    doi = "10.1007/JHEP04(2019)100",
    journal = "JHEP",
    volume = "04",
    pages = "100",
    year = "2019"
}

@article{Castro:2016jjv,
    author = {Castro, Nuno and Erdmann, Johannes and Grunwald, Cornelius and Kr\"oninger, Kevin and Rosien, Nils-Arne},
    title = "{EFTfitter: A tool for interpreting measurements in the context of effective field theories}",
    eprint = "1605.05585",
    archivePrefix = "arXiv",
    primaryClass = "hep-ex",
    doi = "10.1140/epjc/s10052-016-4280-9",
    journal = "Eur. Phys. J. C",
    volume = "76",
    number = "8",
    pages = "432",
    year = "2016"
}

@article{Ethier:2021bye,
    author = "Ethier, Jacob J. and Magni, Giacomo and Maltoni, Fabio and Mantani, Luca and Nocera, Emanuele R. and Rojo, Juan and Slade, Emma and Vryonidou, Eleni and Zhang, Cen",
    collaboration = "SMEFiT",
    title = "{Combined SMEFT interpretation of Higgs, diboson, and top quark data from the LHC}",
    eprint = "2105.00006",
    archivePrefix = "arXiv",
    primaryClass = "hep-ph",
    reportNumber = "OUTP-20-05P, Nikhef-2020-020, CP3-21-12, MCNET-21-07,
  MAN/HEP/2021/004",
    doi = "10.1007/JHEP11(2021)089",
    journal = "JHEP",
    volume = "11",
    pages = "089",
    year = "2021"
}

@string{jhep = {JHEP}}

@article{PhysRevLett.81.2642,
 title = {Evidence for Parton ${k}_{T}$ Effects in High- ${p}_{T}$ Particle Production},
 author = {Apanasevich, L. and Bacigalupi, J. and Baker, W. and Begel, M. and Blusk, S. and Bromberg, C. and Chang, P. and Choudhary, B. and Chung, W. H. and de Barbaro, L. and DeSoi, W. and D\l{}ugosz, W. and Dunlea, J. and Engels, E. and Fanourakis, G. and Ferbel, T. and Ftacnik, J. and Garelick, D. and Ginther, G. and Glaubman, M. and Gutierrez, P. and Hartman, K. and Huston, J. and Johnstone, C. and Kapoor, V. and Kuehler, J. and Lirakis, C. and Lobkowicz, F. and Lukens, P. and Mani, S. and Mansour, J. and Maul, A. and Miller, R. and Oh, B. Y. and Osborne, G. and Pellett, D. and Prebys, E. and Roser, R. and Shepard, P. and Shivpuri, R. and Skow, D. and Slattery, P. and Sorrell, L. and Striley, D. and Toothacker, W. and Varelas, N. and Weerasundara, D. and Whitmore, J. J. and Yasuda, T. and Yosef, C. and Zieli\ifmmode \acute{n}\else \'{n}\fi{}ski, M. and Zutshi, V.},
 journal = {Phys. Rev. Lett.},
 volume = {81},
 pages = {2642--2645},
 numpages = {0},
 year = {1998},
 month = {Sep},
 publisher = {American Physical Society},
 doi = {10.1103/PhysRevLett.81.2642},
}

@article{Cacciari:2008gn,
      author         = "Cacciari, Matteo and Salam, Gavin P. and Soyez, Gregory",
      title          = "{The catchment area of jets}",
      journal        = jhep,
      volume         = "04",
      pages          = "005",
      doi            = "10.1088/1126-6708/2008/04/005",
      year           = "2008",
      eprint         = "0802.1188",
      archivePrefix  = "arXiv",
      primaryClass   = "hep-ph",
      reportNumber   = "LPTHE-07-02",
      SLACcitation   = "%%CITATION = ARXIV:0802.1188;%%",
}

@article{Czakon:2017wor,
	author = "Czakon, Michal and Heymes, David and Mitov, Alexander and Pagani, Davide and Tsinikos, Ioannis and Zaro, Marco",
	title = "{Top-pair production at the LHC through NNLO QCD and NLO EW}",
	eprint = "1705.04105",
	archivePrefix = "arXiv",
	primaryClass = "hep-ph",
	reportNumber = "CAVENDISH-HEP-17-07, CP3-17-12, TUM-HEP-1084-17, TTK-17-15",
	doi = "10.1007/JHEP10(2017)186",
	journal = "JHEP",
	volume = "10",
	pages = "186",
	year = "2017"
}

@article{NNLO_calc,
      author         = "Kidonakis, Nikolaos",
      title          = "{Next-to-next-to-leading soft-gluon corrections for the
                        top quark cross section and transverse momentum
                        distribution}",
      journal        = "Phys. Rev.",
      volume         = "\textmd{D} 82", 
      pages	     = "114030",
      doi            = "10.1103/PhysRevD.82.114030",
      year           = "2010",
      eprint         = "1009.4935",
      archivePrefix  = "arXiv",
      primaryClass   = "hep-ph",
}

@article{nnloMtt,
      author         = "Ahrens, Valentin and Ferroglia, Andrea and Neubert,
                        Matthias and Pecjak, Ben D. and Yang, Li Lin",
      title          = "{Renormalization-group improved predictions for top-quark
                        pair production at hadron colliders}",
      journal        = jhep,
      volume         = "09",
      pages          = "097",
      doi            = "10.1007/JHEP09(2010)097",
      year           = "2010",
      eprint         = "1003.5827",
      archivePrefix  = "arXiv",
      primaryClass   = "hep-ph",
}

@article{PDG,
    author        = "{Particle Data Group} and C. Patrignani and others",
    journal       = "{Chin. Phys.}",
	pages         = "{100001}",
	doi           = "10.1088/1674-1137/40/10/100001",
	volume        = "\textmd{C} 40",
	year          = "2016",
	url           = {http://pdg.lbl.gov/}
}

@article{Frederix:2009,
	author	= "R. Frederix and F. Maltoni",
	title		= "Top pair invariant mass distribution: a window on new physics",
	journal	= jhep,
	volume	= "01",
	pages	= "047",
    doi            = "10.1088/1126-6708/2009/01/047",
	year		= "2009",
    archivePrefix  = "arXiv",
	eprint         = "0712.2355",
    primaryClass   = "hep-ph",
}

@article{Hill:1993hs,
      author         = "{C. T. Hill and S. J. Parke}",
      title          = "{Top production: Sensitivity to new physics}",
      journal        = "Phys. Rev. D",
      volume         = "49",
      year           = "1994",
      pages          = "4454-4462",
      doi            = "10.1103/PhysRevD.49.4454",
      eprint         = "hep-ph/9312324",
      archivePrefix  = "arXiv",
      primaryClass   = "hep-ph",
      reportNumber   = "FERMILAB-PUB-93-397-T",
      SLACcitation   = "%%CITATION = HEP-PH/9312324;%%"
}

@article{top_bkg_interference,
      author         = "{B. Hespel, F. Maltoni and E. Vryonidou}",
      title          = "{Signal background interference effects in heavy scalar
                        production and decay to a top-anti-top pair}",
      journal        = "JHEP",
      volume         = "10",
      year           = "2016",
      pages          = "016",
      doi            = "10.1007/JHEP10(2016)016",
      eprint         = "1606.04149",
      archivePrefix  = "arXiv",
      primaryClass   = "hep-ph",
      reportNumber   = "CP3-16-30, MCNET-16-19",
      SLACcitation   = "%%CITATION = ARXIV:1606.04149;%%"
}

@article{Bevilacqua:2010qb,
      author         = "Bevilacqua, Giuseppe and Czakon, Michal and van Hameren,
                        Andreas and Papadopoulos, Costas G. and Worek, Malgorzata",
      title          = "{Complete off-shell effects in top quark pair
                        hadroproduction with leptonic decay at next-to-leading
                        order}",
      journal        = "JHEP",
      volume         = "02",
      year           = "2011",
      pages          = "083",
      doi            = "10.1007/JHEP02(2011)083",
      eprint         = "1012.4230",
      archivePrefix  = "arXiv",
      primaryClass   = "hep-ph",
      reportNumber   = "TTK-10-56, IFJPAN-IV-2010-9, WUB-10-26",
      SLACcitation   = "%%CITATION = ARXIV:1012.4230;%%"
}

@article{Denner:2010jp,
      author         = "Denner, A. and Dittmaier, S. and Kallweit, S. and
                        Pozzorini, S.",
      title          = "{Next-to-Leading-Order QCD Corrections to $W^{+}W^{-}b\bar{b}$ production at Hadron Colliders}",
      journal        = "Phys. Rev. Lett.",
      volume         = "106",
      year           = "2011",
      pages          = "052001",
      doi            = "10.1103/PhysRevLett.106.052001",
      eprint         = "1012.3975",
      archivePrefix  = "arXiv",
      primaryClass   = "hep-ph",
      reportNumber   = "FR-PHENO-2010-041, PSI-PR-10-14, ZH-TH-19-10",
      SLACcitation   = "%%CITATION = ARXIV:1012.3975;%%"
}

@article{trimming,
      author         = "Krohn, D. and Thaler, J. and Wang, L.-T.",
      title          = "{Jet trimming}",
      year           = "2010",
      journal        = "JHEP",
      volume         = "02",
      pages          = "084",
      doi            = "10.1007/JHEP02(2010)084",
      eprint         = "0912.1342",
      archivePrefix  = "arXiv",
      primaryClass   = "hep-ph",
}

@article{Kogler:2018hem,
    author = "Kogler, Roman and others",
    title = "{Jet Substructure at the Large Hadron Collider}",
    eprint = "1803.06991",
    archivePrefix = "arXiv",
    primaryClass = "hep-ex",
    reportNumber = "FERMILAB-PUB-18-123-PPD",
    doi = "10.1103/RevModPhys.91.045003",
    journal = "Rev. Mod. Phys.",
    volume = "91",
    number = "4",
    pages = "045003",
    year = "2019"
}

@Article{Collins:1977iv, 
	author = "Collins, John C. and Soper, Davison E.", 
	title = "{Angular Distribution of Dileptons in High-Energy Hadron Collisions}", 
	reportNumber = "Print-77-0288 (PRINCETON)", 
	doi = "10.1103/PhysRevD.16.2219", 
	journal = "Phys. Rev. D", 
	volume = "16", 
	pages = "2219", 
	year = "1977" }

@article{Buchmuller:1985jz,
    author = "Buchmuller, W. and Wyler, D.",
    title = "{Effective lagrangian analysis of new interactions and flavor conservation}",
    reportNumber = "CERN-TH-4254/85",
    doi = "10.1016/0550-3213(86)90262-2",
    journal = "Nucl. Phys. B",
    volume = "268",
    pages = "621",
    year = "1986"
}

@article{Behring:2019iiv,
	author = "Behring, Arnd and Czakon, Michal and Mitov, Alexander and Papanastasiou, Andrew S. and Poncelet, Rene",
	title = "{Higher order corrections to spin correlations in top quark pair production at the LHC}",
	eprint = "1901.05407",
	archivePrefix = "arXiv",
	primaryClass = "hep-ph",
	reportNumber = "Cavendish-HEP-19/02, TTK-19-02, TTP19-003",
	doi = "10.1103/PhysRevLett.123.082001",
	journal = "Phys. Rev. Lett.",
	volume = "123",
	number = "8",
	pages = "082001",
	year = "2019"
}

@article{Czakon:2015owf,
	author = "Czakon, Michal and Heymes, David and Mitov, Alexander",
	title = "{High-precision differential predictions for top-quark pairs at the LHC}",
	eprint = "1511.00549",
	archivePrefix = "arXiv",
	primaryClass = "hep-ph",
	reportNumber = "CAVENDISH-HEP-15-10, TTK-15-34",
	doi = "10.1103/PhysRevLett.116.082003",
	journal = "Phys. Rev. Lett.",
	volume = "116",
	number = "8",
	pages = "082003",
	year = "2016"
}

@article{Czakon:2016dgf,
	author = "Czakon, Michal and Heymes, David and Mitov, Alexander",
	title = "{Dynamical scales for multi-TeV top-pair production at the LHC}",
	eprint = "1606.03350",
	archivePrefix = "arXiv",
	primaryClass = "hep-ph",
	reportNumber = "CAVENDISH-HEP-16-08, TTK-16-21",
	doi = "10.1007/JHEP04(2017)071",
	journal = "JHEP",
	volume = "04",
	pages = "071",
	year = "2017"
}

@Article{Cacciari:2008gp,
     author    = "Cacciari, Matteo and Salam, Gavin P. and Soyez, Gregory",
     title     = "{The anti-\(k_{t}\) jet clustering algorithm}",
     journal   = "JHEP",
     volume    = "04",
     year      = "2008",
     pages     = "063",
     eprint    = "0802.1189",
     archivePrefix = "arXiv",
     primaryClass  =  "hep-ph",
     doi       = "10.1088/1126-6708/2008/04/063",
     SLACcitation  = "%%CITATION = 0802.1189;%%"
}

@Article{Fastjet,
      author         = "Cacciari, Matteo and Salam, Gavin P. and Soyez, Gregory",
      title          = "{FastJet user manual}",
      journal        = "Eur. Phys. J. C",
      volume         = "72",
      year           = "2012",
      pages          = "1896",
      doi            = "10.1140/epjc/s10052-012-1896-2",
      eprint         = "1111.6097",
      archivePrefix  = "arXiv",
      primaryClass   = "hep-ph",
      reportNumber   = "CERN-PH-TH-2011-297",
      SLACcitation   = "%%CITATION = ARXIV:1111.6097;%%"
}

@Article{Butterworth:2015oua,
      author         = "Butterworth, Jon and others",
      title          = "{PDF4LHC recommendations for LHC Run II}",
      journal        = "J. Phys. G",
      volume         = "43",
      year           = "2016",
      pages          = "023001",
      doi            = "10.1088/0954-3899/43/2/023001",
      eprint         = "1510.03865",
      archivePrefix  = "arXiv",
      primaryClass   = "hep-ph",
      reportNumber   = "OUTP-15-17P, SMU-HEP-15-12, TIF-UNIMI-2015-14,
                        LCTS-2015-27, CERN-PH-TH-2015-249",
      SLACcitation   = "%%CITATION = ARXIV:1510.03865;%%"
}

@article{Lai:2010vv,
      author         = "Lai, H.-L. and others",
      title          = "{New parton distributions for collider physics}",
      journal        = "Phys. Rev. D",
      volume         = "82",
      pages          = "074024",
      doi            = "10.1103/PhysRevD.82.074024",
      year           = "2010",
      eprint         = "1007.2241",
      archivePrefix  = "arXiv",
      primaryClass   = "hep-ph",
      reportNumber   = "MSUHEP-100707, SMU-HEP-10-10",
      SLACcitation   = "%%CITATION = ARXIV:1007.2241;%%",
}

@article{Gao:2013xoa,
      author         = "Gao, J. and others",
      title          = "{CT10 next-to-next-to-leading order global analysis of QCD}",
      journal        = "Phys. Rev. D",
      volume         = "89",
      pages          = "033009",
      doi            = "10.1103/PhysRevD.89.033009",
      year           = "2014",
      eprint         = "1302.6246",
      archivePrefix  = "arXiv",
      primaryClass   = "hep-ph",
      reportNumber   = "SMU-HEP-12-23",
      SLACcitation   = "%%CITATION = ARXIV:1302.6246;%%",
}

@article{Martin:2009iq,
      author         = "Martin, A. D. and Stirling, W. J. and Thorne, R. S. and Watt, G.",
      title          = "{Parton distributions for the LHC}",
      journal        = "Eur. Phys. J. C",
      volume         = "63",
      year           = "2009",
      pages          = "189",
      doi            = "10.1140/epjc/s10052-009-1072-5",
      eprint         = "0901.0002",
      archivePrefix  = "arXiv",
      primaryClass   = "hep-ph",
      reportNumber   = "IPPP-08-95, DCPT-08-190, CAVENDISH-HEP-08-16",
      SLACcitation   = "%%CITATION = ARXIV:0901.0002;%%"
}

@article{Martin:2009bu,
      author         = "Martin, A. D. and Stirling, W. J. and Thorne, R. S. and Watt, G.",
      title          = "{Uncertainties on \(\alpha_S\) in global PDF analyses and
                        implications for predicted hadronic cross sections}",
      journal        = "Eur. Phys. J. C",
      volume         = "64",
      pages          = "653-680",
      doi            = "10.1140/epjc/s10052-009-1164-2",
      year           = "2009",
      eprint         = "0905.3531",
      archivePrefix  = "arXiv",
      primaryClass   = "hep-ph",
      reportNumber   = "IPPP-09-33, DCPT-09-66, CAVENDISH-HEP-09-06",
      SLACcitation   = "%%CITATION = ARXIV:0905.3531;%%",
}

@article{Harland-Lang:2014zoa,
      author         = "Harland-Lang, L. A. and Martin, A. D. and Motylinski, P.
                        and Thorne, R. S.",
      title          = "{Parton distributions in the LHC era: MMHT 2014 PDFs}",
      journal        = "Eur. Phys. J. C",
      number         = "5",
      volume         = "75",
      pages          = "204",
      doi            = "10.1140/epjc/s10052-015-3397-6",
      year           = "2015",
      eprint         = "1412.3989",
      archivePrefix  = "arXiv",
      primaryClass   = "hep-ph",
      reportNumber   = "LCTS-2014-47, IPPP-14-97, DCPT-14-194",
      SLACcitation   = "%%CITATION = ARXIV:1412.3989;%%",
}

@article{Ball:2012cx,
      author         = "Ball, Richard D. and others",
      title          = "{Parton distributions with LHC data}",
      journal        = "Nucl. Phys. B",
      volume         = "867",
      year           = "2013",
      pages          = "244",
      doi            = "10.1016/j.nuclphysb.2012.10.003",
      eprint         = "1207.1303",
      archivePrefix  = "arXiv",
      primaryClass   = "hep-ph",
      reportNumber   = "EDINBURGH-2012-08, IFUM-FT-997, FR-PHENO-2012-014,
                        RWTH-TTK-12-25, CERN-PH-TH-2012-037, SFB-CPP-12-47\,
                        --CERN-PH-TH-2012-037",
      SLACcitation   = "%%CITATION = ARXIV:1207.1303;%%"
}

@article{Ball:2014uwa,
      author         = "Ball, Richard D. and others",
      title          = "{Parton distributions for the LHC run II}",
      collaboration  = "NNPDF",
      journal        = "JHEP",
      volume         = "04",
      year           = "2015",
      pages          = "040",
      doi            = "10.1007/JHEP04(2015)040",
      eprint         = "1410.8849",
      archivePrefix  = "arXiv",
      primaryClass   = "hep-ph",
      reportNumber   = "EDINBURGH-2014-15, IFUM-1034-FT, CERN-PH-TH-2013-253,
                        OUTP-14-11P, CAVENDISH-HEP-14-11",
      SLACcitation   = "%%CITATION = ARXIV:1410.8849;%%"
}

@article{Czakon:2012pz,
      author         = "Czakon, Michal and Mitov, Alexander",
      title          = "{NNLO corrections to top pair production at hadron
                        colliders: the quark-gluon reaction}",
      journal        = "JHEP",
      volume         = "01",
      pages          = "080",
      doi            = "10.1007/JHEP01(2013)080",
      year           = "2013",
      eprint         = "1210.6832",
      archivePrefix  = "arXiv",
      primaryClass   = "hep-ph",
      SLACcitation   = "%%CITATION = ARXIV:1210.6832;%%",
}

@Article{Sjostrand:2007gs,
	author    = "Sj{\"o}strand, T. and Mrenna, S. and Skands, P.",
	title     = "{A brief introduction to PYTHIA 8.1}",
	journal   = "Comput. Phys. Commun.",
	volume    = "178",
	year      = "2008",
	pages     = "852-867",
	eprint    = "0710.3820",
	archivePrefix = "arXiv",
	primaryClass  =  "hep-ph",
	doi       = "10.1016/j.cpc.2008.01.036",
	SLACcitation  = "%%CITATION = 0710.3820;%%"
}

@article{Sjostrand:2014zea,
      author         = "Sj{\"o}strand, Torbj{\"o}rn and Ask, Stefan and Christiansen,
                        Jesper R. and Corke, Richard and Desai, Nishita and Ilten,
                        Philip and Mrenna, Stephen and Prestel, Stefan and
                        Rasmussen, Christine O. and Skands, Peter Z.",
      title          = "{An introduction to PYTHIA 8.2}",
      journal        = "Comput. Phys. Commun.",
      volume         = "191",
      year           = "2015",
      pages          = "159",
      doi            = "10.1016/j.cpc.2015.01.024",
      eprint         = "1410.3012",
      archivePrefix  = "arXiv",
      primaryClass   = "hep-ph",
      reportNumber   = "LU-TP-14-36, MCNET-14-22, CERN-PH-TH-2014-190,
                        FERMILAB-PUB-14-316-CD, DESY-14-178, SLAC-PUB-16122,
                        --FERMILAB-PUB-14-316-CD",
      SLACcitation   = "%%CITATION = ARXIV:1410.3012;%%"
}

@Article{Lange:2001uf,
      author         = "Lange, D. J.",
      title          = "{The EvtGen particle decay simulation package}",
      booktitle      = "{Proceedings, 7th International Conference on B physics
                        at hadron machines (BEAUTY 2000)}",
      journal        = "Nucl. Instrum. Meth. A",
      volume         = "462",
      year           = "2001",
      pages          = "152",
      doi            = "10.1016/S0168-9002(01)00089-4",
      SLACcitation   = "%%CITATION = NUIMA,A462,152;%%"
}

@Article{Frixione:2008yi,
     author    = "Frixione, Stefano and Laenen, Eric and Motylinski, Patrick and White, Chris and Webber, Bryan R.",
     title     = "{Single-top hadroproduction in association with a \(W\) boson}",
     journal   = "JHEP",
     volume    = "07",
     year      = "2008",
     pages     = "029",
     eprint    = "0805.3067",
     archivePrefix = "arXiv",
     primaryClass  =  "hep-ph",
     doi       = "10.1088/1126-6708/2008/07/029"
}

@Article{Alwall:2014hca,
      author         = "Alwall, J. and Frederix, R. and Frixione, S. and Hirschi,
                        V. and Maltoni, F. and Mattelaer, O. and Shao, H. -S. and
                        Stelzer, T. and Torrielli, P. and Zaro, M.",
      title          = "{The automated computation of tree-level and
                        next-to-leading order differential cross sections, and
                        their matching to parton shower simulations}",
      journal        = "JHEP",
      volume         = "07",
      year           = "2014",
      pages          = "079",
      doi            = "10.1007/JHEP07(2014)079",
      eprint         = "1405.0301",
      archivePrefix  = "arXiv",
      primaryClass   = "hep-ph",
      reportNumber   = "CERN-PH-TH-2014-064, CP3-14-18, LPN14-066, MCNET-14-09,
                        ZU-TH-14-14",
      SLACcitation   = "%%CITATION = ARXIV:1405.0301;%%"
}

@article{Artoisenet:2012st,
      author         = "Artoisenet, Pierre and Frederix, Rikkert and Mattelaer,
                        Olivier and Rietkerk, Robbert",
      title          = "{Automatic spin-entangled decays of heavy resonances in
                        Monte Carlo simulations}",
      journal        = "JHEP",
      volume         = "03",
      pages          = "015",
      doi            = "10.1007/JHEP03(2013)015",
      year           = "2013",
      eprint         = "1212.3460",
      archivePrefix  = "arXiv",
      primaryClass   = "hep-ph",
      reportNumber   = "NIKHEF-2012-021, CERN-PH-TH-2012-329",
      SLACcitation   = "%%CITATION = ARXIV:1212.3460;%%",
}

@article{Frixione:2007zp,
      author         = "Frixione, Stefano and Laenen, Eric and Motylinski,
                        Patrick and Webber, Bryan R.",
      title          = "{Angular correlations of lepton pairs from vector boson
                        and top quark decays in Monte Carlo simulations}",
      journal        = "JHEP",
      volume         = "04",
      year           = "2007",
      pages          = "081",
      doi            = "10.1088/1126-6708/2007/04/081",
      eprint         = "hep-ph/0702198",
      archivePrefix  = "arXiv",
      OPTprimaryClass   = "HEP-PH",
      reportNumber   = "CAVENDISH-HEP-07-01, GEF-TH-09-2007, ITP-UU-07-10,
                        NIKHEF-2007-004",
      SLACcitation   = "%%CITATION = HEP-PH/0702198;%%"
}

@Article{Bahr:2008pv,
      author         = "B{\"a}hr, M. and others",
      title          = "{Herwig++ physics and manual}",
      journal        = "Eur. Phys. J. C",
      volume         = "58",
      year           = "2008",
      pages          = "639",
      doi            = "10.1140/epjc/s10052-008-0798-9",
      eprint         = "0803.0883",
      archivePrefix  = "arXiv",
      primaryClass   = "hep-ph",
      reportNumber   = "CERN-PH-TH-2008-038, CAVENDISH-HEP-08-03, KA-TP-05-2008,
                        DCPT-08-22, IPPP-08-11, CP3-08-05",
      SLACcitation   = "%%CITATION = ARXIV:0803.0883;%%"
}

@Article{Bellm:2015jjp,
      author         = "Bellm, Johannes and others",
      title          = "{Herwig 7.0/Herwig++ 3.0 release note}",
      journal        = "Eur. Phys. J. C",
      volume         = "76",
      year           = "2016",
      number         = "4",
      pages          = "196",
      doi            = "10.1140/epjc/s10052-016-4018-8",
      eprint         = "1512.01178",
      archivePrefix  = "arXiv",
      primaryClass   = "hep-ph",
      reportNumber   = "CERN-PH-TH-2015-289, MAN-HEP-2015-15, IFJPAN-IV-2015-13,
                        HERWIG-2015-01, KA-TP-18-2015, DCPT-15-142, MCNET-15-28,
                        IPPP-15-71, --HERWIG-2015-01",
      SLACcitation   = "%%CITATION = ARXIV:1512.01178;%%"
}

@article{Bellm:2017jjp,
      author         = "Bellm, Johannes and others",
      title          = "{Herwig 7.1 Release Note}",
      eprint         = "1705.06919",
      year           = "2017",
      archivePrefix  = "arXiv",
      primaryClass   = "hep-ph",
      reportNumber   = "CERN-PH-TH-2017-109, MAN/HEP/2017/08, UWTHPH-2017-10, 
                        IFJPAN-IV-2017-7, NIKHEF 2017-026, HERWIG-2017-02, 
                        KA-TP-19-2017, MCnet-17-08, IPPP/17/40",
      SLACcitation   = "%%CITATION = ARXIV:1705.06919;%%"
}

@Article{Nason:2004rx,
      author         = "Nason, Paolo",
      title          = "{A new method for combining NLO QCD with shower Monte Carlo algorithms}",
      journal        = "JHEP",
      volume         = "11",
      pages          = "040",
      doi            = "10.1088/1126-6708/2004/11/040",
      year           = "2004",
      eprint         = "hep-ph/0409146",
      archivePrefix  = "arXiv",
}

@Article{Frixione:2007vw,
      author         = "Frixione, Stefano and Nason, Paolo and Oleari, Carlo",
      title          = "{Matching NLO QCD computations with parton shower
                        simulations: the POWHEG method}",
      journal        = "JHEP",
      volume         = "11",
      pages          = "070",
      doi            = "10.1088/1126-6708/2007/11/070",
      year           = "2007",
      eprint         = "0709.2092",
      archivePrefix  = "arXiv",
      primaryClass   = "hep-ph",
}

@Article{Alioli:2010xd,
      author         = "Alioli, Simone and Nason, Paolo and Oleari, Carlo and Re,
                        Emanuele",
      title          = "{A general framework for implementing NLO calculations in
                        shower Monte Carlo programs: the POWHEG BOX}",
      journal        = "JHEP",
      volume         = "06",
      pages          = "043",
      doi            = "10.1007/JHEP06(2010)043",
      year           = "2010",
      eprint         = "1002.2581",
      archivePrefix  = "arXiv",
      primaryClass   = "hep-ph",
}

@article{Hartanto:2015uka,
      author         = "Hartanto, Heribertus B. and J{\"a}ger, Barbara and Reina,
                        Laura and Wackeroth, Doreen",
      title          = "{Higgs boson production in association with top quarks in
                        the POWHEG BOX}",
      journal        = "Phys. Rev. D",
      volume         = "91",
      year           = "2015",
      number         = "9",
      pages          = "094003",
      doi            = "10.1103/PhysRevD.91.094003",
      eprint         = "1501.04498",
      archivePrefix  = "arXiv",
      primaryClass   = "hep-ph",
      SLACcitation   = "%%CITATION = ARXIV:1501.04498;%%"
}

@article{Aliev:2010zk,
      author         = "Aliev, M. and Lacker, H. and Langenfeld, U. and Moch, S.
                        and Uwer, P. and Wiedermann, M.",
      title          = "{HATHOR -- HAdronic Top and Heavy quarks crOss section
                        calculatoR}",
      journal        = "Comput. Phys. Commun.",
      volume         = "182",
      year           = "2011",
      pages          = "1034-1046",
      doi            = "10.1016/j.cpc.2010.12.040",
      eprint         = "1007.1327",
      archivePrefix  = "arXiv",
      primaryClass   = "hep-ph",
      reportNumber   = "DESY-10-091, HU-EP-10-33, SFB-CPP-10-60",
      SLACcitation   = "%%CITATION = ARXIV:1007.1327;%%"
}

@article{Re:2010bp,
      author         = "Re, Emanuele",
      title          = "{Single-top \(Wt\)-channel production matched with parton
                        showers using the POWHEG method}",
      journal        = "Eur. Phys. J. C",
      volume         = "71",
      year           = "2011",
      pages          = "1547",
      doi            = "10.1140/epjc/s10052-011-1547-z",
      eprint         = "1009.2450",
      archivePrefix  = "arXiv",
      primaryClass   = "hep-ph",
      reportNumber   = "IPPP-10-74, DCPT-10-148",
      SLACcitation   = "%%CITATION = ARXIV:1009.2450;%%"
}

@article{Kant:2014oha,
      author         = "Kant, P. and Kind, O. M. and Kintscher, T. and Lohse, T.
                        and Martini, T. and Mölbitz, S. and Rieck, P. and Uwer, P.",
      title          = "{HatHor for single top-quark production: Updated
                        predictions and uncertainty estimates for single top-quark
                        production in hadronic collisions}",
      journal        = "Comput. Phys. Commun.",
      volume         = "191",
      year           = "2015",
      pages          = "74-89",
      doi            = "10.1016/j.cpc.2015.02.001",
      eprint         = "1406.4403",
      archivePrefix  = "arXiv",
      primaryClass   = "hep-ph",
      reportNumber   = "HU-EP-14-22",
      SLACcitation   = "%%CITATION = ARXIV:1406.4403;%%"
}

@article{Kidonakis:2010ux,
      author         = "Kidonakis, Nikolaos",
      title          = "{Two-loop soft anomalous dimensions for single top quark
                        associated production with a \(W^{-}\) or \(H^{-}\)}",
      journal        = "Phys. Rev. D",
      volume         = "82",
      year           = "2010",
      pages          = "054018",
      doi            = "10.1103/PhysRevD.82.054018",
      eprint         = "1005.4451",
      archivePrefix  = "arXiv",
      primaryClass   = "hep-ph",
      SLACcitation   = "%%CITATION = ARXIV:1005.4451;%%"
}

@inproceedings{Kidonakis:2013zqa,
      author         = "Kidonakis, Nikolaos",
      title          = "{Top Quark Production}",
      booktitle      = "{Proceedings, Helmholtz International Summer School on
                        Physics of Heavy Quarks and Hadrons (HQ 2013)}",
      eventdate      = "2013-07-15/2013-07-28",
      venue          = "JINR, Dubna, Russia",
      pages          = "139-168",
      doi            = "10.3204/DESY-PROC-2013-03/Kidonakis",
      eprint         = "1311.0283",
      archivePrefix  = "arXiv",
      primaryClass   = "hep-ph",
      reportNumber   = "KSU-HEP-110113",
      SLACcitation   = "%%CITATION = ARXIV:1311.0283;%%"
}

@article{Beneke:2011mq,
      author         = "Beneke, M. and Falgari, P. and Klein, S. and Schwinn, C.",
      title          = "{Hadronic top-quark pair production with NNLL threshold
                        resummation}",
      journal        = "Nucl. Phys. B",
      volume         = "855",
      year           = "2012",
      pages          = "695-741",
      doi            = "10.1016/j.nuclphysb.2011.10.021",
      eprint         = "1109.1536",
      archivePrefix  = "arXiv",
      primaryClass   = "hep-ph",
      reportNumber   = "TTK-11-38, ITP-UU-11-26, SPIN-11-19, FR-PHENO-2011-015,
                        SFB-CPP-11-49",
      SLACcitation   = "%%CITATION = ARXIV:1109.1536;%%"
}

@article{Cacciari:2011hy,
      author         = "Cacciari, Matteo and Czakon, Michal and Mangano,
                        Michelangelo and Mitov, Alexander and Nason, Paolo",
      title          = "{Top-pair production at hadron colliders with
                        next-to-next-to-leading logarithmic soft-gluon
                        resummation}",
      journal        = "Phys. Lett. B",
      volume         = "710",
      pages          = "612-622",
      doi            = "10.1016/j.physletb.2012.03.013",
      year           = "2012",
      eprint         = "1111.5869",
      primaryClass   = "hep-ph",
      archivePrefix  = "arXiv",
      reportNumber   = "CERN-PH-TH-2011-277, TTK-11-54",
      SLACcitation   = "%%CITATION = ARXIV:1111.5869;%%",
}

@article{Czakon:2012zr,
      author         = "Czakon, Michal and Mitov, Alexander",
      title          = "{NNLO corrections to top-pair production at hadron
                        colliders: the all-fermionic scattering channels}",
      journal        = "JHEP",
      volume         = "12",
      pages          = "054",
      doi            = "10.1007/JHEP12(2012)054",
      year           = "2012",
      eprint         = "1207.0236",
      archivePrefix  = "arXiv",
      primaryClass   = "hep-ph",
      SLACcitation   = "%%CITATION = ARXIV:1207.0236;%%",
}

@article{Czakon:2013goa,
      author         = "Czakon, Michal and Fiedler, Paul and Mitov, Alexander",
      title          = "{Total Top-Quark Pair-Production Cross Section at
                        Hadron Colliders Through \(O(\alpha_S^4)\)}",
      journal        = "Phys. Rev. Lett.",
      volume         = "110",
      pages          = "252004",
      doi            = "10.1103/PhysRevLett.110.252004",
      year           = "2013",
      eprint         = "1303.6254",
      archivePrefix  = "arXiv",
      primaryClass   = "hep-ph",
      reportNumber   = "CERN-PH-TH-2013-056, TTK-13-08",
      SLACcitation   = "%%CITATION = ARXIV:1303.6254;%%",
}

@article{Baernreuther:2012ws,
      author         = "B{\"a}rnreuther, Peter and Czakon, Michal and Mitov,
                        Alexander",
      title          = "{Percent-Level-Precision Physics at the
                  Tevatron: Next-to-Next-to-Leading Order QCD
                  Corrections to \(q \bar{q} \to t \bar{t} + X\)}",
      journal        = "Phys. Rev. Lett.",
      volume         = "109",
      pages          = "132001",
      doi            = "10.1103/PhysRevLett.109.132001",
      year           = "2012",
      eprint         = "1204.5201",
      archivePrefix  = "arXiv",
      primaryClass   = "hep-ph",
      SLACcitation   = "%%CITATION = ARXIV:1204.5201;%%",
}

@article{Frixione:2007nw,
      author         = "Frixione, Stefano and Ridolfi, Giovanni and Nason, Paolo",
      title          = "{A positive-weight next-to-leading-order Monte Carlo for
                        heavy flavour hadroproduction}",
      journal        = "JHEP",
      volume         = "09",
      pages          = "126",
      doi            = "10.1088/1126-6708/2007/09/126",
      year           = "2007",
      eprint         = "0707.3088",
      archivePrefix  = "arXiv",
      primaryClass   = "hep-ph",
      SLACcitation   = "%%CITATION = ARXIV:0707.3088;%%",
}

@article{deFlorian:2016spz,
      author         = "de Florian, D. and others",
      title          = "{Handbook of LHC Higgs Cross Sections: 4. Deciphering the
                        Nature of the Higgs Sector}",
      collaboration  = "LHC Higgs Cross Section Working Group",
      doi            = "10.23731/CYRM-2017-002",
      year           = "2016",
      eprint         = "1610.07922",
      archivePrefix  = "arXiv",
      primaryClass   = "hep-ph",
      reportNumber   = "FERMILAB-FN-1025-T, CERN-2017-002-M",
      SLACcitation   = "%%CITATION = ARXIV:1610.07922;%%"
}

@article{Czakon:2011xx,
  author         = "Czakon, Michal and Mitov, Alexander",
  title          = "{Top++: A program for the calculation of the top-pair cross-section at hadron colliders}",
  journal        = "Comput. Phys. Commun.",
  volume         = "185",
  year           = "2014",
  pages          = "2930",
  doi            = "10.1016/j.cpc.2014.06.021",
  eprint         = "1112.5675",
  archivePrefix  = "arXiv",
  primaryClass   = "hep-ph",
  reportNumber   = "CERN-PH-TH-2011-303, TTK-11-58",
}

@article{Agostinelli:2002hh,
      author         = "{GEANT4 Collaboration} and Agostinelli, S. and others",
      title          = "{\textsc{Geant4} -- a simulation toolkit}",
      journal        = "Nucl. Instrum. Meth. A",
      volume         = "506",
      year           = "2003",
      pages          = "250",
      doi            = "10.1016/S0168-9002(03)01368-8",
      reportNumber   = "SLAC-PUB-9350, FERMILAB-PUB-03-339",
      SLACcitation   = "%%CITATION = NUIMA,A506,250;%%"
}

@Inproceedings{Adye:2011gm,
    author = "Adye, Tim",
    title = "{Unfolding algorithms and tests using RooUnfold}",
    booktitle = "{Proceedings, 2011 Workshop on Statistical Issues Related to Discovery Claims in Search Experiments and Unfolding (PHYSTAT 2011)}",
    venue = "CERN, Geneva, Switzerland",
    eventdate = "2011-01-17/2011-01-20",
    pages = "313-318",
    eprint = "1105.1160",
    archivePrefix = "arXiv",
    primaryClass = "physics.data-an",
    doi = "10.5170/CERN-2011-006.313",
}

@article{DAgostini:1995,
  author = "Giulio D'Agostini",
  title = "A multidimensional unfolding method based on Bayes' theorem",
  journal = "Nucl. Instrum. Meth. A",
  volume = "362",
  number = "2",
  pages = "487 - 498",
  year = "1995",
  issn = "0168-9002",
  doi = "10.1016/0168-9002(95)00274-X",
}

@Article{PERF-2007-01,
    author         = "{ATLAS Collaboration}",
    title          = "{The ATLAS Experiment at the CERN Large Hadron Collider}",
    journal        = "JINST",
    volume         = "3",
    year           = "2008",
    pages          = "S08003",
    doi            = "10.1088/1748-0221/3/08/S08003",
    primaryClass   = "hep-ex",
}

@Article{SOFT-2010-01,
    author         = "{ATLAS Collaboration}",
    title          = "{The ATLAS Simulation Infrastructure}",
    journal        = "Eur. Phys. J. C",
    volume         = "70",
    year           = "2010",
    pages          = "823",
    doi            = "10.1140/epjc/s10052-010-1429-9",
    eprint         = "1005.4568",
    archivePrefix  = "arXiv",
    primaryClass   = "physics.ins-det",
}

@Article{STDM-2011-38,
    author         = "{ATLAS Collaboration}",
    title          = "{ATLAS measurements of the properties of jets for boosted particle searches}",
    journal        = "Phys. Rev. D",
    volume         = "86",
    year           = "2012",
    pages          = "072006",
    doi            = "10.1103/PhysRevD.86.072006",
    reportNumber   = "CERN-PH-EP-2012-149",
    eprint         = "1206.5369",
    archivePrefix  = "arXiv",
    primaryClass   = "hep-ex",
}

@Article{TOPQ-2011-07,
    author         = "{ATLAS Collaboration}",
    title          = "{Measurements of top quark pair relative differential cross-sections with ATLAS in \(pp\) collisions at \(\sqrt{s} = 7\,\text{TeV}\)}",
    journal        = "Eur. Phys. J. C",
    volume         = "73",
    year           = "2013",
    pages          = "2261",
    doi            = "10.1140/epjc/s10052-012-2261-1",
    reportNumber   = "CERN-PH-EP-2012-165",
    eprint         = "1207.5644",
    archivePrefix  = "arXiv",
    primaryClass   = "hep-ex",
}

@Article{TOPQ-2012-08,
    author         = "{ATLAS Collaboration}",
    title          = "{Measurements of normalized differential cross-sections for  \(t\bar{t}\)  production in \(pp\) collisions at \(\sqrt{s} = 7\,\text{TeV}\) using the ATLAS detector}",
    journal        = "Phys. Rev. D",
    volume         = "90",
    year           = "2014",
    pages          = "072004",
    doi            = "10.1103/PhysRevD.90.072004",
    reportNumber   = "CERN-PH-EP-2014-099",
    eprint         = "1407.0371",
    archivePrefix  = "arXiv",
    primaryClass   = "hep-ex",
}

@Article{TOPQ-2012-15,
    author         = "{ATLAS Collaboration}",
    title          = "{Search for resonances decaying into top-quark pairs using fully hadronic decays in \(pp\) collisions with ATLAS at \(\sqrt{s} = 7\,\text{TeV}\)}",
    journal        = "JHEP",
    volume         = "01",
    year           = "2013",
    pages          = "116",
    doi            = "10.1007/JHEP01(2013)116",
    reportNumber   = "CERN-PH-EP-2012-291",
    eprint         = "1211.2202",
    archivePrefix  = "arXiv",
    primaryClass   = "hep-ex",
}

@Article{STDM-2013-11,
    author         = "{ATLAS Collaboration}",
    title          = "{Measurement of the inclusive jet cross-section in proton--proton collisions at \(\sqrt{s} = 7\,\text{TeV}\) using \(4.5\,\text{fb}^{-1}\) of data with the ATLAS detector}",
    journal        = "JHEP",
    volume         = "02",
    year           = "2015",
    pages          = "153",
    doi            = "10.1007/JHEP02(2015)153",
    reportNumber   = "CERN-PH-EP-2014-155",
    eprint         = "1410.8857",
    archivePrefix  = "arXiv",
    primaryClass   = "hep-ex",
    related        = "STDM-2013-11-err",
    relatedstring  = "Erratum:",
}

@Article{TOPQ-2013-07,
    author         = "{ATLAS Collaboration}",
    title          = "{Differential top-antitop cross-section measurements as a function of observables constructed from final-state particles using \(pp\) collisions at \(\sqrt{s} = 7\,\text{TeV}\) in the ATLAS detector}",
    journal        = "JHEP",
    volume         = "06",
    year           = "2015",
    pages          = "100",
    doi            = "10.1007/JHEP06(2015)100",
    reportNumber   = "CERN-PH-EP-2014-295",
    eprint         = "1502.05923",
    archivePrefix  = "arXiv",
    primaryClass   = "hep-ex",
}

@Article{EXOT-2014-15,
    author         = "{ATLAS Collaboration}",
    title          = "{Search for New Phenomena in Dijet Angular Distributions in Proton--Proton Collisions at \(\sqrt{s} = 8\,\text{TeV}\) Measured with the ATLAS Detector}",
    journal        = "Phys. Rev. Lett.",
    volume         = "114",
    year           = "2015",
    pages          = "221802",
    doi            = "10.1103/PhysRevLett.114.221802",
    reportNumber   = "CERN-PH-EP-2015-067",
    eprint         = "1504.00357",
    archivePrefix  = "arXiv",
    primaryClass   = "hep-ex",
}

@Article{PERF-2014-03,
    author         = "{ATLAS Collaboration}",
    title          = "{Performance of pile-up mitigation techniques for jets in \(pp\) collisions at \(\sqrt{s} = 8\,\text{TeV}\) using the ATLAS detector}",
    journal        = "Eur. Phys. J. C",
    volume         = "76",
    year           = "2016",
    pages          = "581",
    doi            = "10.1140/epjc/s10052-016-4395-z",
    reportNumber   = "CERN-PH-EP-2015-206",
    eprint         = "1510.03823",
    archivePrefix  = "arXiv",
    primaryClass   = "hep-ex",
}

@Article{PERF-2014-07,
    author         = "{ATLAS Collaboration}",
    title          = "{Topological cell clustering in the ATLAS calorimeters and its performance in LHC Run~1}",
    journal        = "Eur. Phys. J. C",
    volume         = "77",
    year           = "2017",
    pages          = "490",
    doi            = "10.1140/epjc/s10052-017-5004-5",
    reportNumber   = "CERN-PH-EP-2015-304",
    eprint         = "1603.02934",
    archivePrefix  = "arXiv",
    primaryClass   = "hep-ex",
}

@Article{TOPQ-2014-15,
    author         = "{ATLAS Collaboration}",
    title          = "{Measurement of the differential cross-section of highly boosted top quarks as a function of their transverse momentum in \(\sqrt{s} = 8\,\text{TeV}\) proton--proton collisions using the ATLAS detector}",
    journal        = "Phys. Rev. D",
    volume         = "93",
    year           = "2016",
    pages          = "032009",
    doi            = "10.1103/PhysRevD.93.032009",
    reportNumber   = "CERN-PH-EP-2015-237",
    eprint         = "1510.03818",
    archivePrefix  = "arXiv",
    primaryClass   = "hep-ex",
}

@Article{TOPQ-2015-06,
    author         = "{ATLAS Collaboration}",
    title          = "{Measurements of top-quark pair differential cross-sections in the lepton+jets channel in \(pp\) collisions at \(\sqrt{s} = 8\,\text{TeV}\) using the ATLAS detector}",
    journal        = "Eur. Phys. J. C",
    volume         = "76",
    year           = "2016",
    pages          = "538",
    doi            = "10.1140/epjc/s10052-016-4366-4",
    reportNumber   = "CERN-PH-EP-2015-239",
    eprint         = "1511.04716",
    archivePrefix  = "arXiv",
    primaryClass   = "hep-ex",
}

@Article{TOPQ-2015-07,
    author         = "{ATLAS Collaboration}",
    title          = "{Measurement of top quark pair differential cross sections in the dilepton channel in \(pp\) collisions at \(\sqrt{s} = 7\) and \(8\,\text{TeV}\) with ATLAS}",
    journal        = "Phys. Rev. D",
    volume         = "94",
    year           = "2016",
    pages          = "092003",
    doi            = "10.1103/PhysRevD.94.092003",
    reportNumber   = "CERN-EP-2016-144",
    eprint         = "1607.07281",
    archivePrefix  = "arXiv",
    primaryClass   = "hep-ex",
}

@Article{TOPQ-2016-01,
    author         = "{ATLAS Collaboration}",
    title          = "{Measurements of top-quark pair differential cross-sections in the lepton+jets channel in \(pp\) collisions at \(\sqrt{s} = 13\,\text{TeV}\) using the ATLAS detector}",
    journal        = "JHEP",
    volume         = "11",
    year           = "2017",
    pages          = "191",
    doi            = "10.1007/JHEP11(2017)191",
    reportNumber   = "CERN-EP-2017-058",
    eprint         = "1708.00727",
    archivePrefix  = "arXiv",
    primaryClass   = "hep-ex",
}

@Article{TOPQ-2016-04,
    author         = "{ATLAS Collaboration}",
    title          = "{Measurements of top-quark pair differential cross-sections in the \(e\mu\) channel in \(pp\) collisions at \(\sqrt{s} = 13\,\text{TeV}\) using the ATLAS detector}",
    journal        = "Eur. Phys. J. C",
    volume         = "77",
    year           = "2017",
    pages          = "292",
    doi            = "10.1140/epjc/s10052-017-4821-x",
    eprint         = "1612.05220",
    archivePrefix  = "arXiv",
    primaryClass   = "hep-ex",
}

@Article{TOPQ-2016-09,
    author         = "{ATLAS Collaboration}",
    title          = "{Measurements of \(t\bar{t}\) differential cross-sections of highly boosted top quarks decaying to all-hadronic final states in \(pp\) collisions at \(\sqrt{s} = 13\,\text{TeV}\) using the ATLAS detector}",
    journal        = "Phys. Rev. D",
    volume         = "98",
    year           = "2018",
    pages          = "012003",
    doi            = "10.1103/PhysRevD.98.012003",
    reportNumber   = "CERN-EP-2017-226",
    eprint         = "1801.02052",
    archivePrefix  = "arXiv",
    primaryClass   = "hep-ex",
}

@Article{TRIG-2016-01,
    author         = "{ATLAS Collaboration}",
    title          = "{Performance of the ATLAS trigger system in 2015}",
    journal        = "Eur. Phys. J. C",
    volume         = "77",
    year           = "2017",
    pages          = "317",
    doi            = "10.1140/epjc/s10052-017-4852-3",
    reportNumber   = "CERN-EP-2016-241",
    eprint         = "1611.09661",
    archivePrefix  = "arXiv",
    primaryClass   = "hep-ex",
}

@Article{DAPR-2018-01,
    author         = "{ATLAS Collaboration}",
    title          = "{ATLAS data quality operations and performance for 2015--2018 data-taking}",
    journal        = "JINST",
    volume         = "15",
    year           = "2020",
    pages          = "P04003",
    doi            = "10.1088/1748-0221/15/04/P04003",
    reportNumber   = "CERN-EP-2019-207",
    eprint         = "1911.04632",
    archivePrefix  = "arXiv",
    primaryClass   = "physics.ins-det",
}

@Article{EGAM-2018-01,
    author         = "{ATLAS Collaboration}",
    title          = "{Electron and photon performance measurements with the ATLAS detector using the 2015--2017 LHC proton--proton collision data}",
    journal        = "JINST",
    volume         = "14",
    year           = "2019",
    pages          = "P12006",
    doi            = "10.1088/1748-0221/14/12/P12006",
    reportNumber   = "CERN-EP-2019-145",
    eprint         = "1908.00005",
    archivePrefix  = "arXiv",
    primaryClass   = "hep-ex",
}

@Article{FTAG-2018-01,
    author         = "{ATLAS Collaboration}",
    title          = "{ATLAS \(b\)-jet identification performance and efficiency measurement with \(t\bar{t}\) events in \(pp\) collisions at \(\sqrt{s} = 13\,\text{TeV}\)}",
    journal        = "Eur. Phys. J. C",
    volume         = "79",
    year           = "2019",
    pages          = "970",
    doi            = "10.1140/epjc/s10052-019-7450-8",
    reportNumber   = "CERN-EP-2019-132",
    eprint         = "1907.05120",
    archivePrefix  = "arXiv",
    primaryClass   = "hep-ex",
}

@Article{JETM-2018-02,
    author         = "{ATLAS Collaboration}",
    title          = "{In situ calibration of large-radius jet energy and mass in \(13\,\text{TeV}\) proton--proton collisions with the ATLAS detector}",
    journal        = "Eur. Phys. J. C",
    volume         = "79",
    year           = "2019",
    pages          = "135",
    doi            = "10.1140/epjc/s10052-019-6632-8",
    reportNumber   = "CERN-EP-2018-191",
    eprint         = "1807.09477",
    archivePrefix  = "arXiv",
    primaryClass   = "hep-ex",
}

@Article{JETM-2018-03,
    author         = "{ATLAS Collaboration}",
    title          = "{Performance of top-quark and \(W\)-boson tagging with ATLAS in Run~2 of the LHC}",
    journal        = "Eur. Phys. J. C",
    volume         = "79",
    year           = "2019",
    pages          = "375",
    doi            = "10.1140/epjc/s10052-019-6847-8",
    reportNumber   = "CERN-EP-2018-192",
    eprint         = "1808.07858",
    archivePrefix  = "arXiv",
    primaryClass   = "hep-ex",
}

@Article{JETM-2018-05,
    author         = "{ATLAS Collaboration}",
    title          = "{Jet energy scale and resolution measured in proton--proton collisions at \(\sqrt{s} = 13\,\text{TeV}\) with the ATLAS detector}",
    journal        = "Eur. Phys. J. C",
    volume         = "81",
    year           = "2020",
    pages          = "689",
    doi            = "10.1140/epjc/s10052-021-09402-3",
    reportNumber   = "CERN-EP-2020-083",
    eprint         = "2007.02645",
    archivePrefix  = "arXiv",
    primaryClass   = "hep-ex",
}

@Article{JETM-2018-06,
    author         = "{ATLAS Collaboration}",
    title          = "{Optimisation of large-radius jet reconstruction for the ATLAS detector in \(13\,\text{TeV}\) proton--proton collisions}",
    journal        = "Eur. Phys. J. C",
    volume         = "81",
    year           = "2020",
    pages          = "334",
    doi            = "10.1140/epjc/s10052-021-09054-3",
    reportNumber   = "CERN-EP-2020-134",
    eprint         = "2009.04986",
    archivePrefix  = "arXiv",
    primaryClass   = "hep-ex",
}

@Article{MUON-2018-03,
    author         = "{ATLAS Collaboration}",
    title          = "{Muon reconstruction and identification efficiency in ATLAS using the full Run~2 \(pp\) collision data set at \(\sqrt{s} = 13\,\text{TeV}\)}",
    journal        = "Eur. Phys. J. C",
    volume         = "81",
    year           = "2021",
    pages          = "578",
    doi            = "10.1140/epjc/s10052-021-09233-2",
    reportNumber   = "CERN-EP-2020-199",
    eprint         = "2012.00578",
    archivePrefix  = "arXiv",
    primaryClass   = "hep-ex",
}

@Article{PIX-2018-001,
    author         = "Abbott, B. and others",
    title          = "{Production and integration of the ATLAS Insertable B-Layer}",
    journal        = "JINST",
    volume         = "13",
    year           = "2018",
    pages          = "T05008",
    doi            = "10.1088/1748-0221/13/05/T05008",
    eprint         = "1803.00844",
    archivePrefix  = "arXiv",
    primaryClass   = "physics.ins-det",
}

@Article{TOPQ-2018-11,
    author         = "{ATLAS Collaboration}",
    title          = "{Differential \(t\bar{t}\) cross-section measurements using boosted top quarks in the all-hadronic final state with \(139\,\text{fb}^{-1}\) of ATLAS data}",
    year           = "2022",
    reportNumber   = "CERN-EP-2022-026",
    eprint         = "2205.02817",
    archivePrefix  = "arXiv",
    primaryClass   = "hep-ex",
}

@Article{TOPQ-2019-28,
    author         = "{ATLAS Collaboration}",
    title          = "{Measurement of the energy asymmetry in \(t\bar{t}j\) production at \(13\,\text{TeV}\) with the ATLAS experiment and interpretation in the SMEFT framework}",
    journal        = "Eur. Phys. J. C",
    volume         = "82",
    year           = "2022",
    pages          = "374",
    doi            = "10.1140/epjc/s10052-022-10101-w",
    reportNumber   = "CERN-EP-2021-181",
    eprint         = "2110.05453",
    archivePrefix  = "arXiv",
    primaryClass   = "hep-ex",
}

@Booklet{ATL-SOFT-PUB-2021-001,
    author         = "{ATLAS Collaboration}",
    title          = "{The ATLAS Collaboration Software and Firmware}",
    howpublished   = "{ATL-SOFT-PUB-2021-001}",
    url            = "https://cds.cern.ch/record/2767187",
    year           = "2021",
}

@Booklet{ATL-SOFT-PUB-2021-003,
    author         = "{ATLAS Collaboration}",
    title          = "{ATLAS Computing Acknowledgements}",
    howpublished   = "{ATL-SOFT-PUB-2021-003}",
    url            = "https://cds.cern.ch/record/2776662",
    year           = "2021",
}

@Report{ATLAS-TDR-19,
    author         = "{ATLAS Collaboration}",
    title          = "{ATLAS Insertable B-Layer: Technical Design Report}",
    type           = "ATLAS-TDR-19; CERN-LHCC-2010-013",
    year           = "2010",
    url            = "https://cds.cern.ch/record/1291633",
    related        = "ATLAS-TDR-19-addm",
    relatedstring  = "Addendum:",
}

@Article{CMS-TOP-11-013,
    author         = "{CMS Collaboration}",
    title          = "{Measurement of differential top-quark-pair production cross sections in \(pp\) colisions at \(\sqrt{s} = 7\,\text{TeV}\)}",
    journal        = "Eur. Phys. J. C",
    volume         = "73",
    year           = "2013",
    pages          = "2339",
    doi            = "10.1140/epjc/s10052-013-2339-4",
    reportNumber   = "CERN-PH-EP-2012-322",
    eprint         = "1211.2220",
    archivePrefix  = "arXiv",
    primaryClass   = "hep-ex",
}

@Article{CMS-TOP-12-028,
    author         = "{CMS Collaboration}",
    title          = "{Measurement of the differential cross section for top quark pair production in \(pp\) collisions at \(\sqrt{s} = 8\,\text{TeV}\)}",
    journal        = "Eur. Phys. J. C",
    volume         = "75",
    year           = "2015",
    pages          = "542",
    doi            = "10.1140/epjc/s10052-015-3709-x",
    reportNumber   = "CERN-PH-EP-2015-117",
    eprint         = "1505.04480",
    archivePrefix  = "arXiv",
    primaryClass   = "hep-ex",
}

@Article{CMS-TOP-14-012,
    author         = "{CMS Collaboration}",
    title          = "{Measurement of the integrated and differential \(t\bar{t}\) production cross sections for high-\(p_{\text{T}}\) top quarks in \(pp\) collisions at \(\sqrt{s} = 8\,\text{TeV}\)}",
    journal        = "Phys. Rev. D",
    volume         = "94",
    year           = "2016",
    pages          = "072002",
    doi            = "10.1103/PhysRevD.94.072002",
    reportNumber   = "CERN-EP-2016-078",
    eprint         = "1605.00116",
    archivePrefix  = "arXiv",
    primaryClass   = "hep-ex",
}

@Article{CMS-TOP-14-013,
    author         = "{CMS Collaboration}",
    title          = "{Measurement of double-differential cross sections for top quark pair production in \(pp\) collisions at \(\sqrt{s} = 8\,\text{TeV}\) and impact on parton distribution functions}",
    journal        = "Eur. Phys. J. C",
    volume         = "77",
    year           = "2017",
    pages          = "459",
    doi            = "10.1140/epjc/s10052-017-4984-5",
    reportNumber   = "CERN-EP-2017-030",
    eprint         = "1703.01630",
    archivePrefix  = "arXiv",
    primaryClass   = "hep-ex",
}

@Article{CMS-TOP-14-018,
    author         = "{CMS Collaboration}",
    title          = "{Measurement of the \(t\bar{t}\)  production cross section in the all-jets final state in \(pp\) collisions at \(\sqrt{s} = 8\,\text{TeV}\)}",
    journal        = "Eur. Phys. J. C",
    volume         = "76",
    year           = "2016",
    pages          = "128",
    doi            = "10.1140/epjc/s10052-016-3956-5",
    reportNumber   = "CERN-PH-EP-2015-243",
    eprint         = "1509.06076",
    archivePrefix  = "arXiv",
    primaryClass   = "hep-ex",
}

@Article{CMS-TOP-15-015,
    author         = "{CMS Collaboration}",
    title          = "{Measurement of the jet mass in highly boosted \(t\bar{t}\) events from \(pp\) collisions at \(\sqrt{s} = 8\,\text{TeV}\)}",
    journal        = "Eur. Phys. J. C",
    volume         = "77",
    year           = "2017",
    pages          = "467",
    doi            = "10.1140/epjc/s10052-017-5030-3",
    reportNumber   = "CERN-EP-2017-018",
    eprint         = "1703.06330",
    archivePrefix  = "arXiv",
    primaryClass   = "hep-ex",
}

@Article{CMS-TOP-16-007,
    author         = "{CMS Collaboration}",
    title          = "{Measurement of normalized differential \(t\bar{t}\) cross sections in the dilepton channel from \(pp\) collisions at \(\sqrt{s} = 13\,\text{TeV}\)}",
    journal        = "JHEP",
    volume         = "04",
    year           = "2018",
    pages          = "060",
    doi            = "10.1007/JHEP04(2018)060",
    reportNumber   = "CERN-EP-2017-120",
    eprint         = "1708.07638",
    archivePrefix  = "arXiv",
    primaryClass   = "hep-ex",
}

@Article{CMS-TOP-16-008,
    author         = "{CMS Collaboration}",
    title          = "{Measurement of differential cross sections for top quark pair production using the lepton+jets final state in proton--proton collisions at \(13\,\text{TeV}\)}",
    journal        = "Phys. Rev. D",
    volume         = "95",
    year           = "2017",
    pages          = "092001",
    doi            = "10.1103/PhysRevD.95.092001",
    reportNumber   = "CERN-EP-2016-227",
    eprint         = "1610.04191",
    archivePrefix  = "arXiv",
    primaryClass   = "hep-ex",
}

@Article{CMS-TOP-16-014,
    author         = "{CMS Collaboration}",
    title          = "{Measurements of differential cross sections of top quark pair production as a function of kinematic event variables in proton--proton collisions at \(\sqrt{s} = 13\,\text{TeV}\)}",
    journal        = "JHEP",
    volume         = "06",
    year           = "2018",
    pages          = "002",
    doi            = "10.1007/JHEP06(2018)002",
    reportNumber   = "CERN-EP-2018-013",
    eprint         = "1803.03991",
    archivePrefix  = "arXiv",
    primaryClass   = "hep-ex",
}

@Article{CMS-TOP-17-002,
    author         = "{CMS Collaboration}",
    title          = "{Measurement of differential cross sections for the production of top quark pairs and of additional jets in lepton+jets events from \(pp\) collisions at \(\sqrt{s} = 13\,\text{TeV}\)}",
    journal        = "Phys. Rev. D",
    volume         = "97",
    year           = "2018",
    pages          = "112003",
    doi            = "10.1103/PhysRevD.97.112003",
    reportNumber   = "CERN-EP-2018-039",
    eprint         = "1803.08856",
    archivePrefix  = "arXiv",
    primaryClass   = "hep-ex",
}

@Article{CMS-TOP-18-004,
    author         = "{CMS Collaboration}",
    title          = "{Measurement of \(t\bar{t}\) normalised multi-differential cross sections in \(pp\) collisions at \(\sqrt{s} = 13\,\text{TeV}\), and simultaneous determination of the strong coupling strength, top quark pole mass, and parton distribution functions}",
    journal        = "Eur. Phys. J. C",
    volume         = "80",
    year           = "2020",
    pages          = "658",
    doi            = "10.1140/epjc/s10052-020-7917-7",
    reportNumber   = "CERN-EP-2019-028",
    eprint         = "1904.05237",
    archivePrefix  = "arXiv",
    primaryClass   = "hep-ex",
}

@Article{CMS-TOP-18-013,
    author         = "{CMS Collaboration}",
    title          = "{Measurement of differential \(t\bar{t}\) production cross sections using top quarks at large transverse momenta in \(pp\) collisions at \(\sqrt{s} = 13\,\text{TeV}\)}",
    journal        = "Phys. Rev. D",
    volume         = "103",
    year           = "2021",
    pages          = "052008",
    doi            = "10.1103/PhysRevD.103.052008",
    reportNumber   = "CERN-EP-2020-121",
    eprint         = "2008.07860",
    archivePrefix  = "arXiv",
    primaryClass   = "hep-ex",
}

@Article{CMS-TOP-19-005,
    author         = "{CMS Collaboration}",
    title          = "{Measurement of the jet mass distribution and top quark mass in hadronic decays of boosted top quarks in \(pp\) collisions at \(\sqrt{s} = 13\,\text{TeV}\)}",
    journal        = "Phys. Rev. Lett.",
    volume         = "124",
    year           = "2020",
    pages          = "202001",
    doi            = "10.1103/PhysRevLett.124.202001",
    reportNumber   = "CERN-EP-2019-226",
    eprint         = "1911.03800",
    archivePrefix  = "arXiv",
    primaryClass   = "hep-ex",
}

@Article{CMS-TOP-20-001,
    author         = "{CMS Collaboration}",
    title          = "{Measurement of differential \(t\bar{t}\) production cross sections in the full kinematic range using lepton+jets events from proton--proton collisions at \(\sqrt{s} = 13\,\text{TeV}\)}",
    journal        = "Phys. Rev. D",
    volume         = "104",
    year           = "2021",
    pages          = "092013",
    doi            = "10.1103/PhysRevD.104.092013",
    reportNumber   = "CERN-EP-2021-135",
    eprint         = "2108.02803",
    archivePrefix  = "arXiv",
    primaryClass   = "hep-ex",
}

@Booklet{ATLAS-CONF-2019-021,
    author         = "{ATLAS Collaboration}",
    title          = "{Luminosity determination in \(pp\) collisions at \(\sqrt{s} = 13\,\text{TeV}\) using the ATLAS detector at the LHC}",
    howpublished   = "{ATLAS-CONF-2019-021}",
    url            = "https://cds.cern.ch/record/2677054",
    year           = "2019",
}

@Booklet{ATLAS-CONF-2020-022,
    author         = "{ATLAS Collaboration}",
    title          = "{Measurement of the ATLAS Detector Jet Mass Response using Forward Folding with \(80\,\text{fb}^{-1}\) of \(\sqrt{s} = 13\,\text{TeV}\) \(pp\) data}",
    howpublished   = "{ATLAS-CONF-2020-022}",
    url            = "https://cds.cern.ch/record/2724442",
    year           = "2020",
}

@Booklet{ATL-PHYS-PUB-2014-021,
    author         = "{ATLAS Collaboration}",
    title          = "{ATLAS Pythia~8 tunes to \(7~\text{TeV}\) data}",
    howpublished   = "{ATL-PHYS-PUB-2014-021}",
    url            = "https://cds.cern.ch/record/1966419",
    year           = "2014",
}

@Booklet{ATL-PHYS-PUB-2015-013,
    author         = "{ATLAS Collaboration}",
    title          = "{Proposal for particle-level object and observable definitions for use in physics measurements at the LHC}",
    howpublished   = "{ATL-PHYS-PUB-2015-013}",
    url            = "https://cds.cern.ch/record/2022743",
    year           = "2015",
}

@Booklet{ATL-PHYS-PUB-2016-004,
    author         = "{ATLAS Collaboration}",
    title          = "{Simulation of top-quark production for the ATLAS experiment at \(\sqrt{s} = 13~\text{TeV}\)}",
    howpublished   = "{ATL-PHYS-PUB-2016-004}",
    url            = "https://cds.cern.ch/record/2120417",
    year           = "2016",
}

@Booklet{ATL-PHYS-PUB-2016-017,
    author         = "{ATLAS Collaboration}",
    title          = "{The Pythia~8 A3 tune description of ATLAS minimum bias and inelastic measurements incorporating the Donnachie--Landshoff diffractive model}",
    howpublished   = "{ATL-PHYS-PUB-2016-017}",
    url            = "https://cds.cern.ch/record/2206965",
    year           = "2016",
}

@Booklet{ATL-PHYS-PUB-2016-020,
    author         = "{ATLAS Collaboration}",
    title          = "{Studies on top-quark Monte Carlo modelling for Top2016}",
    howpublished   = "{ATL-PHYS-PUB-2016-020}",
    url            = "https://cds.cern.ch/record/2216168",
    year           = "2016",
}

@Booklet{ATL-PHYS-PUB-2017-007,
    author         = "{ATLAS Collaboration}",
    title          = "{Studies on top-quark Monte Carlo modelling with Sherpa and MG5\_aMC@NLO}",
    howpublished   = "{ATL-PHYS-PUB-2017-007}",
    url            = "https://cds.cern.ch/record/2261938",
    year           = "2017",
}

@Booklet{ATL-PHYS-PUB-2017-013,
    author         = "{ATLAS Collaboration}",
    title          = "{Optimisation and performance studies of the ATLAS \(b\)-tagging algorithms for the 2017-18 LHC run}",
    howpublished   = "{ATL-PHYS-PUB-2017-013}",
    url            = "https://cds.cern.ch/record/2273281",
    year           = "2017",
}

@Booklet{ATL-PHYS-PUB-2018-009,
    author         = "{ATLAS Collaboration}",
    title          = "{Improvements in \(t\bar{t}\) modelling using NLO+PS Monte Carlo generators for Run~2}",
    howpublished   = "{ATL-PHYS-PUB-2018-009}",
    url            = "https://cds.cern.ch/record/2630327",
    year           = "2018",
}

@Booklet{ATL-PHYS-PUB-2020-017,
    author         = "{ATLAS Collaboration}",
    title          = "{Boosted hadronic vector boson and top quark tagging with ATLAS using Run 2 data}",
    howpublished   = "{ATL-PHYS-PUB-2020-017}",
    url            = "https://cds.cern.ch/record/2724149",
    year           = "2020",
}

@Booklet{ATL-PHYS-PUB-2020-019,
    author         = "{ATLAS Collaboration}",
    title          = "{Identification of Boosted Higgs Bosons Decaying Into \(b\bar{b}\) With Neural Networks and Variable Radius Subjets in ATLAS}",
    howpublished   = "{ATL-PHYS-PUB-2020-019}",
    url            = "https://cds.cern.ch/record/2724739",
    year           = "2020",
}

@Booklet{ATL-PHYS-PUB-2020-023,
    author         = "{ATLAS Collaboration}",
    title          = "{Study of top-quark pair modelling and uncertainties using ATLAS measurements at \(\sqrt{s} = 13\,\text{TeV}\)}",
    howpublished   = "{ATL-PHYS-PUB-2020-023}",
    url            = "https://cds.cern.ch/record/2730443",
    year           = "2020",
}

@Booklet{ATL-PHYS-PUB-2021-003,
    author         = "{ATLAS Collaboration}",
    title          = "{Simulation-based extrapolation of \(b\)-tagging calibrations towards high transverse momenta in the ATLAS experiment}",
    howpublished   = "{ATL-PHYS-PUB-2021-003}",
    url            = "https://cds.cern.ch/record/2753444",
    year           = "2021",
}
 
\clearpage
 
\begin{flushleft}
\hypersetup{urlcolor=black}
{\Large The ATLAS Collaboration}

\bigskip

\AtlasOrcid[0000-0002-6665-4934]{G.~Aad}$^\textrm{\scriptsize 100}$,
\AtlasOrcid[0000-0002-5888-2734]{B.~Abbott}$^\textrm{\scriptsize 118}$,
\AtlasOrcid[0000-0002-7248-3203]{D.C.~Abbott}$^\textrm{\scriptsize 101}$,
\AtlasOrcid[0000-0002-2788-3822]{A.~Abed~Abud}$^\textrm{\scriptsize 35}$,
\AtlasOrcid[0000-0002-1002-1652]{K.~Abeling}$^\textrm{\scriptsize 54}$,
\AtlasOrcid[0000-0002-2987-4006]{D.K.~Abhayasinghe}$^\textrm{\scriptsize 93}$,
\AtlasOrcid[0000-0002-8496-9294]{S.H.~Abidi}$^\textrm{\scriptsize 28}$,
\AtlasOrcid[0000-0002-9987-2292]{A.~Aboulhorma}$^\textrm{\scriptsize 34e}$,
\AtlasOrcid[0000-0001-5329-6640]{H.~Abramowicz}$^\textrm{\scriptsize 149}$,
\AtlasOrcid[0000-0002-1599-2896]{H.~Abreu}$^\textrm{\scriptsize 148}$
% ...
\bigskip
\\

% ...
$^{28}$Physics Department, Brookhaven National Laboratory, Upton NY; United States of America.\\
% ...
$^{34}$$^{(a)}$Facult\'e des Sciences Ain Chock, R\'eseau Universitaire de Physique des Hautes Energies - Universit\'e Hassan II, Casablanca;$^{(b)}$Facult\'{e} des Sciences, Universit\'{e} Ibn-Tofail, K\'{e}nitra;$^{(c)}$Facult\'e des Sciences Semlalia, Universit\'e Cadi Ayyad, LPHEA-Marrakech;$^{(d)}$LPMR, Facult\'e des Sciences, Universit\'e Mohamed Premier, Oujda;$^{(e)}$Facult\'e des sciences, Universit\'e Mohammed V, Rabat;$^{(f)}$Institute of Applied Physics, Mohammed VI Polytechnic University, Ben Guerir; Morocco.\\
$^{35}$CERN, Geneva; Switzerland.\\
% ...
$^{54}$II. Physikalisches Institut, Georg-August-Universit\"{a}t G\"ottingen, G\"ottingen; Germany.\\
% ...
$^{93}$Department of Physics, Royal Holloway University of London, Egham; United Kingdom.\\
% ...
$^{100}$CPPM, Aix-Marseille Universit\'e, CNRS/IN2P3, Marseille; France.\\
$^{101}$Department of Physics, University of Massachusetts, Amherst MA; United States of America.\\
% ...
$^{118}$Homer L. Dodge Department of Physics and Astronomy, University of Oklahoma, Norman OK; United States of America.\\
% ...
$^{148}$Department of Physics, Technion, Israel Institute of Technology, Haifa; Israel.\\
$^{149}$Raymond and Beverly Sackler School of Physics and Astronomy, Tel Aviv University, Tel Aviv; Israel.\\
% ...
$^{a}$ Also Affiliated with an institute covered by a cooperation agreement with CERN.\\
$^{b}$ Also at Borough of Manhattan Community College, City University of New York, New York NY; United States of America.\\
$^{c}$ Also at Bruno Kessler Foundation, Trento; Italy.\\
$^{d}$ Also at Center for High Energy Physics, Peking University; China.\\
$^{e}$ Also at Centro Studi e Ricerche Enrico Fermi; Italy.\\
$^{f}$ Also at CERN, Geneva; Switzerland.\\
% ...
$^{*}$ Deceased

\end{flushleft}


\end{document}